\documentclass[a4paper,11pt]{article}
\usepackage{jheppub} 

\usepackage{lineno}
\usepackage[T1]{fontenc}
\usepackage{  amsthm, amssymb,latexsym,amsfonts}
\usepackage[final]{pdfpages}
\usepackage{float}
\usepackage{tikz}
\usetikzlibrary{matrix,calc,fit,intersections}
\usepackage{gensymb}
\usepackage{tikz-feynman}
\usepackage{relsize}
\usepackage{graphicx}
\tikzfeynmanset{compat=1.1.0}
\usepackage{circuitikz}
\usepackage{relsize}
\usepackage{xcolor}
\usetikzlibrary{arrows.meta,calc,decorations.markings,math,arrows.meta}
\usepackage{mathtools}
 \usepackage[shortlabels]{enumitem}
\usepackage{afterpage}
\usepackage{enumitem}
\usepackage{tensor}
\usepackage[cal=cm,calscaled=.96]{mathalpha}
\usepackage{hyperref}
 \usepackage{graphicx}
\usepackage{tikz-network}
\usepackage{textgreek,stackengine}
\newcommand\textdot[1]{\stackon[1pt]{\csname text#1\endcsname}{.}}

\newcommand{\overbar}[1]{\mkern 1mu\overline{\mkern-1mu#1\mkern-1mu}\mkern 1mu}

\theoremstyle{plain}
\newtheorem{thm}{Theorem}[section]

\theoremstyle{definition}
\newtheorem{dfn}[thm]{Definition}

\def\be{\begin{equation}}
\def\ee{\end{equation}}
\def\bea{\begin{eqnarray}}
\def\eea{\end{eqnarray}}

\definecolor{baby}{rgb}{0.96, 0.76, 0.76}
\newcommand{\mysetminusD}{\hbox{\tikz{\draw[line width=0.6pt,line cap=round] (3pt,0) -- (0,6pt);}}}
\newcommand{\mysetminusT}{\mysetminusD}
\newcommand{\mysetminusS}{\hbox{\tikz{\draw[line width=0.45pt,line cap=round] (2pt,0) -- (0,4pt);}}}
\newcommand{\mysetminusSS}{\hbox{\tikz{\draw[line width=0.4pt,line cap=round] (1.5pt,0) -- (0,3pt);}}}

\newcommand{\mysetminus}{\mathbin{\mathchoice{\mysetminusD}{\mysetminusT}{\mysetminusS}{\mysetminusSS}}}

\usepackage{amstext} 
\usepackage{array}   
\newcolumntype{L}{>{$}l<{$}} 
\usepackage{colortbl}
\definecolor{babypink}{rgb}{0.96, 0.76, 0.76}

\newcommand{\mycomment}[1]{}

\title{\boldmath On fusing matrices associated with conformal boundary conditions}

\author{Anatoly Konechny}
 \author{and Vasileios Vergioglou}
\affiliation{Department of Mathematics, Heriot-Watt University,\\
 Edinburgh EH14 4AS, U.K.}
\affiliation{Maxwell Institute for Mathematical Sciences,\\
Edinburgh, U.K.}

\emailAdd{A.Konechny@hw.ac.uk, vv2004@hw.ac.uk}

\abstract{In the context of rational conformal field theories (RCFT) we look at the fusing matrices that arise when a topological defect is 
attached to a  conformal boundary condition. We call such junctions  open topological defects. One type of fusing matrices 
arises when two open defects fuse while another arises when an open defect passes through a boundary operator. 
We use the topological field  theory approach to RCFTs based on Frobenius algebra objects in modular tensor categories
 to describe the general structure associated with such matrices and 
how to compute them from a given  Frobenius algebra object and its representation theory. We illustrate the computational process on the   rational free boson theories. Applications to boundary renormalisation group flows are briefly discussed. }


\definecolor{ashgrey}{rgb}{0.7, 0.75, 0.71}

\DeclareMathOperator{\Hom}{Hom}
\definecolor{mycolor}{HTML}{960018}
\definecolor{coralred}{rgb}{1.0, 0.25, 0.25}
\definecolor{pansypurple}{rgb}{0.47, 0.09, 0.29}
\definecolor{scarlet}{rgb}{1.0, 0.13, 0.0}
\definecolor{baby}{rgb}{0.96, 0.76, 0.76}
\definecolor{aqua}{rgb}{0.5, 1.0, 0.83}
\definecolor{fuc}{rgb}{0.76, 0.33, 0.76}
\newcommand{\dc}{\prescript{}{A}{\mathcal{C}}_B}
\newcommand{\objc}{\operatorname{Obj}(\mathcal{C})}
\newcommand{\id}{\operatorname{id}}

\newcommand{\dimh}{\operatorname{dim}\operatorname{Hom}}
\newcommand{\affu}{\hat{\mathfrak{u}}}

\begin{document}

\tikzset{myptr/.style={
        decoration={markings,
            mark= at position 0.7 with {\arrow{#1}} ,
        },
        postaction={decorate}
    }
}

\tikzset{myptr1/.style={
        decoration={markings,
            mark= at position 0.4 with {\arrow{#1}} ,
        },
        postaction={decorate}
    }
}
\tikzset{middlearrow/.style={
        decoration={markings,
            mark= at position 0.5 with {\arrow{#1}} ,
        },
        postaction={decorate}
    }
}

\tikzset{myptr2/.style={
        decoration={markings,
            mark= at position 0.9 with {\arrow{#1}} ,
        },
        postaction={decorate}
    }
}

\tikzfeynmanset{
myblob/.style={
shape=circle,
draw=black}
}

\newcolumntype{a}{>{\columncolor{babypink}}c}

\maketitle
\flushbottom

\section{Introduction}
\label{sec:intro}
The crucial algebraic structure underlying  rational two-dimensional conformal field theories (RCFT) is that of a modular tensor category \cite{MS1,MS2}. The approach to  RCFTs based on modular tensor categories and 
Frobenius algebra objects was developed in \cite{FFFS1,FFFS2, FS, FRS0, FRS1, FRS2, FRS3, FRS4, FRS5}. 
Structure constants in this approach are calculated from ribbon graphs using a three-dimensional topological field theory (TFT).
We will refer to this approach as the TFT approach. A review  can be found in  \cite{RFS_review}, \cite{review2} and a historical overview of RCFT that puts the TFT approach into perspective can be found in  \cite{25years}. 

 In this introductory section we will sketch the main structures 
we will be working with leaving the precise definitions to the later sections.
At the chiral symmetry level an RCFT can be described in terms of a modular tensor category ${\mathcal C}$. One of the key data specifying  ${\mathcal C}$ 
is the associator morphisms spaces for the tensor products of objects. Choosing bases for the relevant morphisms these associators are represented by the 
fusing matrices also called $\mathrm{F}$-symbols or 6j-symbols.  Such matrices satisfy a system of non-linear equations called the pentagon equations, which in general are very difficult to solve.  The defining observables of a  full RCFT are local bulk operators which combine a holomorphic and anti-holomorphic representation content of ${\mathcal C}$. It was shown in the papers cited above  that to construct such operators and their correlators it suffices to have a Frobenius algebra object $A$ 
in ${\mathcal C}$ with certain additional properties.   Such an algebra object can be constructed  from a conformal boundary condition. All conformal boundary conditions can then be associated with modules over $A$ and form a module category over $\mathcal{C}$. The multiplication structure of $A$, in addition to the fusing and braiding matrices of ${\mathcal C}$, provides further data satisfying non-trivial non-linear constraints. 
One of the important  computational advantages of the TFT approach is 
that   all structure constants of the CFT associated with the pair ${\mathcal C}, A$ 
can be found solving some linear problems. It was shown that the structure constants  will satisfy all the necessary non-linear constraints by virtue  
of the non-linear equations satisfied by the data defining ${\mathcal C}$ and $ A$. 

\begin{figure}[pt]
    \centering
    \begin{equation*}
        \vcenter{\hbox{\begin{tikzpicture}[font=\scriptsize,inner sep=2pt]
  \begin{feynman}
\vertex (i1) at (0,0) ;
\vertex [right=1cm of i1] (j1);
\vertex [above=0.5cm of i1] (i2) [label=left:\(X\)];
\vertex [above=0.5cm of j1] (j2) [label=left:\(M\)];
\vertex [above=3cm of i1] (i3);
\vertex [right =1cm of i3] (j3) ;
\vertex  [left=0.5cm of i3] (a1)  ;
\vertex [below= 0.4cm of a1] (a2);
 \vertex [right=1cm of a2] (b1);
 \vertex[right=0.5cm of j3] (top);
 \path[pattern=north east lines,pattern color=ashgrey,very thin] (top) rectangle (j1);
   \diagram*{
(i1)--[thick,middlearrow={latex}] (i3),
(j1)--[thick] (j3)
  };
  \end{feynman}
\end{tikzpicture}}}
~ \;\xrightarrow{\text{fuse}}\;~
\vcenter{\hbox{\begin{tikzpicture}[font=\scriptsize,inner sep=2pt]
  \begin{feynman}
\vertex (i1) at (0,0) ;
\vertex [above=0.5cm of i1] (i2) [label=left:\(X\otimes_B\!M\)];
\vertex [above=3cm of i1] (i3);
\vertex  [left=0.5cm of i3] (a1)  ;
\vertex [below= 0.4cm of a1] (a2);
\vertex[right=0.5cm of i3] (top);
\path[pattern=north east lines,pattern color=ashgrey,very thin] (top) rectangle (i1);
   \diagram*{
(i1)--[thick] (i3)
  };
  \end{feynman}
\end{tikzpicture}}}
    \end{equation*}
    \caption{Fusion of a topological interface with a conformal boundary condition.}\label{figintro1}
\end{figure}
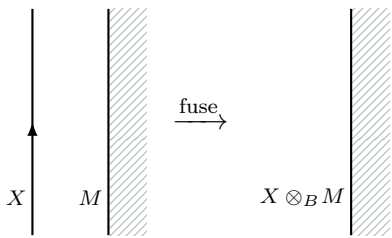

Besides the conformal boundary conditions, described as modules over $A$, another important ingredient of the categorical approach are the topological interfaces.   Given two RCFTs built on the same chiral symmetry category ${\mathcal C}$ from two Frobenius algebra objects: $A$ and $B$, a topological interface between them can be described as an $A$-$B$-bimodule object in ${\mathcal C}$. For the case $A=B$ such bimodules correspond to  topological defects. A new tensor product structure arises on the category of all $A$-$A$-bimodules giving rise to a fusion category. It has its own fusing matrices but no natural braiding. Furthermore, a number of non-trivial operations arise involving both conformal boundary conditions and topological defects (or more generally interfaces).
Firstly, we can define a tensor product (fusion) of a topological interface labelled by an $A$-$B$-bimodule  $X$ and a conformal boundary condition labelled by a $B$-module $M$. This tensor product   gives a conformal boundary condition described by the $A$-module $X\otimes_BM$, see Figure \ref{figintro1}.

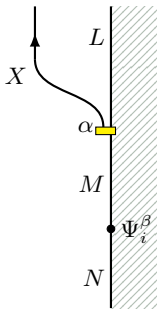
\begin{figure}
    \centering
   \begin{equation*}
        \vcenter{\hbox{\begin{tikzpicture}[font=\footnotesize,inner sep=2pt]
  \begin{feynman}
  \path[pattern=north east lines,pattern color=ashgrey,very thin] (1.01,4) rectangle (1.6,0);
  \vertex  (i1) at (0,0)  ;
  \vertex (x1) at (0,3.3);
  \vertex  (x2) at (0.75, 0.65);
  \vertex [above=4cm of i1] (i2);
  \vertex [right=1cm of i1] (j1);
  \vertex[ above=0.4cm of j1] (j3) [label=left:\(N\)];
  \vertex [right=1cm of i2] (j2);
  \vertex[ below=0.4cm of j2] (j4) [label=left:\(L\)];
  \vertex [below=1.6cm of j2] (j6);
  \vertex   [above=2.3cm of j1] (j5);
  \vertex [small,dot] [below=1.2cm of j5] (p1) [label=right:\(\Psi_i^\beta\)] {};
  \vertex[above=0.6cm of p1] (p2) [label=left:\(M\)];
  \vertex [left=0.15cm of j6] [label=left:\(\alpha\)];
  \vertex [left=0.1cm of j6] (j7);
  \vertex (x) at (-0.25,2.9) [label=above:\(X\)];
\draw [thick]    (j7) to[out=90,in=-90] (x1);
   \draw [fill=yellow] (1.06,2.4) rectangle (0.8, 2.3);
  \diagram*{
  (j1)--[ thick] (j5),
  (j2)--[ thick] (j6),
  (x1)--[thick,middlearrow={latex}] (i2)
    };
  \end{feynman}
\end{tikzpicture}}}
    \end{equation*}
    \caption{Open defect junction and an insertion of a boundary  operator.}\label{klap}
\end{figure}

Secondly, we can define a junction of a defect with a boundary. 
More precisely,  for a given RCFT a topological defect can always end on a conformal boundary condition with the end point in general being a defect operator of non-trivial dimension. In particular cases, when this dimension is zero, the junction is topological. We will often refer to such topological junctions as open topological defects.
In correlation functions involving bulk and boundary operators an open defect  can be freely moved both in the bulk and on the boundary, as long as the defect does not pass through an insertion. We depict a topological junction  on Figure \ref{klap} where we also show an insertion of a boundary operator. 

\begin{figure}
  \begin{equation*}
    \vcenter{\hbox{\begin{tikzpicture}[font=\footnotesize,inner sep=2pt]
  \begin{feynman}
  \path[pattern=north east lines,pattern color=ashgrey,very thin] (1.01,4) rectangle (1.6,0);
  \vertex  (i1) at (0,0)  ;
  \vertex (x1) at (0.2,3.5) ;
  \vertex[above=0.5cm of x1] (x11);
  \vertex[below left=0.001cm of x1] (x5) [label=below left:\(X_y\)];
  \vertex[left=0.75cm of x5] (l1) [label=below left:\(X_x\)];
  \vertex  (x2) at (0.75, 0.65);
  \vertex [above=4cm of i1] (i2);
  \vertex [right=1cm of i1] (j1);
  \vertex[ above=0.4cm of j1] (j3) [label=right:\(M_c\)];
  \vertex [right=1cm of i2] (j2);
  \vertex[ below=0.4cm of j2] (j4) [label=right:\(M_a\)];
  \vertex [below=1.4cm of j2] (j6);
  \vertex   [above=1cm of j1] (j5) ;
  \vertex[ above =1.1cm of j1] (j55);
  \vertex [left=0.15cm of j5]  [label=left:\(\alpha\)];
    \vertex [left=0.1cm of j55]  (j);
  \vertex [left=0.1cm of j6] (j7);
  \vertex (y) at (-0.5,2.6) ;
  \vertex[above=1.4cm of y] (y11);
  \vertex [below left=0.001cm of y] (y1) ;
  \vertex [below=1.4cm of j2] (k1);
  \vertex [below=1.5cm of j2] (k2);
  \vertex [left=0.15cm of k1] (k3);
  \vertex [left=0.15cm of k2] (k4) [label=left:\(\beta\)];
\draw [thick]   (j7) to[out=90,in=-90] (x1);
   \draw  (1.06,1.1) rectangle (0.8, 1);
   \draw (1.06,2.6) rectangle (0.8, 2.5);
   \draw [thick]    (j) to[out=90,in=-90] (y);
   \vertex[above=0.65 of j55] [label=right:\(M_b\)];
  \diagram*{
  (j2)--[thick] (k1),
  (k2)--[thick] (j55),
  (j5)--[thick] (j1),
  (x1)--[thick,middlearrow={latex}] (x11),
  (y)--[thick,middlearrow={latex}] (y11)
    };
  \end{feynman}
\end{tikzpicture}}} \hspace{1mm}~= \mathlarger{\sum}_{v,\gamma,\delta} \mathrm{T}^{(x\,y\,a)c}_{\alpha b\beta,\gamma v\delta} ~
\vcenter{\hbox{\hspace{1mm}\begin{tikzpicture}[font=\footnotesize,inner sep=2pt]
  \begin{feynman}
   \path[pattern=north east lines,pattern color=ashgrey,very thin] (1.01,4) rectangle (1.6,0);
  \vertex  (i1) at (0,0)  ;
  \vertex [ empty dot] (x1) at (-0.5,2.3) [label=left:\(\gamma\)] {};
  \vertex  (x2) at (0.75, 0.65);
  \vertex [above=4cm of i1] (i2);
  \vertex [right=1cm of i1] (j1);
  \vertex[ above=0.4cm of j1] (j3) [label=right:\(M_c\)];
  \vertex [right=1cm of i2] (j2);
  \vertex[ below=0.4cm of j2] (j4) [label=right:\(M_a\)];
  \vertex   [above=1cm of j1] (j5);
  \vertex   [above=1.1cm of j1] (j6);
  \vertex [left=0.15cm of j5] [label=left:\(\delta\)];
  \vertex [left=0.1cm of j6] (j7);
  \vertex (x) at (-0.3,1.4) [label=right:\(X_v\)];
\draw [thick] [middlearrow={latex}]   (j7) to[out=90,in=-90] (x1);
   \draw  (1.06,1) rectangle (0.8, 1.1);
   \vertex (k1) at (-1,2.8) ;
   \vertex (k2) at (-1,3.4)  [label=left:\(X_x\)];
   \vertex (k3) at (0,2.8) ;
   \vertex (k4) at (0,3.4) [label=right:\(X_y\)];
   \vertex[above=1.1cm of k2] (k5) ;
   \vertex[above=0.6cm of k4] (k6);
   \vertex[left=0.25cm of k5] (k55);
   \vertex[right=0.25cm of j2] (j11);
    \vertex[below=0.5cm of k5] (k7) ;
   \draw [myptr={latex},thick,rounded corners=1mm]  (x1) -- (k1) --(k7);
   \draw [myptr={latex},thick,rounded corners=1mm]  (x1) -- (k3) --(k6);
  \diagram*{
  (j1)--[ thick] (j5),
  (j2)--[ thick] (j6)
    };
  \end{feynman}
\end{tikzpicture}}}
   \end{equation*}
    \caption{Fusion of open defects.}
    \label{figintro3}
\end{figure}
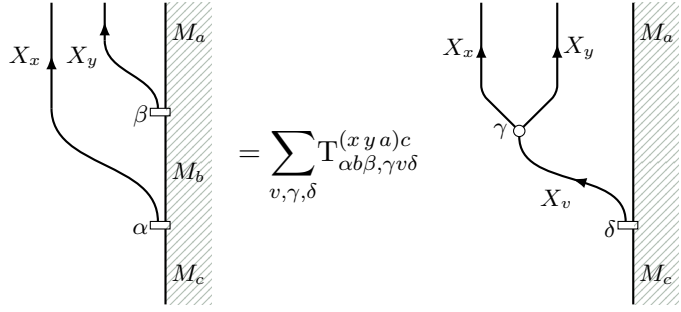

A nice piece of terminology related to the two constructions was proposed in \cite{weak_strong}. If a boundary condition is a simultaneous eigenstate 
under fusion with a subring of topological defects we say that this boundary condition is strongly symmetric with respect to this subring. 
If a  subset of topological defects each admits a  topological junction  with a boundary condition we say that that boundary condition is weakly symmetric with respect to  this subset. In the latter case one can fuse such junctions by putting them on top of each other. Such a fusion is governed by a new set of $\mathrm{F}$-matrices which we denote as $\mathrm{T}$ and  show schematically on  Figure \ref{figintro3}.

The rules for passing a topological defect through a bulk operator were formulated in \cite{FFRS} (see  \cite{KW} for a nice discussion in the context of the Ising model). In general this will result in a defect operator but in some special cases the passing through  just multiplies the bulk  
operator by a constant. As discussed in \cite{5A} such a situation has non-trivial implications for bulk RG flows triggered by the bulk operator at hand. 
If the constant is 1 the bulk operator commutes with some topological defects which have to survive the flow. If the constant is -1  the bulk operator anticommutes with a topological defect and such a defect must be present between the CFTs (possibly trivial) obtained as the end points with the opposite signs of the coupling. A slightly different take on the role of defects was presented in \cite{weak_strong}. For a massive flow the infrared vacuum can be described by 
a conformal boundary state \cite{Kon2}, \cite{Cardy_vac} specifying an RG boundary. In the presence of topological defects commuting with a perturbation 
the RG boundary must be strongly symmetric. Recently non-invertible and categorical symmetries received a lot of attention, see e.g. \cite{Sakura_lect}, \cite{Shao_lect}, \cite{Brennan_lect} for a review. The constraints from topological defects on RG flows were discussed in this broader framework in 
  \cite{Sakura_etal1}, \cite{Sakura_etal2}, \cite{Sakura_etal3}, \cite{SymTree}. 

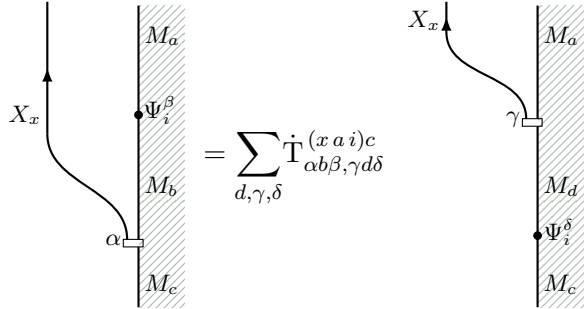
\begin{figure}
    \centering
    \begin{equation*}
     \vcenter{\hbox{\hspace{-10mm}\begin{tikzpicture}[font=\footnotesize,inner sep=2pt]
  \begin{feynman}
   \path[pattern=north east lines,pattern color=ashgrey,very thin] (0.01,4) rectangle (0.6,0);
  \vertex (j1) at (0,0);
  \vertex [above=0.8 of j1] (j2);
  \vertex[ above =0.9 of j1] (j3);
  \vertex [small,dot] [ above=2.5cm of j1] (psi)  {};
  \vertex[right=0.2cm of psi] (psi1) ;
  \vertex[below=0.2cm of psi1] (psi2) [label=\(\,\,\,\Psi_i^\beta\)];
  \vertex [above=3.4cm of j1] (j4) ;
  \vertex[above=3.5cm of j1] (j5)  ;
  \vertex[ left=0.15cm of  j4] (j7);
  \vertex [left=0.15cm of j3] (j8) [label=left:\(\alpha\)]; 
  \vertex [above =4cm of j1] (j6);
  \vertex [below=0.4cm of j6] (p1)  [label=right:\(M_a\)];
  \draw  (0.06,0.8) rectangle (-0.2, 0.9);
    \vertex[below=0.1cm of j6] (m1) ;
    \vertex[above=1.6cm of j1] (m3) [label=right:\(M_b\)];
    \vertex[above=0.9cm of psi] (m2);
    \vertex[left=1.2cm of psi] (x) [label=left:\(X_x\)];
    \vertex[above=0.3 of j1] (jj) [label=right:\(M_c\)];
    \vertex [below=0.25cm of x] (x1);
    \vertex [above=0.25cm of x] (x2);
    \vertex[above=1.5cm of x] (xtop);
    \draw [thick]    (j8) to[out=90,in=-90] (x1);
    \draw [thick] [middlearrow={latex}]   (x1) to (xtop);
   \diagram*{
    (j1) --[thick] (j2),
    (j3) --[thick] (psi),
    (psi) --[thick] (j6)
  };
  \end{feynman}
\end{tikzpicture}}}
~=\mathlarger{\sum}_{d,\gamma,\delta}{\mathrm{\dot{T}}}^{\,(x\,a\,i)c}_{\alpha b\beta,\gamma d\delta}~
\vcenter{\hbox{\hspace{1mm}\begin{tikzpicture}[font=\footnotesize,inner sep=2pt]
  \begin{feynman}
   \path[pattern=north east lines,pattern color=ashgrey,very thin] (0.01,4) rectangle (0.6,0);
  \vertex (j1) at (0,0);
  \vertex [above=2.4 of j1] (j2);
  \vertex[ above =2.5 of j1] (j3);
  \vertex [small,dot] [ above=0.9cm of j1] (psi)  {};
  \vertex[right=0.2cm of psi] (psi1);
  \vertex[below=0.2cm of psi1] (psi2) [label=\(\,\,\,\Psi_i^\delta\)];
  \vertex [above=3.4cm of j1] (j4) ;
  \vertex[above=3.5cm of j1] (j5)  ;
  \vertex[ left=0.15cm of  j4] (j7);
  \vertex [left=0.15cm of j3] (j8) [label=left:\(\gamma\)]; 
  \vertex [above =4cm of j1] (j6);
  \vertex[below=0.4cm of j6] (p2) [label=right:\(M_a\)];
  \draw (0.06,2.4) rectangle (-0.2, 2.5);
    \vertex[below=0.1cm of j6] (m1) ;
    \vertex[above=1.6cm of j1] (m3) [label=right:\(M_d\)];
    \vertex[above=0.9cm of j2] (m2) ;
    \vertex[above=0.5cm of j4] (xx);
    \vertex[left=1.2cm of xx] (x);
    \vertex[below=0.1cm of x] (x100)  [label=left:\(X_x\)];
    \vertex[above=0.3 of j1] (jj) [label=right:\(M_c\)];
    \vertex [below=0.25cm of x] (x1);
    \vertex [above=0.15cm of x] (x2);
    \vertex[left=0.25cm of x2] (x22);
    \vertex[above=0.6cm of j6] (j7);
    \vertex[right=0.25cm of j7] (j77);
    \vertex[left=0.25cm of x2] (x22);
    \vertex[above=0.05cm of x] (x3);
    \vertex[left=0.25cm of x3] (x33);
    \vertex[above=0.1cm of x33] (x333);
    \vertex[above=0.4cm of x2] (x4);
    \draw [thick]    (j8) to[out=90,in=-90] (x1);
    \draw [thick] [middlearrow={latex}]   (x1) to (x2);
   \diagram*{
    (j1) --[thick] (j2),
    (j3) --[thick] (j6)
  };
  \end{feynman}
\end{tikzpicture}}}
\end{equation*}
    \caption{Moving an open topological defect past a boundary field.}
    \label{figintro4}
\end{figure}

Another arena of applications of topological defects is the boundary RG flows which are triggered by a boundary deformation switched on a conformal boundary condition. Both the fusion of Figure \ref{figintro1} and the junctions on Figure \ref{klap} proved to be useful. As shown in \cite{Graham_Watts} any boundary RG flow 
can be fused with a topological defect to give another RG flow triggered by an operator of the same weight. Constraints on the same flow arise in the situation when one has a topological junction that either commutes or anti-commutes with a relevant boundary field.  Similarly to  the bulk case discussed above moving a topological junction past a boundary insertion in general introduces  a new boundary condition between the insertion and the junction.
Such a move is depicted on Figure \ref{figintro4}. 

It is governed by a new set of fusing matrices which we denote\footnote{In this paper we introduced a new notation for three kinds of fusing matrices: $\mathrm{T}$, $\dot {\mathrm{T}}$ and the fusing matrices for junctions of topological defects which we denote as $\mathrm{Y}$. We decided to break with the tradition of keeping the letter $\mathrm{F}$ in the notation in favour of some mnemonic rule that would make  it easier to remember which situation the fusing matrix describes. Thus, the  triple junctions of topological defects resemble the letter Y while the junctions of defects with boundaries are depicted as letter-T-shaped junctions. The dot in the notation  $\dot {\mathrm{T}}$ is meant to remind of an insertion of a boundary operator.  } by $\dot{\mathrm{T}}$. In the case when the boundary condition remains the same and the boundary field is left intact we say that the open defect commutes with the boundary operator. When 
moving past the insertion amounts only to multiplication of the boundary field by $-1$ we say that it anti-commutes. Concrete examples of such situations were presented in \cite{Kon1} for the Virasoro minimal models. The commuting situation means that the topological junction survives into the infrared fixed point. When there are several such topological junctions their fusion algebra must also be an invariant under the RG flow. This means that the relevant 
fusing matrices $\mathrm{T}$  defined on Figure \ref{figintro3}  must be invariant (up to a change of basis). 

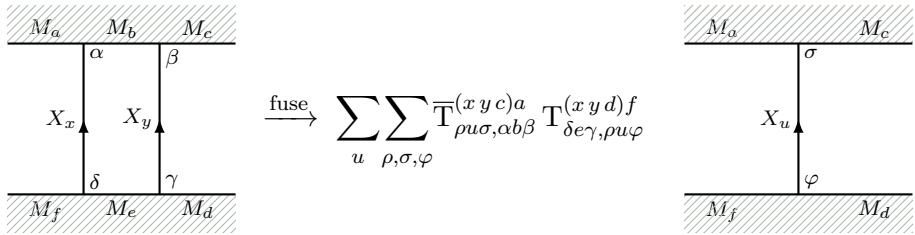
\begin{figure}
    \centering
    \begin{equation*}
        \vcenter{\hbox{\begin{tikzpicture}[font=\scriptsize,inner sep=2pt]
  \begin{feynman}
\vertex (i1) at (0,0) ;
\vertex[right=3cm of i1] (j1);
\vertex[above=0.5cm of i1] (i2);
\vertex[above=0.28cm of i1] (i22) ;
\vertex[right=0.2cm of i22] (i222) ;
\vertex[below=0.08cm of i222] (i2222) [label=right:\(M_a\)];
 \path[pattern=north east lines,pattern color=ashgrey,very thin] (i2) rectangle (j1);
 \vertex[below=2cm of i1] (i3);
 \vertex[right=3cm of i3] (j2);
 \vertex[below=0.5cm of i3] (i4);
 \vertex[below=0.35cm of i3] (i33) ;
 \vertex[right=0.2cm of i33] (i333);
 \vertex[above=0.12cm of i333] (i3333)  [label=right:\(M_f\)];
  \path[pattern=north east lines,pattern color=ashgrey,very thin] (i4) rectangle (j2);
  \vertex[right=1cm of i1] (d1)[label=below right:\(\alpha\)];
  \vertex[right=1cm of d1] (d2)[label=below right:\(\beta\)];
  \vertex[below=2cm of d1] (d3)[label=above right:\(\delta\)];
  \vertex[right=1cm of d3] (d4)[label=above right:\(\gamma\)];
  \vertex[right=0.5cm of d1] (m1) [label=above:\(M_b\)];
   \vertex[right=0.5cm of d2] (m1) [label=above:\(M_c\)];
    \vertex[right=0.5cm of d3] (m1) [label=below:\(M_e\)];
     \vertex[right=0.5cm of d4] (m1) [label=below:\(M_d\)];
   \diagram*{
(i1)--[thick] (j1),
(j2)--[thick] (i3),
(d3)--[thick,middlearrow={latex},edge label=$X_x$] (d1),
(d4)--[thick,middlearrow={latex},edge label=$X_y$] (d2)
  };
  \end{feynman}
\end{tikzpicture}}}
~ \;\xrightarrow{\text{fuse}}\;\mathlarger{\sum}_u\mathlarger{\sum}_{\rho,\sigma,\varphi}\overline{\mathrm{T}}^{(x\,y\,c)a}_{\rho u\sigma,\alpha b\beta}\,\mathrm{T}^{(x\,y\,d)f}_{\delta e\gamma,\rho u\varphi}~
 \vcenter{\hbox{\quad\begin{tikzpicture}[font=\scriptsize,inner sep=2pt]
  \begin{feynman}
\vertex (i1) at (0,0) ;
\vertex[right=3cm of i1] (j1);
\vertex[above=0.5cm of i1] (i2);
\vertex[above=0.28cm of i1] (i22);
\vertex[right=0.2cm of i22] (i222)  ;
\vertex[below=0.08cm of i222] (i2222) [label=right:\(M_a\)];
 \path[pattern=north east lines,pattern color=ashgrey,very thin] (i2) rectangle (j1);
 \vertex[below=2cm of i1] (i3);
 \vertex[right=3cm of i3] (j2);
 \vertex[below=0.5cm of i3] (i4);
 \vertex[below=0.35cm of i3] (i33) ;
 \vertex[right=0.2cm of i33] (i333) ;
 \vertex[above=0.12cm of i333] (i3333)  [label=right:\(M_f\)];
  \path[pattern=north east lines,pattern color=ashgrey,very thin] (i4) rectangle (j2);
  \vertex[right=1.5cm of i1] (d1)[label=below right:\(\sigma\)];
  \vertex[below=2cm of d1] (d3)[label=above right:\(\varphi\)];
   \vertex[right=0.5cm of d2] (m1) [label=above:\(M_c\)];
     \vertex[right=0.5cm of d4] (m1) [label=below:\(M_d\)];
   \diagram*{
(i1)--[thick] (j1),
(j2)--[thick] (i3),
(d3)--[thick,middlearrow={latex},edge label=$X_u$] (d1)
  };
  \end{feynman}
\end{tikzpicture}}}
    \end{equation*}
    \caption{Fusion of compact open defects.}
    \label{figintro5}
\end{figure}

Another fusion algebra is obtained when we look at defects with both ends attached to the UV boundary condition. On Figure \ref{figintro5} this is depicted as defect lines stretched across a strip. 
It is not hard to show  that the fusion algebra of such {\it compact open defects} is associative. This was demonstrated in \cite{Koj} for the case of charge-conjugation modular invariant (i.e. the trivial Frobenius algebra $A={\bf 1}$). We do this in the general case in section \ref{sec_tube}. When the boundary condition is the direct sum of all irreducible boundary conditions with multiplicity 1 the corresponding associative algebra  was considered in \cite{BBW} under the name of the {\it tube algebra}. This notion   goes back to \cite{Ocn1} and also emerges in bulk theories, see \cite{5A}, \cite{Lin_Tach}, 
 \cite{BullBar}, \cite{Weak_Hopf} for recent discussions of the tube algebras arising in bulk QFTs and lattice theories. 
 The structure constants of the  algebras 
 associated with compact open defects are expressed in terms of the matrices $\mathrm{T}$ and the dual matrices $\overline{ \mathrm{T}}$ as on 
 Figure \ref{figintro5}. 
 
 When there is a boundary operator that commutes with some subset of topological junctions the compact open defects from this  subset generate an associative algebra which is an invariant of the RG flow triggered by this operator.
Examples of such algebras and boundary RG flows were considered in  \cite{Kon1} in the context of $A$-series Virasoro minimal models. 
When we have an operator that anti-commutes with some topological junctions similar conclusions apply giving defects linking the two infrared boundary conditions obtained for the opposite signs of the boundary coupling. More generally we can consider boundary RG flows in a multi-coupling space which are triggered by switching on several boundary operators. In such cases topological junctions arise between the boundary conditions on different RG trajectories. 
The assignment of a junction to pairs of RG trajectories can be expressed in terms of the fusing matrix $\dot {\mathrm{T}}$. 
We will consider concrete examples of such situations in section  \ref{RG_section}. 

We see that for the applications to RG flows one needs to know the matrices ${\mathrm T}$ and $\dot {\mathrm{T}}$. When $A=\bf {1}$ each irreducible module and bimodule is described by an irreducible object in $\mathcal{C}$. It follows from this that in an appropriate basis the matrices $\mathrm{T}$ and $\dot {\mathrm{T}}$ are given by the fusing matrices of  $\mathcal{C}$. For a non-trivial $A$ this is no longer the case and the new fusing matrices  need to be calculated. The main purpose of this paper is to lay out in detail the general structure associated with these matrices for the  case of $A\ne {\bf 1}$ and in the presence of multiplicities in the fusion rules. Furthermore we work out in detail how these matrices can be computed from a given representation theoretic data for the pair $(\mathcal{C}, A)$. In the TFT approach, as with all structure constants, this task reduces to solving some linear algebra problems. While conceptually we don't offer anything new to the computational part, we work out the details of the linear problems to the level at which one can use software. We illustrate the  method on particular examples of the rational free boson theories. An altogether different approach to calculating the fusing matrices associated with 
fusion categories which are Morita equivalent to the one with the known fusing matrices is developed in \cite{BBW}. Another recent calculation of fusing matrices arising in a CFT with a triality defect was performed in \cite{triality}.



The main body of the paper is organised as follows. In section \ref{sec_tensor} we introduce the tensor product structures related to boundaries and interfaces arising in RCFT. We formally introduce the associators and the fusing matrices for those tensor structures and discuss the related pentagon identities. In section 
\ref{sec_tube} we introduce the compact open defects and their fusion algebras. In section \ref{sec_calculation}  we work out the general formulae for the new fusing matrices in terms of the data in $(\mathcal{C}, A)$. In section \ref{sec_boson} we apply our general formulae to obtain the fusing matrices for 
some rational free boson theories. In section \ref{RG_section} we describe some concrete applications to boundary RG flows in rational free boson examples. The appendices contain a detailed discussion of associators with proofs of pentagon equations, as well as calculational details related to the free boson examples.

\section{Tensor product structures related to RCFTs}\label{sec_tensor}

In this paper we are following the TFT approach to RCFT (see  e.g. \cite{RFS_review}, \cite{review2} for a review). The  relevant mathematical structures used in this approach are well exposed in the literature (see e.g. \cite{BK,Etin}).   We will thus focus on conventions and notations providing only certain definitions that will be of particular importance for our calculations. Throughout this section we will assume that the modular tensor category $\mathcal{C}$ that encodes the chiral symmetry algebra of our RCFT is fixed.

\subsection{\texorpdfstring{Morphisms in $\mathcal{C}$}{Conventions for morphisms in C}}
 
We represent morphisms in  $\mathcal{C}$ as graphs in which lines stand for identity morphisms. For example an identity morphism $\operatorname{id}_U$ and $f\in \operatorname{Hom}(U,V)$ are depicted as
\begin{equation}
\operatorname{id}_U=~
\vcenter{\hbox{\hspace{1mm}\begin{tikzpicture}[font=\footnotesize,inner sep=2pt]
  \begin{feynman}
  \vertex (j1) at (0,0) [label=below:\(U\)];
  \vertex [above=2cm of j1] (j2) [label=above:\(U\)];
   \diagram*{
    (j1)--[thick] (j2)
  };
  \end{feynman}
\end{tikzpicture}}}
~\hspace{25mm}f=~ 
\vcenter{\hbox{\hspace{1mm}\begin{tikzpicture}[font=\footnotesize,inner sep=2pt]
  \begin{feynman}
  \vertex (j1) at (0,0) [label=below:\(U\)];
  \vertex [above=2cm of j1] (j2) [label=above:\(U\)];
  \vertex [above=1.2cm of j1] (j4);
  \draw [fill=yellow] (-0.3,0.8) rectangle (0.3, 1.2);
  \vertex [above=0.8cm of j1] (j3) [label=above:\scriptsize\(f\)];
   \diagram*{
    (j1)--[thick] (j3),
    (j4)--[thick] (j2)
  };
  \end{feynman}
\end{tikzpicture}}}
\end{equation}
Here the convention is that such pictures are read from bottom to top. Since $\operatorname{id}_{\mathbf{1}}=1\in\mathbb{C}$, lines labelled by $\mathbf{1}$ will be omitted, where $\mathbf{1}$ is the tensor unit of $\mathcal{C}$ (see definition \ref{def1}).  Composition of morphisms corresponds to concatenation of lines, while the tensor product to juxtaposition.

To extract information from the axioms of $\mathcal{C}$ we usually need to make an explicit  choice of basis in the three-point coupling spaces. 
To this end we choose representatives $U_i,\,i\in\mathcal{I}$ of the isomorphisms classes of the simple objects in $\mathcal{C}$ where  $\mathcal{I}$ is a finite indexing set. For these simple objects we fix  bases $\{\lambda^\alpha_{(i,j)k}\}$ in the hom-spaces $\operatorname{Hom}(U_i\otimes U_j,U_k)$ as well as dual bases $\{\bar{\lambda}_{\alpha}^{(i,j)k}\}$ in $\operatorname{Hom}(U_k,U_i\otimes U_j)$. These bases will be depicted as
\begin{equation}\label{bases1}
    \vcenter{\hbox{\hspace{2mm}\begin{tikzpicture}[font=\footnotesize,inner sep=2pt]
  \begin{feynman}
  \vertex (j1) at (0,0) [label=below:\(U_i\)];
  \vertex [above=1.7cm of j1] (j4);
  \draw [fill=yellow] (-0.25,1.8) rectangle (0.75, 1.3);
  \vertex [above=1.3cm of j1] (j3) [label=above:\( \hspace{5mm}\lambda^\alpha_{(i,j)k}\)];
  \vertex [right=0.5cm of j1] (i1) [label=below:\(U_j\)];
  \vertex (x) at (0.25,3) [label=above:\(U_k\)];
  \vertex [below=1.2cm of x] (x2);
  \vertex[above=1.7 cm of i1] (i4);
  \vertex[ above=1.3 cm of i1] (i3);
   \diagram*{
    (j1)--[thick] (j3),
    (i1)--[thick] (i3),
    (x2)--[thick] (x)
  };
  \end{feynman}
\end{tikzpicture}}}
~=~
  \vcenter{\hbox{\hspace{2mm}\begin{tikzpicture}[font=\footnotesize,inner sep=2pt]
  \begin{feynman}
\vertex (i1) at (0,0) [label=below:\(U_i\)];
\vertex [right=1cm of i1] (j1) [label=below:\(U_j\)];
\vertex [above=0.5cm of i1] (i2);
\vertex [above =0.5cm of j1] (j2);
\vertex  [right=0.5cm of i1] (k1) ;
\vertex [small,orange, dot][above =1.5cm of k1] (k2)[label=right:\(\alpha\)] {};
\vertex [above=1.5cm of k2] (k3) [label=above:\(U_k\)];
   \draw [thick,rounded corners=1mm] (i1)--(i2)--(k2);
      \draw [thick,rounded corners=1mm] (j1)--(j2)--(k2);
   \diagram*{
(k2)--[thick] (k3)
  };
  \end{feynman}
\end{tikzpicture}}}
\hspace{20mm}
 \vcenter{\hbox{\hspace{2mm}\begin{tikzpicture}[font=\footnotesize,inner sep=2pt]
  \begin{feynman}
  \vertex (j1) at (0,0) [label=above:\(U_i\)];
  \vertex [below=1.3cm of j1] (j2);
  \draw [fill=yellow] (-0.25,-1.8) rectangle (0.75, -1.3);
  \vertex [below=1.8cm of j1] (j3) [label=\( \hspace{5mm}\bar{\lambda}_{\alpha}^{(i,j)k}\)];
  \vertex [right=0.5cm of j1] (i1) ;
  \vertex[below=0.04cm of i1] (xx) [label=above:\(U_j\)];
  \vertex (x) at (0.25,-3) [label=below:\(U_k\)];
  \vertex [above=1.2cm of x] (x2);
  \vertex[below=1.3cm of i1] (i2);
   \diagram*{
    (j1)--[thick] (j2),
    (i1)--[thick] (i2),
    (x)--[thick] (x2)
  };
  \end{feynman}
\end{tikzpicture}}}
~=~
  \vcenter{\hbox{\hspace{2mm}\begin{tikzpicture}[font=\footnotesize,inner sep=2pt]
  \begin{feynman}
\vertex (i1) at (0,0) [label=below:\(U_k\)];
\vertex [small,orange, dot][above=1.5cm of i1] (i2) [label=right:\(\bar{\alpha}\)]{};
\vertex [above=3cm of i1] (k1);
\vertex [right =0.5cm of k1] (k2) ;
\vertex[below=0.04cm of k2] (xx) [label=above:\(U_j\)];
\vertex  [left=0.5cm of k1] (k3) [label=above:\(U_i\)] ;
\vertex [below=0.5cm of k3] (k4);
\vertex [below=0.5cm of k2] (k5);
   \draw [thick,rounded corners=1mm] (k2)--(k5)--(i2);
      \draw [thick,rounded corners=1mm] (k3)--(k4)--(i2);
   \diagram*{
(i2)--[thick] (i1)
  };
  \end{feynman}
\end{tikzpicture}}}
\end{equation}
where $\alpha\in \{ 1,\ldots,\tensor{N}{_{ij}}{^k}\}$ with $\tensor{N}{_{ij}}{^k}$ being the fusion coefficients defined as 
    \begin{equation}
    \tensor{N}{_{ij}}{^k}:=\operatorname{dim}\operatorname{Hom}(U_i\otimes U_j,U_k) \, .
    \end{equation}
    Duality of these bases means that the following normalisation holds
\begin{equation} \label{e5}
\vcenter{\hbox{\hspace{-10mm}\begin{tikzpicture}[font=\footnotesize,inner sep=2pt]
  \begin{feynman}
  \vertex  [small,orange, dot] (del) at (0,0) [label=below:\(\alpha\)] {} ;
  \vertex  [small,orange, dot] (gam) at (0,-2) [label=above:\(\beta\)] {} ;
 \vertex [below=1cm of gam] (j1) [label=below:\(U_j\)];
  \vertex [above=1cm of del] (j2) [label=above:\(U_m\)];
   \diagram*{
     (j2)--[thick] (del),
     (j1)--[thick] (gam),
     (del) --[half left, thick, looseness=1.2,edge label=$U_k$] (gam),
     (gam) --[half left, thick, looseness=1.2, edge label=$U_i$] (del)
  };
  \end{feynman}
\end{tikzpicture}}}
~\hspace{5mm}=\delta_{m,j}\delta_{\alpha,\beta} 
~ 
\vcenter{\hbox{\hspace{5mm}\begin{tikzpicture}[font=\footnotesize,inner sep=2pt]
  \begin{feynman}
  \vertex (1) at (0,1) [label=above:\(U_j\)];
  \vertex (2) at (0,-3) [label=below:\(U_j\)];
   \diagram*{
     (1)--[thick] (2)
  };
  \end{feynman}
\end{tikzpicture}}}\end{equation}
These bases also satisfy a completeness relation due to the semisimplicity of the category $\mathcal{C}$ \cite{FRS1}:
\begin{equation}\label{completelambda}
    \vcenter{\hbox{\hspace{-10mm}\begin{tikzpicture}[font=\footnotesize,inner sep=2pt]
  \begin{feynman}
  \vertex  (i1) at (0,0) [label=below:\(U_i\)]  ;
  \vertex [above=3cm of i1] (i2) [label=above:\(U_i\)];
  \vertex [right=1cm of i1] (j1) [label=below:\(U_j\)];
  \vertex [right=1cm of i2] (j2) [label=above:\(U_j\)];
  \diagram*{
  (i1)--[thick] (i2),
  (j1)--[thick] (j2)
    };
  \end{feynman}
\end{tikzpicture}}}
~\hspace{5mm}=\mathlarger{\sum}_{k\in\mathcal{I}}\mathlarger{\sum}_\gamma\,
~ 
\vcenter{\hbox{\hspace{1mm}\begin{tikzpicture}[font=\footnotesize,inner sep=2pt]
  \begin{feynman}
  \vertex  [small,orange,dot] (up) at (0,0) [label=above:\(\gamma\)] {};
  \vertex [small,orange,dot] [below=1cm of up] (down) [label=below:\(\gamma\)] {};
  \vertex [above left=0.75cm of up] (i1);
  \vertex [above=0.6cm of i1] (i2) [label=above:\(U_i\)];
  \vertex [above right=0.75cm of up] (j1);
  \vertex [above=0.6cm of j1] (j2) [label=above:\(U_j\)];
  \draw [thick,rounded corners=1mm] (up) -- (i1) --(i2);
  \draw [thick,rounded corners=1mm] (up) -- (j1) --(j2);
  \vertex [below left=0.75cm of down] (i3) ;
  \vertex [below=0.6cm of i3] (i4) [label=below:\(U_i\)];
  \vertex [below right=0.75cm of down] (j3);
  \vertex [below=0.6cm of j3] (j4) [label=below:\(U_j\)];
  \draw [thick,rounded corners=1mm] (down) -- (j3) --(j4);
   \draw [thick,rounded corners=1mm] (down) -- (i3) --(i4);
   \diagram*{
      (down)--[thick,edge label=$U_k$] (up)
  };
  \end{feynman}
\end{tikzpicture}}}
\end{equation}
In the literature one finds other normalisation choices than (\ref{e5}). In particular the choice made in  \cite{Koj}, for theories with  ${\tensor{N}{_{ij}}{^k}} \in\{0,1\}$, can be characterised by the equation 
\begin{equation}\label{koj1}
\vcenter{\hbox{\hspace{-10mm}\begin{tikzpicture}[inner sep=2pt,font=\footnotesize]
  \begin{feynman}
  \vertex  [small,orange, dot] (l) at (0,0)  {} ;
  \vertex [above right=1.25cm of l] [small,orange, dot] (alpha)  {};
  \vertex [above=1cm of alpha] (i) [label=above:\(U_i\)];
  \vertex [below right=1cm of alpha] (j1);
  \vertex [below right=0.3cm of j1] (j2);
  \vertex [below=0.6cm of j2] (j) [label=below:\(U_j\)];
  \vertex [below=0.7cm of l] (1) [label=below:\(\mathbf{1}\)];
  \vertex [above left=0.8cm of l] (k1) ;
  \vertex [above left=0.3cm of k1] (k2);
  \vertex [above=1.1cm of k2] (k) [label=above:\(U_k\)];
   \draw [thick,rounded corners=1mm] (alpha) -- (j1)  -- (j2)--(j);
   \draw [thick,rounded corners=1mm] (l) -- (k1)  -- (k2)--(k);
    \diagram*{
     (l) --[thick,edge label= $U_{\bar{k}}$] (alpha),
     (alpha) --[thick] (i),
     (l) --[scalar] (1)
    };
  \end{feynman}
\end{tikzpicture}}}
~\hspace{5mm}=
~ 
\vcenter{\hbox{\hspace{2mm}\begin{tikzpicture}[font=\footnotesize,inner sep=2pt]
  \begin{feynman}
\vertex (i1) at (0,0) [label=below:\(U_j\)];
\vertex [small,orange, dot][above=1.5cm of i1] (i2) {};
\vertex [above=3cm of i1] (k1);
\vertex [right =0.5cm of k1] (k2) ;
\vertex[below=0.04cm of k2] (xx) [label=above:\(U_i\)];
\vertex  [left=0.5cm of k1] (k3) [label=above:\(U_k\)] ;
\vertex [below=0.5cm of k3] (k4);
\vertex [below=0.5cm of k2] (k5);
   \draw [thick,rounded corners=1mm] (k2)--(k5)--(i2);
      \draw [thick,rounded corners=1mm] (k3)--(k4)--(i2);
   \diagram*{
(i2)--[thick] (i1)
  };
  \end{feynman}
\end{tikzpicture}}}\end{equation}
where $\bar{k}$  is the label of the dual object of $U_k$.


\subsection{Frobenius algebras and  modules over them}

Let $A$ be a symmetric special Frobenius algebra in $\mathcal{C}$. We  introduce  the following graphical notation for the multiplication, unit, comultiplication and counit morphisms of  $A$
\begin{equation}
   m
~=~
  \vcenter{\hbox{\hspace{2mm}\begin{tikzpicture}[font=\footnotesize,inner sep=2pt]
  \begin{feynman}
\vertex (i1) at (0,0) [label=below:\(A\)];
\vertex [right=1cm of i1] (j1) [label=below:\(A\)];
\vertex [above=0.5cm of i1] (i2);
\vertex [above =0.5cm of j1] (j2);
\vertex  [right=0.5cm of i1] (k1) ;
\vertex [small, scarlet, dot][above =1.5cm of k1] (k2) {};
\vertex [above=1.5cm of k2] (k3) [label=above:\(A\)];
   \draw [thick,rounded corners=1mm] (i1)--(i2)--(k2);
      \draw [thick,rounded corners=1mm] (j1)--(j2)--(k2);
   \diagram*{
(k2)--[thick] (k3)
  };
  \end{feynman}
\end{tikzpicture}}}
\hspace{12mm}
\eta~=~
 \vcenter{\hbox{\begin{tikzpicture}[font=\footnotesize,inner sep=2pt]
  \begin{feynman}
\vertex(i1) at (0,0) [label=below:\(\mathbf{1}\)]{};
\draw[black,fill=baby] (0,0) circle (.5ex);
\vertex [above=1cm of i1] (i2) [label=above:\(A\)];
   \diagram*{
(i1)--[thick] (i2)
  };
  \end{feynman}
\end{tikzpicture}}}
\hspace{12mm} \Delta ~=~
      \vcenter{\hbox{\begin{tikzpicture}[font=\footnotesize,inner sep=2pt]
  \begin{feynman}
\vertex (i1) at (0,0) [label=below:\(A\)];
\vertex [small,blue, dot][above=1.5cm of i1] (i2){};
\vertex [above=3cm of i1] (k1);
\vertex [right =0.5cm of k1] (k2) [label=above:\(A\)];
\vertex  [left=0.5cm of k1] (k3) [label=above:\(A\)] ;
\vertex [below=0.5cm of k3] (k4);
\vertex [below=0.5cm of k2] (k5);
   \draw [thick,rounded corners=1mm] (k2)--(k5)--(i2);
      \draw [thick,rounded corners=1mm] (k3)--(k4)--(i2);
   \diagram*{
(i2)--[thick] (i1)
  };
  \end{feynman}
\end{tikzpicture}}}
\hspace{12mm}
\varepsilon~=~ 
\vcenter{\hbox{\begin{tikzpicture}[font=\footnotesize,inner sep=2pt]
  \begin{feynman}
\vertex(i1) at (0,0) [label=above:\(\mathbf{1}\)]{};
\draw[black,fill=aqua] (0,0) circle (.5ex);
\vertex [below=1cm of i1] (i2) [label=below:\(A\)];
   \diagram*{
(i1)--[thick] (i2)
  };
  \end{feynman}
\end{tikzpicture}}}
\end{equation}
As $\mathcal{C}$ is semisimple,   $A$ is a finite direct sum of  $U_i$. We fix  all bases $\{\imath^A_{i,\alpha}\}$ in the embedding spaces $\operatorname{Hom}(U_i,A)$, as well as dual bases $\{\jmath_A^{i,\alpha}\}$ in $\operatorname{Hom}(A,U_i)$ which we draw as
\begin{equation}\label{embed0}
    \vcenter{\hbox{\hspace{1mm}\begin{tikzpicture}[font=\footnotesize,inner sep=2pt]
  \begin{feynman}
  \vertex (j1) at (0,0) [label=below:\(U_i\)];
  \vertex [above=2.1cm of j1] (j2) [label=above:\(A\)];
  \vertex [above=1.3cm of j1] (j4);
  \draw [fill=yellow] (-0.3,0.8) rectangle (0.3, 1.3);
  \vertex [above=0.8cm of j1] (j3) [label=above:\scriptsize\(\imath^A_{i,\alpha}\)];
   \diagram*{
    (j1)--[thick] (j3),
    (j4)--[thick] (j2)
  };
  \end{feynman}
\end{tikzpicture}}}
~=:~ 
 \vcenter{\hbox{\hspace{1mm}\begin{tikzpicture}[font=\footnotesize,inner sep=2pt]
  \begin{feynman}
  \vertex (j1) at (0,0) [label=below:\(U_i\)];
  \vertex [above=2.1cm of j1] (j2) [label=above:\(A\)];
  \vertex [above=0.9cm of j1] (j4);
 \node[isosceles triangle,
    draw,
    rotate=90,
    fill=yellow,
    minimum size =0.4cm] (T1)at (j4){};
    \vertex [below=0.1cm of j4] (j5) [label=above:\tiny\(\alpha\)];
   \diagram*{
    (j1)--[thick] (j5),
    (T1)--[thick] (j2)
  };
  \end{feynman}
\end{tikzpicture}}}
\qquad\quad \text{and}\qquad \quad
 \vcenter{\hbox{\hspace{1mm}\begin{tikzpicture}[font=\footnotesize,inner sep=2pt]
  \begin{feynman}
  \vertex (j1) at (0,0) [label=below:\(A\)];
  \vertex [above=2.1cm of j1] (j2) [label=above:\(U_i\)];
  \vertex [above=1.3cm of j1] (j4);
  \draw [fill=yellow] (-0.3,0.8) rectangle (0.3, 1.3);
  \vertex [above=0.8cm of j1] (j3) [label=above:\scriptsize\(\jmath_A^{i,\alpha}\)];
   \diagram*{
    (j1)--[thick] (j3),
    (j4)--[thick] (j2)
  };
  \end{feynman}
\end{tikzpicture}}}
~=:~ 
 \vcenter{\hbox{\hspace{1mm}\begin{tikzpicture}[font=\footnotesize,inner sep=2pt]
  \begin{feynman}
  \vertex (j1) at (0,0) [label=below:\(A\)];
  \vertex [above=2.1cm of j1] (j2) [label=above:\(U_i\)];
  \vertex [above=1.1cm of j1] (j4);
 \node[isosceles triangle,
    draw,
    rotate=270,
    fill=yellow,
    minimum size =0.4cm] (T1)at (j4){};
    \vertex [below=0.2cm of j4] (j5) [label=\tiny\(\alpha\)];
   \diagram*{
    (j1)--[thick] (T1),
    (T1)--[thick] (j2)
  };
  \end{feynman}
\end{tikzpicture}}}
\end{equation}
They satisfy the following duality and completeness relations:
\begin{align}
    &\jmath_A^{(j,\beta)}\circ \imath^A_{i,\alpha}=\delta_{i,j}\delta_{\alpha,\beta}\id_{U_i} \\
    & \sum_{i, \alpha}  \imath^A_{i,\alpha} \circ  \jmath_A^{i,\alpha}  = \id_{M}\,.
\end{align}
Using these bases we can expand the comultiplication morphism in terms of $\bar{\lambda}^{(i,j)k}_{\bar{\alpha}}$ as 
\begin{equation}\label{comultiplication}
     \vcenter{\hbox{\hspace{2mm}\begin{tikzpicture}[font=\footnotesize,inner sep=2pt]
   \begin{feynman}
  \vertex (l) at (0,0)  ;
  \vertex[below=0.8cm of l] (w1);
  \vertex[above=0.5cm of l] ;
   \vertex[right=0.1cm of l] (phi) ;
  \vertex  [above=0.8cm of l] (gam) {};
  \vertex [small,blue,dot] [above left=1.25cm of gam] (del) {} ;
  \vertex [above right=1cm of gam] (k1);
  \vertex [above=1.4cm of k1] (k) ;
  \vertex [above left=0.7cm of del] (i1);
  \vertex [above=0.7cm of i1] (i) ;
   \vertex [above right=0.7cm of del] (j1);
  \vertex [above=0.7cm of j1] (j) ;
     \draw [thick,rounded corners=1mm] (del) -- (j1) --(j);
       \draw [thick,rounded corners=1mm] (del) -- (i1) --(i);
       \vertex[above=0.8cm of i] (i3) [label=above:\(U_i\)];
       \vertex[above=0.8cm of j] (j3) [label=above:\(U_j\)];
       \vertex[above=0.8cm of k] (k3) ;
        \vertex[above right=0.1cm of k] (tau);
        \vertex[below=0.5cm of i] (i4) [label=left:\(A\)];
         \vertex[below=0.5cm of j] (j4) [label=right:\(A\)];
         \vertex[below=0.8cm of k] (k4);
    \node[isosceles triangle,
    draw,
    rotate=270,
    fill=yellow,
    minimum size =0.35cm] (T3)at (j){};
    \node[isosceles triangle,
    draw,
    rotate=270,
    fill=yellow,
    minimum size =0.35cm] (T2)at (i){};
    \vertex[below=0.24cm of j] (b1) [label=:\tiny\(\beta\)];
    \vertex[below=0.17cm of i] (a1) [label=:\tiny \(\alpha\)];
    \vertex[below=1cm of del] (c1) ;
    \vertex[above=0.4cm of c1] (c3) [label=above right:\(A\)];
    \node[isosceles triangle,
    draw,
    rotate=90,
    fill=yellow,
    minimum size =0.35cm] (T1)at (c1){};
    \vertex[below=0.12cm of c1] (c2) [label=:\tiny\(\hspace{-0.5mm}\gamma\)];
    \vertex[below=0.7cm of c1] (bot) [label=below:\(U_k\)];
    \diagram*{
     (T2)--[thick] (i3),
     (T3)--[thick] (j3),
     (T1)--[thick] (del),
     (T1)--[thick] (bot)
    };
  \end{feynman}
\end{tikzpicture}}}
=\mathlarger{\sum}_{\delta=1}^{\tensor{N}{_{ij}}{^k}} \Delta_{k\gamma;\delta}^{i\alpha,j\beta} \vcenter{\hbox{\hspace{2mm}\begin{tikzpicture}[font=\footnotesize,inner sep=2pt]
  \begin{feynman}
\vertex (i1) at (0,0) [label=below:\(U_k\)];
\vertex [small,orange, dot][above=2cm of i1] (i2) [label=right:\(\bar{\delta}\)]{};
\vertex [above=3.8cm of i1] (k1);
\vertex [right =0.5cm of k1] (k2) ;
\vertex[below=0.04cm of k2] (xx) [label=above:\(U_j\)];
\vertex  [left=0.5cm of k1] (k3) [label=above:\(U_i\)] ;
\vertex [below=0.65cm of k3] (k4);
\vertex [below=0.65cm of k2] (k5);
   \draw [thick,rounded corners=1mm] (k2)--(k5)--(i2);
      \draw [thick,rounded corners=1mm] (k3)--(k4)--(i2);
   \diagram*{
(i2)--[thick] (i1)
  };
  \end{feynman}
\end{tikzpicture}}}
\end{equation}
The numbers $\Delta_{k\gamma;\delta}^{i\alpha,j\beta}\in{\mathbb C}$ are called the \textit{comultiplication numbers} of $A$.
Similarly we can introduce the \textit{multiplication numbers} by
\begin{equation}
     \vcenter{\hbox{\hspace{1mm}\begin{tikzpicture}[font=\footnotesize,inner sep=2pt]
  \begin{feynman}
  \vertex (a1) at (0,0) [label=below:\(U_i\)];
  \vertex [right=1cm of a1] (b1) [label=below:\(U_j\)];
  \vertex[above=0.8cm of a1] (a2) ;
  \vertex [above=0.8cm of b1] (b2) ;
  \node[isosceles triangle,
    draw,
    rotate=90,
    fill=yellow,
    minimum size =0.4cm] (T1)at (a2){};
   \node[isosceles triangle,
    draw,
    rotate=90,
    fill=yellow,
    minimum size =0.4cm] (T2)at (b2){};
     \vertex[above=0.7cm of a1] (a2) [label=\tiny\(\alpha\)];
     \vertex [right=0.5cm of a1] (x);
     \vertex [small,scarlet, dot][above=1.8cm of x] (x1){};
  \vertex [above=0.7cm of b1] (b2) [label=\tiny\(\beta\)];
  \vertex[above=0.4cm of a2] (a3);
  \vertex[above=0.4cm of b2] (b3);
  \vertex[above=0.8cm of x1] (c1);
  \node[isosceles triangle,
    draw,
    rotate=270,
    fill=yellow,
    minimum size =0.4cm] (T3)at (c1){};
    \vertex[above=0.6cm of x1] (c11) [label=above:\tiny\(\gamma\)];
    \vertex [above=0.6cm of T3] (c) [label=above:\(U_k\)];
   \diagram*{
    (a1)--[thick] (T1),
    (b1)--[thick] (T2),
    (a3) --[quarter left, thick, looseness=0.9] (x1),
     (b3) --[quarter right, thick, looseness=0.9] (x1),
     (T3)--[thick] (c),
     (x1)--[thick] (T3)
  };
  \end{feynman}
\end{tikzpicture}}}
~=\mathlarger{\sum}_{\delta=1}^{\tensor{N}{_{ij}}{^k} }m_{i\alpha,j\beta}^{k\gamma;\delta} ~
\vcenter{\hbox{\hspace{2mm}\begin{tikzpicture}[font=\footnotesize,inner sep=2pt]
  \begin{feynman}
\vertex (i1) at (0,0) [label=below:\(U_i\)];
\vertex [right=1cm of i1] (j1) [label=below:\(U_j\)];
\vertex [above=0.5cm of i1] (i2);
\vertex [above =0.5cm of j1] (j2);
\vertex  [right=0.5cm of i1] (k1) ;
\vertex [small,orange, dot][above =1.5cm of k1] (k2)[label=right:\(\delta\)] {};
\vertex [above=1.5cm of k2] (k3) [label=above:\(U_k\)];
   \draw [thick,rounded corners=1mm] (i1)--(i2)--(k2);
      \draw [thick,rounded corners=1mm] (j1)--(j2)--(k2);
   \diagram*{
(k2)--[thick] (k3)
  };
  \end{feynman}
\end{tikzpicture}}}
\end{equation}
As shown in  \cite{FRS1} in the case $\tensor{N}{_{ij}}{^k}\in\{0,1\}$ and $\dimh(U_k,A)\in\{0,1\}$ the following relation between these two sets of numbers holds:
\begin{equation}\label{comultnumbers}
    \Delta_k^{i,j}=\frac{1}{\operatorname{dim}(A)} \cdot \frac{m_{\bar{i}, k}^j}{m_{\bar{i}, i}{ }^0} \cdot \frac{\mathrm{F}_{j, 0}^{(i\, \bar{i}\, k) k}}{\mathrm{F}_{0,0}^{(i\, \bar{i}\, i) i}} \, .
\end{equation}

Conformal boundary conditions for an RCFT are labelled by modules of the algebra $A$. We remind the reader the following two definitions  (see e.g. \cite{FRS1}) we will be using later. 
\begin{dfn} \label{dmod}
    Let $(A,m,\eta)$ be an algebra in a tensor category $\mathcal{C}$. A \textit{left} $A$-\textit{module} is a pair $N=(\dot{N},\rho^N)$ where $\dot{N}\in\operatorname{Obj}(\mathcal{C})$ and $\rho^N\in\Hom(A\otimes \dot{N},\dot{N})$ such that
    \begin{equation}
        \rho^N \circ\left(m \otimes \operatorname{id}_{\dot{N}}\right)=\rho^N \circ\left(\operatorname{id}_A \otimes \rho^N\right) \qquad \text { and } \qquad \rho^N \circ\left(\eta \otimes \operatorname{id}_{\dot{N}}\right)=\operatorname{id}_{\dot{N}}\,.
    \end{equation}
\end{dfn}
Pictorially:
\begin{equation}
     \vcenter{\hbox{\hspace{2mm}\begin{tikzpicture}[font=\footnotesize,inner sep=2pt]
  \begin{feynman}
  \vertex (j1) at (0,0) [label=below:\(A\)];
  \vertex [above=1.7cm of j1] (j4);
  \draw [fill=yellow] (-0.25,1.8) rectangle (0.75, 1.3);
  \vertex [above=1.3cm of j1] (j3) [label=above:\( \hspace{5mm}\rho^N\)];
  \vertex [right=0.5cm of j1] (i1) [label=below:\(\dot{N}\)];
  \vertex (x) at (0.25,3) [label=above:\(\dot{N}\)];
  \vertex [below=1.2cm of x] (x2);
  \vertex[above=1.7 cm of i1] (i4);
  \vertex[ above=1.3 cm of i1] (i3);
   \diagram*{
    (j1)--[thick] (j3),
    (i1)--[thick] (i3),
    (x2)--[thick] (x)
  };
  \end{feynman}
\end{tikzpicture}}}
~=:~
 \vcenter{\hbox{\hspace{2mm}\begin{tikzpicture}[font=\footnotesize,inner sep=2pt]
  \begin{feynman}
 \vertex (a1) at (0,0) [label=below:\(A\)];
 \vertex [right=1cm of a1] (n1) [label=below:\(\dot{N}\)];
 \vertex [above=1.4cm of n1] (r) ;
 \vertex [above=1.5cm of r] (n2) [label=above:\(\dot{N}\)];
 \vertex [above=0.5cm of a1] (a2);
 \vertex [left=0.3cm of r] (r1);
 \vertex [above=0.2cm of r] (r2);
 \draw[fuc, thick] (r) -- (r1) -- (r2) -- cycle;
     \draw [thick,rounded corners=1mm] (a1) -- (a2) --(r1);
     \fill[fuc] (r) -- (r1) -- (r2) -- cycle;
   \diagram*{
    (n1)--[thick] (r),
    (r)--[thick] (n2)
  };
  \end{feynman}
\end{tikzpicture}}}
\hspace{17mm}
 \vcenter{\hbox{\hspace{2mm}\begin{tikzpicture}[font=\footnotesize,inner sep=2pt]
  \begin{feynman}
 \vertex (a1) at (0,0) [label=below:\(A\)];
 \vertex [right=0.8cm of a1] (n1) [label=below:\(\dot{N}\)];
 \vertex [above=1.3cm of n1] (r) ;
 \vertex [above=3cm of n1] (n2) [label=above:\(\dot{N}\)];
 \vertex [above=0.5cm of a1] (a2);
 \vertex [left=0.3cm of r] (r1);
 \vertex [above=0.2cm of r] (r2);
 \draw[fuc, thick] (r) -- (r1) -- (r2) -- cycle;
     \draw [thick,rounded corners=1mm] (a1) -- (a2) --(r1);
     \fill[fuc] (r) -- (r1) -- (r2) -- cycle;
     \vertex [above=0.6cm of r2] (s1) ;
     \vertex [above=0.2cm of s1] (s2);
     \vertex [left=0.3cm of s1] (s3);
     \draw[fuc, thick] (s1) -- (s2) -- (s3) -- cycle;
     \fill[fuc] (s1) -- (s2) -- (s3) -- cycle;
     \vertex[left=0.7cm of a1] (aa1) [label=below:\(A\)];
     \vertex [above =0.5cm of aa1] (aa2);
     \draw [thick,rounded corners=1mm] (aa1) -- (aa2) --(s3);
   \diagram*{
    (n1)--[thick] (n2)
  };
  \end{feynman}
\end{tikzpicture}}}
~=~
 \vcenter{\hbox{\hspace{2mm}\begin{tikzpicture}[font=\footnotesize,inner sep=2pt]
  \begin{feynman}
 \vertex (a1) at (0,0) [label=below:\(A\)];
 \vertex [right=0.8cm of a1] (n1) [label=below:\(\dot{N}\)];
 \vertex [above=2cm of n1] (r) ;
 \vertex [above=3cm of n1] (n2) [label=above:\(\dot{N}\)];
 \vertex [above=0.5cm of a1] (a2);
 \vertex [left=0.3cm of r] (r1);
 \vertex [above=0.2cm of r] (r2);
 \vertex [left=0.8cm of a1] (aa1) [label=below:\(A\)];
 \vertex[ left=0.4cm of a1] (b1);
 \vertex[above=0.5cm of aa1] (aa2);
 \vertex [small,red, dot][above=0.7cm of b1] (b2) {};
 \vertex[above=0.5cm of b2] (b3);
 \draw[fuc, thick] (r) -- (r1) -- (r2) -- cycle;
     \draw [thick,rounded corners=1mm] (a1) -- (a2) --(b2);
       \draw [thick,rounded corners=1mm] (aa1) -- (aa2) --(b2);
        \draw [thick,rounded corners=1mm] (b2) -- (b3) --(r1);
     \fill[fuc] (r) -- (r1) -- (r2) -- cycle;
   \diagram*{
    (n1)--[thick] (r),
    (r)--[thick] (n2)
  };
  \end{feynman}
\end{tikzpicture}}}
\hspace{17mm}
\vcenter{\hbox{\hspace{2mm}\begin{tikzpicture}[font=\footnotesize,inner sep=2pt]
  \begin{feynman}
 \vertex (a1) at (0,0) ;
 \vertex [above=0.4cm of a1] (aa1);
 \vertex[above=0.5cm of aa1] (aa2);
 \vertex [right=1cm of a1] (n1) [label=below:\(\dot{N}\)];
 \vertex [above=1.4cm of n1] (r) ;
 \vertex [above=1.5cm of r] (n2) [label=above:\(\dot{N}\)];
 \vertex [above=0.5cm of a1] (a2);
 \vertex [left=0.3cm of r] (r1);
 \vertex [above=0.2cm of r] (r2);
 \draw[fuc, thick] (r) -- (r1) -- (r2) -- cycle;
     \draw [thick,rounded corners=1mm] (aa1) -- (aa2) --(r1);
     \fill[fuc] (r) -- (r1) -- (r2) -- cycle;
     \draw[black,fill=baby] (aa1) circle (.5ex);
   \diagram*{
    (n1)--[thick] (r),
    (r)--[thick] (n2)
  };
  \end{feynman}
\end{tikzpicture}}}
~=~
\vcenter{\hbox{\hspace{2mm}\begin{tikzpicture}[font=\footnotesize,inner sep=2pt]
  \begin{feynman}
 \vertex (n1) at (0,0) [label=below:\(\dot{N}\)] ;
\vertex[above=3cm of n1] (n2) [label=above:\(\dot{N}\)];
   \diagram*{
    (n1)--[thick] (n2)    
  };
  \end{feynman}
\end{tikzpicture}}}
\end{equation}
\begin{dfn}
    Let  $(\dot{N},\rho^N)$ and $(\dot{M},\rho^M)$ be left $A$-modules. An $A$-\textit{module} \textit{intertwiner} is a morphism $f\in\Hom(\dot{N},\dot{M})$ such that
    \begin{equation}
        f\circ \rho^N=\rho^M\circ \left(\operatorname{id}_A\otimes f\right)\,.
    \end{equation}
    Pictorially:
    \begin{equation}
        \vcenter{\hbox{\hspace{2mm}\begin{tikzpicture}[font=\footnotesize,inner sep=2pt]
  \begin{feynman}
 \vertex (a1) at (0,0) [label=below:\(A\)];
 \vertex [right=1cm of a1] (n1) [label=below:\(\dot{N}\)];
 \vertex [above=1.4cm of n1] (r) ;
 \vertex [above=3cm of n1] (n2) [label=above:\(\dot{N}\)];
 \vertex [above=0.5cm of a1] (a2);
 \vertex [left=0.3cm of r] (r1);
 \vertex [above=0.2cm of r] (r2);
 \vertex [above=0.5cm of r2] (f1) [label=above:\( f\)];
 \vertex[above=0.5cm of f1] (f2);
 \draw[fuc, thick] (r) -- (r1) -- (r2) -- cycle;
     \draw [thick,rounded corners=1mm] (a1) -- (a2) --(r1);
     \fill[fuc] (r) -- (r1) -- (r2) -- cycle;
     \draw [fill=yellow] (0.75,2.1) rectangle (1.25, 2.6);
      \vertex [above=0.5cm of r2] (f1) [label=above:\(f\)];
   \diagram*{
    (n1)--[thick] (r),
    (r)--[thick] (f1),
    (f2)--[thick] (n2)
  };
  \end{feynman}
\end{tikzpicture}}}
~=~
 \vcenter{\hbox{\hspace{2mm}\begin{tikzpicture}[font=\footnotesize,inner sep=2pt]
  \begin{feynman}
 \vertex (a1) at (0,0) [label=below:\(A\)];
 \vertex [right=1cm of a1] (n1) [label=below:\(\dot{N}\)];
 \vertex [above=2.1cm of n1] (r) ;
 \vertex [above=3cm of n1] (n2) [label=above:\(\dot{N}\)];
 \vertex [above=0.6cm of a1] (a2);
 \vertex [left=0.3cm of r] (r1);
 \vertex [above=0.2cm of r] (r2);
 \vertex [above=0.9cm of n1] (f1) [label=above:\(f\)];
 \vertex[above=0.5cm of f1] (f2);
 \draw[fuc, thick] (r) -- (r1) -- (r2) -- cycle;
     \draw [thick,rounded corners=1mm] (a1) -- (a2) --(r1);
     \fill[fuc] (r) -- (r1) -- (r2) -- cycle;
     \draw [fill=yellow] (0.75,0.9) rectangle (1.25, 1.4);
      \vertex [above=0.9cm of n1] (f1) [label=above:\(f\)];
   \diagram*{
    (n1)--[thick] (f1),
    (f2)--[thick] (r),
    (r)--[thick] (n2)
  };
  \end{feynman}
\end{tikzpicture}}}
    \end{equation}
\end{dfn}
\noindent Given an algebra $A$ in $\mathcal{C}$ we can obtain the \textit{category of left $A$-modules} $\mathcal{C}_A$ whose objects are the $A$-modules and whose morphisms are the $A$-module intertwiners, i.e., elements of the subspaces
\begin{equation}
    \operatorname{Hom}_A(N, M):=\left\{f \in \operatorname{Hom}(\dot{N}, \dot{M}) \mid f \circ \rho^N=\rho^M \circ\left(\mathrm{id}_A \otimes f\right)\right\}\,.
\end{equation}
A simple object in  $\mathcal{C}_A$ is called a \textit{simple $A$-module}. We will denote the representatives for the isomorphism classes of simple $A$-modules by $M_a$ and the corresponding indexing set by $\mathcal{J}_A$. Some of the properties of $\mathcal{C}$ are inherited by $\mathcal{C}_A$ . For instance in our case $\mathcal{C}_A$ is semisimple, see \cite{FS}.

Recall that a \textit{module category} $\mathcal{M}$ over a tensor category $\mathcal{C}$ is a category together with a bifunctor \footnote{The usual convention is to denote this functor by the same symbol as the tensor product bifunctor of $\mathcal{C}$.}
$
    \otimes: \mathcal{M}\times\mathcal{C} \to 
    \mathcal{M}
$ 
that obeys generalised unit and associativity constraints. More specifically, $\mathcal{M}$ is equipped with a natural isomorphism with components
\begin{equation}
    m_{M,U,V}: M\otimes (U\otimes V)\to (M\otimes U)\otimes V, \qquad U,V\in\operatorname{Obj}(\mathcal{C}),\, M\in\operatorname{Obj}(\mathcal{M})
\end{equation}
 satisfying the module pentagon identity \eqref{modpent1}.

The category $\mathcal{C}_A$ is a module category over $\mathcal{C}$  as for any $U\in\objc$ and $M\in\operatorname{Obj}(\mathcal{C}_A)$ the object $M\otimes U$ has a natural left $A$-module structure. In particular, choosing $M=A$ we obtain:
\begin{dfn}
    Let $U\in\objc$ and $A$ an algebra in $\mathcal{C}$. The \textit{induced left module} $\operatorname{Ind}_A(U)$ is the left $A$-module whose underlying object is $A\otimes U$ and whose representation morphism is $m\otimes \operatorname{id}_U$, that is
    \begin{equation}
        \operatorname{Ind}_A(U):=\left( A\otimes U,\,m\otimes \operatorname{id}_U\right)\,.
    \end{equation}
\end{dfn}

Since $M\otimes U_i$ is a left $A$-module and thus labels a conformal boundary condition, we can  define boundary fields as follows. \textit{Boundary fields} which join a segment of boundary labelled by $M$ to a segment labelled by $N$ and that transform in the representation $U_i$ of $\mathfrak{V}$ are elements of $\Hom_{A}(M\otimes U_i,N)$.

For computational purposes, it is useful to express the representation map $\rho^M$ introduced in \eqref{dmod}  with respect to a  basis. To that end, for any given module $(\dot{M},\rho^M)$ we choose bases $b^M_{i,\alpha}$ for $\Hom(U_i,\dot{M})$ and dual bases $b_M^{j,\beta}$ for $\Hom(\dot{M},U_j)$ 
depicted as 
\begin{equation}\label{emb1}
   b^M_{i,\alpha}~=:~ \vcenter{\hbox{\hspace{1mm}\begin{tikzpicture}[font=\footnotesize,inner sep=2pt]
  \begin{feynman}
  \vertex (j1) at (0,0) [label=below:\(U_i\)];
  \vertex [above=2.1cm of j1] (j2) [label=above:\(\dot{M}\)];
  \vertex [above=0.9cm of j1] (j4);
 \node[isosceles triangle,
    draw,
    rotate=90,
    fill=aqua,
    minimum size =0.3cm] (T1)at (j4) {};
    \vertex [below=0.1cm of j4] (j5) ;
    \vertex [above right=0.12 of j5] (jj5) [label=above right:\scriptsize\(\alpha\)];
   \diagram*{
    (j1)--[thick] (j5),
    (T1)--[thick] (j2)
  };
  \end{feynman}
\end{tikzpicture}}}
\hspace{25mm}
b_M^{j,\beta}~=:~
\vcenter{\hbox{\hspace{1mm}\begin{tikzpicture}[font=\footnotesize,inner sep=2pt]
  \begin{feynman}
  \vertex (j1) at (0,0) [label=below:\(\dot{M}\)];
  \vertex [above=2.1cm of j1] (j2) [label=above:\(U_j\)];
  \vertex [above=1.1cm of j1] (j4);
 \node[isosceles triangle,
    draw,
    rotate=270,
    fill=aqua,
    minimum size =0.3cm] (T1)at (j4){};
    \vertex [below=0.25cm of j4] (j5) ;
    \vertex [right=0.23cm of j5] (jj5) [label=\scriptsize\(\beta\)];
   \diagram*{
    (j1)--[thick] (T1),
    (T1)--[thick] (j2)
  };
  \end{feynman}
\end{tikzpicture}}}
\end{equation}
These morphisms are 
normalised so  that
\begin{equation}
    b_M^{j,\beta}\circ b^M_{i,\alpha}=\delta_{i,j}\delta_{\alpha,\beta}\operatorname{id}_{U_i}\,
   \end{equation}
and satisfy the completeness condition 
\be
\sum_{i, \alpha}  b^M_{i,\alpha} \circ  b_M^{i,\alpha}  = \mathrm{id}_{M} \, .
\ee
Using these bases we have the decomposition
    \begin{equation}\label{rep2}
        \vcenter{\hbox{\hspace{1mm}\begin{tikzpicture}[font=\footnotesize,inner sep=2pt]
  \begin{feynman}
  \vertex (j1) at (0,0) [label=below:\(U_i\)];
  \vertex [above=2.8cm of j1] (j2) ;
  \vertex [above=1.4cm of j1] (j4);
  \vertex [above=2.1cm of j1] (j5) [label=right:\(\dot{M}\)];
 \node[isosceles triangle,
    draw,
    rotate=90,
    fill=aqua,
    minimum size =0.3cm] (T1)at (j4)  {};
    \vertex [below=0.1cm of j4] (j5) ;
    \vertex [above right=0.15cm of j5] (jj5) [label=right:\scriptsize\(\beta\)];
    \node[isosceles triangle,
    draw,
    rotate=270,
    fill=aqua,
    minimum size =0.3cm] (T2)at (j2){};
    \vertex [below right=0.1cm of j2] (jj2) [label= right:\scriptsize\(\gamma\)];
    \vertex [above=0.8cm of j2] (top) [label=above:\(U_j\)];
    \vertex [left=1cm of j1] (a1) [label=below:\(U_a\)];
    \vertex [above=0.6cm of a1] (a2) [label=above:\scriptsize\(\alpha\)];
    \vertex [above=0.6cm of T1] (a3);
    \vertex [left=0.3cm of a3] (a4);
    \vertex [above=0.2cm of a3] (a5);
     \draw[fuc, thick] (a3) -- (a4) -- (a5) -- cycle;
     \fill[fuc] (a3) -- (a4) -- (a5) -- cycle;
     \node[isosceles triangle,
    draw,
    rotate=90,
    fill=yellow,
    minimum size =0.4cm] (T3)at (a2)  {};
\vertex[above=0.6cm of T3] (a5);
    \draw [thick,rounded corners=1mm] (T3) -- (a5) --(a4);
     \vertex [above=0.5cm of a1] (a2) [label=above:\scriptsize\(\alpha\)];
   \diagram*{
    (j1)--[thick] (j5),
    (T1)--[thick] (T2),
    (T2)--[thick] (top),
    (a1)--[thick] (T3)
  };
  \end{feynman}
\end{tikzpicture}}}~\;=\mathlarger{\sum}_{\delta=1}^{\tensor{N}{_{ai}}{^j}} \,\mathlarger{\rho}^{M\;(j\gamma);\delta}_{(a\alpha)(i\beta)} \;
 \vcenter{\hbox{\hspace{2mm}\begin{tikzpicture}[font=\footnotesize,inner sep=2pt]
  \begin{feynman}
\vertex (i1) at (0,0) [label=below:\(U_a\)];
\vertex [right=1cm of i1] (j1) [label=below:\(U_i\)];
\vertex [above=0.7cm of i1] (i2);
\vertex [above =0.7cm of j1] (j2);
\vertex  [right=0.5cm of i1] (k1) ;
\vertex [small,orange, dot][above =1.8cm of k1] (k2)[label=right:\(\delta\)] {};
\vertex [above=1.8cm of k2] (k3) [label=above:\(U_j\)];
   \draw [thick,rounded corners=1mm] (i1)--(i2)--(k2);
      \draw [thick,rounded corners=1mm] (j1)--(j2)--(k2);
   \diagram*{
(k2)--[thick] (k3)
  };
  \end{feynman}
\end{tikzpicture}}}
    \end{equation}
  where $\rho^{M\;(j\gamma);\delta}_{(a\alpha)(i\beta)}\in{\mathbb C}$ are the \textit{representation numbers} of $M$.  Similarly we can define representation numbers in the case where we have a right module over $A$.

\subsection{Bimodules over Frobenius algebras and their fusion}
In this subsection we consider topological interfaces between two RCFTs sharing the same chiral symmetry algebra described by ${\mathcal C}$. 
In the TFT approach such interfaces are described  by $A$-$B$-bimodules where the RCFTs on each side of the interface are the ones obtained from the Frobenius algebras $A$ and $B$ respectively \cite{FRS1,FFRS}.

\begin{dfn} \label{bimoduledef}
\begin{enumerate}[label=(\roman*)]
   \item Let $A,B$  be algebra objects in a tensor category $\mathcal{C}$. An $A$-$B$-\textit{bimodule} $M$ is a triple $(\dot{M},\rho^M,\tilde{\rho}^M)$ where $\dot{M}\in\operatorname{Obj}(\mathcal{C})$, $\rho^M\in\Hom(A\otimes\dot{M},\dot{M})$ and $\tilde{\rho}^M\in\Hom(\dot{M}\otimes B,\dot{M})$ such that $(\dot{M},\rho^M)$ is a left $A$-module, $(\dot{M},\tilde{\rho}^M)$ is a right $B$-module and
    two actions commute:
    \begin{equation}
        \rho^M\circ(\operatorname{id}_A\otimes \tilde{\rho}^M)=\tilde{\rho}^M\circ(\rho^M\otimes \operatorname{id}_B)\,.
    \end{equation}
    \item Given two $A$-$B$-bimodules $X,Y$, the space of \textit{bimodule intertwiners} is defined as the subspace
    \begin{equation}
        \Hom_{A|B}(X,Y):=\{f\in\Hom(\dot{X},\dot{Y})\,|\,f\circ \rho^X=\rho^Y\circ(\operatorname{id}_A\otimes f),\; f\circ \Tilde{\rho}_X=\tilde{\rho}^Y\circ (f\otimes \operatorname{id}_B)\}\,.
    \end{equation}
    \end{enumerate}
\end{dfn} 
Thus, we obtain the category $\prescript{}{A}{\mathcal{C}}_B$ whose objects are $A$-$B$-bimodules and whose morphisms are $A$-$B$-bimodule intertwiners. Certain properties of $\mathcal{C}$ are inherited by  $\prescript{}{A}{\mathcal{C}}_B$. For example, if $\mathcal{C}$ has a finite number of isomorphism classes of simple objects, then so has $\dc$. In the following, we shall choose particular representatives of the isomorphism classes of simple $A$-$B$-bimodules denoted $\{X_x\}$ for $x$ in an indexing set $\mathcal{K}_{AB}$. An object $X\in\operatorname{Obj}(\prescript{}{A}{\mathcal{C}}_B)$ will  be represented graphically as
\begin{equation}
\vcenter{\hbox{\begin{tikzpicture}[block/.style ={rectangle, draw=black, thick, text width=5em,align=center, minimum height=1em},font=\scriptsize,inner sep=2pt]
  \begin{feynman}
\vertex (i1) at (0,0) ;
\vertex [right=1cm of i1] (j1);
\vertex [above=0.5cm of i1] (i2) [label=right:\(X\)];
\vertex [above=0.5cm of j1] (j2) ;
\vertex [above=3cm of i1] (i3);
\vertex [right =1cm of i3] (j3) ;
\vertex  [left=0.5cm of i3] (a1)  ;
\vertex [below= 0.4cm of a1] (a2);
 \node at (a2) [rectangle,draw] (a3) {$A$};
 \vertex [right=1cm of a2] (b1);
 \node at (b1) [rectangle,draw] (b3) {$B$};
 \vertex [right=1cm of b1] (c1);
   \diagram*{
(i1)--[thick,middlearrow={latex}] (i3)
  };
  \end{feynman}
\end{tikzpicture}}}
\end{equation}
While we will only work with the bimodules in $\prescript{}{A}{\mathcal{C}}_B$ with the arrow directed upwards, the opposite arrow represents the dual objects in $\prescript{}{B}{\mathcal{C}}_A$.

 Unlike $\mathcal{C}_A$, $\prescript{}{A}{\mathcal{C}}_A$ has a structure of a tensor category. To see this, consider two simple defects  running parallel to each other. In the limit of vanishing distance they fuse to a single defect. Thus emerges the notion of a fusion algebra of defects \cite{PZ}. In the TFT framework  this  fusion corresponds to the tensor product of bimodules over an algebra. Before we formally define this we need to introduce the tensor product of a right $A$-module $M$ and a left $A$-module $N$. 
\begin{dfn}\label{tensorover}
    Let $(\dot{M},\rho^M)$ be a right $A$-module and $(\dot{N},\rho^N)$ a left $A$-module where $A$ is a Frobenius algebra in a modular tensor category $\mathcal{C}$. The \textit{tensor product over $A$} denoted $M\otimes_AN$ is the object
    \begin{equation}\label{coker}
        M\otimes_AN:=\operatorname{Coker}\left[(\rho^M\otimes\operatorname{id}_N)-(\operatorname{id}_M\otimes\rho^N)\right]\,.
    \end{equation}
    The corresponding idempotent $P_{M,N}\in\operatorname{End}(M\otimes N)$ with $\operatorname{Im}(P_{M,N})=M\otimes_AN$ is found to be \footnote{Note that in the following, in order to lighten the notation in diagrams we will often identify the module with its underlying object $M\equiv \dot{M}$.}
    \begin{equation}\label{proj}
        P_{M,N}~=~ 
         \vcenter{\hbox{\hspace{2mm}\begin{tikzpicture}[font=\footnotesize,inner sep=2pt]
  \begin{feynman}
 \vertex (m1) at (0,0) [label=below:\(M\)];
 \vertex [above=2.1cm of m1] (m2);
 \vertex [right=0.3cm of m2] (m3);
 \vertex [above=0.2cm of m2] (m4);
 \draw[fuc, thick] (m2) -- (m3) -- (m4) -- cycle;
     \fill[fuc] (m2) -- (m3) -- (m4) -- cycle;
     \vertex[above=3cm of m1] (m5) [label=above:\(M\)];
     \vertex [right=1.5cm of m1] (n1) [label=below:\(N\)];
     \vertex [above=2.1cm of n1] (n2);
     \vertex [left=0.3cm of n2] (n3);
     \vertex [above=0.2cm of n2] (n4);
      \draw[fuc, thick] (n2) -- (n3) -- (n4) -- cycle;
     \fill[fuc] (n2) -- (n3) -- (n4) -- cycle;
     \vertex [above=3cm of n1] (n5) [label=above:\(N\)];
     \vertex[right=0.75cm of m1] (a1);
     \vertex [above=0.6cm of a1] (a2);
     \vertex  [small,blue, dot][above=0.5cm of a2] (a3) {};
     \vertex [above left=0.5cm of a3] (a4);
     \vertex[above right=0.5cm of a3] (a5);
      \draw [thick,rounded corners=1mm,looseness=3] (a3) -- (a4) --(m3);
       \draw [thick,rounded corners=1mm,looseness=3] (a3) -- (a5) --(n3);
   \diagram*{
   (m1)--[thick] (m5),
   (n1) --[thick] (n5),
   (a2)--[thick] (a3)
  };
   \draw[black,fill=baby] (a2) circle (.5ex);
  \end{feynman}
\end{tikzpicture}}}
    \end{equation}
\end{dfn}
The tensor product $M\otimes_AN$ is not in general an $A$-module. Interesting situations occur when the initial objects have bimodule structures. More specifically, consider an $A$-$B$-bimodule $X$ and a $B$-$C$-bimodule $Y$ where $A,B$ and $C$ are Frobenius algebras in $\mathcal{C}$. Then $X\otimes_ BY$ is an $A$-$C$-bimodule. Also, noticing that $X\otimes Y$ is an $A$-$C$-bimodule one can show that $P_{X,Y}$ is a bimodule intertwiner $P_{X,Y}\in\Hom_{A|C}(X\otimes Y,X\otimes Y)$. Since $\mathcal{C}$ is abelian, there exists an object $\operatorname{Im}(P_{X,Y})$ in $\prescript{}{A}{\mathcal{C}}_C$, a monomorphism $e_{X,Y}\in\Hom_{A|C}(\operatorname{Im}(P_{X,Y}),X\otimes Y)$ and an epimorphism $r_{X,Y}\in\Hom_{A|C}(X\otimes Y,\operatorname{Im}(P_{X,Y}))$ such that
\begin{equation}\label{proj11}
    P_{X,Y}=e_{X,Y}\circ r_{X,Y}\,.
\end{equation}
Then we define
\begin{equation}
    X\otimes_BY:=\operatorname{Im}(P_{X,Y})\,.
\end{equation}
Furthermore, defining the action on morphisms $f\in\Hom_{A|B}(X,X^\prime)$ and $g\in \Hom_{B|C}(Y,Y^\prime)$ as
\begin{equation}\label{newprod}
    f \otimes_B g:=r_{X^{\prime}, Y^{\prime}} \circ(f \otimes g) \circ e_{X, Y}
\end{equation}
we obtain a bifunctor $\otimes_B:\mathcal{C}_{A|B}\times\mathcal{C}_{B|C}\to\mathcal{C}_{A|C}$.

We still have to find an appropriate associator for $\prescript{}{A}{\mathcal{C}}_A$ to define a tensor category structure. It will prove useful to start from the most general tensor structure that can appear between three bimodules. Let $X\in\operatorname{Obj}(\prescript{}{A}{\mathcal{C}}_B),\, Y\in\operatorname{Obj}(\prescript{}{B}{\mathcal{C}}_C),\,Z\in\operatorname{Obj}( \prescript{}{C}{\mathcal{C}}_D)$. We define  the associator maps
\begin{equation}
    \check{\alpha}_{X,Y,Z}:X\otimes_B(Y\otimes_CZ)\to(X\otimes_BY)\otimes_CZ
\end{equation}
 by
\begin{equation}\label{associator2}
     \check{\alpha}_{X,Y,Z}=r_{X\otimes_BY,Z}\circ\left(r_{X,Y}\otimes \operatorname{id}_{Z}\right)\circ\left(\operatorname{id}_{X}\otimes e_{Y,Z}\right)\circ e_{X,Y\otimes_CZ}\,.
\end{equation}
Specialising to the case  $A=B=C=D$ we  introduce the associator 
\begin{equation}\label{associator11}
\check{\alpha}_{X,Y,Z}:X\otimes_A(Y\otimes_AZ)\to(X\otimes_AY)\otimes_AZ
\end{equation}
of $\prescript{}{A}{\mathcal{C}}_A$ defined by
\begin{equation}\label{associator1}
    \check{\alpha}_{X,Y,Z}=r_{X\otimes_AY,Z}\circ\left(r_{X,Y}\otimes \operatorname{id}_Z\right)\circ\left(\operatorname{id}_X\otimes e_{Y,Z}\right)\circ e_{X,Y\otimes_AZ}\,.
\end{equation}
 It is shown in appendix  \ref{app3} that this associator satisfies  the pentagon identity turning $\prescript{}{A}{\mathcal{C}}_A$ into a tensor category.

Similarly to the fusion of two interfaces described above, we can define a fusion of an interface and a conformal boundary condition. 
Consider an $A$-$B$-interface described by a $A$-$B$-bimodule $X$ running close to a $B$-boundary described by a left $B$-module $M$. 
The result of the fusion is the object $X\otimes_BM$ defined in \eqref{coker} which has a left $A$-module structure and thus represents a new boundary condition.

\subsection{Junctions}
 In this subsection we introduce various types of junctions between interfaces and boundaries. Such junctions appear in a network of topological interfaces and conformal boundary conditions and have been studied in \cite{FFRS,5A,Koj}.

The first type of junctions we consider joins  three topological interfaces.  At this point we choose the conventions of \cite{FFRS} so that the morphisms in the diagrams which involve interfaces flow from top to bottom. Then the junctions labelled by morphisms $\alpha\in \operatorname{Hom}_{A|B}(X^\prime,X)$ and $\beta \in \Hom_{A|C}(Z,X\otimes_BY)$ are represented graphically as
\begin{equation}\label{jun1}
    \vcenter{\hbox{\begin{tikzpicture}[font=\footnotesize,inner sep=2pt]
  \begin{feynman}
\vertex (i1) at (0,0) [label=above right:\(X\)] ;
\vertex [dot,yellow] [above =1.5cm of i1] (i2) [label=right:\(\alpha\)] {};
\vertex[above=0.2cm of i2] (i4);
\vertex[above=0.4cm of i4] (i6) [label=right:\(X^\prime\)];
\vertex [above=3cm of i1] (top) ;
\vertex [left=0.5cm of top] (a1);
\vertex [below=0.3cm of a1] (a2);
 \node at (a2) [rectangle,draw] (a3) {$A$};
\vertex [right=0.5cm of top] (b1);
\vertex [below=0.3cm of b1] (b2);
 \node at (b2) [rectangle,draw] (b3) {$B$};
   \diagram*{
(i1)--[thick,middlearrow={latex}] (i2),
(i2)--[thick,middlearrow={latex}] (top)
  };
  \end{feynman}
\end{tikzpicture}}}
~\qquad \text{and}\qquad~
 \vcenter{\hbox{\begin{tikzpicture}[font=\footnotesize,inner sep=2pt]
  \begin{feynman}
\vertex (i1) at (0,0) [label=above left:\(X\)] ;
\vertex[right=1.3cm of i1] (j1)[label=above right:\(Y\)];
\vertex[right=0.65cm of i1] (b1);
\node at (b1) [rectangle,draw] (b2) {$B$};
\vertex [dot,yellow] [above=1.5cm of b1] (b3) [label=right:\(\beta\)] {};
\vertex[left=0.2cm of b3] (b5);
\vertex[above=0.2cm of b5] (b6);
 \vertex[above=3cm of b1] (top);
 \vertex[above=0.2cm of b3] (b7);
 \vertex[above=0.6cm of b7] (b8) [label=right:\(Z\)];
 \vertex[left=0.7cm of top] (a1);
 \vertex[below=0.4cm of a1] (a2);
  \node at (a2) [rectangle,draw] (a3) {$A$};
  \vertex[right=0.9cm of top] (c1);
 \vertex[below=0.4cm of c1] (c2);
  \node at (c2) [rectangle,draw] (c3) {$C$};
   \diagram*{
(i1)--[thick,middlearrow={latex}] (b3),
(j1)--[thick,middlearrow={latex}] (b3),
(b3)--[thick,middlearrow={latex}] (top)
  };
  \end{feynman}
\end{tikzpicture}}}
\end{equation}
For  calculations  we introduce  bases in  morphism spaces:
\begin{equation}\label{base2}
    \Lambda^\alpha_{(x,y)z}\in\Hom_{A|C}(X_x\otimes_BX_y,X_z)\qquad \text{and}\qquad \bar{\Lambda}^\alpha_{(x,y)z}\in\Hom_{A|C}(X_z,X_x\otimes_BX_y)
\end{equation}
where $\alpha=1,\ldots,\operatorname{dim}\Hom_{A|C}(X_x\otimes_BX_y,X_z,)$ and $x\in\mathcal{K}_{AB},\;y\in\mathcal{K}_{BC},\;z\in\mathcal{K}_{AC}$. We fix their normalisation by
\begin{equation}\label{normLambda}
    \Lambda^\alpha_{(x,y)z}\circ \bar{\Lambda}^\beta_{(x,y)w}=\delta_{z,w}\delta_{\alpha,\beta}\operatorname{id}_{X_z}
\end{equation}
or in graphical notation
\begin{equation} \label{defnorm}
\vcenter{\hbox{\hspace{-10mm}\begin{tikzpicture}[font=\footnotesize,inner sep=2pt]
  \begin{feynman}
  \vertex  [small,yellow, dot] (del) at (0,0) [label=below:\(\beta\)] {} ;
  \vertex  [small,yellow, dot] (gam) at (0,-2) [label=above:\(\alpha\)] {} ;
 \vertex [below=1cm of gam] (j1) [label=below:\(X_z\)];
  \vertex [above=1cm of del] (j2) [label=above:\(X_w\)];
  \vertex (kek) at (0.8,-1) [label=right:\(X_y\)];
   \diagram*{
     (del)--[thick,middlearrow={latex}] (j2),
     (j1)--[thick,middlearrow={latex}] (gam),
     (gam) --[half right, thick, looseness=1.2,middlearrow={latex}] (del),
     (gam) --[half left, thick, looseness=1.2, edge label=$X_x$,middlearrow={latex}] (del)
  };
  \end{feynman}
\end{tikzpicture}}}
~\hspace{5mm}=\delta_{z,w}\delta_{\alpha,\beta} 
~ 
\vcenter{\hbox{\hspace{5mm}\begin{tikzpicture}[font=\footnotesize,inner sep=2pt]
  \begin{feynman}
  \vertex (1) at (0,1) [label=above:\(X_z\)];
  \vertex (2) at (0,-3) [label=below:\(X_z\)];
   \diagram*{
     (2)--[thick,middlearrow={latex}] (1)
  };
  \end{feynman}
\end{tikzpicture}}}\end{equation}
Let us note that the basis morphisms \eqref{base2} satisfy a completeness relation analogous to \eqref{completelambda}:
\begin{equation}\label{complete3def}
    \vcenter{\hbox{\hspace{-10mm}\begin{tikzpicture}[font=\footnotesize,inner sep=2pt]
  \begin{feynman}
  \vertex  (i1) at (0,0) [label=below:\(X_x\)]  ;
  \vertex [above=3cm of i1] (i2) [label=above:\(X_x\)];
  \vertex [right=1cm of i1] (j1) [label=below:\(X_y\)];
  \vertex [right=1cm of i2] (j2) [label=above:\(X_y\)];
  \diagram*{
  (i1)--[middlearrow={latex},thick] (i2),
  (j1)--[middlearrow={latex},thick] (j2)
    };
  \end{feynman}
\end{tikzpicture}}}
~\hspace{5mm}=\mathlarger{\sum}_{z\in\mathcal{K}_{AC}}\mathlarger{\sum}_\gamma\,
~ 
\vcenter{\hbox{\hspace{1mm}\begin{tikzpicture}[font=\footnotesize,inner sep=2pt]
  \begin{feynman}
  \vertex  [small,yellow,dot] (up) at (0,0) [label=above:\(\gamma\)] {};
  \vertex [small,yellow,dot] [below=1cm of up] (down) [label=below:\(\gamma\)] {};
  \vertex [above left=0.75cm of up] (i1);
  \vertex [above=0.6cm of i1] (i2) [label=above:\(X_x\)];
  \vertex [above right=0.75cm of up] (j1);
  \vertex [above=0.6cm of j1] (j2) [label=above:\(X_y\)];
  \draw [middlearrow={latex},thick] (up) -- (i1) --(i2);
  \draw [middlearrow={latex},thick] (up) -- (j1) --(j2);
  \vertex [below left=0.75cm of down] (i3) ;
  \vertex [below=0.6cm of i3] (i4) [label=below:\(X_x\)];
  \vertex [below right=0.75cm of down] (j3);
  \vertex [below=0.6cm of j3] (j4) [label=below:\(X_y\)];
  \draw [myptr={latex},thick] (j4) -- (j3) --(down);
   \draw [myptr={latex},thick] (i4) -- (i3) --(down);
   \diagram*{
      (down)--[middlearrow={latex},thick,edge label=$X_z$] (up)
  };
  \end{feynman}
\end{tikzpicture}}}
\end{equation}
due to the semisimplicity of the category $_A\mathcal{C}_C$ as established in \cite{FFRS}.

The hom-spaces introduced in \eqref{base2} can be described as subspaces of larger hom-spaces -- the ones defined using the tensor product in $\mathcal{C}$. Namely, the following two isomorphisms can be defined
\begin{align}\label{iso1}
\phi_{X,Y}^Z:{\Hom_{A|C}}_{(P_{X,Y})}(X\otimes Y,Z) &\stackrel{\cong}{\longrightarrow} \Hom_{A|C}(X\otimes_BY,Z) \quad \text{and}\nonumber\\
\bar{\phi}_{X,Y}^Z:\Hom_{A|C}^{\;\;\;\;\;(P_{X,Y})}(Z,X\otimes Y) &\stackrel{\cong}{\longrightarrow} \operatorname{Hom}_{A \mid C}\left(Z, X \otimes_B Y\right) 
\end{align}
where 
\begin{align}\label{spaces}
 {\Hom_{A|C}}_{(P_{X,Y})}(X\otimes Y,Z)    &:=\{f \in  \Hom_{A|C}(X\otimes Y,Z) \mid f \circ P_{X,Y}=f\} \, ,  \nonumber\\ \Hom_{A|C}^{\;\;\;\;\;(P_{X,Y})}(Z,X\otimes Y) &:=\{g \in \Hom_{A|C}(Z,X\otimes Y) \mid P_{X,Y} \circ g=g\}
    \end{align}
 and the isomorphisms in (\ref{iso1}) act as 
\begin{equation} \label{phi_action}
\phi_{X,Y}^Z (f) = f\circ e_{X,Y} \, , \qquad \bar{\phi}_{X,Y}^Z(g)=r_{X,Y}\circ g \, .
\end{equation}

The images of  the bases \eqref{base2} under the isomorphisms \eqref{iso1} are
\begin{equation} \label{base3}
    \Tilde{\Lambda}^\alpha_{(x,y)z}\in{\Hom_{A|C}}_{(P_{X_x,X_y})}(X_x\otimes X_y,X_z) \quad \text{and} \quad \bar{\Tilde{\Lambda}}^\alpha_{(x,y)z}\in\Hom_{A|C}^{\;\;\;\;\;(P_{X_x,X_y})}(X_z,X_x\otimes X_y)
\end{equation}
defined by 
\begin{equation}
    \Tilde{\Lambda}^\alpha_{(x,y)z}=  \Lambda^\alpha_{(x,y)z}\circ r_{X_x,X_y} \qquad \text{and} \qquad \bar{ \Tilde{\Lambda}^\alpha_{(x,y)z}} =e_{X_x,X_y}\circ \bar{\Lambda}^\alpha_{(x,y)z}\,.
\end{equation}
The inverse formula reads 
\begin{equation}\label{baserel1}
    \Lambda^\alpha_{(x,y)z}=\Tilde{\Lambda}^\alpha_{(x,y)z}\circ e_{X_x,X_y} \qquad \text{and} \qquad \bar{\Lambda}^\alpha_{(x,y)z}=r_{X_x,X_y}\circ \bar{\Tilde{\Lambda}}^\alpha_{(x,y)z} \, .
\end{equation}
On the diagrams we will depict the $ \Tilde{\Lambda}$ morphisms and their duals as hollow circles.

Next we consider junctions of topological interfaces with conformal boundaries   where an $A$-$B$-interface $X$ can start or end on the boundary \cite{FFRS}, which we call \textit{open defects}. These will correspond respectively to morphisms $\alpha\in\Hom_A(M,X\otimes_BN)$ and $\beta\in\Hom_A(X\otimes_BN,M)$ where $M$ is an $A$-module while $N$ is a $B$-module. Graphically
 \begin{equation}\label{jun2}
  \vcenter{\hbox{\hspace{-10mm}\begin{tikzpicture}[font=\footnotesize,inner sep=2pt]
  \begin{feynman}
  \vertex  (i1) at (0,0)  ;
  \vertex (x1) at (-0.25,0.9);
  \vertex  (x2) at (0.75, 0.65);
  \vertex[below=0.9cm of x1] (xbot);
  \vertex [above=3cm of i1] (i2);
  \vertex [right=1cm of i1] (j1);
  \vertex[ above=0.4cm of j1] (j3) [label=left:\(N\)];
  \vertex [right=1cm of i2] (j2);
  \vertex[ below=0.4cm of j2] (j4) [label=left:\(M\)];
  \vertex [below=1.1cm of j2] (j6) ;
  \vertex[left=0.15cm of j6] [label=left:\(\alpha\)];
  \vertex   [above=1.8cm of j1] (j5) {};
  \vertex [left=0.1cm of j5] (j7);
  \vertex (x) at (-0.5,0.6) [label=above:\(X\)];
\draw [thick]    (x1) to[out=90,in=-90] (j7);
   \path[pattern=north east lines,pattern color=ashgrey,very thin] (1.01,3) rectangle (1.6,0);
   \draw [fill=yellow] (1.06,1.8) rectangle (0.8, 1.9);
    \draw [thick] (-0.1,1.4) rectangle (-0.5, 1.8);
    \vertex at (-0.3,1.4) [label=\(A\)];
      \draw [thick] (0.1,0.4) rectangle (0.5, 0);
    \vertex at (0.3,0) [label=\(B\)];
  \diagram*{
  (j1)--[ thick] (j5),
  (j2)--[ thick] (j6),
      (xbot)--[thick,middlearrow={latex}] (x1)
    };
  \end{feynman}
\end{tikzpicture}}}
~\hspace{14mm} \text{and} \hspace{14mm}~
 \vcenter{\hbox{\begin{tikzpicture}[font=\footnotesize,inner sep=2pt]
  \begin{feynman}
  \vertex  (i1) at (0,0)  ;
  \vertex (x1) at (-0.25,2.2);
  \vertex  (x2) at (0.75, 0.65);
  \vertex [above=3cm of i1] (i2);
  \vertex [right=1cm of i1] (j1);
  \vertex[ above=0.4cm of j1] (j3) [label=left:\(M\)];
  \vertex [right=1cm of i2] (j2);
  \vertex[ below=0.4cm of j2] (j4) [label=left:\(N\)];
  \vertex [below=1.7cm of j2] (j6);
  \vertex   [above=1.2cm of j1] (j5);
  \vertex [left=0.15cm of j5] [label=left:\(\beta\)];
  \vertex [left=0.1cm of j6] (j7);
  \vertex (x) at (-0.5,2.1) [label=above:\(X\)];
\draw [thick]    (j7) to[out=90,in=-90] (x1);
\vertex[above=0.8cm of x1] (xtop);
   \path[pattern=north east lines,pattern color=ashgrey,very thin] (1.01,3) rectangle (1.6,0);
   \draw [fill=yellow] (1.06,1.2) rectangle (0.8, 1.3);
   \draw [thick] (0.1,2.6) rectangle (0.5, 3);
    \vertex at (0.3,2.6) [label=\(B\)];
    \draw [thick] (-0.1,1.2) rectangle (-0.5, 1.6);
    \vertex at (-0.3,1.2) [label=\(A\)];
  \diagram*{
  (j1)--[ thick] (j5),
  (j2)--[ thick] (j6),
  (x1)--[thick,middlearrow={latex}] (xtop)
    };
  \end{feynman}
\end{tikzpicture}}}
  \end{equation}
 We choose bases for these junction spaces 
 \begin{equation}\label{base4}
    \Omega^\beta_{(x,a)b}\in\Hom_A(X_x\otimes_BM_a,M_b)\qquad \text{and}\qquad \bar{\Omega}^\alpha_{(x,a)b}\in\Hom_A(M_b,X_x\otimes_BM_a)
\end{equation}
where $X_x$, $M_{a}$, $M_{b}$ are a simple bimodule and two simple modules respectively. 
The basis elements satisfy the normalisation condition
\begin{equation} \label{junctions_norm}
     \vcenter{\hbox{\hspace{-10mm}\begin{tikzpicture}[font=\footnotesize,inner sep=2pt]
  \begin{feynman}
  \vertex (j1) at (0,0);
  \vertex [above=0.5 of j1] (j2);
  \vertex[ above =0.6 of j1] (j3);
  \vertex  [ above=2cm of j1] (psi) {};
  \vertex [above=3.4cm of j1] (j4) ;
  \vertex[above=3.5cm of j1] (j5)  ;
  \vertex[ left=0.15cm of  j4] (j7) [label=left:\(\alpha\)];
  \vertex [left=0.15cm of j3] (j8) [label=left:\(\beta\)]; 
  \vertex [above =4cm of j1] (j6);
  \path[pattern=north east lines,pattern color=ashgrey,very thin] (0.01,4) rectangle (0.6,0);
  \draw [fill=yellow] (0.06,0.5) rectangle (-0.2, 0.6);
    \draw [fill=yellow] (0.06,3.4) rectangle (-0.2, 3.5);
    \vertex[below=0.1cm of j6] (m1) [label=left:\(M_a\)];
    \vertex[ above=0.1cm of j1] (m4) [label=left:\(M_c\)];
    \vertex[above=1.3cm of j1] (m3) ;
    \vertex[above=0.01cm of psi] (m2) [label=right:\(M_b\)];
    \vertex[left=1.2cm of psi] (x) [label=left:\(X_x\)];
    \vertex [below=0.25cm of x] (x1);
    \vertex [above=0.25cm of x] (x2);
    \draw [thick]    (j8) to[out=90,in=-90] (x1);
    \draw [thick]    (x2) to[out=90,in=-90] (j7);
    \draw [thick] [middlearrow={latex}]   (x1) to (x2);
     \draw [thick] (-1.3,2.8) rectangle (-1.7, 3.2);
    \vertex at (-1.5,2.8) [label=\(A\)];
    \draw [thick] (-0.3,1.8) rectangle (-0.7, 2.2);
    \vertex at (-0.5,1.8) [label=\(B\)];
   \diagram*{
    (j1) --[thick] (j2),
    (j3) --[thick] (j4),
    (j5)--[thick] (j6)
  };
  \end{feynman}
\end{tikzpicture}}}
~= \delta_{a,c}\delta_{\alpha,\beta} ~
\vcenter{\hbox{\hspace{2mm}\begin{tikzpicture}[font=\footnotesize,inner sep=2pt]
  \begin{feynman}
  \vertex (1) at (0,0) [label=above left:\(M_a\)];
  \vertex (2) at (0,4);
  \path[pattern=north east lines,pattern color=ashgrey,very thin] (0.01,4) rectangle (0.6,0);
  \draw [thick] (-0.4,1.8) rectangle (-0.8, 2.2);
    \vertex at (-0.6,1.8) [label=\(A\)];
   \diagram*{
     (1)--[thick] (2)
  };
  \end{feynman}
\end{tikzpicture}}}
\end{equation}
The basis morphisms $\Omega$ also satisfy the following completeness relation due to the semisimplicity of the category $\mathcal{C}_A$ \cite{FS}:
\begin{equation}
    \vcenter{\hbox{\begin{tikzpicture}[font=\footnotesize,inner sep=2pt]
  \begin{feynman}
\vertex (i1) at (0,0) ;
\vertex [right=1cm of i1] (j1);
\vertex [above=0.5cm of i1] (i2) [label=left:\(X_x\)];
\vertex [above=0.5cm of j1] (j2) [label=left:\(M_a\)];
\vertex [above=4cm of i1] (i3);
\vertex [right =1cm of i3] (j3) ;
\vertex  [left=0.5cm of i3] (a1)  ;
\vertex [below= 0.4cm of a1] (a2);
 \vertex [right=1cm of a2] (b1);
 \vertex[right=0.5cm of j3] (top);
 \path[pattern=north east lines,pattern color=ashgrey,very thin] (top) rectangle (j1);
   \diagram*{
(i1)--[thick,middlearrow={latex}] (i3),
(j1)--[thick] (j3)
  };
  \end{feynman}
\end{tikzpicture}}} = \mathlarger{\sum}_{b\in\mathcal{J}_A}\mathlarger{\sum}_\gamma   \vcenter{\hbox{\hspace{5mm}\begin{tikzpicture}[font=\footnotesize,inner sep=2pt]
  \begin{feynman}
   \path[pattern=north east lines,pattern color=ashgrey,very thin] (1.01,4) rectangle (1.6,0);
  \vertex  (i1) at (0,0)  ;
  \vertex (x1) at (0,3.5);
  \vertex  (x2) at (0, 0.6);
  \vertex [above=4cm of i1] (i2);
  \vertex [right=1cm of i1] (j1);
  \vertex[ above=0.1cm of j1] (j3) [label=left:\(M_a\)];
  \vertex [right=1cm of i2] (j2);
  \vertex[ below=0.1cm of j2] (j4) [label=left:\(M_a\)];
  \vertex [below=1.4cm of j2] (j6) ;
  \vertex[left=0.2cm of j6] (jjj) ;
  \vertex[below=0.1cm of jjj] (jjjj) [label=left:\(\gamma\)];
  \vertex   [above=1.5cm of j1] (j5);
  \vertex [left=0.2cm of j5] [label=left:\(\gamma\)];
  \vertex [left=0.1cm of j6] (j7);
  \vertex[left=0.1cm of j5] (j0);
  \vertex (x) at (-0.25,2.9) [label=above:\(X_x\)];
  \vertex (yy) at (-0.25,0.7) [label=above:\(X_x\)];
  \vertex[below=0.1cm of j6] (g1);
  \vertex[above=0.1cm of j5] (g2);
  \vertex[above=0.5cm of g2] (g3) [label=right:\(M_b\)];
\draw [thick]    (j7) to[out=90,in=-90] (x1);
\draw [thick]    (x2) to[out=90,in=-90] (j0);
   \draw [fill=yellow] (1.06,2.5) rectangle (0.8, 2.6);
    \draw [fill=yellow] (1.06,1.5) rectangle (0.8, 1.6);
  \diagram*{
  (j1)--[ thick] (j5),
  (j2)--[ thick] (j6),
  (g1)--[thick] (g2),
  (x1)--[thick,middlearrow={latex}] (i2),
   (i1)--[thick,myptr={latex}] (x2)
    };
  \end{feynman}
\end{tikzpicture}}}
\end{equation}

Similarly to the morphisms $\Tilde{\Lambda}$ introduced above we introduce  morphisms $\tilde \Omega$ using the isomorphism  \eqref{iso1}.
In more detail, consider an isomorphism
\begin{equation}\label{iso5}
\operatorname{Hom}_{A|\mathbf{1}}(X_x\otimes_BM_b,M_c)\cong  {\Hom_{A|\mathbf{1}}}_{(P_{X_x,M_b})}(X_x\otimes M_b,M_c)
\end{equation}
given by
\be \label{fff}
\phi_{x,b}^c:\;{\Hom_{A|\mathbf{1}}}_{(P_{X_x,M_b})}(X_x\otimes M_b,M_c) \to \operatorname{Hom}_{A|\mathbf{1}}(X_x\otimes_BM_b,M_c)
\ee
as defined in (\ref{phi_action}).
Here we have used $\mathbf{1}$ which is the tensor unit of $\mathcal{C}$  thought of as the trivial algebra. 
This isomorphism introduces a basis  in the hom-space on the left of (\ref{fff}): 
\begin{equation}\label{omegatilde}
    \tilde{\Omega}^\alpha_{(x,b)c} = \Omega^\alpha_{(x,b)c} \circ r_{x,b}
\end{equation}
and similarly the dual basis $ \bar{\Tilde{\Omega}}^\alpha_{(x,b)c} $. On the diagrams we will depict the morphisms $ \tilde{\Omega}$ and their 
duals as hollow rectangles.

The final type of junction that we consider in this paper is a boundary field. As mentioned earlier, these morphisms in general are elements of $\operatorname{Hom}_A(M\otimes U,N)$, where $M,N$ are left $A$-modules and $U\in\objc$. We introduce the bases 
\begin{equation}\label{bcco}
    \Psi^\alpha_{(a,i)b}\in\operatorname{Hom}_A(M_a\otimes U_i,M_b) \qquad \text{and}\qquad \bar{\Psi}^\alpha _{(a,i)b}\in \operatorname{Hom}_A(M_b,M_a\otimes U_i)
\end{equation} 
where  all the objects involved are simple. 
We represent these bases of  boundary  fields  by black blobs on the boundary, as on the following picture 
\begin{equation}\label{bccopic}
   \Psi^\alpha_{(a,i)b}= \vcenter{\hbox{\hspace{5mm}\begin{tikzpicture}[font=\footnotesize,inner sep=2pt]
  \begin{feynman}
  \path[pattern=north east lines,pattern color=ashgrey,very thin] (0.01,4) rectangle (0.6,0);
  \vertex (a) at (0,0);
  \vertex [above=4cm of a] (b);
  \vertex [dot] [above=2cm of a] (psi) {};
  \vertex[right=0.2cm of psi] (i) ;
  \vertex[below=0.2cm of i] (ii) [label=:\(\;\;\,\Psi_i^\alpha\)];
   \vertex [above=1.4cm of psi] (b1) [label=left:\(M_a\)];
   \vertex [below=1.4cm of psi] (b2) [label=left:\(M_b\)];
   \draw [thick] (-1.1,2.2) rectangle (-1.5, 1.8);
    \vertex at (-1.3,1.8) [label=\(A\)];
   \diagram*{
     (a)--[thick] (psi),
     (psi)--[thick] (b)
    };
  \end{feynman}
\end{tikzpicture}}}
\end{equation}
We further  choose the normalisation condition 
\begin{equation}
    \Psi^\alpha_{(a,i)b} \circ \bar{\Psi}^\beta _{(a,i)c}=\delta_{b,c}\delta_{\alpha,\beta}\operatorname{id}_{M_b}
\end{equation}
that  graphically looks like
\begin{equation}
   \vcenter{\hbox{\hspace{5mm}\begin{tikzpicture}[font=\footnotesize,inner sep=2pt]
  \begin{feynman}
   \path[pattern=north east lines,pattern color=ashgrey,very thin] (0.01,4) rectangle (0.6,0);
  \vertex (a) at (0,0);
  \vertex [above=4cm of a] (b);
  \vertex [dot] [above=1.4cm of a] (psi) {};
  \vertex[above=0.6cm of psi] (ok) [label=left:\(M_a\)];
  \vertex [dot] [above=2.6cm of a] (psi1){};
  \vertex[right=0.2cm of psi] (i) ;
  \vertex[below=0.2cm of i] (ii) [label=:\(\;\;\,\Psi_i^\alpha\)];
   \vertex[right=0.2cm of psi1] (d1) ;
  \vertex[below=0.2cm of d1] (d2) [label=:\(\;\;\,\bar{\Psi}_i^{\beta}\)];
   \vertex [above=3.8cm of a] (b1) [label=left:\(M_c\)];
   \vertex [above=0.2cm of a] (b2) [label=left:\(M_b\)];
   \draw [thick] (-1.1,2.2) rectangle (-1.5, 1.8);
    \vertex at (-1.3,1.8) [label=\(A\)];
   \diagram*{
     (a)--[thick] (psi),
     (psi)--[thick] (b)
    };
  \end{feynman}
\end{tikzpicture}}}
= \delta_{b,c}\delta_{\alpha,\beta}\vcenter{\hbox{\hspace{2mm}\begin{tikzpicture}[font=\footnotesize,inner sep=2pt]
  \begin{feynman}
  \vertex (1) at (0,0) [label=above left:\(M_b\)];
  \vertex (2) at (0,4);
  \path[pattern=north east lines,pattern color=ashgrey,very thin] (0.01,4) rectangle (0.6,0);
  \draw [thick] (-0.4,1.8) rectangle (-0.8, 2.2);
    \vertex at (-0.6,1.8) [label=\(A\)];
   \diagram*{
     (1)--[thick] (2)
  };
  \end{feynman}
\end{tikzpicture}}}
\end{equation}
The basis morphisms $\Psi$ satisfy the following completeness relation due to the semisimplicity of $\mathcal{C}_A$:
\begin{equation}
    \id_{M_a\otimes U_i} =\sum_{b\in\mathcal{J}_A}\sum_\gamma \bar{\Psi}^\gamma_{(a,i)b}\circ \Psi^\gamma_{(a,i)b}
\end{equation}

\subsection{\texorpdfstring{Fusing matrices}{Fusing matrices}} \label{sec2.5}
In this section we will introduce various fusing matrices related to the tensor structures  introduced in the previous sections.

\subsubsection{Fusing matrix of \texorpdfstring{$\mathcal{C}$}{C} }

Recall that we assume $\mathcal{C}$ to be a strict category with trivial associator morphisms. 
The fusing matrices of  $\mathcal{C}$ arise from two different representations of $\operatorname{Hom}(U_i\otimes U_j\otimes U_k,U_l)$:
as 
$$\bigoplus_q \operatorname{Hom}(U_i\otimes U_j,U_q)\otimes \operatorname{Hom}(U_q\otimes U_k,U_l)$$ and as 
$$\bigoplus_p \operatorname{Hom}(U_j\otimes U_k,U_p)\otimes \operatorname{Hom}(U_i\otimes U_p,U_l)\, .$$
 Each representation plus the choice of basis (\ref{bases1}) gives $\operatorname{Hom}(U_i\otimes U_j\otimes U_k,U_l)$ 
a distinguished basis. 
The matrices that perform the transformation between the two bases are the fusion matrices of $\mathcal{C}$. Their entries are defined by
\begin{equation} \label{fmove1}
\vcenter{\hbox{\hspace{-10mm}\begin{tikzpicture}[font=\footnotesize,inner sep=2pt]
  \begin{feynman}
  \vertex (l) at (0,0) [label=above:\(U_l\)] ;
  \vertex [small,orange, dot] (alpha) at (0,-1) [label=left:\footnotesize\(\alpha\)] {};
 \vertex  [small,orange, dot] (beta) at (1,-2) [label=right:\footnotesize\(\beta\)] {};
 \vertex  (i) at (-1,-3) [label=below:\(U_i\)];
 \vertex  (left) at (-0.7,-2);
 \vertex [above=0.5cm of i] (i2);
 \vertex [below left=0.042cm of left] (i3);
 \vertex (j) at (0.4,-3) [label=below:\(U_j\)];
  \vertex (k) at (1.6,-3) [label=below:\(U_k\)];
  \vertex  (j2) at (0.45,-2.6);
  \vertex [above right=0.15cm of j2] (j3);
  \vertex [above right=0.15cm of j3] (j4);
  \vertex  (k2) at (1.55,-2.6);
  \vertex [above left=0.3cm of k2] (k3);
  \draw [thick,rounded corners=1mm] (alpha) -- (left) -- (i2) -- (i);
   \draw [thick,rounded corners=1mm] (beta)  -- (j2) -- (j);
   \draw [thick,rounded corners=1mm] (beta)  -- (k2) -- (k);
    \diagram*{
      (l) --[thick]  (alpha)   ,
      (alpha) -- [thick,edge label=$U_p$] (beta)
    };
  \end{feynman}
\end{tikzpicture}}}
~\hspace{5mm}=\mathlarger{\sum}_{q\in\mathcal{I}}\mathlarger{\sum}_{\gamma,\delta}\,\mathrm{F}_{\alpha p \beta,\gamma q\delta}^{(i\,j\,k)l}
~ 
\vcenter{\hbox{\hspace{1mm}\begin{tikzpicture}[font=\footnotesize,inner sep=2pt]
  \begin{feynman} 
  \vertex (l) at (0,0) [label=above:\(U_l\)] ;
  \vertex [small,orange, dot] (del) at (0,-1) [label=left:\footnotesize\(\delta\)] {};
 \vertex  [small, orange, dot] (gamma) at (-1,-2) [label=right:\footnotesize\(\gamma\)] {};
 \vertex  (i) at (-1.6,-3) [label=below:\(U_i\)];
 \vertex  (i2) at (-1.55,-2.6);
 \vertex [below left=0.03cm of i2] (i3);
 \vertex (j) at (-0.4,-3) [label=below:\(U_j\)];
  \vertex (k) at (1,-3) [label=below:\(U_k\)];
  \vertex (j2) at (-0.45,-2.6) ;
  \vertex [below right=0.03cm of j2] (j3) ;
  \vertex [above=0.5cm of k] (k2);
  \vertex [above left=0.005cm of k2] (k3);
  \vertex (right) at (0.7,-2);
   \draw [thick,rounded corners=1mm] (del) -- (right) -- (k2) -- (k);
    \draw [thick,rounded corners=1mm] (gamma) -- (i2)  -- (i);
     \draw [thick,rounded corners=1mm] (gamma) -- (j2) -- (j);
    \diagram*{
      (l) --[thick]  (del)   ,
      (del) -- [thick, edge label=$U_q$] (gamma)  
  };
  \end{feynman}
\end{tikzpicture}}}\end{equation}
The same matrix appears in  an analogous relation involving dual morphisms
\begin{equation}
   \vcenter{\hbox{\hspace{2mm}\begin{tikzpicture}[font=\footnotesize,inner sep=2pt]
   \begin{feynman}
  \vertex (l) at (0,0) [label=below:\(U_l\)] ;
  \vertex [small,orange,dot] [above=1cm of l] (gam) [label=above:\footnotesize\(\delta\)] {};
  \vertex [small,orange,dot] [above left=1.25cm of gam] (del) [label=above:\footnotesize\(\gamma\)] {} ;
  \vertex [above right=1cm of gam] (k1);
  \vertex [above=1.4cm of k1] (k) [label=above:\(U_k\)];
  \vertex [above left=0.7cm of del] (i1);
  \vertex [above=0.7cm of i1] (i) [label=above: \(U_i\)];
   \vertex [above right=0.7cm of del] (j1);
  \vertex [above=0.7cm of j1] (j) [label=above: \(U_j\)];
   \draw [thick,rounded corners=1mm] (gam) -- (k1) --(k);
     \draw [thick,rounded corners=1mm] (del) -- (j1) --(j);
       \draw [thick,rounded corners=1mm] (del) -- (i1) --(i);
    \diagram*{
     (l)--[thick] (gam),
     (gam) --[thick,edge label= $U_q$] (del)
    };
  \end{feynman}
\end{tikzpicture}}}
~=\;\mathlarger{\sum}_{p\in\mathcal{I}}\mathlarger{\sum}_{\alpha,\beta}\mathrm{F}^{(i\,j\,k)l}_{\alpha p\beta,\gamma q\delta}~
\vcenter{\hbox{\begin{tikzpicture}[font=\footnotesize,inner sep=2pt]
  \begin{feynman}
  \vertex (l) at (0,0) [label=below:\(U_l\)] ;
  \vertex [small,orange,dot] [above=1cm of l] (beta) [label=above:\footnotesize\(\alpha\)] {};
  \vertex [above left=1cm of beta] (i1);
  \vertex [above=1.4cm of i1] (i) [label=above:\(U_i\)];
  \vertex [small,orange,dot] [above right=1.25cm of beta] (alpha) [label=right:\footnotesize\(\,\beta\)] {};
  \vertex[above left=0.7cm of alpha] (j1);
  \vertex [above=0.7cm of j1] (j) [label=above:\(U_j\)];
  \vertex [above right=0.7cm of alpha] (k1);
    \vertex [above=0.7cm of k1] (k) [label=above:\(U_k\)];
    \draw [thick,rounded corners=1mm] (alpha) -- (j1) --(j);
    \draw [thick,rounded corners=1mm] (alpha) -- (k1) --(k);
    \draw [thick,rounded corners=1mm] (beta) -- (i1) --(i);
    \diagram*{
      (alpha) --[thick,edge label= $U_p$] (beta),
      (l)--[thick] (beta)
    };
  \end{feynman}
\end{tikzpicture}}}
\end{equation}
The inverse of $\mathrm{F}$  is defined as
\begin{equation}
    \vcenter{\hbox{\hspace{1mm}\begin{tikzpicture}[font=\footnotesize,inner sep=2pt]
  \begin{feynman} 
  \vertex (l) at (0,0) [label=above:\(U_l\)] ;
  \vertex [small,orange, dot] (del) at (0,-1) [label=left:\footnotesize\(\beta\)] {};
 \vertex  [small, orange, dot] (gamma) at (-1,-2) [label=right:\footnotesize\(\alpha\)] {};
 \vertex  (i) at (-1.6,-3) [label=below:\(U_i\)];
 \vertex  (i2) at (-1.55,-2.6);
 \vertex [below left=0.03cm of i2] (i3);
 \vertex (j) at (-0.4,-3) [label=below:\(U_j\)];
  \vertex (k) at (1,-3) [label=below:\(U_k\)];
  \vertex (j2) at (-0.45,-2.6) ;
  \vertex [below right=0.03cm of j2] (j3) ;
  \vertex [above=0.5cm of k] (k2);
  \vertex [above left=0.005cm of k2] (k3);
  \vertex (right) at (0.7,-2);
   \draw [thick,rounded corners=1mm] (del) -- (right) -- (k2) -- (k);
    \draw [thick,rounded corners=1mm] (gamma) -- (i2)  -- (i);
     \draw [thick,rounded corners=1mm] (gamma) -- (j2) -- (j);
    \diagram*{
      (l) --[thick]  (del)   ,
      (del) -- [thick, edge label=$U_p$] (gamma)  
  };
  \end{feynman}
\end{tikzpicture}}}
~=\; \mathlarger{\sum}_{q,\gamma,\delta} \mathrm{G}^{(i\,j\,k)l}_{\alpha p\beta,\gamma q\delta}~
\vcenter{\hbox{\begin{tikzpicture}[font=\footnotesize,inner sep=2pt]
  \begin{feynman}
  \vertex (l) at (0,0) [label=above:\(U_l\)] ;
  \vertex [small,orange, dot] (alpha) at (0,-1) [label=left:\footnotesize\(\gamma\)] {};
 \vertex  [small,orange, dot] (beta) at (1,-2) [label=right:\footnotesize\(\delta\)] {};
 \vertex  (i) at (-1,-3) [label=below:\(U_i\)];
 \vertex  (left) at (-0.7,-2);
 \vertex [above=0.5cm of i] (i2);
 \vertex [below left=0.042cm of left] (i3);
 \vertex (j) at (0.4,-3) [label=below:\(U_j\)];
  \vertex (k) at (1.6,-3) [label=below:\(U_k\)];
  \vertex  (j2) at (0.45,-2.6);
  \vertex [above right=0.15cm of j2] (j3);
  \vertex [above right=0.15cm of j3] (j4);
  \vertex  (k2) at (1.55,-2.6);
  \vertex [above left=0.3cm of k2] (k3);
  \draw [thick,rounded corners=1mm] (alpha) -- (left) -- (i2) -- (i);
   \draw [thick,rounded corners=1mm] (beta)  -- (j2) -- (j);
   \draw [thick,rounded corners=1mm] (beta)  -- (k2) -- (k);
    \diagram*{
      (l) --[thick]  (alpha)   ,
      (alpha) -- [thick,edge label=$U_q$] (beta)
    };
  \end{feynman}
\end{tikzpicture}}}
\end{equation}

\subsubsection{Fusing matrix \texorpdfstring{$\mathrm{Y}$}{Y} for bimodules} \label{sec252}
Next we   define  fusing matrices for the case when all the objects involved are bimodules. More specifically, consider equation \eqref{fmove1} with all simple objects $U_i$ of $\mathcal{C}$ replaced with simple objects $X_x$ of  bimodule categories. Taking into account the construction of the fusion of bimodules, the most general definition  includes three initial simple bimodules: $X_x\in\operatorname{Obj}(\prescript{}{A}{\mathcal{C}}_B),\, X_y\in\operatorname{Obj}(\prescript{}{B}{\mathcal{C}}_C),\,X_z\in\operatorname{Obj}( \prescript{}{C}{\mathcal{C}}_D)$ while the  simple bimodule at the top of the diagram must be $X_w\in \operatorname{Obj}(\prescript{}{A}{\mathcal{C}}_D)$. The final ingredient needed for the resulting equation to make sense  is the associator 
\begin{equation}
    \check{\alpha}_{x,y,z}:X_x\otimes_B(X_y\otimes_CX_z)\to(X_x\otimes_BX_y)\otimes_CX_z
\end{equation}
already introduced in \eqref{associator2}. In the general case of distinct algebras $A$, $B$, $C$, $D$ it is defined as
\begin{equation}\label{associator22}
     \check{\alpha}_{x,y,z}=r_{X_x\otimes_BX_y,X_z}\circ\left(r_{X_x,X_y}\otimes \operatorname{id}_{X_z}\right)\circ\left(\operatorname{id}_{X_x}\otimes e_{X_y,X_z}\right)\circ e_{X_x,X_y\otimes_CX_z}\,.
\end{equation}
We can now obtain a matrix $\mathrm{Y}$ that furnishes the vector space isomorphism  
\begin{equation}\label{Ymatrixiso}
\Hom_{A|D}\left(X_x\otimes_B(X_y\otimes_C X_z),X_w\right)
\cong \Hom_{A|D}\left((X_x\otimes_BX_y)\otimes_C X_z,X_w\right)
\end{equation}
induced by the associator \eqref{associator22}. The entries of this matrix are defined by the equation
\begin{equation} \label{cross1}
\vcenter{\hbox{\hspace{-10mm}\begin{tikzpicture}[font=\footnotesize,inner sep=2pt]
  \begin{feynman}
  \vertex (l) at (0,0) [label=above:\(X_w\)] ;
  \vertex [yellow, dot] (alpha) at (0,-1) [label=left:\(\alpha\)] {};
 \vertex  [yellow, dot] (beta) at (1,-2) [label=right:\(\beta\)] {};
 \vertex  (i) at (-1,-4.5) [label=below:\(X_x\)];
 \vertex  (left) at (-0.7,-3);
 \vertex [above=1cm of i] (i2);
 \vertex [below left=0.042cm of left] (i3);
 \vertex (j) at (0.4,-4.5) [label=below:\(X_y\)];
  \vertex (k) at (1.6,-4.5) [label=below:\(X_z\)];
  \vertex [above=0.8cm of j] (j11);
  \vertex[above=0.8cm of k] (k11);
  \vertex  (j2) at (0.45,-3.6);
  \vertex [above right=0.15cm of j2] (j3);
  \vertex [above right=0.15cm of j3] (j4);
  \vertex  (k2) at (1.55,-3.6);
  \vertex [above left=0.3cm of k2] (k3);
  \draw [middlearrow={latex},thick,rounded corners=1mm] (i) -- (i2) -- (alpha);
   \draw [middlearrow={latex},thick,rounded corners=1mm] (j)  -- (j11) -- (beta);
   \draw [middlearrow={latex},thick,rounded corners=1mm] (k)  -- (k11) -- (beta);
    \diagram*{
      (alpha) --[middlearrow={latex},thick]  (l)   ,
      (beta) -- [middlearrow={latex},thick,edge label=$X_u$] (alpha)
    };
  \end{feynman}
\end{tikzpicture}}}
~\hspace{5mm}=\mathlarger{\sum}_{v\in\mathcal{K}_{AC}}\mathlarger{\sum}_{\gamma,\delta}\,\mathrm{Y}_{\alpha u\beta,\gamma v\delta}^{(x\,y\,z)w}
~ 
\vcenter{\hbox{\hspace{1mm}\begin{tikzpicture}[font=\footnotesize,inner sep=2pt]
  \begin{feynman}
  \vertex (l) at (0,0) [label=above:\(X_w\)] ;
  \vertex [yellow, dot] (del) at (0,-1) [label=left:\(\delta\)] {};
 \vertex  [yellow, dot] (gamma) at (-1,-2) [label=right:\(\gamma\)] {};
 \vertex  (i) at (-1.6,-3.5);
 \vertex  (i2) at (-1.55,-2.6);
 \vertex [below left=0.03cm of i2] (i3);
 \vertex (j) at (-0.4,-3.5) ;
  \vertex (k) at (1,-3.5) ;
  \vertex (j2) at (-0.45,-2.6) ;
  \vertex [below right=0.03cm of j2] (j3) ;
  \vertex [above=0.5cm of k] (k2);
  \vertex [above left=0.005cm of k2] (k3);
  \vertex (right) at (0.7,-2);
  \vertex[above=0.7cm of i] (i11);
  \vertex[above=0.7cm of j] (j11);
  \vertex[above=1cm of k] (k11);
  \vertex[below=0.4cm of i] (a1);
  \vertex[left=0.25cm of a1] (a2);
  \vertex[right=0.25cm of k] (c1);
  \draw [fill=yellow] (a2) rectangle (c1);
   \vertex[below=0.4cm of i] (a1) ;
   \vertex[below=0.46cm of i] (iii) [label=above right:\(\quad\quad\check{\alpha}_{x,y,z}\)];
   \vertex[below=0.6cm of a1] (a3) [label=below:\(X_x\)];
   \vertex[below=1cm of j] (b1) [label=below:\(X_y\)];
   \vertex[below=1cm of k] (c2) [label=below:\(X_z\)];
   \vertex[below=0.4cm of j] (b2);
   \vertex[below=0.4cm of k] (c3);
    \draw [middlearrow={latex},thick,rounded corners=1mm] (a3) -- (a1) ;
    \draw [middlearrow={latex},thick,rounded corners=1mm] (b1) -- (b2)  ;
     \draw [middlearrow={latex},thick,rounded corners=1mm] (c2) -- (c3) ;
   \draw [middlearrow={latex},thick,rounded corners=1mm] (k) -- (k11)  -- (del);
    \draw [middlearrow={latex},thick,rounded corners=1mm] (i) -- (i11)  -- (gamma);
     \draw [middlearrow={latex},thick,rounded corners=1mm] (j) -- (j11) -- (gamma);
    \diagram*{
      (del) --[middlearrow={latex},thick]  (l)   ,
      (gamma) -- [middlearrow={latex},thick, edge label=$X_v$] (del)  
  };
  \end{feynman}
\end{tikzpicture}}}
\end{equation}
or in algebraic  form 
\begin{equation}
\Lambda^\alpha_{(x,u)w}\circ\left(\operatorname{id}_{X_x}\otimes_B \Lambda^\beta_{(y,z)u}\right)=\sum_{v\in\mathcal{K}_{AC}}\sum_{\gamma,\delta}\,\mathrm{Y}_{\alpha u \beta,\gamma v\delta}^{(x\,y\,z)w} \,\Lambda^\delta_{(v,z)w}\circ\left(\Lambda^\gamma_{(x,y)v}\otimes_C\operatorname{id}_{X_z}\right)\circ \check{\alpha}_{x,y,z}\,.
\end{equation}
Notice that \eqref{cross1} uses the convention of the diagrams flowing from bottom to top. Furthermore, to simplify notation in some places we will just write the labels $x,y,z$ instead of the corresponding bimodules $X_x,X_y,X_z$. To obtain a simpler expression for the matrix $\mathrm{Y}$ we compose \eqref{cross1} with the epimorphism $r_{x,y\otimes_Cz}\circ\left(\id_x\otimes r_{y,z}\right)$ which leads to the equivalent equation
\begin{equation} \label{cross11}
\vcenter{\hbox{\hspace{-10mm}\begin{tikzpicture}[font=\footnotesize,inner sep=2pt]
  \begin{feynman}
  \vertex (l) at (0,0) [label=above:\(X_w\)] ;
  \vertex [yellow, dot] (alpha) at (0,-1) [label=left:\(\alpha\)] {};
 \vertex  [yellow, dot] (beta) at (1,-2) [label=right:\(\beta\)] {};
 \vertex  (i) at (-1,-3) [label=below:\(X_x\)];
 \vertex  (left) at (-0.7,-3);
 \vertex [above=1cm of i] (i2);
 \vertex [below left=0.042cm of left] (i3);
 \vertex (j) at (0.4,-3) [label=below:\(X_y\)];
  \vertex (k) at (1.6,-3) [label=below:\(X_z\)];
  \vertex [above=0.5cm of j] (j11);
  \vertex[above=0.5cm of k] (k11);
  \vertex  (j2) at (0.45,-3.6);
  \vertex [above right=0.15cm of j2] (j3);
  \vertex [above right=0.15cm of j3] (j4);
  \vertex  (k2) at (1.55,-3.6);
  \vertex [above left=0.3cm of k2] (k3);
  \vertex[below=0.4cm of i] (a1);
  \vertex[below=0.4cm of j] (b1);
  \vertex[below=0.4cm of k] (c1);
  \vertex[left=0.25cm of i] (a11);
  \vertex[right=0.25cm of c1] (c11);
  \draw [fill=yellow] (c11) rectangle (a11);
  \vertex[left=0.25cm of i] (a11) [label=below right:\(\qquad \;r_{x,y\otimes_Cz}\)];
  \vertex[below=1.6cm of a1] (a2) [label=below:\(X_x\)];
  \vertex[below=0.5cm of b1] (b2);
  \vertex[below=0.5cm of c1] (c2);
  \vertex[left=0.25cm of b2] (b22);
  \vertex[below=0.4cm of c2] (c3);
  \vertex[right=0.25cm of c3] (c33);
   \draw [fill=yellow] (c33) rectangle (b22);
   \vertex[below=0.4cm of b2] (b3);
   \vertex[below=0.7cm of c3] (c4) [label=below:\(X_z\)];
   \vertex[below=0.7cm of b3] (b4) [label=below:\(X_y\)];
    \draw [middlearrow={latex},thick,rounded corners=1mm] (b4) -- (b3) ;
     \draw [middlearrow={latex},thick,rounded corners=1mm] (c4) -- (c3) ;
   \vertex[left=0.25cm of b2] (b22) [label=below right:\(\quad r_{y,z}\)];
   \draw [middlearrow={latex},thick,rounded corners=1mm] (a2) -- (a1) ;
    \draw [middlearrow={latex},thick,rounded corners=1mm] (b2) -- (b1) ;
     \draw [middlearrow={latex},thick,rounded corners=1mm] (c2) -- (c1) ;
  \draw [middlearrow={latex},thick,rounded corners=1mm] (i) -- (i2) -- (alpha);
   \draw [middlearrow={latex},thick,rounded corners=1mm] (j)  -- (j11) -- (beta);
   \draw [middlearrow={latex},thick,rounded corners=1mm] (k)  -- (k11) -- (beta);
    \diagram*{
      (alpha) --[middlearrow={latex},thick]  (l)   ,
      (beta) -- [middlearrow={latex},thick,edge label=$X_u$] (alpha)
    };
  \end{feynman}
\end{tikzpicture}}}
~\hspace{5mm}=\mathlarger{\sum}_{v\in\mathcal{K}_{AC}}\mathlarger{\sum}_{\gamma,\delta}\,\mathrm{Y}_{\alpha u\beta,\gamma v\delta}^{(x\,y\,z)w}
~ 
\vcenter{\hbox{\hspace{1mm}\begin{tikzpicture}[font=\footnotesize,inner sep=2pt]
  \begin{feynman}
  \vertex (l) at (0,0) [label=above:\(X_w\)] ;
  \vertex [yellow, dot] (del) at (0,-1) [label=left:\(\delta\)] {};
 \vertex  [yellow, dot] (gamma) at (-1,-2) [label=right:\(\gamma\)] {};
 \vertex  (i) at (-1.6,-3);
 \vertex  (i2) at (-1.55,-2.6);
 \vertex [below left=0.03cm of i2] (i3);
 \vertex (j) at (-0.4,-3) ;
  \vertex (k) at (1,-3) ;
  \vertex (j2) at (-0.45,-2.6) ;
  \vertex [below right=0.03cm of j2] (j3) ;
  \vertex [above=0.5cm of k] (k2);
  \vertex [above left=0.005cm of k2] (k3);
  \vertex (right) at (0.7,-2);
  \vertex[above=0.7cm of i] (i11);
  \vertex[above=0.7cm of j] (j11);
  \vertex[above=1cm of k] (k11);
  \vertex[below=0.4cm of i] (a1);
  \vertex[left=0.25cm of a1] (a2);
  \vertex[right=0.25cm of k] (c1);
  \draw [fill=yellow] (a2) rectangle (c1);
   \vertex[below=0.4cm of i] (a1) ;
   \vertex[below=0.46cm of i] (iii) [label=above right:\(\quad\quad\check{\alpha}_{x,y,z}\)];
   \vertex[below=0.6cm of a1] (a3) [label=below:\(X_x\)]; 
   \vertex[below=1cm of j] (b1) [label=below:\(X_y\)]; 
   \vertex[below=1cm of k] (c2) [label=below:\(X_z\)];
   \vertex[below=0.4cm of j] (b2);
   \vertex[below=0.4cm of k] (c3);
   \vertex[below=0.4cm of c2] (z1); 
   \vertex[left=0.25cm of a3] (x11);
   \vertex[right=0.25cm of z1] (z11);
    \draw [fill=yellow] (x11) rectangle (z11);
     \vertex[left=0.25cm of a3] (x11) [label=below right:\(\qquad\;r_{x,y\otimes_Cz}\)];
     \vertex[below=0.4cm of a3] (a4); 
     \vertex[below=0.4cm of b1] (b3); 
     \vertex[below=1.3cm of a4] (a5) [label=below:\(X_x\)]; 
     \vertex[below=0.5cm of b3] (b4); 
     \vertex[below=0.5cm of z1] (z2); 
     \vertex[below=0.4cm of z2] (z3); 
     \vertex[left=0.25cm of b4] (b44);
     \vertex[right=0.25cm of z3] (z33);
      \draw [fill=yellow] (b44) rectangle (z33);
      \vertex[left=0.25cm of b4] (b44) [label=below right:\(\qquad r_{y,z}\)];
      \vertex[below=0.4cm of b4] (b5);
      \vertex[below=0.4cm of b5] (b6) [label=below:\(X_y\)];
      \vertex[below=0.4cm of z3] (z4) [label=below:\(X_z\)];
      \draw [middlearrow={latex},thick,rounded corners=1mm] (a5) -- (a4) ;
      \draw [middlearrow={latex},thick,rounded corners=1mm] (b6) -- (b5) ;
      \draw [middlearrow={latex},thick,rounded corners=1mm] (z4) -- (z3) ;
    \draw [middlearrow={latex},thick,rounded corners=1mm] (a3) -- (a1) ;
     \draw [middlearrow={latex},thick,rounded corners=1mm] (b4) -- (b3) ;
      \draw [middlearrow={latex},thick,rounded corners=1mm] (z2) -- (z1) ;
    \draw [middlearrow={latex},thick,rounded corners=1mm] (b1) -- (b2)  ;
     \draw [middlearrow={latex},thick,rounded corners=1mm] (c2) -- (c3) ;
   \draw [middlearrow={latex},thick,rounded corners=1mm] (k) -- (k11)  -- (del);
    \draw [middlearrow={latex},thick,rounded corners=1mm] (i) -- (i11)  -- (gamma);
     \draw [middlearrow={latex},thick,rounded corners=1mm] (j) -- (j11) -- (gamma);
    \diagram*{
      (del) --[middlearrow={latex},thick]  (l)   ,
      (gamma) -- [middlearrow={latex},thick, edge label=$X_v$] (del)  
  };
  \end{feynman}
\end{tikzpicture}}}\end{equation}
Using \eqref{newprod} in the top part of \eqref{cross11} as well as the explicit form of $\check{\alpha}_{x,y,z}$ leads to the following equation
\begin{equation} 
\vcenter{\hbox{\hspace{-10mm}\begin{tikzpicture}[font=\footnotesize,inner sep=2pt]
  \begin{feynman}
  \vertex (l) at (0,0) [label=above:\(X_w\)] ;
  \vertex [empty dot] (alpha) at (0,-1) [label=left:\(\alpha\)] {};
 \vertex  [empty dot] (beta) at (1,-2) [label=right:\(\beta\)] {};
 \vertex  (i) at (-1,-3) [label=below:\(X_x\)];
 \vertex  (left) at (-0.7,-3);
 \vertex [above=1cm of i] (i2);
 \vertex [below left=0.042cm of left] (i3);
 \vertex (j) at (0.4,-3) [label=below:\(X_y\)];
  \vertex (k) at (1.6,-3) [label=below:\(X_z\)];
  \vertex [above=0.5cm of j] (j11);
  \vertex[above=0.5cm of k] (k11);
  \vertex  (j2) at (0.45,-3.6);
  \vertex [above right=0.15cm of j2] (j3);
  \vertex [above right=0.15cm of j3] (j4);
  \vertex  (k2) at (1.55,-3.6);
  \vertex [above left=0.3cm of k2] (k3);
  \vertex[below=0.4cm of i] (a1);
  \vertex[below=0.4cm of j] (b1);
  \vertex[below=0.4cm of k] (c1);
  \vertex[left=0.25cm of i] (a11);
  \vertex[right=0.25cm of c1] (c11);
  \draw [fill=yellow] (c11) rectangle (a11);
  \vertex[left=0.25cm of i] (a11) [label=below right:\(\qquad \;\;P_{x,y\otimes z}\)];
  \vertex[below=0.9cm of a1] (a2) [label=below:\(X_x\)];
  \vertex[below=0.5cm of b1] (b2);
  \vertex[below=0.5cm of c1] (c2);
  \vertex[left=0.25cm of b2] (b22);
  \vertex[below=0.4cm of c2] (c3);
  \vertex[right=0.25cm of c3] (c33);
   \vertex[below=0.4cm of b2] (b3);
   \vertex[below=0.00001cm of c3] (c4) [label=below:\(X_z\)];
   \vertex[below=0.00001cm of b3] (b4) [label=below:\(X_y\)];
    \draw [middlearrow={latex},thick,rounded corners=1mm] (b4) -- (b1) ;
     \draw [middlearrow={latex},thick,rounded corners=1mm] (c4) -- (c1) ;
   \draw [middlearrow={latex},thick,rounded corners=1mm] (a2) -- (a1) ;
  \draw [middlearrow={latex},thick,rounded corners=1mm] (i) -- (i2) -- (alpha);
   \draw [middlearrow={latex},thick,rounded corners=1mm] (j)  -- (j11) -- (beta);
   \draw [middlearrow={latex},thick,rounded corners=1mm] (k)  -- (k11) -- (beta);
    \diagram*{
      (alpha) --[middlearrow={latex},thick]  (l)   ,
      (beta) -- [middlearrow={latex},thick,edge label=$X_u$] (alpha)
    };
  \end{feynman}
\end{tikzpicture}}}
~\hspace{5mm}=\mathlarger{\sum}_{v\in\mathcal{K}_{AC}}\mathlarger{\sum}_{\gamma,\delta}\,\mathrm{Y}_{\alpha u\beta,\gamma v\delta}^{(x\,y\,z)w}
~ 
\vcenter{\hbox{\hspace{1mm}\begin{tikzpicture}[font=\footnotesize,inner sep=2pt]
  \begin{feynman}
  \vertex (l) at (0,0) [label=above:\(X_w\)] ;
  \vertex [empty dot] (del) at (0,-1) [label=left:\(\delta\)] {};
 \vertex  [empty dot] (gamma) at (-1,-2) [label=right:\(\gamma\)] {};
 \vertex  (i) at (-1.6,-3.2);
 \vertex  (i2) at (-1.55,-2.6);
 \vertex [below left=0.03cm of i2] (i3);
 \vertex (j) at (-0.4,-3.2) ;
  \vertex (k) at (1,-3.2) ;
  \vertex (j2) at (-0.45,-2.6) ;
  \vertex [below right=0.03cm of j2] (j3) ;
  \vertex [above=0.5cm of k] (k2);
  \vertex [above left=0.005cm of k2] (k3);
  \vertex (right) at (0.7,-2);
  \vertex[above=0.7cm of i] (i11);
  \vertex[above=0.7cm of j] (j11);
  \vertex[above=1cm of k] (k11);
  \vertex[below=0.4cm of i] (a1);
  \vertex[left=0.25cm of a1] (a2);
  \vertex[right=0.25cm of k] (c1);
  \draw [fill=yellow] (a2) rectangle (c1);
   \vertex[below=0.4cm of i] (a1) ;
   \vertex[below=0.8cm of a1] (a3) [label=below:\(X_x\)];
   \vertex[below=1.2cm of j] (b1) [label=below:\(X_y\)];
   \vertex[below=1.2cm of k] (c2) [label=below:\(X_z\)];
   \vertex[below=0.4cm of j] (b2);
   \vertex[below=0.4cm of k] (c3);
   \vertex (j) at (-0.4,-3.2) [label=below:\(P_{x\otimes y,z}\)];
    \draw [middlearrow={latex},thick,rounded corners=1mm] (a3) -- (a1) ;
    \draw [middlearrow={latex},thick,rounded corners=1mm] (b1) -- (b2)  ;
     \draw [middlearrow={latex},thick,rounded corners=1mm] (c2) -- (c3) ;
   \draw [middlearrow={latex},thick,rounded corners=1mm] (k) -- (k11)  -- (del);
    \draw [middlearrow={latex},thick,rounded corners=1mm] (i) -- (i11)  -- (gamma);
     \draw [middlearrow={latex},thick,rounded corners=1mm] (j) -- (j11) -- (gamma);
    \diagram*{
      (del) --[middlearrow={latex},thick]  (l)   ,
      (gamma) -- [middlearrow={latex},thick, edge label=$X_v$] (del)  
  };
  \end{feynman}
\end{tikzpicture}}}\end{equation}
where the  junctions indicated by hollow black blobs represent morphisms of the form $\tilde{\Lambda}$ as in the first equation in \eqref{base3}. For the left-hand side we notice that the morphisms in the top part satisfy
\begin{equation}\tilde{\Lambda}^\alpha_{(x,u)w}\circ\left(\id_x\otimes\tilde{\Lambda}^\beta_{(y,z)u}\right)\in {\Hom_{A|D}}_{(P_{X_x,X_y\otimes X_z})}(X_x\otimes( X_y\otimes X_z),X_w)\end{equation} while for the right-hand side we see that  \begin{equation}\tilde{\Lambda}^\delta_{(v,z)w}\circ\left(\tilde{\Lambda}^\gamma_{(x,y)v}\otimes\id_z\right)\in {\Hom_{A|D}}_{(P_{X_x\otimes X_y,X_z})}((X_x\otimes X_y)\otimes X_z),X_w)\,.\end{equation}

\begin{figure}[pt]
    \centering
    \includegraphics{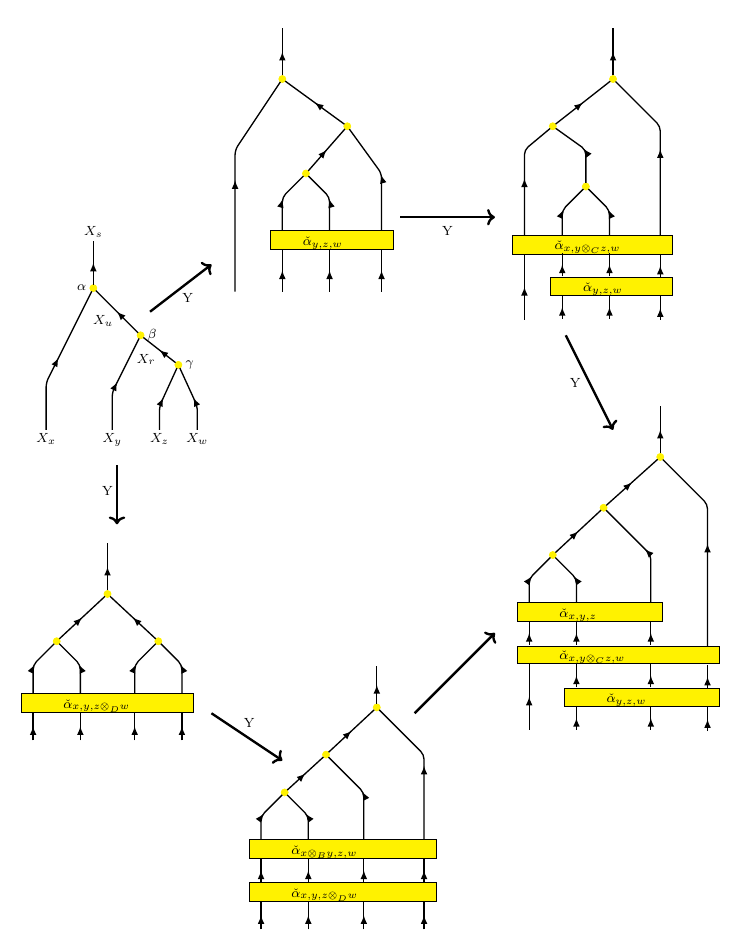}
    \caption{Pentagon identity for the $\mathrm{Y}$-matrices.}\label{figure1}
\end{figure}

Using the definition \eqref{spaces} of these spaces we find  that the projection morphisms can be removed yielding an equation in which 
all tensor products are those of the objects in  $\mathcal{C}$:
\begin{equation}\label{crossf1}
    \vcenter{\hbox{\hspace{-10mm}\begin{tikzpicture}[font=\footnotesize,inner sep=2pt]
  \begin{feynman}
  \vertex (l) at (0,0) [label=above:\(X_w\)] ;
  \vertex [empty dot] (alpha) at (0,-1) [label=left:\(\alpha\)] {};
 \vertex  [empty dot] (beta) at (1,-2) [label=right:\(\beta\)] {};
 \vertex  (i) at (-1,-4) [label=below:\(X_x\)];
 \vertex  (left) at (-0.7,-3);
 \vertex [above=1cm of i] (i2);
 \vertex [below left=0.042cm of left] (i3);
 \vertex (j) at (0.4,-4) [label=below:\(X_y\)];
  \vertex (k) at (1.6,-4) [label=below:\(X_z\)];
  \vertex [above=0.8cm of j] (j11);
  \vertex[above=0.8cm of k] (k11);
  \vertex  (j2) at (0.45,-3.6);
  \vertex [above right=0.15cm of j2] (j3);
  \vertex [above right=0.15cm of j3] (j4);
  \vertex  (k2) at (1.55,-3.6);
  \vertex [above left=0.3cm of k2] (k3);
  \draw [middlearrow={latex},thick,rounded corners=1mm] (i) -- (i2) -- (alpha);
   \draw [middlearrow={latex},thick,rounded corners=1mm] (j)  -- (j11) -- (beta);
   \draw [middlearrow={latex},thick,rounded corners=1mm] (k)  -- (k11) -- (beta);
    \diagram*{
      (alpha) --[middlearrow={latex},thick]  (l)   ,
      (beta) -- [middlearrow={latex},thick,edge label=$X_u$] (alpha)
    };
  \end{feynman}
\end{tikzpicture}}}
~\hspace{5mm}=\mathlarger{\sum}_{v\in\mathcal{K}_{AC}}\mathlarger{\sum}_{\gamma,\delta}\,\mathrm{Y}_{\alpha u\beta,\gamma v\delta}^{(x\,y\,z)w}
~ 
\vcenter{\hbox{\hspace{1mm}\begin{tikzpicture}[font=\footnotesize,inner sep=2pt]
  \begin{feynman}
  \vertex (l) at (0,0) [label=above:\(X_w\)] ;
  \vertex [empty dot] (del) at (0,-1) [label=left:\(\delta\)] {};
 \vertex  [empty dot] (gamma) at (-1,-2) [label=right:\(\gamma\)] {};
 \vertex  (i) at (-1.6,-4) [label=below:\(X_x\)];
 \vertex  (i2) at (-1.55,-2.6);
 \vertex [below left=0.03cm of i2] (i3);
 \vertex (j) at (-0.4,-4) [label=below:\(X_y\)];
  \vertex (k) at (1,-4) [label=below:\(X_z\)];
  \vertex (j2) at (-0.45,-2.6) ;
  \vertex [below right=0.03cm of j2] (j3) ;
  \vertex [above=0.5cm of k] (k2);
  \vertex [above left=0.005cm of k2] (k3);
  \vertex (right) at (0.7,-2);
  \vertex[above=0.7cm of i] (i11);
  \vertex[above=0.7cm of j] (j11);
  \vertex[above=1cm of k] (k11);
   \draw [middlearrow={latex},thick,rounded corners=1mm] (k) -- (k11)  -- (del);
    \draw [middlearrow={latex},thick,rounded corners=1mm] (i) -- (i11)  -- (gamma);
     \draw [middlearrow={latex},thick,rounded corners=1mm] (j) -- (j11) -- (gamma);
    \diagram*{
      (del) --[middlearrow={latex},thick]  (l)   ,
      (gamma) -- [middlearrow={latex},thick, edge label=$X_v$] (del)  
  };
  \end{feynman}
\end{tikzpicture}}}
\end{equation}
The last equation is equivalent to the definition \eqref{cross1} of the $\mathrm{Y}$ but it is in a form more suitable for calculations which we will later use to construct an explicit expression for $\mathrm{Y}$.

 The associator \eqref{associator22} obeys the pentagon identity
\begin{equation}\label{genpent1} \check{\alpha}_{x\otimes_By,z,w}\circ\check{\alpha}_{x,y,z\otimes_Dw}=\left(\check{\alpha}_{x,y,z}\otimes_D\operatorname{id}_{w}\right)\circ \check{\alpha}_{x,y\otimes_Cz,w}\circ\left(\id_{x}\otimes_B\check{\alpha}_{y,z,w}\right)
\end{equation}
which is a generalised version of \eqref{a1}. Using this pentagon identity as well as the definition of the $\mathrm{Y}$-matrices \eqref{cross1} 
we   obtain a new pentagon identity that the $\mathrm{Y}$-matrices must obey. Graphically the sequence of moves for the new pentagon 
is represented on figure \ref{figure1}. On that figure between the bottom graph and the one to its right   we have used the pentagon equation for the associator $\check{\alpha}$ \eqref{genpent1}. Demanding that these two ways lead to the same result gives the following pentagon identity for the $\mathrm{Y}$-matrices
\begin{equation}\label{pentaY}
    \sum_{q}\sum_{\delta,\varepsilon,\sigma} \mathrm{Y}^{(y\, z\,w)u}_{\beta r\gamma,\delta q\varepsilon}\,\mathrm{Y}^{(x\,q\,w)s}_{\alpha u\varepsilon,\sigma v\tau}\,\mathrm{Y}^{(x\,y\,z)v}_{\sigma q\delta,\theta p\lambda}=\sum_{\eta}\mathrm{Y}^{(x\,y\,r)s}_{\alpha u\beta,\theta p\eta}\,\mathrm{Y}^{(p\,z\,w)s}_{\eta r\gamma,\lambda v\tau}\,.
\end{equation}

\subsubsection{Fusing matrix \texorpdfstring{$\mathrm{T}$}{T} for open defects} \label{Tmatrix}

Next, we would like to define the fusing matrix for open topological interfaces. 
 To this end we consider the  associator
 $\tilde{\alpha}:\prescript{}{A}{\mathcal{C}}_B\times\prescript{}{B}{\mathcal{C}}_C\times\mathcal{C}_C\to \prescript{}{A}{\mathcal{C}}_B\times\prescript{}{B}{\mathcal{C}}_C\times\mathcal{C}_C$  with components
\begin{align}\label{associator3}
    & \tilde{\alpha}_{X,Y,M}:X\otimes_B(Y\otimes_CM)\to (X\otimes_BY)\otimes_CM, \nonumber \\
    & \tilde{\alpha}_{X,Y,M}= r_{X\otimes_BY,M}\circ\left(r_{X,Y}\otimes\id_M\right)\circ\left(\id_X\otimes e_{Y,M}\right)\circ e_{X,Y\otimes_CM}\,.
\end{align}
This associator is required to obey a certain mixed pentagon identity the proof of which is presented in appendix \ref{app4}. Notice that for the case $A=B=C$  this associator is exactly the module associativity constraint for the left module category $\mathcal{C}_A$ over the tensor category $\prescript{}{A}{\mathcal{C}}_A$ (see  \ref{modcat} for the definition). 

Taking $X,Y$ and $M$ to be simple objects in their corresponding categories as well as fixing the bases for the relevant junction spaces we obtain the matrix $\mathrm{T}$ that furnishes the vector space isomorphism 
\begin{equation}
\Hom_A\left(X_x\otimes_B(X_y\otimes_C M_a),M_c\right)\cong\Hom_A\left((X_x\otimes_BX_y)\otimes_C M_a,M_c\right)
\end{equation}
induced by the associator \eqref{associator3}. The entries of this matrix are defined by
\begin{equation}\label{cross4}
   \vcenter{\hbox{\begin{tikzpicture}[font=\footnotesize,inner sep=2pt]
  \begin{feynman}
  \path[pattern=north east lines,pattern color=ashgrey,very thin] (1.01,4) rectangle (1.6,0);
  \vertex  (i1) at (0,0)  ;
  \vertex (x1) at (0,3.5) ;
  \vertex[above=0.5cm of x1] (top1);
  \vertex[below right=0.1cm of x1] (x5) [label=below left:\(X_y\)];
  \vertex  (x2) at (0.75, 0.65);
  \vertex [above=4cm of i1] (i2);
  \vertex [right=1cm of i1] (j1);
  \vertex[ above=0.4cm of j1] (j3) [label=right:\(M_c\)];
  \vertex [right=1cm of i2] (j2);
  \vertex[ below=0.2cm of j2] (j4) [label=right:\(M_a\)];
  \vertex [below=1.4cm of j2] (j6);
  \vertex   [above=1cm of j1] (j5) ;
  \vertex[ above =1.1cm of j1] (j55);
  \vertex [left=0.15cm of j5]  [label=left:\(\alpha\)];
    \vertex [left=0.1cm of j55]  (j);
  \vertex [left=0.1cm of j6] (j7);
  \vertex (y) at (-0.7,2.6) ;
  \vertex[above=1.4cm of y] (top2);
  \vertex [below right=0.1cm of y] (y1) ;
  \vertex[left=0.8cm of x5] (xx) [label=below left:\(X_x\)];
  \vertex [below=1.4cm of j2] (k1);
  \vertex [below=1.5cm of j2] (k2);
  \vertex [left=0.15cm of k1] (k3);
  \vertex [left=0.15cm of k2] (k4) [label=left:\(\beta\)];
\draw [thick]   (j7) to[out=90,in=-90] (x1);
   \draw [fill=yellow] (1.06,1.1) rectangle (0.8, 1);
   \draw [fill=yellow] (1.06,2.6) rectangle (0.8, 2.5);
   \draw [thick]  (j) to[out=90,in=-90] (y);
   \vertex[above=0.65 of j55] [label=right:\(M_b\)];
    \draw [thick] (0,0.8) rectangle (0.4, 1.2);
    \vertex at (0.2,0.8) [label=\(A\)];
     \draw [thick] (0,2.3) rectangle (0.4, 2.7);
    \vertex at (0.2,2.3) [label=\(B\)];
     \draw [thick] (0.3,3.6) rectangle (0.7, 4);
    \vertex at (0.5,3.6) [label=\(C\)];
  \diagram*{
  (j2)--[thick] (k1),
  (k2)--[thick] (j55),
  (j5)--[thick] (j1),
  (x1)--[thick,myptr1={latex}] (top1),
   (y)--[thick,myptr1={latex}] (top2)
    };
  \end{feynman}
\end{tikzpicture}}} \hspace{1mm}~= \mathlarger{\sum}_{v\in\mathcal{K}_{AC}}\mathlarger{\sum}_{\gamma=1}^{\tensor{\check{N}}{_{yx}}{^v}}\mathlarger{\sum}_{\delta=1}^{\tensor{B}{_{va}}{^c}} \mathrm{T}^{(x\,y\,a)c}_{\alpha b\beta,\gamma v\delta} ~
\vcenter{\hbox{\hspace{5mm}\begin{tikzpicture}[font=\footnotesize,inner sep=2pt]
  \begin{feynman}
   \path[pattern=north east lines,pattern color=ashgrey,very thin] (1.01,4) rectangle (1.6,0);
  \vertex  (i1) at (0,0)  ;
  \vertex [dot,yellow] (x1) at (-0.5,2.3) [label=left:\(\gamma\)] {};
  \vertex  (x2) at (0.75, 0.65);
  \vertex [above=4cm of i1] (i2);
  \vertex [right=1cm of i1] (j1);
  \vertex[ above=0.4cm of j1] (j3) [label=right:\(M_c\)];
  \vertex [right=1cm of i2] (j2);
  \vertex[ below=0.4cm of j2] (j4) [label=right:\(M_a\)];
  \vertex   [above=1cm of j1] (j5);
  \vertex   [above=1.1cm of j1] (j6);
  \vertex [left=0.15cm of j5] [label=left:\(\delta\)];
  \vertex [left=0.1cm of j6] (j7);
  \vertex (x) at (-0.15,1.9) [label=right:\(X_v\)];
\draw [thick] [middlearrow={latex}]   (j7) to[out=90,in=-90] (x1);
   \draw [fill=yellow] (1.06,1) rectangle (0.8, 1.1);
   \vertex (k1) at (-1,2.8) ;
   \vertex (k2) at (-1,3.4)  [label=left:\(X_x\)];
   \vertex (k3) at (0,2.8) ;
   \vertex (k4) at (0,3.4) [label=right:\(X_y\)]; 
   \vertex[above=1.1cm of k2] (k5) ;
   \vertex[above=0.6cm of k4] (k6);
   \vertex[left=0.25cm of k5] (k55);
   \vertex[right=0.25cm of j2] (j11);
   \draw [fill=yellow] (j11) rectangle (k55);
    \vertex[below=0.5cm of k5] (k7) [label=above right:\(\quad\tilde{\alpha}_{x,y,a}\)];
   \draw [myptr={latex},thick,rounded corners=1mm]  (x1) -- (k1) --(k7);
   \draw [myptr={latex},thick,rounded corners=1mm]  (x1) -- (k3) --(k6);
    \draw [thick] (-0.1,0.8) rectangle (-0.5, 1.2);
    \vertex at (-0.3,0.8) [label=\(A\)];
     \draw [thick] (-0.3,2.8) rectangle (-0.7, 3.2);
    \vertex at (-0.5,2.8) [label=\(B\)];
     \draw [thick] (0.4,2.6) rectangle (0.8, 3);
    \vertex at (0.6,2.6) [label=\(C\)];
  \diagram*{
  (j1)--[ thick] (j5),
  (j2)--[ thick] (j6)
    };
  \end{feynman}
\end{tikzpicture}}}
\end{equation}
where $\tensor{\check{N}}{_{xy}}{^u}:=\operatorname{dim}\text{Hom}_{A|C}(X_x\otimes_BX_y,X_u)$ and $\tensor{B}{_{va}}{^c}:=\operatorname{dim}\text{Hom}_A(X_v\otimes_CM_a,M_c)$. To avoid  cluttering the morphism graphs we will sometimes omit the algebra labels $A,B,C,\ldots$ when they are clear from the context.

To clarify the notation used in \eqref{cross4} we also rewrite this equation using the first graphical convention where the diagrams flow from bottom to top
\begin{equation}\label{cross4.1}
    \vcenter{\hbox{\hspace{-10mm}\begin{tikzpicture}[font=\footnotesize,inner sep=2pt]
  \begin{feynman}
  \vertex (l) at (0,0) [label=above:\(M_c\)] ;
  \vertex  (alpha) at (0,-1) [label=left:\(\alpha\)] {};
  \Vertex[x=0,y=-1,color=yellow,shape=rectangle,size=0.15]{A};
 \vertex  (beta) at (1,-2) [label=right:\(\beta\)] {};
 \Vertex[x=1,y=-2,color=yellow,shape=rectangle,size=0.15]{A};
 \vertex  (i) at (-1,-4.5) [label=below:\(X_x\)];
 \vertex  (left) at (-0.7,-3);
 \vertex [above=1cm of i] (i2);
 \vertex [below left=0.042cm of left] (i3);
 \vertex (j) at (0.4,-4.5) [label=below:\(X_y\)];
  \vertex (k) at (1.6,-4.5) [label=below:\(M_a\)];
  \vertex [above=0.8cm of j] (j11);
  \vertex[above=0.8cm of k] (k11);
  \vertex  (j2) at (0.45,-3.6);
  \vertex [above right=0.15cm of j2] (j3);
  \vertex [above right=0.15cm of j3] (j4);
  \vertex  (k2) at (1.55,-3.6);
  \vertex [above left=0.3cm of k2] (k3);
  \draw [middlearrow={latex},thick,rounded corners=1mm] (i) -- (i2) -- (alpha);
   \draw [middlearrow={latex},thick,rounded corners=1mm] (j)  -- (j11) -- (beta);
   \draw [thick,rounded corners=1mm] (k)  -- (k11) -- (beta);
    \diagram*{
      (alpha) --[thick]  (l)   ,
      (beta) -- [thick,edge label=$M_b$] (alpha)
    };
  \end{feynman}
\end{tikzpicture}}}
\hspace{1mm}~= \mathlarger{\sum}_{v\in\mathcal{K}_{AC}}\mathlarger{\sum}_{\gamma=1}^{\tensor{\check{N}}{_{yx}}{^v}}\mathlarger{\sum}_{\delta=1}^{\tensor{B}{_{va}}{^c}} \mathrm{T}^{(x\,y\,a)c}_{\alpha b\beta,\gamma v\delta} ~
\vcenter{\hbox{\hspace{1mm}\begin{tikzpicture}[font=\footnotesize,inner sep=2pt]
  \begin{feynman}
  \vertex (l) at (0,0) [label=above:\(M_c\)] ;
  \vertex  (del) at (0,-1) [label=left:\(\delta\)] {};
 \vertex  [yellow,dot] (gamma) at (-1,-2) [label=right:\(\gamma\)] {};
 \Vertex[x=0,y=-1,color=yellow,shape=rectangle,size=0.15] {G};
 \vertex  (i) at (-1.6,-3.5);
 \vertex  (i2) at (-1.55,-2.6);
 \vertex [below left=0.03cm of i2] (i3);
 \vertex (j) at (-0.4,-3.5) ;
  \vertex (k) at (1,-3.5) ;
  \vertex (j2) at (-0.45,-2.6) ;
  \vertex [below right=0.03cm of j2] (j3) ;
  \vertex [above=0.5cm of k] (k2);
  \vertex [above left=0.005cm of k2] (k3);
  \vertex (right) at (0.7,-2);
  \vertex[above=0.7cm of i] (i11);
  \vertex[above=0.7cm of j] (j11);
  \vertex[above=1cm of k] (k11);
  \vertex[below=0.4cm of i] (a1);
  \vertex[left=0.25cm of a1] (a2);
  \vertex[right=0.25cm of k] (c1);
  \draw [fill=yellow] (a2) rectangle (c1);
   \vertex[below=0.4cm of i] (a1) ;
   \vertex[below=0.46cm of i] (iii) [label=above right:\(\quad\quad\tilde{\alpha}_{x,y,a}\)];
   \vertex[below=0.6cm of a1] (a3) [label=below:\(X_x\)];
   \vertex[below=1cm of j] (b1) [label=below:\(X_y\)];
   \vertex[below=1cm of k] (c2) [label=below:\(M_a\)];
   \vertex[below=0.4cm of j] (b2);
   \vertex[below=0.4cm of k] (c3);
    \draw [middlearrow={latex},thick,rounded corners=1mm] (a3) -- (a1) ;
    \draw [middlearrow={latex},thick,rounded corners=1mm] (b1) -- (b2)  ;
     \draw [thick,rounded corners=1mm] (c2) -- (c3) ;
   \draw [thick,rounded corners=1mm] (k) -- (k11)  -- (del);
    \draw [myptr={latex},thick,rounded corners=1mm] (i) -- (i11)  -- (gamma);
     \draw [myptr={latex},thick,rounded corners=1mm] (j) -- (j11) -- (gamma);
    \diagram*{
      (del) --[thick]  (l)   ,
      (gamma) -- [middlearrow={latex},thick, edge label=$X_v$] (del)  
  };
  \end{feynman}
\end{tikzpicture}}}
\end{equation}
To obtain a simpler expression in which the associator is absent and the tensor products are those in $\mathcal{C}$ we proceed exactly as we did in subsection \ref{sec252}. We note that the only difference between \eqref{cross1} and \eqref{cross4.1} is that $X_z$ is a bimodule while $M_a$ is a module. However, the fact that $X_z$ has a right-module structure was never used in the derivation of the equivalent definition \eqref{crossf1}. Therefore, we arrive at 
\begin{equation}\label{cross4.2}
    \vcenter{\hbox{\hspace{-10mm}\begin{tikzpicture}[font=\footnotesize,inner sep=2pt]
  \begin{feynman}
  \vertex (l) at (0,0) [label=above:\(M_c\)] ;
  \vertex (alpha) at (0,-1) [label=left:\(\alpha\)] {};
 \vertex   (beta) at (1,-2) [label=right:\(\beta\)] {};
 \path ( 0,-1) node [shape=rectangle,draw] {};
 \path ( 1,-2) node [shape=rectangle,draw] {};
 \vertex  (i) at (-1,-4) [label=below:\(X_x\)];
 \vertex  (left) at (-0.7,-3);
 \vertex [above=1cm of i] (i2);
 \vertex [below left=0.042cm of left] (i3);
 \vertex (j) at (0.4,-4) [label=below:\(X_y\)];
  \vertex (k) at (1.6,-4) [label=below:\(M_a\)];
  \vertex [above=0.8cm of j] (j11);
  \vertex[above=0.8cm of k] (k11);
  \vertex  (j2) at (0.45,-3.6);
  \vertex [above right=0.15cm of j2] (j3);
  \vertex [above right=0.15cm of j3] (j4);
  \vertex  (k2) at (1.55,-3.6);
  \vertex [above left=0.3cm of k2] (k3);
  \draw [middlearrow={latex},thick,rounded corners=1mm] (i) -- (i2) -- (alpha);
   \draw [middlearrow={latex},thick,rounded corners=1mm] (j)  -- (j11) -- (beta);
   \draw [thick,rounded corners=1mm] (k)  -- (k11) -- (beta);
    \diagram*{
      (alpha) --[thick]  (l)   ,
      (beta) -- [thick,edge label=$M_b$] (alpha)
    };
  \end{feynman}
\end{tikzpicture}}}
\hspace{1mm}~= \mathlarger{\sum}_{v\in\mathcal{K}_{AC}}\mathlarger{\sum}_{\gamma=1}^{\tensor{\check{N}}{_{yx}}{^v}}\mathlarger{\sum}_{\delta=1}^{\tensor{B}{_{va}}{^c}} \mathrm{T}^{(x\,y\,a)c}_{\alpha b\beta,\gamma v\delta} ~
\vcenter{\hbox{\hspace{1mm}\begin{tikzpicture}[font=\footnotesize,inner sep=2pt]
  \begin{feynman}
  \vertex (l) at (0,0) [label=above:\(M_c\)] ;
  \vertex  (del) at (0,-1) [label=left:\(\delta\)] {};
 \vertex [empty dot]  (gamma) at (-1,-2) [label=right:\(\gamma\)] {};
 \path ( 0,-1) node [shape=rectangle,draw] {};
 \vertex  (i) at (-1.6,-4) [label=below:\(X_x\)];
 \vertex  (i2) at (-1.55,-2.6);
 \vertex [below left=0.03cm of i2] (i3);
 \vertex (j) at (-0.4,-4) [label=below:\(X_y\)];
  \vertex (k) at (1,-4) [label=below:\(M_a\)];
  \vertex (j2) at (-0.45,-2.6) ;
  \vertex [below right=0.03cm of j2] (j3) ;
  \vertex [above=0.5cm of k] (k2);
  \vertex [above left=0.005cm of k2] (k3);
  \vertex (right) at (0.7,-2);
  \vertex[above=0.7cm of i] (i11);
  \vertex[above=0.7cm of j] (j11);
  \vertex[above=1cm of k] (k11);
   \draw [thick,rounded corners=1mm] (k) -- (k11)  -- (del);
    \draw [middlearrow={latex},thick,rounded corners=1mm] (i) -- (i11)  -- (gamma);
     \draw [middlearrow={latex},thick,rounded corners=1mm] (j) -- (j11) -- (gamma);
    \diagram*{
      (del) --[thick]  (l)   ,
      (gamma) -- [middlearrow={latex},thick, edge label=$X_v$] (del)  
  };
  \end{feynman}
\end{tikzpicture}}}
\end{equation}
where the junctions indicated by hollow black squares represent morphisms $\Tilde{\Omega}$ defined in (\ref{omegatilde}).

\begin{figure}[pt]
    \centering
    \includegraphics{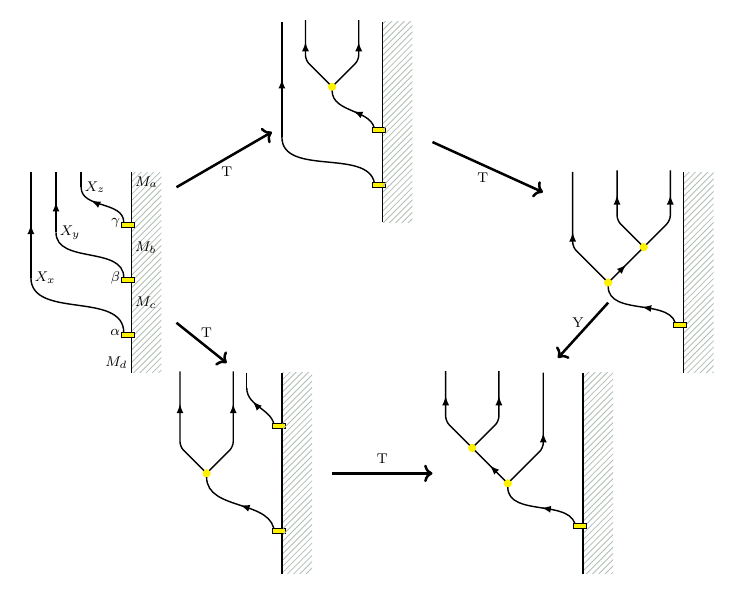}
    \caption{Pentagon identity for the $\mathrm{T}$-matrices.}\label{figure2}
\end{figure}

The fusing matrix $\mathrm{T}$ satisfies a mixed pentagon identity which can be derived either by starting with \eqref{cross4.1} and using the pentagon identity for $\tilde{\alpha}$  or by starting with the equivalent definition \eqref{cross4.2}. 
Proceeding either way 
we obtain the following mixed pentagon identity for the $\mathrm{T}$-matrices
\begin{equation}\label{penT1}
     \sum_v\sum_{\varepsilon,\zeta,\xi} \mathrm{T}^{(y\,z\,a)c}_{\beta b\alpha,\varepsilon v\zeta}\,\mathrm{T}^{(x\,v\,a)d}_{\gamma c\zeta,\xi w \nu} \,\mathrm{Y}^{(x\,y\,z)w}_{\xi v \varepsilon,\rho u \mu}= \sum_\sigma \mathrm{T}^{(x\,y\,b)d}_{\gamma c\beta,\rho u \sigma} \,\mathrm{T}^{(u \,z\,a)d}_{\sigma b \alpha, \mu w \nu}\,.
\end{equation}
This identity is illustrated on figure \ref{figure2}
where to avoid clutter we omitted the associators which  drop out from the final result as  in section \ref{sec252}.

Finally, we introduce the inverse $\overline{\mathrm{T}}$ of the $\mathrm{T}$-matrix
\begin{equation}\label{Tinv}
    \vcenter{\hbox{\hspace{1mm}\begin{tikzpicture}[font=\footnotesize,inner sep=2pt]
  \begin{feynman}
  \path[pattern=north east lines,pattern color=ashgrey,very thin] (1.01,4) rectangle (1.6,0);
  \vertex  (i1) at (0,0)  ;
  \vertex (y1) at (-0.7,1.5);
  \vertex[below=1.5cm of y1] (bot1);
  \vertex  (y2) at (0.75, 0.65);
  \vertex [above=4cm of i1] (i2);
  \vertex [right=1cm of i1] (j1);
  \vertex[ above=0.4cm of j1] (j3) [label=right:\(M_c\)];
  \vertex [right=1cm of i2] (j2);
  \vertex[ below=0.4cm of j2] (j4) [label=right:\(M_a\)];
  \vertex [below=1.4cm of j2] (j6) ;
  \vertex   [above=1.5cm of j1] (j5) {};
  \vertex [left=0.1cm of j5] (j7);
   \vertex[below=1cm of j2] (k1);
   \vertex[below=1.1cm of j2] (k2);
   \vertex[left=0.15cm of k1] (k3) [label=left:\(\alpha\)];
   \vertex[left=0.15cm of k2] (k4);
  \vertex (y) at (-0.25,2.1) [label=above:\(X_x\)];
\draw [thick]    (y1) to[out=90,in=-90] (k4);
   \draw [fill=yellow] (1.06,2.9) rectangle (0.8,3);
   \vertex (x) at (-0.25,0.6) [label=\(X_y\)];
    \vertex (x1) at (0,0.5);
    \vertex[below=0.5cm of x1] (bot2);
    \vertex[above=1.4cm of j1] (k5);
    \vertex[above=1.5cm of j1] (k6);
    \vertex[left=0.15cm of k5] (k7);
    \vertex[left=0.15cm of k6] (k8) [label=left:\(\beta\)];
    \draw [fill=yellow] (1.06,1.4) rectangle (0.8,1.5);
    \draw [thick]    (x1) to[out=90,in=-90] (k7);
     \vertex[above=0.65cm of k6] (b) [label=right:\(M_b\)];
      \draw [thick] (-0.1,2.8) rectangle (0.3, 3.2);
    \vertex at (0.1,2.8) [label=\(A\)];
     \draw [thick] (-0.1,1.3) rectangle (0.3, 1.7);
    \vertex at (0.1,1.3) [label=\(B\)];
     \draw [thick] (0.3,0) rectangle (0.7, 0.4);
    \vertex at (0.5,0) [label=\(C\)];
  \diagram*{
  (j2)--[ thick] (k1),
  (k2) --[thick] (k6),
  (k5)--[thick] (j1) ,
  (bot1)--[thick,middlearrow={latex}] (y1),
  (bot2)--[thick,myptr2={latex}] (x1)
    };
  \end{feynman}
\end{tikzpicture}}}
\hspace{1mm}~=\mathlarger{\sum}_{u\in\mathcal{K}_{AC}}\mathlarger{\sum}_{\gamma=1}^{\tensor{\check{N}}{_{u}}{^xy}}\mathlarger{\sum}_{\delta=1}^{\tensor{B}{_a}{^{uc}}}\,\overline{\mathrm{T}}^{(x\,y\,c)a}_{\gamma u\delta,\alpha b\beta}~
  \vcenter{\hbox{\hspace{5mm}\begin{tikzpicture}[font=\footnotesize,inner sep=2pt]
  \begin{feynman}
  \path[pattern=north east lines,pattern color=ashgrey,very thin] (1.01,4) rectangle (1.6,0);
  \vertex  (i1) at (0,0)  ;
  \vertex [yellow,dot] (y1) at (-0.5,1.7) [label=left:\(\gamma\)] {};
  \vertex  (y2) at (-1, 0.65) [label=left:\(X_x\)];
  \vertex (y3) at (-1,1.2);
  \vertex  (y4) at (0, 0.65) [label=right :\(X_y\)];
  \vertex[below=0.65cm of y2] (y22);
  \vertex[below=0.65cm of y4] (y44);
  \vertex (y5) at (0,1.2);
  \vertex [above=4cm of i1] (i2);
  \vertex [right=1cm of i1] (j1);
  \vertex[ above=0.4cm of j1] (j3) [label=right:\(M_c\)];
  \vertex [right=1cm of i2] (j2);
  \vertex[ below=0.4cm of j2] (j4) [label=right:\(M_a\)];
  \vertex [below=1.4cm of j2] (j6) ;
  \vertex   [above=1.5cm of j1] (j5) {};
  \vertex [left=0.1cm of j5] (j7);
   \vertex[below=1cm of j2] (k1);
   \vertex[below=1.1cm of j2] (k2);
   \vertex[left=0.15cm of k1] (k3) [label=left:\(\delta\)];
   \vertex[left=0.15cm of k2] (k4);
  \vertex (y) at (-0.25,2.4) [label=above right:\(X_u\)];
\draw [thick] [middlearrow={latex}]   (y1) to[out=90,in=-90] (k4);
   \draw [fill=yellow] (1.06,2.9) rectangle (0.8,3);
    \vertex (x1) at (-0.5,0.5);
    \vertex[above=1.4cm of j1] (k5);
    \vertex[above=1.5cm of j1] (k6);
    \vertex[left=0.15cm of k5] (k7);
    \vertex[right=0.25cm of j1] (j11);
    \vertex[below=0.5cm of y22] (y222);
    \vertex[left=0.25cm of y222] (y0);
    \draw [fill=yellow] (y0) rectangle (j11);
    \vertex[right=1.2cm of y0] (y00) [label=above:\(\Tilde{\alpha}^{-1}_{x,y,c}\)];
     \draw [middlearrow={latex},thick,rounded corners=1mm] (y22) -- (y3) --(y1);
      \draw [middlearrow={latex},thick,rounded corners=1mm]  (y44) -- (y5) --(y1);
       \draw [thick] (-0.1,3) rectangle (-0.5, 3.4);
    \vertex at (-0.3,3) [label=\(A\)];
     \draw [thick] (-0.3,0.5) rectangle (-0.7, 0.9);
    \vertex at (-0.5,0.5) [label=\(B\)];
     \draw [thick] (0.2,1.1) rectangle (0.6, 1.5);
    \vertex at (0.4,1.1) [label=\(C\)];
  \diagram*{
  (j2)--[ thick] (k1),
  (k2) --[thick] (j1)
    };
  \end{feynman}
\end{tikzpicture}}}
\end{equation}
where  
$$\tensor{\check{N}}{_{u}}{^{xy}}:=\operatorname{dim}\operatorname{Hom}_{A|C}(X_u,X_x\otimes_BX_y),\, \quad 
\tensor{B}{_a}{^{uc}}:=\operatorname{dim}\text{Hom}_A(M_a,X_u\!\otimes_C\!M_c)$$ 
and $X_x\in\operatorname{Obj}(\prescript{}{A}{\mathcal{C}}_B),\,X_y\in\operatorname{Obj}(\prescript{}{B}{\mathcal{C}}_C)$ while $M_a\in \operatorname{Obj}(\mathcal{C}_A)$, $M_b\in\operatorname{Obj}(\mathcal{C}_B)$, $M_c\in\operatorname{Obj}(\mathcal{C}_C)$. The matrix $\overline{\mathrm{T}}$ satisfies a mixed pentagon identity similar to \eqref{penT1}:
\begin{equation}\label{penT2}
     \sum_u\sum_{\rho,\sigma,\mu}\overline{\mathrm{T}}^{(y\,z\,d)b}_{\rho u\sigma,\beta c\gamma}\,\overline{\mathrm{T}}^{(x\,u\,d)a}_{\mu w\nu,\alpha b \sigma} 
    \,\overline{\mathrm{Y}}^{(x\,y\,z)w}_{\tau v\varphi,\mu u\rho}=\sum_\zeta \overline{\mathrm{T}}^{(x\,y\,c)a}_{\zeta v\tau,\alpha b\beta}\overline{\mathrm{T}}^{(v\,z\,d)a}_{\varphi w\nu,\zeta c\gamma}
\end{equation}
where $\overline{\mathrm{Y}}$ is the inverse of the matrix $\mathrm{Y}$.

\subsubsection{Fusing matrix \texorpdfstring{$\dot{\mathrm{T}}$}{\.T} involving an open defect and  a boundary field}\label{dotTmatrix}

In this subsection we define the fusing matrices for open defects passing through a boundary field. Consider the associator $\dot{\alpha}:\prescript{}{A}{\mathcal{C}}_B\times \mathcal{C}_B\times\mathcal{C}\to\prescript{}{A}{\mathcal{C}}_B\times \mathcal{C}_B\times\mathcal{C}$ with components given by
 \begin{align}\label{associator4}
    & \dot{\alpha}_{X,M,U}:X\otimes_B\left(M\otimes U\right)\to \left(X\otimes_BM\right)\otimes U, \nonumber \\
    & \dot{\alpha}_{X,M,U}=\left(r_{X,M}\otimes\id_U\right)\circ \alpha_{\dot{X},\dot{M},U}\circ e_{X,M\otimes U}
 \end{align}
 where $\alpha$ denotes the associator of the category $\mathcal{C}$ which will be taken to be the identity map since $\mathcal{C}$ is  a strict category. To be consistent, the associator \eqref{associator4} is required to obey a certain  pentagon identity the proof of which can be found in appendix \ref{app5}.

 Taking $X,M$ and $U$ to be simple objects and fixing the bases for the relevant junction spaces we obtain the matrix $\dot{\mathrm{T}}$ that furnishes the vector space isomorphism 
 \begin{equation}
\Hom_A\left(X_x\otimes_B(M_a\otimes U_i),M_c\right)\cong\Hom_A\left((X_x\otimes_BM_a)\otimes U_i,M_c\right)
\end{equation}
induced by the associator \eqref{associator4}. The entries of this matrix are defined by
\begin{equation}\label{cross2}
     \vcenter{\hbox{\hspace{-10mm}\begin{tikzpicture}[font=\footnotesize,inner sep=2pt]
  \begin{feynman}
  \path[pattern=north east lines,pattern color=ashgrey,very thin] (0.01,4) rectangle (0.6,0);
  \vertex (j1) at (0,0);
  \vertex [above=0.8 of j1] (j2);
  \vertex[ above =0.9 of j1] (j3);
  \vertex [small,dot] [ above=2.5cm of j1] (psi)  {};
  \vertex[right=0.2cm of psi] (psi1) ;
  \vertex[below=0.2cm of psi1] (psi2) [label=\(\,\,\,\Psi_i^\beta\)];
  \vertex [above=3.4cm of j1] (j4) ;
  \vertex[above=3.5cm of j1] (j5)  ;
  \vertex[ left=0.15cm of  j4] (j7);
  \vertex [left=0.15cm of j3] (j8) [label=left:\(\alpha\)]; 
  \vertex [above =4cm of j1] (j6);
  \draw [fill=yellow] (0.06,0.8) rectangle (-0.2, 0.9);
    \vertex[below=0.1cm of j6] (m1) ;
    \vertex[above=1.6cm of j1] (m3) [label=right:\(M_b\)];
    \vertex[below=0.4cm of j6] (m2) [label=right:\(M_a\)];
    \vertex[left=1.2cm of psi] (x) [label=left:\(X_x\)];
    \vertex[above=0.3 of j1] (jj) [label=right:\(M_c\)];
    \vertex [below=0.25cm of x] (x1);
    \vertex [above=1.45cm of x] (x2);
    \draw [thick]    (j8) to[out=90,in=-90] (x1);
    \draw [thick] [middlearrow={latex}]   (x1) to (x2);
    \draw [thick] (-1.5,1.5) rectangle (-1.9, 1.9);
    \vertex at (-1.7,1.5) [label=\(A\)];
    \draw [thick] (-0.3,2) rectangle (-0.7, 2.4);
    \vertex at (-0.5,2) [label=\(B\)];
   \diagram*{
    (j1) --[thick] (j2),
    (j3) --[thick] (psi),
    (psi) --[thick] (j6)
  };
  \end{feynman}
\end{tikzpicture}}}
~=\mathlarger{\sum}_{d\in\mathcal{J}_A}\mathlarger{\sum}_{\gamma=1}^{\tensor{B}{_{xa}}{^d}}\mathlarger{\sum}_{\delta=1}^{\tensor{A}{_{di}}{^c}}\,{\mathrm{\dot{T}}}^{\,(x\,a\,i)c}_{\alpha b\beta,\gamma d\delta}~
\vcenter{\hbox{\hspace{1mm}\begin{tikzpicture}[font=\footnotesize,inner sep=2pt]
  \begin{feynman}
  \path[pattern=north east lines,pattern color=ashgrey,very thin] (0.01,4) rectangle (0.6,0);
  \vertex (j1) at (0,0);
  \vertex [above=2.4 of j1] (j2);
  \vertex[ above =2.5 of j1] (j3);
  \vertex [small,dot] [ above=0.9cm of j1] (psi)  {};
  \vertex[right=0.2cm of psi] (psi1);
  \vertex[below=0.21cm of psi1] (psi2) [label=\(\,\,\,\Psi_i^\delta\)];
  \vertex [above=3.4cm of j1] (j4) ;
  \vertex[above=3.5cm of j1] (j5)  ;
  \vertex[ left=0.15cm of  j4] (j7);
  \vertex [left=0.15cm of j3] (j8) [label=left:\(\gamma\)]; 
  \vertex [above =4cm of j1] (j6);
  \draw [fill=yellow] (0.06,2.4) rectangle (-0.2, 2.5);
    \vertex[below=0.1cm of j6] (m1) ;
    \vertex[above=1.7cm of j1] (m3) [label=right:\(M_d\)];
    \vertex[below=0.4cm of j6] (m2) [label=right:\(M_a\)];
    \vertex[above=0.5cm of j4] (xx);
    \vertex[left=1.2cm of xx] (x);
    \vertex[below=0.1cm of x] (x100)  [label=left:\(X_x\)];
    \vertex[above=0.3 of j1] (jj) [label=right:\(M_c\)];
    \vertex [below=0.25cm of x] (x1);
    \vertex [above=0.15cm of x] (x2);
    \vertex[left=0.25cm of x2] (x22);
    \vertex[above=0.6cm of j6] (j7);
    \vertex[right=0.25cm of j7] (j77);
    \draw [fill=yellow] (x22) rectangle (j77);
    \vertex[left=0.25cm of x2] (x22);
    \vertex[above=0.05cm of x] (x3);
    \vertex[left=0.25cm of x3] (x33);
    \vertex[above=0.1cm of x33] (x333) [label=above right:\(\quad\dot{\alpha}_{x,a,i}\)];
    \vertex[above=0.4cm of x2] (x4);
    \draw [thick]    (j8) to[out=90,in=-90] (x1);
    \draw [thick] [middlearrow={latex}]   (x1) to (x2);
    \draw [thick] (-1.4,2.8) rectangle (-1.8, 3.2);
    \vertex at (-1.6,2.8) [label=\(A\)];
    \draw [thick] (-0.2,3.3) rectangle (-0.6, 3.7);
    \vertex at (-0.4,3.3) [label=\(B\)];
   \diagram*{
    (j1) --[thick] (j2),
    (j3) --[thick] (j6)
  };
  \end{feynman}
\end{tikzpicture}}}
\end{equation}
where  $\tensor{A}{_{di}}{^c}:=\operatorname{dim}\text{Hom}_A(M_d\!\otimes\!U_i,M_c)$. Notice that in (\ref{cross2}) the diagram flows from top to bottom. 
For clarity  we  also rewrite  \eqref{cross2}  using the first graphical convention where the diagrams flow from bottom to top
\begin{equation}\label{cross2.1}
    \vcenter{\hbox{\hspace{-10mm}\begin{tikzpicture}[font=\footnotesize,inner sep=2pt]
  \begin{feynman}
  \vertex (l) at (0,0) [label=above:\(M_c\)] ;
  \vertex  (alpha) at (0,-1) [label=left:\(\alpha\)] {};
  \Vertex[x=0,y=-1,color=yellow,shape=rectangle,size=0.15]{A};
 \vertex  [black, dot] (beta) at (1,-2) [label=right:\(\beta\)] {};
 \vertex  (i) at (-1,-4.5) [label=below:\(X_x\)];
 \vertex  (left) at (-0.7,-3);
 \vertex [above=1cm of i] (i2);
 \vertex [below left=0.042cm of left] (i3);
 \vertex (j) at (0.4,-4.5) [label=below:\(M_a\)];
  \vertex (k) at (1.6,-4.5) [label=below:\(U_i\)];
  \vertex [above=0.8cm of j] (j11);
  \vertex[above=0.8cm of k] (k11);
  \vertex  (j2) at (0.45,-3.6);
  \vertex [above right=0.15cm of j2] (j3);
  \vertex [above right=0.15cm of j3] (j4);
  \vertex  (k2) at (1.55,-3.6);
  \vertex [above left=0.3cm of k2] (k3);
  \draw [middlearrow={latex},thick,rounded corners=1mm] (i) -- (i2) -- (alpha);
   \draw [thick,rounded corners=1mm] (j)  -- (j11) -- (beta);
   \draw [thick,rounded corners=1mm] (k)  -- (k11) -- (beta);
    \diagram*{
      (alpha) --[thick]  (l)   ,
      (beta) -- [thick,edge label=$M_b$] (alpha)
    };
  \end{feynman}
\end{tikzpicture}}}
~\hspace{5mm}=\mathlarger{\sum}_{d\in\mathcal{J}_A}\mathlarger{\sum}_{\gamma=1}^{\tensor{B}{_{xa}}{^d}}\mathlarger{\sum}_{\delta=1}^{\tensor{A}{_{di}}{^c}}\,\mathrm{\dot{T}}_{\alpha b\beta,\gamma d\delta}^{(x\,a\,i)c}
~ 
\vcenter{\hbox{\hspace{1mm}\begin{tikzpicture}[font=\footnotesize,inner sep=2pt]
  \begin{feynman}
  \vertex (l) at (0,0) [label=above:\(M_c\)] ;
  \vertex [ dot] (del) at (0,-1) [label=left:\(\delta\)] {};
 \vertex  [yellow, dot] (gamma) at (-1,-2) [label=right:\(\gamma\)] {};
 \Vertex[x=-1,y=-2,color=yellow,shape=rectangle,size=0.15] {G};
 \vertex  (i) at (-1.6,-3.5);
 \vertex  (i2) at (-1.55,-2.6);
 \vertex [below left=0.03cm of i2] (i3);
 \vertex (j) at (-0.4,-3.5) ;
  \vertex (k) at (1,-3.5) ;
  \vertex (j2) at (-0.45,-2.6) ;
  \vertex [below right=0.03cm of j2] (j3) ;
  \vertex [above=0.5cm of k] (k2);
  \vertex [above left=0.005cm of k2] (k3);
  \vertex (right) at (0.7,-2);
  \vertex[above=0.7cm of i] (i11);
  \vertex[above=0.7cm of j] (j11);
  \vertex[above=1cm of k] (k11);
  \vertex[below=0.4cm of i] (a1);
  \vertex[left=0.25cm of a1] (a2);
  \vertex[right=0.25cm of k] (c1);
  \draw [fill=yellow] (a2) rectangle (c1);
   \vertex[below=0.4cm of i] (a1) ;
   \vertex[below=0.46cm of i] (iii) [label=above right:\(\quad\quad\dot{\alpha}_{x,a,i}\)];
   \vertex[below=0.6cm of a1] (a3) [label=below:\(X_x\)];
   \vertex[below=1cm of j] (b1) [label=below:\(M_a\)];
   \vertex[below=1cm of k] (c2) [label=below:\(U_i\)];
   \vertex[below=0.4cm of j] (b2);
   \vertex[below=0.4cm of k] (c3);
    \draw [middlearrow={latex},thick,rounded corners=1mm] (a3) -- (a1) ;
    \draw [thick,rounded corners=1mm] (b1) -- (b2)  ;
     \draw [thick,rounded corners=1mm] (c2) -- (c3) ;
   \draw [thick,rounded corners=1mm] (k) -- (k11)  -- (del);
    \draw [middlearrow={latex},thick,rounded corners=1mm] (i) -- (i11)  -- (gamma);
     \draw [thick,rounded corners=1mm] (j) -- (j11) -- (gamma);
    \diagram*{
      (del) --[thick]  (l)   ,
      (gamma) -- [thick, edge label=$M_d$] (del)  
  };
  \end{feynman}
\end{tikzpicture}}}
\end{equation}
where the yellow square vertices represent morphisms of type \eqref{jun2} while the black dot vertices are boundary fields as introduced in \eqref{bcco} and \eqref{bccopic}.

\begin{figure}[pt]
\[
     \vcenter{\hbox{\hspace{1mm}\scalebox{0.8}{\begin{tikzpicture}[font=\footnotesize,inner sep=2pt]
  \begin{feynman}
   \path[pattern=north east lines,pattern color=ashgrey,very thin] (0.01,4) rectangle (0.6,0);
 \vertex (j1) at (0,0);
  \vertex [above=0.6 of j1] (j2);
  \vertex[ above =0.7 of j1] (j3);
  \vertex [small,dot] [ above=3cm of j1] (psi)  {};
  \vertex[right=0.2cm of psi] (psi1) ;
  \vertex[below=0.2cm of psi1] (psi2) [label=\(\,\,\,\,\Psi_i^\alpha\)];
  \vertex [above=3.4cm of j1] (j4) ;
  \vertex[above=3.5cm of j1] (j5)  ;
  \vertex[ left=0.15cm of  j4] (j7);
  \vertex [left=0.15cm of j3] (j8) [label=left:\(\gamma\)]; 
  \vertex [above =4cm of j1] (j6);
  \draw [fill=yellow] (0.06,0.6) rectangle (-0.2, 0.7);
    \vertex[below=0.1cm of j6] (m1) ;
    \vertex[above=2.4cm of j1] (m3) [label=right:\(M_b\)];
    \vertex[above=0.7cm of psi] (m2) [label=right:\(M_a\)];
    \vertex (x) at (-1.4,1.8) [label=right:\(X_x\)];
    \vertex[above=0.2 of j1] (jj) [label=left:\(M_d\)];
    \vertex[above=1.2cm of j1] (jj0) [label=right:\(M_c\)];
    \vertex [below=0.25cm of x] (x1);
    \vertex [above=2.2cm of x] (x2);
    \draw [thick]    (j8) to[out=90,in=-90] (x1);
    \draw [thick] [middlearrow={latex}]   (x1) to (x2);
    \vertex[above=1.6cm of j1] (jj1);
    \vertex[above=1.7cm of j1] (jj2);
    \vertex[left=0.15cm of jj2] (jj3)[label=left:\(\beta\)];
     \draw [fill=yellow] (0.06,1.6) rectangle (-0.2, 1.7);
 \vertex (y) at (-0.7,2.5) [label=right:\(X_y\)];
 \vertex [below=0.25cm of y] (y1);
    \vertex [above=1.4cm of y] (y2);
\draw [thick]    (jj3) to[out=90,in=-90] (y1);
 \draw [thick] [middlearrow={latex}]   (y1) to (y2);
      \draw[->, line width=0.5mm] (1.2,3.7) -- (3.3,4.7);
      \node at (2.4,4) (eq1) {$\mathlarger{\dot{\mathrm{T}}}$};
      \vertex (l1) at (5,3); 
     \vertex[above=0.8cm of l1] (l2);
     \vertex[above=0.9cm of l1] (l3);
     \vertex[small,dot][above=0.8cm of l3] (psi3){};
   \vertex[above=2.7cm of l1] (l4);
   \vertex[above=2.8cm of l1] (l5);
   \vertex[left=0.15cm of l5] (l55);
   \vertex[left=0.15cm of l3] (l33);
   \vertex[above=4cm of l1] (l6);
\path[pattern=north east lines,pattern color=ashgrey,very thin] (5.01,7) rectangle (5.6,3);
 \draw [fill=yellow] (5.06,3.8) rectangle (4.8, 3.9);
 \draw [fill=yellow] (5.06,5.7) rectangle (4.8, 5.8);
     \vertex (yi1) at (4.3,6.7);
\vertex [below=0.25cm of yi1] (yi2);
    \vertex [above=0.3cm of yi1] (yi3);
\draw [thick]    (l55) to[out=90,in=-90] (yi2);
 \draw [thick] [middlearrow={latex}]   (yi2) to (yi3);
 \vertex (xi1) at (3.6,5);
\vertex [below=0.25cm of xi1] (xi2);
    \vertex [above=2cm of xi1] (xi3);
\draw [thick]    (l33) to[out=90,in=-90] (xi2);
 \draw [thick] [middlearrow={latex}]   (xi2) to (xi3);

      \draw[->, line width=0.5mm] (6.5,4.6) -- (9,3.6);
      \node at (7.5,3.9) (eq1) {$\mathlarger{\dot{\mathrm{T}}}$};
      \vertex (r1) at (11,0); 
 \vertex[above=1.7cm of r1] (r2);
     \vertex[above=1.8cm of r1] (r3);
     \vertex[small,dot][above=0.9cm of r1] (psi4){};
   \vertex[above=2.7cm of r1] (r4);
   \vertex[above=2.8cm of r1] (r5);
   \vertex[left=0.15cm of r5] (r55);
   \vertex[left=0.15cm of r3] (r33);
   \vertex[above=4cm of r1] (r6);
\path[pattern=north east lines,pattern color=ashgrey,very thin] (11.01,4) rectangle (11.6,0);
 \draw [fill=yellow] (11.06,1.7) rectangle (10.8, 1.8);
 \draw [fill=yellow] (11.06,2.7) rectangle (10.8, 2.8);
     \vertex (yi4) at (10.3,3.7);
\vertex [below=0.25cm of yi4] (yi5);
    \vertex [above=0.3cm of yi4] (yi6);
\draw [thick]    (r55) to[out=90,in=-90] (yi5);
 \draw [thick] [middlearrow={latex}]   (yi5) to (yi6);
 \vertex (xi4) at (9.6,2.9);
\vertex [below=0.25cm of xi4] (xi5);
    \vertex [above=1.1cm of xi4] (xi6);
\draw [thick]    (r33) to[out=90,in=-90] (xi5);
 \draw [thick] [middlearrow={latex}]   (xi5) to (xi6);

     \draw[->, line width=0.5mm]  (9.8,1.6) -- (8.7,0.3) ;
      \node at (9.1,1.1) (eq1) {$\mathlarger{\mathrm{T}}$};
      \vertex (s1) at (9,-4); 
     \vertex[small,dot][above=0.8cm of s1] (psi5){};
     \vertex[above=1.7cm of s1] (s2);
     \vertex[above=1.8cm of s1] (s3);
     \vertex[left=0.15cm of s3] (s33);
     \vertex [yellow,dot] (z1) at (8,-1.2){};
\draw [fill=yellow] (9.06,-2.3) rectangle (8.8, -2.2);
\vertex[above=4cm of s1] (s4);
\path[pattern=north east lines,pattern color=ashgrey,very thin] (9.01,0) rectangle (9.6,-4);
 \vertex[above left=0.7cm of z1] (z2);
 \vertex[above right=0.7cm of z1] (z3);
 \vertex[above=0.7cm of z2] (z4);
 \vertex[above=0.7cm of z3] (z5);
 \draw [myptr={latex},thick,rounded corners=1mm]  (z1) -- (z2) --(z4);
 \draw [myptr={latex},thick,rounded corners=1mm]  (z1) -- (z3) --(z5);
\draw [thick,middlearrow={latex}]    (s33) to[out=90,in=-90] (z1);
         
         \vertex (q1) at (3,-4); 
\vertex[small,dot][above=3.1cm of q1] (psi6){};
     \vertex[above=0.8cm of q1] (q2);
     \vertex[above=0.9cm of q1] (q3);
     \vertex[left=0.15cm of q3] (q33);
     \vertex [yellow,dot] (zz1) at (2,-1.9){};
\draw [fill=yellow] (3.06,-3.2) rectangle (2.8, -3.1);
\vertex[above=4cm of q1] (q4);
\path[pattern=north east lines,pattern color=ashgrey,very thin] (3.01,0) rectangle (3.6,-4);
 \vertex[above left=0.7cm of zz1] (zz2);
 \vertex[above right=0.7cm of zz1] (zz3);
 \vertex[above=1.4cm of zz2] (zz4);
 \vertex[above=1.4cm of zz3] (zz5);
 \draw [myptr={latex},thick,rounded corners=1mm]  (zz1) -- (zz2) --(zz4);
 \draw [myptr={latex},thick,rounded corners=1mm]  (zz1) -- (zz3) --(zz5);
\draw [thick,middlearrow={latex}]    (q33) to[out=90,in=-90] (zz1);
             \draw[->, line width=0.5mm] (4,-2) -- (7,-2);
      \node at (5.5,-1.7) (eq1) {$\mathlarger{\dot{\mathrm{T}}}$};
        \draw[->, line width=0.5mm] (1.2,0.9) -- (2.2,0.1);
      \node at (1.8,0.7) (eq1) {$\mathlarger{\mathrm{T}}$};
        \diagram*{
   (j1) --[thick] (j2),
    (j3) --[thick] (jj1),
    (jj2)--[thick] (psi),
    (psi) --[thick] (j6),
    (l1)--[thick] (l2),
    (l3)--[thick] (psi3),
    (psi3)--[thick] (l4),
    (l5)--[thick] (l6),
    (r1)--[thick] (psi4),
    (psi4)--[thick] (r2),
    (r3)--[thick] (r4),
    (r5)--[thick] (r6),
    (s1)--[thick] (psi5),
    (psi5)--[thick] (s2),
    (s3)--[thick] (s4),
    (q1)--[thick] (q2),
    (q3)--[thick] (psi6),
    (psi6)--[thick] (q4)
    };
  \end{feynman}
\end{tikzpicture}}}}
\]\caption{Pentagon identity for the $\dot{\mathrm{T}}$ matrices.}\label{figure3}
\end{figure}
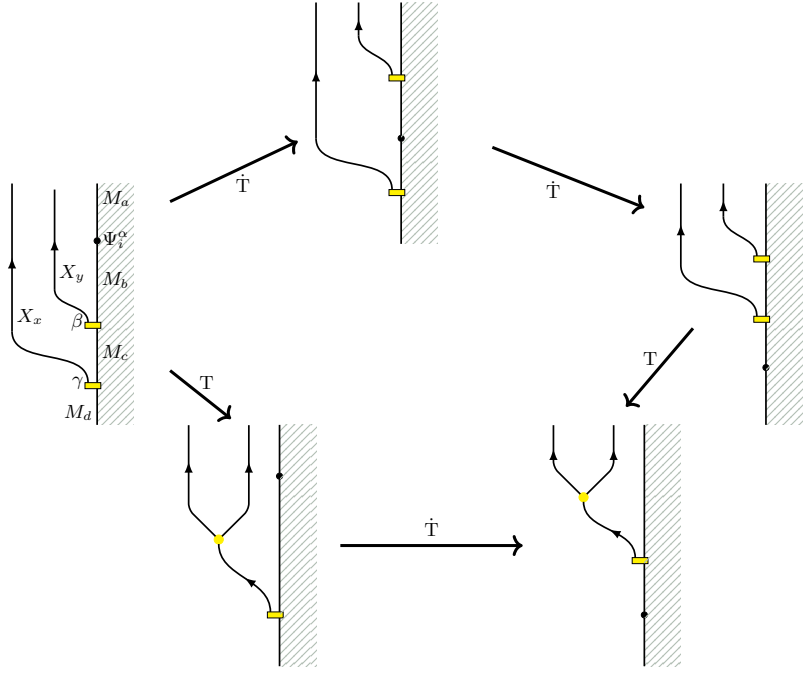

To extract more information from this definition, we proceed in a similar way as we did for the $\mathrm{Y}$-matrices. We compose \eqref{cross2.1} with the epimorphism $r_{x,a\otimes i}$ and also use \eqref{newprod} on the left-hand side. This leads to the equivalent definition for $\mathrm{\dot{T}}$
\begin{equation}\label{cross2.2}
     \vcenter{\hbox{\hspace{-10mm}\begin{tikzpicture}[font=\footnotesize,inner sep=2pt]
  \begin{feynman}
  \vertex (l) at (0,0) [label=above:\(M_c\)] ;
  \vertex  (alpha) at (0,-1) [label=left:\(\alpha\)] {};
 \vertex [dot]  (beta) at (1,-2) [label=right:\(\beta\)] {};
  \path ( 0,-1) node [shape=rectangle,draw] {};
 \vertex  (i) at (-1,-4) [label=below:\(X_x\)];
 \vertex  (left) at (-0.7,-3);
 \vertex [above=1cm of i] (i2);
 \vertex [below left=0.042cm of left] (i3);
 \vertex (j) at (0.4,-4) [label=below:\(M_a\)];
  \vertex (k) at (1.6,-4) [label=below:\(U_i\)];
  \vertex [above=0.8cm of j] (j11);
  \vertex[above=0.8cm of k] (k11);
  \vertex  (j2) at (0.45,-3.6);
  \vertex [above right=0.15cm of j2] (j3);
  \vertex [above right=0.15cm of j3] (j4);
  \vertex  (k2) at (1.55,-3.6);
  \vertex [above left=0.3cm of k2] (k3);
  \draw [middlearrow={latex},thick,rounded corners=1mm] (i) -- (i2) -- (alpha);
   \draw [thick,rounded corners=1mm] (j)  -- (j11) -- (beta);
   \draw [thick,rounded corners=1mm] (k)  -- (k11) -- (beta);
    \diagram*{
      (alpha) --[thick]  (l)   ,
      (beta) -- [thick,edge label=$M_b$] (alpha)
    };
  \end{feynman}
\end{tikzpicture}}}
~\hspace{5mm}=\mathlarger{\sum}_{d\in\mathcal{J}_A}\mathlarger{\sum}_{\gamma,\delta}\,\mathrm{\dot{T}}_{\alpha b\beta,\gamma d\delta}^{(x\,a\,i)c}
~ 
\vcenter{\hbox{\hspace{1mm}\begin{tikzpicture}[font=\footnotesize,inner sep=2pt]
  \begin{feynman}
  \vertex (l) at (0,0) [label=above:\(M_c\)] ;
  \vertex [ dot] (del) at (0,-1) [label=left:\(\delta\)] {};
 \vertex   (gamma) at (-1,-2) [label=right:\(\gamma\)] {};
 \path ( -1,-2) node [shape=rectangle,draw] {};
 \vertex  (i) at (-1.6,-4) [label=below:\(X_x\)];
 \vertex  (i2) at (-1.55,-2.6);
 \vertex [below left=0.03cm of i2] (i3);
 \vertex (j) at (-0.4,-4) [label=below:\(M_a\)];
  \vertex (k) at (1,-4) [label=below:\(U_i\)];
  \vertex (j2) at (-0.45,-2.6) ;
  \vertex [below right=0.03cm of j2] (j3) ;
  \vertex [above=0.5cm of k] (k2);
  \vertex [above left=0.005cm of k2] (k3);
  \vertex (right) at (0.7,-2);
  \vertex[above=0.7cm of i] (i11);
  \vertex[above=0.7cm of j] (j11);
  \vertex[above=1cm of k] (k11);
   \draw [thick,rounded corners=1mm] (k) -- (k11)  -- (del);
    \draw [middlearrow={latex},thick,rounded corners=1mm] (i) -- (i11)  -- (gamma);
     \draw [thick,rounded corners=1mm] (j) -- (j11) -- (gamma);
    \diagram*{
      (del) --[thick]  (l)   ,
      (gamma) -- [thick, edge label=$M_d$] (del)  
  };
  \end{feynman}
\end{tikzpicture}}}
\end{equation}
Using  \eqref{cross2} we can derive a mixed pentagon identity 
\begin{equation} \label{Tdot_pentagon}
  \sum_e\sum_{\delta,\varepsilon,\zeta} \dot{\mathrm{T}}^{(y\,a\,i)c}_{\beta b\alpha,\delta e\varepsilon}\dot{\mathrm{T}}^{(x\,e\,i)d}_{\gamma c\varepsilon,\zeta f\eta}\mathrm{T}^{(x\,y\,a)f}_{\zeta e\delta,\mu u\nu}=\sum_\sigma \mathrm{T}^{(x\,y\,b)d}_{\gamma c\beta,\mu u\sigma}\dot{\mathrm{T}}^{(u\,a\,i)d}_{\sigma b\alpha,\nu f\eta}\,.
\end{equation}
which is illustrated\footnote{To avoid clutter we have omitted the relevant associators on figure \ref{figure3} as they drop from the final result. } 
on figure \ref{figure3}.

\subsubsection{Fusing matrix \texorpdfstring{$\mathrm{F}[A]$ related to boundary OPE}{F[A] related to boundary OPE}} \label{sec_boundaryope}

Here we introduce the fusing matrices corresponding to the tensor structure on $\mathcal{C}_A\times\mathcal{C}\times\mathcal{C}$. The relevant associator
\begin{equation}
    \alpha_{M,U,V}: M\otimes\left(U\otimes V\right)\to \left(M\otimes U\right)\otimes V
\end{equation}
is the associator $\alpha$ of the category $\mathcal{C}$ which is assumed to be strict so that $ \alpha_{M,U,V}$ is trivial. 

Let $M_a$ be a simple object in $\mathcal{C}_A$ and $U_j,U_k$ simple objects in $\mathcal{C}$. We define the fusing matrices $\mathrm{F}[A]$ as the matrices giving the vector space isomorphism 
\begin{equation}
\Hom_A\left(M_a\otimes(U_j\otimes U_k),M_c\right)\cong\Hom_A\left((M_a\otimes U_j)\otimes U_k,M_c\right)
\end{equation}
with the matrix entries defined by
\begin{equation} \label{(1)F}
\vcenter{\hbox{\hspace{-10mm}\begin{tikzpicture}[font=\footnotesize,inner sep=2pt]
  \begin{feynman}
  \vertex (l) at (0,0) [label=above:\(M_c\)] ;
  \vertex [small, dot] (alpha) at (0,-1) [label=left:\footnotesize\(\alpha\)] {};
 \vertex  [small,orange, dot] (beta) at (1,-2) [label=right:\footnotesize\(\beta\)] {};
 \vertex  (i) at (-1,-3) [label=below:\(M_a\)];
 \vertex  (left) at (-0.7,-2);
 \vertex [above=0.5cm of i] (i2);
 \vertex [below left=0.042cm of left] (i3);
 \vertex (j) at (0.4,-3) [label=below:\(U_j\)];
  \vertex (k) at (1.6,-3) [label=below:\(U_k\)];
  \vertex  (j2) at (0.45,-2.6);
  \vertex [above right=0.15cm of j2] (j3);
  \vertex [above right=0.15cm of j3] (j4);
  \vertex  (k2) at (1.55,-2.6);
  \vertex [above left=0.3cm of k2] (k3);
  \draw [thick,rounded corners=1mm] (alpha) -- (left) -- (i2) -- (i);
   \draw [thick,rounded corners=1mm] (beta)  -- (j2) -- (j);
   \draw [thick,rounded corners=1mm] (beta)  -- (k2) -- (k);
    \diagram*{
      (l) --[thick]  (alpha)   ,
      (alpha) -- [thick,edge label=$U_p$] (beta)
    };
  \end{feynman}
\end{tikzpicture}}}
~\hspace{5mm}=\mathlarger{\sum}_{b\in\mathcal{J}_A}\mathlarger{\sum}_{\gamma=1}^{\tensor{A}{_{aj}}{^b}}\mathlarger{\sum}_{\delta=1}^{\tensor{A}{_{bk}}{^c}}\,\mathrm{F}[A]_{\alpha p \beta,\gamma b\delta}^{(a\,j\,k)c}
~ 
\vcenter{\hbox{\hspace{1mm}\begin{tikzpicture}[font=\footnotesize,inner sep=2pt]
  \begin{feynman} 
  \vertex (l) at (0,0) [label=above:\(M_c\)] ;
  \vertex [ small, dot] (del) at (0,-1) [label=left:\(\delta\)] {};
 \vertex  [small,  dot] (gamma) at (-1,-2) [label=right:\(\gamma\)] {};
 \vertex  (i) at (-1.6,-3) [label=below:\(M_a\)];
 \vertex  (i2) at (-1.55,-2.6);
 \vertex [below left=0.03cm of i2] (i3);
 \vertex (j) at (-0.4,-3) [label=below:\(U_j\)];
  \vertex (k) at (1,-3) [label=below:\(U_k\)];
  \vertex (j2) at (-0.45,-2.6) ;
  \vertex [below right=0.03cm of j2] (j3) ;
  \vertex [above=0.5cm of k] (k2);
  \vertex [above left=0.005cm of k2] (k3);
  \vertex (right) at (0.7,-2);
   \draw [thick,rounded corners=1mm] (del) -- (right) -- (k2) -- (k);
    \draw [thick,rounded corners=1mm] (gamma) -- (i2)  -- (i);
     \draw [thick,rounded corners=1mm] (gamma) -- (j2) -- (j);
    \diagram*{
      (l) --[thick]  (del)   ,
      (del) -- [thick, edge label=$M_b$] (gamma)  
  };
  \end{feynman}
\end{tikzpicture}}}\end{equation}
Here the diagrams flow from bottom to top and the black blob vertices denote boundary fields.

We next explain  how this fusing matrix is related to the boundary operator product expansion. Using the notation \eqref{bcco} we define the boundary 
OPE coefficients via the following identity\footnote{As everywhere in the TFT formalism all coordinate dependence is removed and the identities are between morphisms.} between two basis boundary fields $\Psi_{(j,a)b}^\alpha$ and $\Psi_{(k,b)c}^\beta$  
\begin{equation}\label{boundaryOPE}
   \vcenter{\hbox{\hspace{5mm}\begin{tikzpicture}[font=\footnotesize,inner sep=2pt]
  \begin{feynman}
   \path[pattern=north east lines,pattern color=ashgrey,very thin] (0.01,4) rectangle (0.6,0);
  \vertex (a) at (0,0);
  \vertex [above=4cm of a] (b);
  \vertex [small,dot] [above=1.4cm of a] (psi) {};
  \vertex [small,dot] [above=2.6cm of a] (psi1){};
  \vertex[right=0.2cm of psi] (i) ;
  \vertex[below=0.2cm of i] (ii) [label=:\(\;\;\,\Psi_k^\beta\)];
  \vertex[above=0.6cm of i] (p1) [label=left:\(\hspace{-10mm}M_b\)];
   \vertex[right=0.2cm of psi1] (d1) ;
  \vertex[below=0.2cm of d1] (d2) [label=:\(\;\;\,\Psi_j^{\alpha}\)];
   \vertex [above=3.8cm of a] (b1) [label=left:\(M_a\)];
   \vertex [above=0.2cm of a] (b2) [label=left:\(M_c\)];
   \diagram*{
     (a)--[thick] (psi),
     (psi)--[thick] (b)
    };
  \end{feynman}
\end{tikzpicture}}}
= \mathlarger{\sum}_{p\in\mathcal{I}}\mathlarger{\sum}_{\gamma=1}^{\tensor{A}{_{ap}}{^c}}\mathlarger{\sum}_{\delta=1}^{\tensor{N}{_{jk}}{^p}}C^{(a\,b\,c)p\gamma;\delta}_{j\alpha,k\beta}\vcenter{\hbox{\hspace{5mm}\begin{tikzpicture}[font=\footnotesize,inner sep=2pt]
  \begin{feynman}
  \vertex (a) at (0,0);
  \vertex [above=4cm of a] (b);
  \vertex [small,dot] [above=2cm of a] (psi) {};
  \vertex[right=0.2cm of psi] (i) ;
  \vertex[below=0.2cm of i] (ii) [label=:\(\;\;\,\Psi_p^\gamma\)];
   \path[pattern=north east lines,pattern color=ashgrey,very thin] (0.01,4) rectangle (0.6,0);
   \vertex [above=1.6cm of psi] (b1) [label=left:\(M_a\)];
   \vertex [below=1.6cm of psi] (b2) [label=left:\(M_b\)];
   \diagram*{
     (a)--[thick] (psi),
     (psi)--[thick] (b)
    };
  \end{feynman}
\end{tikzpicture}}}
\end{equation}
Here the numbers $C^{(a\,b\,c)p\gamma;\delta}_{j\alpha,k\beta}$ are the boundary OPE coefficients. Comparison with \eqref{(1)F} immediately yields
\begin{equation}\label{OPEcoefgen}
C^{(a\,b\,c)p\gamma;\delta}_{j\alpha,k\beta}= \mathrm{G}[A]^{(a\,j\,k)c}_{\alpha b\beta,\gamma p\delta}
\end{equation}
with $\mathrm{G}[A]$ being the inverse of the $\mathrm{F}[A]$ matrix.
The fusing matrices $\mathrm{F}[A]$ and $\mathrm{G}[A]$ were introduced in \cite{Behrend_etal} where they were denoted as $\prescript{(1)}{}{\mathrm{F}}$ and $\prescript{(3)}{}{\mathrm{F}}$. We use the notation of   \cite{FRS4} where these matrices were introduced in a more general categorical setting 
and the general  relation to boundary OPE coefficients was noted.

\begin{figure}[pt]
\[
     \vcenter{\hbox{\hspace{1mm}\scalebox{0.8}{\begin{tikzpicture}[font=\footnotesize,inner sep=2pt]
  \begin{feynman}
   \path[pattern=north east lines,pattern color=ashgrey,very thin] (0.01,4) rectangle (0.6,0);
 \vertex (j1) at (0,0);
  \vertex [above=0.6 of j1] (j2);
  \vertex[ above =0.7 of j1] (j3);
  \vertex [small,dot] [ above=3cm of j1] (psi)  {};
  \vertex[right=0.2cm of psi] (psi1) ;
  \vertex[below=0.01cm of psi] (psi2) [label=left:\(\,\,\,\,\Psi_i^\alpha\)];
  \vertex [small,dot] [below=1.4cm of psi] (psi4){};
  \vertex[right=0.2cm of psi4] (psi5) ;
  \vertex[below=0.01cm of psi4] (psi6) [label=left:\(\,\,\,\,\Psi_j^\beta\)];
  \vertex [above=3.4cm of j1] (j4) ;
  \vertex[above=3.5cm of j1] (j5)  ;
  \vertex[ left=0.15cm of  j4] (j7);
  \vertex [left=0.15cm of j3] (j8) [label=left:\(\gamma\)]; 
  \vertex [above =4cm of j1] (j6);
  \draw [fill=yellow] (0.06,0.6) rectangle (-0.2, 0.7);
    \vertex[below=0.1cm of j6] (m1) ;
    \vertex[above=2.4cm of j1] (m3) [label=right:\(M_b\)];
    \vertex[above=0.7cm of psi] (m2) [label=right:\(M_a\)];
    \vertex (x) at (-1,1.8) [label=left:\(X_x\)];
    \vertex[above=0.2 of j1] (jj) [label=right:\(M_d\)];
    \vertex[above=1.2cm of j1] (jj0) [label=right:\(M_c\)];
    \vertex [below=0.25cm of x] (x1);
    \vertex [above=2.2cm of x] (x2);
    \draw [thick]    (j8) to[out=90,in=-90] (x1);
    \draw [thick] [middlearrow={latex}]   (x1) to (x2);
    \vertex[above=1.6cm of j1] (jj1);
    \vertex[above=1.7cm of j1] (jj2);
    \vertex[left=0.15cm of jj2] (jj3);
 \vertex [below=0.25cm of y] (y1);
    \vertex [above=0.25cm of y] (y2);
      \draw[->, line width=0.5mm] (1.2,3.7) -- (3.5,4.7);
      \node at (2.5,4) (eq1) {$\mathlarger{\dot{\mathrm{T}}}$};
      \vertex (l1) at (5,3); 
     \vertex[above=0.8cm of l1] (l2);
     \vertex[above=1.9cm of l1] (l3);
     \vertex[small,dot][above=1.2cm of l3] (psi3){};
   \vertex[above=2.7cm of l1] (l4);
   \vertex[above=2.8cm of l1] (l5);
   \vertex[left=0.15cm of l5] (l55);
   \vertex[left=0.15cm of l3] (l33);
   \vertex[below=0.1cm of l3] (l333);
   \vertex[small,dot] [below=0.8cm of l333] (psii){};
   \vertex[above=4cm of l1] (l6);
\path[pattern=north east lines,pattern color=ashgrey,very thin] (5.01,7) rectangle (5.6,3);
 \draw [fill=yellow] (5.06,4.8) rectangle (4.8, 4.9);
     \vertex (yi1) at (4,6.7);
\vertex [below=0.25cm of yi1] (yi2);
    \vertex [above=0.25cm of yi1] (yi3);
 \vertex (xi1) at (4,6);
\vertex [below=0.25cm of xi1] (xi2);
    \vertex [above=1cm of xi1] (xi3);
\draw [thick]    (l33) to[out=90,in=-90] (xi2);
 \draw [thick] [middlearrow={latex}]   (xi2) to (xi3);

      \draw[->, line width=0.5mm] (6.5,4.6) -- (9,3.6);
      \node at (7.5,3.9) (eq1) {$\mathlarger{\dot{\mathrm{T}}}$};
      \vertex (r1) at (11,0); 
 \vertex[above=1.7cm of r1] (r2);
     \vertex[above=1.8cm of r1] (r3);
     \vertex[small,dot][above=0.9cm of r1] (psi4){};
     \vertex[small,dot][above=0.9cm of psi4] (kekw){};
   \vertex[above=2.7cm of r1] (r4);
   \vertex[above=2.8cm of r1] (r5);
   \vertex[left=0.15cm of r5] (r55);
   \vertex[left=0.15cm of r3] (r33);
   \vertex[above=4cm of r1] (r6);
\path[pattern=north east lines,pattern color=ashgrey,very thin] (11.01,4) rectangle (11.6,0);
 \draw [fill=yellow] (11.06,2.7) rectangle (10.8, 2.8);
     \vertex (yi4) at (10,3.7);
\vertex [below=0.25cm of yi4] (yi5);
    \vertex [above=0.3cm of yi4] (yi6);
\draw [thick]    (r55) to[out=90,in=-90] (yi5);
 \draw [thick] [middlearrow={latex}]   (yi5) to (yi6);
 \vertex (xi4) at (10,2.7);
\vertex [below=0.25cm of xi4] (xi5);
    \vertex [above=0.25cm of xi4] (xi6);

     \draw[->, line width=0.5mm]  (9.8,1.6) -- (8.7,0.3) ;
      \node at (8.9,1.1) (eq1) {$\mathlarger{\mathrm{G}[A]}$};
      \vertex (s1) at (9,-4); 
     \vertex[small,dot][above=0.9cm of s1] (psi5){};
     \vertex[above=1.7cm of s1] (s2);
     \vertex[above=2.5cm of s1] (s3);
     \vertex[below=0.1cm of s3] (s333);
     \vertex[left=0.15cm of s3] (s33);
     \vertex  (z1) at (8,-0.7){};
     \vertex[above=0.7cm of z1] (z0);
     \vertex[below=0.15cm of z1] (z111);
\draw [fill=yellow] (9.06,-1.6) rectangle (8.8, -1.5);
\vertex[above=4cm of s1] (s4);
\path[pattern=north east lines,pattern color=ashgrey,very thin] (9.01,0) rectangle (9.6,-4);
 \vertex[above left=0.7cm of z1] (z2);
 \vertex[above right=0.7cm of z1] (z3);
 \vertex[above=0.7cm of z2] (z4);
 \vertex[above=0.7cm of z3] (z5);
\draw [thick]    (s33) to[out=90,in=-90] (z1);
\draw [thick,middlearrow={latex}]    (z111) to (z0);
         
         \vertex (q1) at (3,-4); 
\vertex[small,dot][above=3cm of q1] (psi6){};
     \vertex[above=0.8cm of q1] (q2);
     \vertex[above=0.9cm of q1] (q3);
     \vertex[left=0.15cm of q3] (q33);
     \vertex  (zz1) at (2,-2){};
\draw [fill=yellow] (3.06,-3.2) rectangle (2.8, -3.1);
\vertex[above=4cm of q1] (q4);
\path[pattern=north east lines,pattern color=ashgrey,very thin] (3.01,0) rectangle (3.6,-4);
 \vertex[above left=0.7cm of zz1] (zz2);
 \vertex[above right=0.7cm of zz1] (zz3);
 \vertex[above=0.7cm of zz2] (zz4);
 \vertex[above=0.7cm of zz3] (zz5);
 \vertex[above=2cm of zz1] (zz9);
 \vertex[below=0.15cm of zz1] (zz0);
\draw [thick]    (q33) to[out=90,in=-90] (zz1);
\draw [thick,middlearrow={latex}]    (zz0) to (zz9);
             \draw[->, line width=0.5mm] (4,-2) -- (7,-2);
      \node at (5.5,-1.7) (eq1) {$\mathlarger{\dot{\mathrm{T}}}$};
        \draw[->, line width=0.5mm] (1.2,0.9) -- (2.3,0.1);
      \node at (1.9,0.75) (eq1) {$\mathlarger{\mathrm{G}[B]}$};
        \diagram*{
   (j1) --[thick] (j2),
    (j3) --[thick] (jj2),
    (jj2)--[thick] (psi),
    (psi) --[thick] (j6),
    (l1)--[thick] (l333),
    (l3)--[thick] (psi3),
    (psi3)--[thick] (l4),
    (l5)--[thick] (l6),
    (r1)--[thick] (psi4),
    (psi4)--[thick] (r3),
    (r3)--[thick] (r4),
    (r5)--[thick] (r6),
    (s1)--[thick] (psi5),
    (psi5)--[thick] (s333),
    (s3)--[thick] (s4),
    (q1)--[thick] (q2),
    (q3)--[thick] (psi6),
    (psi6)--[thick] (q4)
    };
  \end{feynman}
\end{tikzpicture}}}}
\]\caption{Mixed pentagon identity for $\mathrm{G}[A],\mathrm{G}[B]$ and $\dot{\mathrm{T}}$ matrices.}\label{mixpentaOPE}
\end{figure}
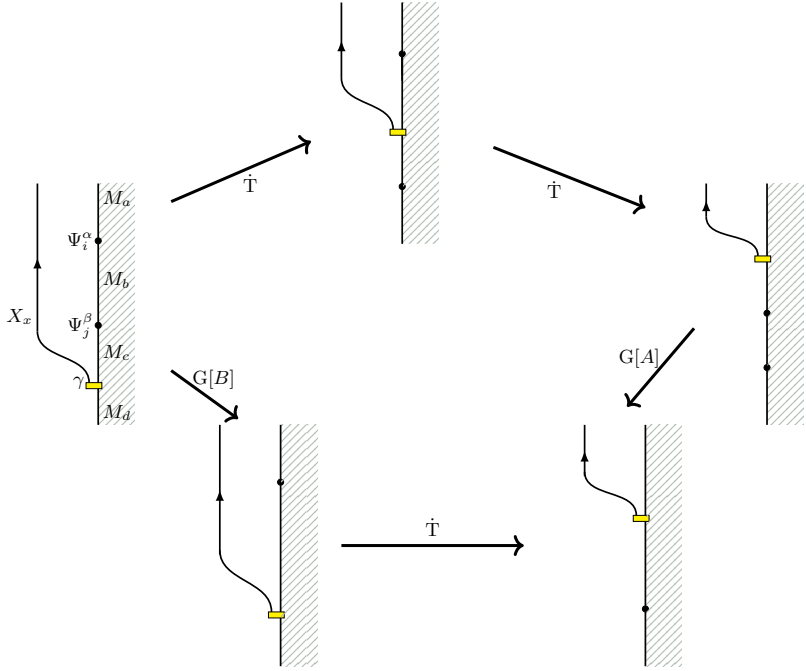

Given a topological  interface between two Frobenius algebras: $A$ and $B$, its junction with a boundary and an insertion of two boundary operators we can pass the interface through the boundary operators before or after the expansion (\ref{boundaryOPE}). The equivalence of these gives rise to the 
mixed pentagon identity depicted in Figure \ref{mixpentaOPE}  between the fusing matrices $\mathrm{G}[A]$,  $\mathrm{G}[B]$ and $\mathrm{\dot{T}}$. In this figure to relief notation we omitted the associators $\dot{\alpha}$ of the form \eqref{associator4}. They drop from the calculation since they obey the corresponding mixed pentagon:
\begin{equation}
\alpha^{-1}_{X\otimes_BM,U,V}\circ\left(\dot{\alpha}_{X,M,U}\otimes\id_V\right)\circ\dot{\alpha}_{X,M\otimes U,V}= \dot{\alpha}_{X,M,U\otimes V}\circ \alpha^{-1}_{M,U,V}
\end{equation}
for any $X\in \operatorname{Obj}(_A\mathcal{C}_B),M\in\operatorname{Obj}(\mathcal{C}_B)$ and $U,V\in\operatorname{Obj}(\mathcal{C})$.
In matrix form the pentagon identity reads 
\begin{equation}\label{mixedpentaOPE2}
    \sum_f\sum_{\zeta,\eta,\nu} \dot{\mathrm{T}}^{(x\,b\,j)d}_{\gamma c\beta,\zeta f\eta}\dot{\mathrm{T}}^{(x\,a\,i)f}_{\zeta b\alpha,\rho e \nu} \mathrm{G}[A]^{(e\,i\,j)d}_{\nu f\eta,\sigma k \varepsilon}=\sum_\delta \mathrm{G}[B] ^{(a\,i\,j)c}_{\alpha b \beta,\delta k \varepsilon}\dot{\mathrm{T}}^{(x\,a\,k)d}_{\gamma c\delta,\rho e \sigma}\,.
\end{equation}

\subsection{Fusing matrices from module and bimodule categories}

In this section we will attempt to bring the  various fusing matrices we introduced  into some system.  To introduce a fusing matrix one has to have a tensor product structure either within a single category or between two or more categories. The tensor product structure comes with   an associator which essentially amounts to an isomorphism between certain morphism spaces  from three objects to one. When the categories at hand have a finite number of isomorphism classes of simple objects,  the associator data can be encoded into a finite matrix which we call the fusing matrix.
We will consider two general classes of examples related to the concepts of a module category and a bimodule category.

Let $(\mathcal{D},\otimes,\mathbf{1})$ be a tensor category. Consider  a right module category $\mathcal{M}$ over $\mathcal{D}$ which comes with a new tensor product
\begin{equation}
    \otimes_{\mathcal{M}}:\;\mathcal{M}\times \mathcal{D}\to \mathcal{M}
\end{equation}
for which  there are  associator isomorphisms defined
\begin{equation}\label{gen1}
     m_{M,U,V}: M\otimes_{\mathcal{M}} (U\otimes V)\to (M\otimes_{\mathcal{M}} U)\otimes_{\mathcal{M}} V, \qquad U,V\in\operatorname{Obj}(\mathcal{D}),\, M\in\operatorname{Obj}(\mathcal{M})
\end{equation}
satisfying the mixed pentagon identity \eqref{modpent1}. We choose representatives for the isomorphism classes of simple objects $U_i,\,i\in\mathcal{I}$ in $\mathcal{D}$ and $M_a,\,a\in\mathcal{J}$ in $\mathcal{M}$ for some finite index sets $\mathcal{I},\mathcal{J}$.  The fusing matrices are then the matrices furnishing the vector space isomorphism
\begin{equation} \label{vsp_is1}
   \Hom_{\mathcal{M}}( (M_a\otimes_{\mathcal{M}} U_i)\otimes_{\mathcal{M}} U_j,M_b)\cong \Hom_{\mathcal{M}}( M_a\otimes_{\mathcal{M}} (U_i\otimes U_j),M_b) \, .
\end{equation}
When $\mathcal{D}$ is a modular tensor category encoding the chiral data of a two-dimensional RCFT and $\mathcal{M}=\mathcal{D}_A$ is the category of left modules over a symmetric special Frobenius algebra $A$ in $\mathcal{D}$ the corresponding  fusing matrices 
are the matrices $\mathrm{F}[A]$ introduced in section \ref{sec_boundaryope}.

 The matrix $\mathrm{T}$ introduced in section \ref{Tmatrix} can be also described as arising from an isomorphism similar to the one  given in (\ref{vsp_is1}).
 To that end we first note  that the construction leading to (\ref{vsp_is1}) can be easily modified to work for the case of a left module category over a tensor category. In the case when we consider topological junctions with defects for a fixed Frobenius algebra $A$ we obtain the fusing matrix  $\mathrm{T}$ 
 by choosing $\mathcal{D}={}_{A}\mathcal{C}_{A}$ and $\mathcal{M}=\mathcal{C}_{A}$ -- the category of left $A$-modules. 
  For the case when we have interfaces between different Frobenius algebras: $A, B,C$ we 
 can replace the tensor category $\mathcal{D}$ by three different 
 abelian categories $\mathcal{D}$, $\mathcal{E}$, $\mathcal{F}$ equipped with a bifunctor 
 \begin{equation}
\otimes_{\mathcal{DE}}^{\mathcal{F}}: \mathcal{D} \times \mathcal{E} \to \mathcal{F}
 \end{equation}
 and replace a single module category $\mathcal{M}$ over $\mathcal{D}$ by three abelian categories  $\mathcal{M}$,  $\mathcal{M}'$,  $\mathcal{M}''$
 and three bifunctors (we work now with left actions) 
 \be
 \otimes_{1}: \mathcal{E} \times \mathcal{M} \to \mathcal{M}' \, , \quad  \otimes_{2}: \mathcal{D} \times \mathcal{M}' \to \mathcal{M}'' \, , \quad 
  \otimes_{3}: \mathcal{F} \times \mathcal{M} \to \mathcal{M}'' \, .
 \ee
  The associator is now a functor 
\begin{equation}\label{gen6}
    \tilde{\alpha}: \;\mathcal{D}\times\mathcal{E} \times \mathcal{M} \to \mathcal{D}\times\mathcal{E} \times \mathcal{M} \, 
\end{equation}
which on the objects is defined by the morphisms 
\be
    m_{X,Y,M}: (X\otimes_{\mathcal{DE}}^{\mathcal{F}} Y)\otimes_{3} \mathcal{M}\to X\otimes_{2} ( Y \otimes_{1} \mathcal{M})
   \ee
    where 
    \be
     X \in\operatorname{Obj}(\mathcal{D}),\, \quad  Y \in\operatorname{Obj}(\mathcal{E}) \, , \quad 
     M\in\operatorname{Obj}(\mathcal{M}) \, .
\ee
To obtain $\mathrm{T}$ we choose  $\mathcal{D}=\prescript{}{A}{\mathcal{C}}_B$, $\mathcal{E}=\prescript{}{B}{\mathcal{C}}_C$, 
$\mathcal{F}=\prescript{}{A}{\mathcal{C}}_C$ and $\mathcal{M}=\mathcal{C}_C$, $\mathcal{M}'=\mathcal{C}_B$, $\mathcal{M}''=\mathcal{C}_A$ where $\mathcal{C}$ is a modular tensor category encoding the chiral data  and $A,B,C$ are symmetric special Frobenius algebras in $\mathcal{C}$. Furthermore, the tensor product structure $\otimes_{\mathcal{DE}}^{\mathcal{F}}$ is given by the tensor product $\otimes_B$ ( see definition  \ref{tensorover}) and the products $\otimes_{1}$, $\otimes_{2}$, $\otimes_{3}$ are given by $\otimes_{C}$, $\otimes_{B}$, $\otimes_{C}$ respectively.

We  proceed by choosing simple objects in the  categories involved. Namely let $X_{x}$, $x\in \mathcal{K}_{AB}$, $Y_{y}$, $y\in \mathcal{K}_{BC}$   be the complete sets 
of simple bimodules in  $\prescript{}{A}{\mathcal{C}}_B$  and $\prescript{}{B}{\mathcal{C}}_C$ respectively. Let $M_{a}$, $a\in \mathcal{J}_C$ be 
a complete set of simple modules in  $\mathcal{C}_C$. Then, using the associator maps and the chosen bases in the relevant morphism spaces, the matrix establishing the 
isomorphism
\begin{equation}
\Hom_A\left((X_x\otimes_BY_y)\otimes_C M_a,M_b\right)\cong\Hom_A\left(X_x\otimes_B(Y_y\otimes_C M_a),M_b\right)
\end{equation}
is the fusing matrix $\mathrm{T}$ (or its inverse, depending on the direction of the isomorphism chosen and the conventions). 

Another general setting in which new fusing matrices arise involves  bimodule categories. Let $(\mathcal{D},\otimes_{\mathcal{D}})$ and  $(\mathcal{E},\otimes_{\mathcal{E}})$ be two tensor categories. A bimodule category $\mathcal{B}$ is a category equipped with both a left action of $\mathcal{D}$ and a right action of $\mathcal{E}$,
\begin{equation}\label{gen3}
    \otimes_{\mathcal{B}}^l:\;\mathcal{D}\times\mathcal{B}\to \mathcal{B} \qquad \text{and}\qquad \otimes_{\mathcal{B}}^r:\; \mathcal{B}\times \mathcal{E}\to \mathcal{B}
\end{equation}
such that $\mathcal{B}$ is a left module category over $\mathcal{D}$, $\mathcal{B}$ is a right module category over $\mathcal{E}$ and the left and right actions commute. The final condition amounts to the existence of an associator  which furnishes the isomorphism
\begin{equation}\label{gen2}
    \dot{\alpha}:\;  \left(U \otimes_{\mathcal{B}}^l X\right) \otimes_{\mathcal{B}}^r R\to U \otimes_{\mathcal{B}}^l\left(X \otimes_{\mathcal{B}}^r R\right) 
\end{equation}
for $U\in\operatorname{Obj}(\mathcal{D}),\,R\in\operatorname{Obj}(\mathcal{E})$ and $X\in\operatorname{Obj}(\mathcal{B})$. To obtain a fusing matrix for this associator $\dot{\alpha}$, one has to choose simple objects in the corresponding categories and then proceed as discussed above . For instance we can choose   $\mathcal{D}=\mathcal{C}, \,\mathcal{E}=\overline{\mathcal{C}}$ (the tensor category dual to $\mathcal{C}$) and $\mathcal{B}=\prescript{}{A}{\mathcal{C}}_A$ with $(\mathcal{C},A)$ being
a pair defining a RCFT in the TFT approach.  From this data we obtain    a fusing matrix $\mathrm{F}[A|A]$ that in a certain basis gives   \cite{FRS4} the OPE coefficients of defect fields including in particular the OPE coefficients of the local bulk fields.

To describe the fusing matrices $\dot{\mathrm{T}}$ introduced in section \ref{dotTmatrix} we first note that in the case when we have defects of a fixed Frobenius algebra $A$ in the symmetry category $\mathcal{C}$ the matrix $\dot{\mathrm{T}}$ corresponds to (\ref{gen2}) with $\mathcal{B}=\mathcal{C}_{A}$, $\mathcal{D}={}_{A}\mathcal{C}_{A}$ 
and $\mathcal{E}=\mathcal{C}$.
For the case of interfaces between different Frobenius algebras we generalise the above setup to the case when 
$ \mathcal{D}$ and  $\mathcal{E}$ are not tensor but some abelian categories which act on a third abelian category $ \mathcal{B}$ by means of two bifunctors: $  \otimes_{\mathcal{B}}^l$ and   $\otimes_{\mathcal{B}}^r$, as in (\ref{gen3}). 
Demanding that the two actions commute up to an isomorphism introduces an associator of the form  \eqref{gen2} which, expressed in the appropriate bases, gives rise to a fusing matrix.

To obtain $\dot {\mathrm{T}}$ we choose  $\mathcal{D}=\prescript{}{A}{\mathcal{C}}_B,\,\mathcal{E}=\mathcal{C}$ and $\mathcal{B}=\mathcal{C}_B$ where $(\mathcal{C},\otimes)$ is the modular tensor category encoding the chiral data of a two-dimensional RCFT while $A,B$ are symmetric special Frobenius algebras in $\mathcal{C}$. Furthermore, choose $\otimes^l_{\mathcal{B}}=\otimes_B$ and $\otimes^r_{\mathcal{B}}=\otimes$. Then the associator $\dot{\alpha}$ introduced in \eqref{associator4} is of the general form \eqref{gen2} for the aforementioned choices and being expressed in bases it gives rise to 
the fusing matrix $\dot {\mathrm{T}}$.

As we have seen above in the case of a single Frobenius algebra $A$ in the symmetry category $\mathcal{C}$ the five fusing matrices: 
$\mathrm{F}$, $\mathrm{F}[A]$, $\mathrm{Y}$, $\mathrm{T}$, $\dot {\mathrm{T}}$ arise from five associativity constraints related to two 
tensor categories $\mathcal{C}$, ${}_{A}\mathcal{C}_{A}$ and the bimodule category $\mathcal{C}_{A}$ over them. One can describe this structure more neatly in terms of two-categories \cite{Picard}.

\subsection{Compact open defects and their fusion algebras} \label{sec_tube}

Let us consider the following composite morphism configuration consisting of two open defect junctions and one boundary field
\begin{equation}\label{tube1}
   \vcenter{\hbox{\hspace{-10mm}\begin{tikzpicture}[font=\footnotesize,inner sep=2pt]
  \begin{feynman}
  \vertex (j1) at (0,0);
  \vertex [above=0.5 of j1] (j2);
  \vertex[ above =0.6 of j1] (j3);
  \vertex [small,dot] [ above=2cm of j1] (psi)  {};
  \vertex[right=0.2cm of psi] (psi1) ;
  \vertex[below=0.2cm of psi1] (psi2) [label=\(\,\,\,\Psi_i^\theta\)];
  \vertex [above=3.4cm of j1] (j4) ;
  \vertex[above=3.5cm of j1] (j5)  ;
  \vertex[ left=0.15cm of  j4] (j7) [label=left:\(\alpha\)];
  \vertex [left=0.15cm of j3] (j8) [label=left:\(\beta\)]; 
  \vertex [above =4cm of j1] (j6);
  \path[pattern=north east lines,pattern color=ashgrey,very thin] (0.01,4) rectangle (0.6,0);
  \draw (0.06,0.5) rectangle (-0.2, 0.6);
    \draw (0.06,3.4) rectangle (-0.2, 3.5);
    \vertex[below=0.1cm of j6] (m1) [label=left:\(M_a\)];
    \vertex[ above=0.1cm of j1] (m4) [label=left:\(M_d\)];
    \vertex[above=1.3cm of j1] (m3) [label=right:\(M_c\)];
    \vertex[above=0.7cm of psi] (m2) [label=right:\(M_b\)];
    \vertex[left=1.2cm of psi] (x) [label=left:\(X_x\)];
    \vertex [below=0.25cm of x] (x1);
    \vertex [above=0.25cm of x] (x2);
    \draw [thick]    (j8) to[out=90,in=-90] (x1);
    \draw [thick]    (x2) to[out=90,in=-90] (j7);
    \draw [thick] [middlearrow={latex}]   (x1) to (x2);
    \draw [thick] (-1.3,2.8) rectangle (-1.7, 3.2);
    \vertex at (-1.5,2.8) [label=\(A\)];
    \draw [thick] (-0.15,1.2) rectangle (-0.55, 1.6);
    \vertex at (-0.35,1.2) [label=\(B\)];
   \diagram*{
    (j1) --[thick] (j2),
    (j3) --[thick] (psi),
    (psi) --[thick] (j4),
    (j5)--[thick] (j6)
  };
  \end{feynman}
\end{tikzpicture}}}
\end{equation}
considered as an element of $\operatorname{Hom}_A(M_a\otimes U_i,M_d)$  where every object is a simple object in its corresponding category. More specifically $X_x\in\prescript{}{A}{\mathcal{C}}_B$, $M_a,M_d\in\mathcal{C}_A$ and $M_b,M_c\in\mathcal{C}_B$. Furthermore, the junctions labelled by $\alpha$ and $\beta$ each represent a basis element in the corresponding morphism spaces, namely $\bar{\Tilde{\Omega}}^\alpha_{(x,b)a}$ and $\tilde \Omega^\beta_{(x,c)d}$ respectively\footnote{We can also consider a similar diagram with the junctions taken as morphisms in the hom-spaces with $\otimes_{B}$ used as in (\ref{jun2}). Such a diagram requires an associator. Using the isomorphisms (\ref{fff})  removes the associator and leads to the diagram in (\ref{tube1}). }.  The sequence of morphisms depicted in  \eqref{tube1}  can be interpreted as a defect acting on a boundary  field which is studied for instance in  \cite{Koj}. 
For a given boundary operator we can calculate this action explicitly by moving the defect past the boundary operator using the matrix $\dot {\mathrm{T}}$ 
and then collapsing the defect bubble using the normalisation equation (\ref{junctions_norm}):
\begin{equation} \label{tube_rep}
    \vcenter{\hbox{\hspace{-10mm}\begin{tikzpicture}[font=\footnotesize,inner sep=2pt]
  \begin{feynman}
   \path[pattern=north east lines,pattern color=ashgrey,very thin] (0.01,4) rectangle (0.6,0);
  \vertex (j1) at (0,0);
  \vertex [above=0.5 of j1] (j2);
  \vertex[ above =0.6 of j1] (j3);
  \vertex [small,dot] [ above=2cm of j1] (psi)  {};
  \vertex[right=0.2cm of psi] (psi1) ;
  \vertex[below=0.2cm of psi1] (psi2) [label=\(\,\,\,\Psi_i^\theta\)];
  \vertex [above=3.4cm of j1] (j4) ;
  \vertex[above=3.5cm of j1] (j5)  ;
  \vertex[ left=0.15cm of  j4] (j7) [label=left:\(\alpha\)];
  \vertex [left=0.15cm of j3] (j8) [label=left:\(\beta\)]; 
  \vertex [above =4cm of j1] (j6);
  \draw (0.06,0.5) rectangle (-0.2, 0.6);
    \draw (0.06,3.4) rectangle (-0.2, 3.5);
    \vertex[below=0.1cm of j6] (m1) [label=left:\(M_a\)];
    \vertex[ above=0.1cm of j1] (m4) [label=left:\(M_d\)];
    \vertex[above=1.3cm of j1] (m3) [label=right:\(M_c\)];
    \vertex[above=0.7cm of psi] (m2) [label=right:\(M_b\)];
    \vertex[left=1.2cm of psi] (x) [label=left:\(X_x\)];
    \vertex [below=0.25cm of x] (x1);
    \vertex [above=0.25cm of x] (x2);
    \draw [thick]    (j8) to[out=90,in=-90] (x1);
    \draw [thick]    (x2) to[out=90,in=-90] (j7);
    \draw [thick] [middlearrow={latex}]   (x1) to (x2);
    \draw [thick] (-1.3,2.8) rectangle (-1.7, 3.2);
    \vertex at (-1.5,2.8) [label=\(A\)];
    \draw [thick] (-0.15,1.2) rectangle (-0.55, 1.6);
    \vertex at (-0.35,1.2) [label=\(B\)];
   \diagram*{
    (j1) --[thick] (j2),
    (j3) --[thick] (psi),
    (psi) --[thick] (j4),
    (j5)--[thick] (j6)
  };
  \end{feynman}
\end{tikzpicture}}}
~=\mathlarger{\sum}_\delta {\dot{\mathrm{T}}}_{\beta c \theta,\alpha a \delta}^{\,(x\,b\, i)d}  ~
\vcenter{\hbox{\hspace{5mm}\begin{tikzpicture}[font=\footnotesize,inner sep=2pt]
  \begin{feynman}
  \path[pattern=north east lines,pattern color=ashgrey,very thin] (0.01,4) rectangle (0.6,0);
  \vertex (a) at (0,0);
  \vertex [above=4cm of a] (b);
  \vertex [small,dot] [above=2cm of a] (psi) {};
  \vertex[right=0.05 of psi] [label=right:\(\Psi_i^\delta\)];
   \vertex [above=1.6cm of psi] (b1) [label=left:\(M_a\)];
   \vertex [below=1.6cm of psi] (b2) [label=left:\(M_d\)];
   \draw [thick] (-1.1,2.2) rectangle (-1.5, 1.8);
    \vertex at (-1.3,1.8) [label=\(A\)];
   \diagram*{
     (a)--[thick] (psi),
     (psi)--[thick] (b)
    };
  \end{feynman}
\end{tikzpicture}}}
\end{equation}

\begin{figure}[H]
\[
     \vcenter{\hbox{\hspace{-10mm}{\scalebox{0.80}{\begin{tikzpicture}[font=\footnotesize,inner sep=2pt]
  \begin{feynman}
  \vertex (j1) at (0,0);
  \vertex [above=0.4 of j1] (j2);
  \vertex[ above =0.5 of j1] (j3);
  \vertex [small,dot] [ above=3cm of j1] (psi) [label=left:\(\Psi_i^\theta\)]  {};
  \vertex[right=0.05 of psi] (kek1) ;
  \vertex [above =6cm of j1] (j6);
  \vertex [below=0.5cm of j6] (j4) ;
  \vertex[below=0.4cm of j6] (j5)  ;
  \vertex[ left=0.15cm of  j4] (j7) ;
  \vertex [left=0.15cm of j3] (j8) ; 
   \path[pattern=north east lines,pattern color=ashgrey,very thin] (0.01,6) rectangle (0.6,0);
  \draw (0.06,0.4) rectangle (-0.2, 0.5);
    \draw  (0.06,5.5) rectangle (-0.2, 5.6);
    \vertex[below=0.1cm of j6] (m1) [label=left:\(M_a\)];
    \vertex[ above=0.1cm of j1] (m4) [label=left:\(M_h\)];
    \vertex[above=2.5cm of j1] (m3) [label=right:\(M_e\)];
    \vertex[above=1.3cm of psi] (m2) [label=right:\(M_c\)];
    \vertex[above=1.7cm of j1] (m5) [label=right:\(M_f\)];
    \vertex[above=0.9cm of j1] (m5) [label=right:\(M_g\)];
    \vertex[left=2.1cm of psi] (x) [label=left:\(X_x\)];
    \vertex [below=0.25cm of x] (x1);
    \vertex [above=0.25cm of x] (x2);
    \draw [thick]    (j8) to[out=90,in=-90] (x1);
    \draw [thick]    (x2) to[out=90,in=-90] (j7);
    \draw [thick] [middlearrow={latex}]   (x1) to (x2);
    \vertex[below=1.2cm of j6] (i1);
    \vertex[below=1.3cm of j6] (i2);
    \vertex [left=0.15cm of i1] (i3) ;
    \vertex [left=0.15cm of i2] (i4) ;
    \vertex [above=1.2cm of j1] (i5) ;
    \vertex[above=1.3cm of j1] (i6) ;
    \vertex[left=0.15cm of i5] (i7) ;
    \vertex[left=0.15cm of i6] (i8);
    \vertex[left=1.4cm of psi] (y) [label=left:\(X_y\)];
    \vertex[below=0.25 of y] (y1);
    \vertex[above=0.25 of y] (y2);
    \vertex[above=2.1cm of psi] (m5) [label=right:\(M_b\)];
    \vertex[above=0.5cm of psi] (m6) [label=right:\(M_d\)];
    \draw (0.06,1.2) rectangle (-0.2, 1.3);
    \draw  (0.06,4.7) rectangle (-0.2, 4.8);
    \draw [thick]    (y2) to[out=90,in=-90] (i4);
    \draw [thick]    (i8) to[out=90,in=-90] (y1);
     \draw [thick] [middlearrow={latex}]   (y1) to (y2);
     \vertex [left=0.8cm of psi] (xx1) [label=left:\(X_z\)];
     \vertex[below=0.25 of xx1] (xx2);
     \vertex[above=0.25 of xx1] (xx3);
     \draw [thick] [middlearrow={latex}]   (xx2) to (xx3);
     \vertex[above=0.9cm of psi] (xx4);
     \vertex[above=1cm of psi] (xx5);
      \vertex[below=0.9cm of psi] (xx6);
      \vertex[below=1cm of psi] (xx7);
     \vertex[left=0.15cm of xx4] (xx8);
     \vertex[left=0.15cm of xx6] (xx9);
     \draw (0.06,3.95) rectangle (-0.2, 4.05);
     \draw  (0.06,2.15) rectangle (-0.2, 2.05);
      \draw [thick]    (xx3) to[out=90,in=-90] (xx8);
       \draw [thick]    (xx9) to[out=90,in=-90] (xx2);
       \draw[->, line width=0.5mm] (0.5,6.5) -- (2,8);
      \node at (1.55,7.1) (eq1) {$\mathlarger{\overline{\mathrm{T}}}\mathrm{T}$};
          \vertex (q1) at (3.5,7.5);
  \vertex [above=0.5 of q1] (q2);
  \vertex[ above =0.6 of q1] (q3);
  \vertex [small,dot] [ above=2.5cm of q1] (psi1) [label=left:\(\Psi_i^\theta\)] {};
  \vertex[right=0.05 of psi1] (kek2);
  \vertex [above =5cm of q1] (q6);
  \vertex [below=0.6cm of q6] (q4) ;
  \vertex[below=0.5cm of q6] (q5)  ;
  \vertex[ left=0.15cm of  q4] (q7) ;
  \vertex [left=0.15cm of q3] (q8) ; 
   \path[pattern=north east lines,pattern color=ashgrey,very thin] (3.51,7.5) rectangle (4.1,12.5);
  \draw (3.56,8) rectangle (3.3, 8.1);
    \draw  (3.56,11.9) rectangle (3.3, 12);
    \vertex[below=0.1cm of q6] (m1) [label=left:\(M_a\)];
    \vertex[ above=0.1cm of q1] (m4) [label=left:\(M_h\)];
    \vertex[above=1.1cm of q1] (m3) [label=right:\(M_f\)];
    \vertex[above=0.5cm of psi1] (m2) [label=right:\(M_d\)];
    \vertex[left=1.8cm of psi1] (xd) [label=left:\(X_u\)];
    \vertex [below=0.25cm of xd] (xd1);
    \vertex [above=0.25cm of xd] (xd2);
    \draw [thick]    (q8) to[out=90,in=-90] (xd1);
    \draw [thick]    (xd2) to[out=90,in=-90] (q7);
    \draw [thick] [middlearrow={latex}]   (xd1) to (xd2);
    \vertex[below=1.5cm of q6] (p1);
    \vertex[below=1.6cm of q6] (p2);
    \vertex [left=0.15cm of p1] (p3) ;
    \vertex [left=0.15cm of p2] (p4) ;
    \vertex [above=1.5cm of q1] (p5) ;
    \vertex[above=1.6cm of q1] (p6) ;
    \vertex[left=0.15cm of p5] (p7) ;
    \vertex[left=0.15cm of p6] (p8);
    \vertex[left=1cm of psi1] (yd) [label=left:\(X_z\)];
    \vertex[below=0.25 of yd] (yd1);
    \vertex[above=0.25 of yd] (yd2);
    \vertex[above=1.5cm of psi1] (m5) [label=right:\(M_c\)];
    \vertex[below=0.5cm of psi1] (m6) [label=right:\(M_e\)];
    \draw (3.56,9) rectangle (3.3, 9.1);
    \draw  (3.56,10.9) rectangle (3.3, 11);
    \draw [thick]    (yd2) to[out=90,in=-90] (p4);
    \draw [thick]    (p8) to[out=90,in=-90] (yd1);
     \draw [thick] [middlearrow={latex}]   (yd1) to (yd2);   
     
      \draw[->, line width=0.5mm] (4.7,7.8) -- (6.2,6.3);
      \node at (5.1,6.9) (eq1) {$\mathlarger{\overline{\mathrm{T}}}\mathrm{T}$};
      \vertex (r1) at (7,0.5);
  \vertex [above=0.5 of r1] (r2);
  \vertex[ above =0.6 of r1] (r3);
  \vertex [small,dot] [ above=2.5cm of r1] (psi2) [label=left:\(\Psi_i^\theta\)] {};
  \vertex[ right=0.05cm of psi2] (kek3) ;
  \vertex [above=4.4cm of r1] (r4) ;
  \vertex[above=4.5cm of r1] (r5)  ;
  \vertex[ left=0.15cm of  r4] (r7);
  \vertex [left=0.15cm of r3] (r8) ; 
  \vertex [above =5cm of r1] (r6);
  \path[pattern=north east lines,pattern color=ashgrey,very thin] (7.01,5.5) rectangle (7.6,0.5);
  \draw  (7.06,1) rectangle (6.8, 1.1);
    \draw  (7.06,4.9) rectangle (6.8, 5);
    \vertex[below=0.1cm of r6] (n1) [label=left:\(M_a\)];
    \vertex[ above=0.1cm of r1] (n4) [label=left:\(M_h\)];
    \vertex[above=1.5cm of r1] (n3) [label=right:\(M_e\)];
    \vertex[above=0.9cm of psi2] (n2) [label=right:\(M_d\)];
    \vertex[left=1.2cm of psi2] (xdd) [label=left:\(X_w\)];
    \vertex [below=0.25cm of xdd] (xdd1);
    \vertex [above=0.25cm of xdd] (xdd2);
    \draw [thick]    (r8) to[out=90,in=-90] (xdd1);
    \draw [thick]    (xdd2) to[out=90,in=-90] (r7);
    \draw [thick] [middlearrow={latex}]   (xdd1) to (xdd2);
     \draw[->, line width=0.5mm] (0.5,-0.5) -- (2,-2);
      \node at (1.4,-0.95) (eq1) {$\mathlarger{\overline{\mathrm{T}}}\mathrm{T}$};
       \vertex (l1) at (3.5,-6.5);
  \vertex [above=0.5 of l1] (l2);
  \vertex[ above =0.6 of l1] (l3);
  \vertex [small,dot] [ above=2.5cm of l1] (psi3) [label=left:\(\Psi_i^\theta\)] {};
  \vertex[ right=0.05cm of psi3] (kek4);
  \vertex [above =5cm of l1] (l6);
  \vertex [below=0.6cm of l6] (l4) ;
  \vertex[below=0.5cm of l6] (l5)  ;
  \vertex[ left=0.15cm of  l4] (l7) ;
  \vertex [left=0.15cm of l3] (l8) ; 
   \path[pattern=north east lines,pattern color=ashgrey,very thin] (3.51,-1.5) rectangle (4.1,-6.5);
  \draw  (3.56,-6) rectangle (3.3, -5.9);
    \draw  (3.56,-2.1) rectangle (3.3, -2);
    \vertex[below=0.1cm of l6] (w1) [label=left:\(M_a\)];
    \vertex[ above=0.1cm of l1] (w4) [label=left:\(M_h\)];
    \vertex[above=1.1cm of l1] (w3) [label=right:\(M_g\)];
    \vertex[above=0.5cm of psi3] (w2) [label=right:\(M_d\)];
    \vertex[left=1.8cm of psi3] (xxd) [label=left:\(X_x\)];
    \vertex [below=0.25cm of xxd] (xxd1);
    \vertex [above=0.25cm of xxd] (xxd2);
    \draw [thick]    (l8) to[out=90,in=-90] (xxd1);
    \draw [thick]    (xxd2) to[out=90,in=-90] (l7);
    \draw [thick] [middlearrow={latex}]   (xxd1) to (xxd2);
    \vertex[below=1.5cm of l6] (s1);
    \vertex[below=1.6cm of l6] (s2);
    \vertex [left=0.15cm of s1] (s3) ;
    \vertex [left=0.15cm of s2] (s4) ;
    \vertex [above=1.5cm of l1] (s5) ;
    \vertex[above=1.6cm of l1] (s6) ;
    \vertex[left=0.15cm of s5] (s7);
    \vertex[left=0.15cm of s6] (s8);
    \vertex[left=1cm of psi3] (ydd) [label=left:\(X_v\)];
    \vertex[below=0.25 of ydd] (ydd1);
    \vertex[above=0.25 of ydd] (ydd2);
    \vertex[above=1.5cm of psi3] (w5) [label=right:\(M_b\)];
    \vertex[below=0.5cm of psi3] (w6) [label=right:\(M_e\)];
    \draw  (3.56,-5) rectangle (3.3, -4.9);
    \draw  (3.56,-3.1) rectangle (3.3, -3);
    \draw [thick]    (ydd2) to[out=90,in=-90] (s4);
    \draw [thick]    (s8) to[out=90,in=-90] (ydd1);
     \draw [thick] [middlearrow={latex}]   (ydd1) to (ydd2);
     \draw[->, line width=0.5mm] (4.7,-1.8) -- (6.2,-0.3);
      \node at (5.25,-0.85) (eq1) {$\mathlarger{\overline{\mathrm{T}}}\mathrm{T}$};
   \diagram*{
    (j1) --[thick] (j2),
    (j3) --[thick] (i5),
    (psi)--[thick] (xx4),
    (i1)--[thick] (j4),
    (j5)--[thick] (j6),
    (i6) --[thick] (xx7),
    (xx6)--[thick] (psi),
    (i2)--[thick] (xx5),
     (q1) --[thick] (q2),
    (q3) --[thick] (p5),
    (p6)--[thick] (psi1),
    (psi1) --[thick] (p2),
    (p1)--[thick] (q4),
    (q5)--[thick] (q6),
    (r1) --[thick] (r2),
    (r3) --[thick] (psi2),
    (psi2) --[thick] (r4),
    (r5)--[thick] (r6),
    (l1) --[thick] (l2),
    (l3) --[thick] (s5),
    (s6)--[thick] (psi3),
    (psi3) --[thick] (s2),
    (s1)--[thick] (l4),
    (l5)--[thick] (l6)
     };
  \end{feynman}
\end{tikzpicture}}}}}
 \]
 \caption{Two distinct ways of calculating the fusion of three compact open defects.}\label{f4}
\end{figure}
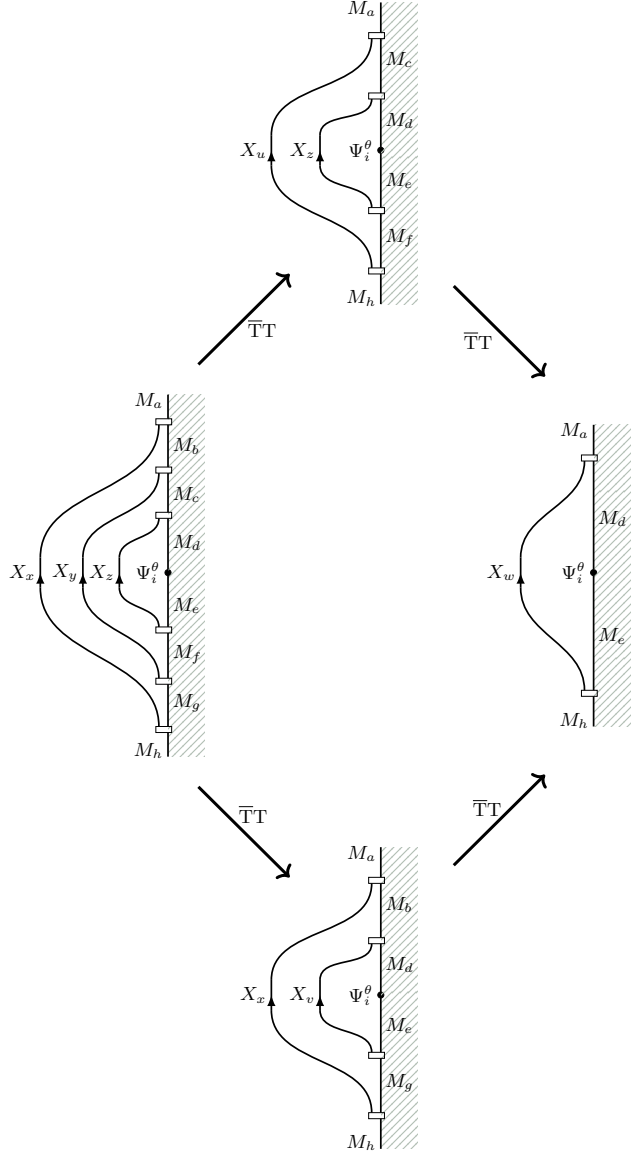

At this stage we are not interested in calculating the action on a boundary field but in the case where we have a second defect line attached to the boundary on both ends. This configuration is depicted graphically as the following morphism
\begin{equation}
     \vcenter{\hbox{\hspace{-10mm}\begin{tikzpicture}[font=\scriptsize,inner sep=2pt]
  \begin{feynman}
  \vertex (j1) at (0,0);
  \vertex [above=0.5 of j1] (j2);
  \vertex[ above =0.6 of j1] (j3);
  \vertex [small,dot] [ above=2.5cm of j1] (psi)[label=left:\(\Psi_i^\theta\)] {};
  \vertex[right=0.05 of psi] (psi1);
  \vertex [above =5cm of j1] (j6);
  \vertex [below=0.6cm of j6] (j4) ;
  \vertex[below=0.5cm of j6] (j5)  ;
  \vertex[ left=0.15cm of  j4] (j7) [label=left:\(\alpha\)];
  \vertex [left=0.15cm of j3] (j8) [label=left:\(\delta\)]; 
   \path[pattern=north east lines,pattern color=ashgrey,very thin] (0.01,5) rectangle (0.6,0);
  \draw  (0.06,0.5) rectangle (-0.2, 0.6);
    \draw  (0.06,4.4) rectangle (-0.2, 4.5);
    \vertex[below=0.1cm of j6] (m1) [label=left:\(M_a\)];
    \vertex[ above=0.1cm of j1] (m4) [label=left:\(M_f\)];
    \vertex[above=1.1cm of j1] (m3) [label=right:\(M_e\)];
    \vertex[above=0.5cm of psi] (m2) [label=right:\(M_c\)];
    \vertex[left=1.8cm of psi] (x) [label=left:\(X_x\)];
    \vertex [below=0.25cm of x] (x1);
    \vertex [above=0.25cm of x] (x2);
    \draw [thick]    (j8) to[out=90,in=-90] (x1);
    \draw [thick]    (x2) to[out=90,in=-90] (j7);
    \draw [thick] [middlearrow={latex}]   (x1) to (x2);
    \vertex[below=1.5cm of j6] (i1);
    \vertex[below=1.6cm of j6] (i2);
    \vertex [left=0.15cm of i1] (i3) [label=left:\(\beta\)];
    \vertex [left=0.15cm of i2] (i4) ;
    \vertex [above=1.5cm of j1] (i5) ;
    \vertex[above=1.6cm of j1] (i6) ;
    \vertex[left=0.15cm of i5] (i7) [label=left:\(\gamma\)];
    \vertex[left=0.15cm of i6] (i8);
    \vertex[left=1cm of psi] (y) [label=left:\(X_y\)];
    \vertex[below=0.25 of y] (y1);
    \vertex[above=0.25 of y] (y2);
    \vertex[above=1.5cm of psi] (m5) [label=right:\(M_b\)];
    \vertex[below=0.5cm of psi] (m6) [label=right:\(M_d\)];
    \draw (0.06,1.5) rectangle (-0.2, 1.6);
    \draw  (0.06,3.4) rectangle (-0.2, 3.5);
    \draw [thick]    (y2) to[out=90,in=-90] (i4);
    \draw [thick]    (i8) to[out=90,in=-90] (y1);
     \draw [thick] [middlearrow={latex}]   (y1) to (y2);
     \draw [thick] (-1.8,1.4) rectangle (-2.2, 1.8);
    \vertex at (-2,1.4) [label=\(A\)];
    \draw [thick] (-1.5,3.7) rectangle (-1.9, 4.1);
    \vertex at (-1.7,3.7) [label=\(B\)];
    \draw [thick] (-0.8,0.6) rectangle (-1.2, 1);
    \vertex at (-1,0.6) [label=\(C\)];
     \draw [dashed,thin] (-1.5,3.7) -- (-0.9,3.3);
      \draw [dashed,thin] (-0.8,1) -- (-0.3,2.2);
   \diagram*{
    (j1) --[thick] (j2),
    (j3) --[thick] (i5),
    (i6)--[thick] (psi),
    (psi) --[thick] (i2),
    (i1)--[thick] (j4),
    (j5)--[thick] (j6)
     };
  \end{feynman}
\end{tikzpicture}}}
\end{equation}
We can represent the last diagram as a linear combination of the diagrams  in (\ref{tube1}) using the $\mathrm{T}$-matrix \eqref{cross4} and its inverse \eqref{Tinv}. This is done via the following diagrammatic equations
\begin{align}\label{fusi1}
   \vcenter{\hbox{\hspace{-10mm}\begin{tikzpicture}[font=\footnotesize,inner sep=2pt]
  \begin{feynman}
   \path[pattern=north east lines,pattern color=ashgrey,very thin] (0.01,5) rectangle (0.6,0);
  \vertex (j1) at (0,0);
  \vertex [above=0.5 of j1] (j2);
  \vertex[ above =0.6 of j1] (j3);
  \vertex [small,dot] [ above=2.5cm of j1] (psi) [label=left:\(\Psi_i^\theta\)] {};
  \vertex[right=0.1 of psi] (psi1);
  \vertex [above =5cm of j1] (j6);
  \vertex [below=0.6cm of j6] (j4) ;
  \vertex[below=0.5cm of j6] (j5)  ;
  \vertex[ left=0.15cm of  j4] (j7) [label=left:\(\alpha\)];
  \vertex [left=0.15cm of j3] (j8) [label=left:\(\delta\)]; 
  \draw  (0.06,0.5) rectangle (-0.2, 0.6);
    \draw  (0.06,4.4) rectangle (-0.2, 4.5);
    \vertex[below=0.1cm of j6] (m1) [label=left:\(M_a\)];
    \vertex[ above=0.1cm of j1] (m4) [label=left:\(M_f\)];
    \vertex[above=1.1cm of j1] (m3) [label=right:\(M_e\)];
    \vertex[above=0.5cm of psi] (m2) [label=right:\(M_c\)];
    \vertex[left=1.8cm of psi] (x) [label=left:\(X_x\)];
    \vertex [below=0.25cm of x] (x1);
    \vertex [above=0.25cm of x] (x2);
    \draw [thick]    (j8) to[out=90,in=-90] (x1);
    \draw [thick]    (x2) to[out=90,in=-90] (j7);
    \draw [thick] [middlearrow={latex}]   (x1) to (x2);
    \vertex[below=1.5cm of j6] (i1);
    \vertex[below=1.6cm of j6] (i2);
    \vertex [left=0.15cm of i1] (i3) [label=left:\(\beta\)];
    \vertex [left=0.15cm of i2] (i4) ;
    \vertex [above=1.5cm of j1] (i5) ;
    \vertex[above=1.6cm of j1] (i6) ;
    \vertex[left=0.15cm of i5] (i7) [label=left:\(\gamma\)];
    \vertex[left=0.15cm of i6] (i8);
    \vertex[left=1cm of psi] (y) [label=left:\(X_y\)];
    \vertex[below=0.25 of y] (y1);
    \vertex[above=0.25 of y] (y2) ;
    \vertex[above=1.5cm of psi] (m5) [label=right:\(M_b\)];
    \vertex[below=0.5cm of psi] (m6) [label=right:\(M_d\)];
    \draw  (0.06,1.5) rectangle (-0.2, 1.6);
    \draw  (0.06,3.4) rectangle (-0.2, 3.5);
    \draw [thick]    (y2) to[out=90,in=-90] (i4);
    \draw [thick]    (i8) to[out=90,in=-90] (y1);
     \draw [thick] [middlearrow={latex}]   (y1) to (y2);
   \diagram*{
    (j1) --[thick] (j2),
    (j3) --[thick] (i5),
    (i6)--[thick] (psi),
    (psi) --[thick] (i2),
    (i1)--[thick] (j4),
    (j5)--[thick] (j6)
     };
  \end{feynman}
\end{tikzpicture}}} ~&=\mathlarger{\sum}_{u,v\in\mathcal{K}_{AC}} \mathlarger{\sum}_{\rho,\sigma,\tau,\varphi} \overline{\mathrm{T}}^{(x\,y\,c)a}_{\rho u\sigma,\alpha b\beta}\,\mathrm{T}^{(x\,y\,d)f}_{\delta e\gamma,\tau v \varphi}~
  \vcenter{\hbox{\hspace{1.5mm}\begin{tikzpicture}[font=\footnotesize,inner sep=2pt]
  \begin{feynman}
   \path[pattern=north east lines,pattern color=ashgrey,very thin] (0.01,5) rectangle (0.6,0);
  \vertex (j1) at (0,0);
  \vertex [above=0.5 of j1] (j2);
  \vertex[ above =0.6 of j1] (j3);
  \vertex [small,dot] [ above=2.5cm of j1] (psi)  {};
  \vertex[right=0.05 of psi] (psi1) [label=right:\(\Psi_i^\theta\)];
  \vertex [above=4.4cm of j1] (j4) ;
  \vertex[above=4.5cm of j1] (j5)  ;
  \vertex[ left=0.15cm of  j4] (j7) [label=left:\(\sigma\)];
  \vertex [left=0.15cm of j3] (j8) [label=left:\(\varphi\)]; 
  \vertex [above =5cm of j1] (j6);
  \draw  (0.06,0.5) rectangle (-0.2, 0.6);
    \draw  (0.06,4.4) rectangle (-0.2, 4.5);
    \vertex[below=0.1cm of j6] (m1) [label=left:\(M_a\)];
    \vertex[ above=0.1cm of j1] (m4) [label=left:\(M_f\)];
    \vertex[above=1.4cm of j1] (m3) [label=right:\(M_d\)];
    \vertex[above=1.1cm of psi] (m2) [label=right:\(M_c\)];
    \vertex[left=1.2cm of psi] (x) ;
    \vertex [empty dot][below=0.75cm of x] (x1)[label=above:\(\tau\)]{};
    \vertex [empty dot][above=0.75cm of x] (x2)[label=below:\(\rho\)]{};
    \draw [thick,middlearrow={latex}]    (j8) to[out=90,in=-90] (x1);
    \draw [thick,middlearrow={latex}]    (x2) to[out=90,in=-90] (j7);
    \vertex (xx) at (-0.55,2.53) [label=left:\(X_y\)];
    \vertex at (-0.85,3.9) [label=\(X_u\)];
    \vertex at (-0.85,0.75) [label=\(X_v\)];
   \diagram*{
    (j1) --[thick] (j2),
    (j3) --[thick] (psi),
    (psi) --[thick] (j4),
    (j5)--[thick] (j6),
    (x1) --[middlearrow={latex},half left,thick, looseness=1.2,edge label=$X_x$] (x2),
     (x1) --[middlearrow={latex},half right,thick, looseness=1.2] (x2)
  };
  \end{feynman}
\end{tikzpicture}}}\nonumber \\[5ex]
~&=\mathlarger{\sum}_{u\in\mathcal{K}_{AC}}\mathlarger{\sum}_{\rho,\sigma,\varphi}\overline{\mathrm{T}}^{(x\,y\,c)a}_{\rho u\sigma,\alpha b\beta}\,\mathrm{T}^{(x\,y\,d)f}_{\delta e\gamma,\rho u\varphi} ~
 \vcenter{\hbox{\hspace{1.5mm}\begin{tikzpicture}[font=\footnotesize,inner sep=2pt]
  \begin{feynman}
   \path[pattern=north east lines,pattern color=ashgrey,very thin] (0.01,5) rectangle (0.6,0);
  \vertex (j1) at (0,0);
  \vertex [above=0.5 of j1] (j2);
  \vertex[ above =0.6 of j1] (j3);
  \vertex [small,dot] [ above=2.5cm of j1] (psi)  {};
  \vertex[right=0.05 of psi] [label=right:\(\Psi_i^\theta\)] (psi1) ;
  \vertex [above=4.4cm of j1] (j4) ;
  \vertex[above=4.5cm of j1] (j5)  ;
  \vertex[ left=0.15cm of  j4] (j7) [label=left:\(\sigma\)];
  \vertex [left=0.15cm of j3] (j8) [label=left:\(\varphi\)]; 
  \vertex [above =5cm of j1] (j6);
  \draw (0.06,0.5) rectangle (-0.2, 0.6);
    \draw  (0.06,4.4) rectangle (-0.2, 4.5);
    \vertex[below=0.1cm of j6] (m1) [label=left:\(M_a\)];
    \vertex[ above=0.1cm of j1] (m4) [label=left:\(M_f\)];
    \vertex[above=1.3cm of j1] (m3) [label=right:\(M_d\)];
    \vertex[above=1.1cm of psi] (m2) [label=right:\(M_c\)];
    \vertex[left=1.2cm of psi] (x) [label=left:\(X_u\)];
    \vertex [below=0.25cm of x] (x1);
    \vertex [above=0.25cm of x] (x2);
    \draw [thick]    (j8) to[out=90,in=-90] (x1);
    \draw [thick]    (x2) to[out=90,in=-90] (j7);
    \draw [thick] [middlearrow={latex}]   (x1) to (x2);
   \diagram*{
    (j1) --[thick] (j2),
    (j3) --[thick] (psi),
    (psi) --[thick] (j4),
    (j5)--[thick] (j6)
  };
  \end{feynman}
\end{tikzpicture}}}
\end{align}
where for the final step we used \eqref{defnorm}. As the boundary field in this relation is arbitrary and the coefficients do not depend on it we can interpret 
this relation as a new kind of fusion of defects attached to a boundary at both ends.
To show  that this fusion  is associative we perform the  fusion  of three such defects via two distinct ways.  This is shown schematically on figure \ref{f4}.

 The diagram on figure \ref{f4} is commutative provided the following tensor identity holds
\begin{equation}\label{asso2}
    \sum_{u}\sum_{\rho,\sigma,\lambda,\varphi} \overline{\mathrm{T}}^{(x\,y\,c)a}_{\rho u\sigma,\alpha b\beta}\,\overline{\mathrm{T}}^{(u\,z\,d)a}_{\varphi w \xi,\sigma c\gamma}\,\mathrm{T}^{(x\,y\,f)h}_{\zeta g\varepsilon,\rho u\lambda}\,\mathrm{T}^{(u\,z\,e)h}_{\lambda f\delta,\varphi w\omega}= \sum_v\sum_{\mu,\nu,\eta,\kappa} \overline{\mathrm{T}}^{(y\,z\,d)b}_{\mu v\nu,\beta c\gamma}\,\overline{\mathrm{T}}^{(x\,v\,d)a}_{\kappa w\xi,\alpha b\nu}\,\mathrm{T}^{(y\,z\,e)g}_{\varepsilon f\delta,\mu v\eta}\,\mathrm{T}^{(x\,v\,e)h}_{\zeta g\eta,\kappa w\omega} \, .
\end{equation}
Using the two pentagon equations \eqref{penT1} and \eqref{penT2} on the left-hand side of \eqref{asso2} we obtain 
\begin{align}\label{asso3}
   \sum_{u}\sum_{\rho,\sigma,\lambda,\varphi} \overline{\mathrm{T}}^{(x\,y\,c)a}_{\rho u\sigma,\alpha b\beta}\,\overline{\mathrm{T}}^{(u\,z\,d)a}_{\varphi w \xi,\sigma c\gamma}\,\mathrm{T}^{(x\,y\,f)h}_{\zeta g\varepsilon,\rho u\lambda}\,\mathrm{T}^{(u\,z\,e)h}_{\lambda f\delta,\varphi w\nu}=& \nonumber\\
    &\hspace{-55mm}= \sum_{\substack{u,\rho,\varphi\\q ,\alpha_1,\alpha_2,\alpha_3\\ p,\beta_1,\beta_2,\beta_3}}\overline{\mathrm{T}}^{(y\,z\,d)b}_{\beta_1 p\beta_2,\beta c\gamma}\overline{\mathrm{T}}^{(x\,p\,d)a}_{\beta_3 w \xi,\alpha b \beta_2}\overline{\mathrm{Y}}^{(x\,y\,z)w}_{\rho u\varphi,\beta_3 p \beta_1}\mathrm{T}^{(y\,z\,e)g}_{\varepsilon f \delta,\alpha_1 q\alpha_2}\mathrm{T}^{(x\,q\,e)h}_{\zeta g \alpha_2,\alpha_3 w \omega}\mathrm{Y}^{(x\,y\,z)w}_{\alpha_3 q\alpha_1,\rho u\varphi} \nonumber \\
 &\hspace{-55mm}= \sum_{\substack{q,\alpha_1,\alpha_2,\alpha_3\\p,\beta_1,\beta_2,\beta_3}} \delta_{\alpha_3,\beta_3}\delta_{p,q}\delta_{\alpha_1,\beta_1}\overline{\mathrm{T}}^{(y\,z\,d)b}_{\beta_1 p\beta_2,\beta c\gamma}\,\overline{\mathrm{T}}^{(x\,p\,d)a}_{\beta_3 w\xi,\alpha b\beta_2}\,\mathrm{T}^{(y\,z\,e)c}_{\varepsilon f\delta,\alpha_1 q\alpha_2}\,\mathrm{T}^{(x\,q\,e)h}_{\zeta g\alpha_2,\alpha_3 w\omega}\nonumber\\
 &\hspace{-55mm}= \sum_{\substack{p,\beta_1,\beta_2\\\beta_3,\alpha_2}} \overline{\mathrm{T}}^{(y\,z\,d)b}_{\beta_1p\beta_2,\beta c\gamma}\,\overline{\mathrm{T}}^{(x\,p\,d)a}_{\beta_3 w\xi,\alpha b \beta_2}\,\mathrm{T}^{(y\,z\,e)g}_{\varepsilon f\delta,\beta_1 p\alpha_2}\,\mathrm{T}^{(x\,p\,e)h}_{\zeta g\alpha_2,\beta_3 w\omega}
\end{align}
which is equal to the right hand side of \eqref{asso2} under suitable renaming of the summation indices, namely $p\rightarrow u,\, \beta_1\rightarrow\mu,\,\beta_2\rightarrow\nu,\,\beta_3\rightarrow \kappa,\,\alpha_2\rightarrow \eta$. In the right-hand side of the first line in \eqref{asso3} we used the fact that the matrices $\mathrm{Y},\overline{\mathrm{Y}}$ are inverses of each other:
\begin{equation}
\sum_q\sum_{\gamma,\delta}\mathrm{Y}^{(x\,y\,z)w}_{\alpha p\beta,\gamma q\delta}\,\overline{\mathrm{Y}}^{(x\,y\,z)w}_{\gamma q\delta,\rho r\sigma}=\delta_{p,r}\delta_{\alpha,\rho}\delta_{\beta,\sigma} \, .
\end{equation}

To lighten the notation we will assume from now on that we work with bimodules over some fixed Frobenius algebra $A$ so that 
$B=C=A$ in \eqref{tube1} and  \eqref{fusi1}. We will later describe how to generalise our considerations to the most general case. 
To proceed note that the boundary field $\Psi^\theta_{(c,i)d}$ plays no role in the fusion calculation  \eqref{fusi1} or in the proof of associativity.
We only needed it to interpret the fusion process as an equation between morphisms.
 This suggests that we can treat    a pair of two open defect junctions
\begin{equation}\label{tube2}
\left(\bar{\Tilde{\Omega}}^\alpha_{(x,b)a},\Tilde{\Omega}^\beta_{(x,c)d}\right) 
\end{equation}
that appears in the configuration represented in \eqref{tube1} as a separate object 
for which, based on \eqref{fusi1}, we  define a fusion product $\ast$  as a  formal linear combination
\begin{equation}\label{tube3}
\left(\bar{\Tilde{\Omega}}^\alpha_{(x,b^\prime)a},\Tilde{\Omega}^\delta_{(x,e^\prime)f}\right)\ast \left(\bar{\Tilde{\Omega}}^\beta_{(y,c)b},\Tilde{\Omega}^\gamma_{(y,d)e}\right)=\delta_{b,b^\prime}\delta_{e,e^\prime}\mathlarger{\sum}_{u\in\mathcal{K}_{\! \!AA}}\mathlarger{\sum}_{\rho,\sigma,\varphi}\overline{\mathrm{T}}^{(x\,y\,c)a}_{\rho u\sigma,\alpha b\beta}\,\mathrm{T}^{(x\,y\,d)f}_{\delta e\gamma,\rho u\varphi} \left(\bar{\Tilde{\Omega}}^\sigma_{(u,c)a},\Tilde{\Omega}^\varphi_{(u,d)f}\right)
\end{equation}
So far we have considered the configurations in which all objects are irreducible. We can replace (\ref{tube1}) by a more general diagram 
\begin{equation}\label{tube1_general}
   \vcenter{\hbox{\hspace{-10mm}\begin{tikzpicture}[font=\footnotesize,inner sep=2pt]
  \begin{feynman}
  \vertex (j1) at (0,0);
  \vertex [above=0.5 of j1] (j2);
  \vertex[ above =0.6 of j1] (j3);
  \vertex [small,dot] [ above=2cm of j1] (psi) [label=left:\( \)] {};
  \vertex[right=0.2cm of psi] (psi1) ;
  \vertex[below=0.15cm of psi1] (psi2) [label=\(\psi\)];
  \vertex [above=3.4cm of j1] (j4) ;
  \vertex[above=3.5cm of j1] (j5)  ;
  \vertex[ left=0.15cm of  j4] (j7) [label=left:\(\bar J\, \)];
  \vertex [left=0.15cm of j3] (j8) [label=left:\(I\, \)]; 
  \vertex [above =4cm of j1] (j6);
  \path[pattern=north east lines,pattern color=ashgrey,very thin] (0.01,4) rectangle (0.6,0);
  \draw  (0.06,0.5) rectangle (-0.2, 0.6);
    \draw  (0.06,3.4) rectangle (-0.2, 3.5);
    \vertex[below=0.1cm of j6] (m1) [label=left:\(Q\)];
    \vertex[ above=0.1cm of j1] (m4) [label=left:\(M\)];
    \vertex[above=1.3cm of j1] (m3) [label=right:\(N\)];
    \vertex[above=0.7cm of psi] (m2) [label=right:\(P\)];
    \vertex[left=1.2cm of psi] (x) [label=left:\(X\)];
    \vertex [below=0.25cm of x] (x1);
    \vertex [above=0.25cm of x] (x2);
    \draw [thick]    (j8) to[out=90,in=-90] (x1);
    \draw [thick]    (x2) to[out=90,in=-90] (j7);
    \draw [thick] [middlearrow={latex}]   (x1) to (x2);
   \diagram*{
    (j1) --[thick] (j2),
    (j3) --[thick] (psi),
    (psi) --[thick] (j4),
    (j5)--[thick] (j6)
  };
  \end{feynman}
\end{tikzpicture}}}
\end{equation}
in which the interface $X$ and  the boundary conditions $P, Q, N, M$ 
are direct sums of the  irreducible ones, the junctions 
\be 
\label{IJ}
\bar J\in {\Hom_{A|\mathbf{1}}}_{(P_{X,P})}(Q,X\otimes P)\, , \quad 
I\in  {\Hom_{A|\mathbf{1}}}_{(P_{X,N})}(X\otimes N,M)
\ee
 and $\psi$ is a generic boundary field. In more detail, let 
\be
 X \cong \bigoplus_{x\in \mathcal{K}_{AA}}n(X,x)X_x
\ee
where $n(X,x) \in {\mathbb Z}$ are the multiplicities of the irreducible components. We fix this isomorphism by choosing the morphisms 
\be
d^{X}_{x, \alpha} \in   \operatorname{Hom}_{A|A}(X, X_{x})\, , \quad d_{X}^{x, \alpha} \in   \operatorname{Hom}_{A|A}( X_{x},X) \, , 
\enspace \alpha=1, \dots , n(X,x)
\ee
that satisfy
\be
d^{X}_{x, \alpha}\circ d_{X}^{y, \beta} = \delta_{x,y} \delta_{\alpha,\beta} {\rm id}_{X_{x}} \, , \qquad 
\sum_{x,\alpha} d_{X}^{x, \alpha}\circ d^{X}_{x, \alpha} = {\rm id}_{X} \, . 
\ee
The junctions (\ref{IJ}) are then represented by the components 
\bea
\bar J_{x,\alpha} &=& \bar J \circ (d_{X}^{x,\alpha}\otimes {\rm id}_{P}) \in   {\Hom_{A|\mathbf{1}}}_{(P_{X_x,P})}(Q,X_{x}\otimes P) \, , \enspace \alpha=1, \dots , n(X,x) \, , 
\nonumber \\
I_{x, \alpha} &=& (d^{X}_{x,\alpha} \otimes {\rm id}_{N})\circ I  \in {\Hom_{A|\mathbf{1}}}_{(P_{X_x,N})}(X_{x}\otimes N,M)  \, , \enspace \alpha=1, \dots , n(X,x) \, .
\eea
It follows from linearity of the compositions and tensor products that the diagrammatic identity \eqref{fusi1} extends  to the case when we have a composition of two configurations of the type depicted in (\ref{tube1_general}). This motivates representing such a configuration by an element 
\bea \label{compact_vs}
&&(\bar J, I) \equiv \sum_{x}\left( \sum_{\alpha} (\bar J_{x,\alpha}, I_{x,\alpha}) \right)\nonumber \\
&& \in \bigoplus_{x\in  \mathcal{K}_{AA}} 
 {\Hom_{A|\mathbf{1}}}_{(P_{X_x,P})}(Q,X_{x}\otimes P)\otimes  {\Hom_{A|\mathbf{1}}}_{(P_{X_x,N})}(X_{x}\otimes N,M) \, .
\eea
We call elements $(\bar J, I)$ defined above   \textit{compact open defects}. 
The space on the right hand side of (\ref{compact_vs}) is a vector space. Moreover, in the case  when the modules $Q,P,N,M$ are direct sums of the irreducible modules $M_{a}$ with multiplicities 1 or 0  the corresponding  space of compact open defects has a basis made of  elementary pairs 
(\ref{tube2}).  Vice versa, the linear combination on the right hand side of 
(\ref{tube3}) can be interpreted as a compact open interface. 
The fusion product $\ast$ defined in (\ref{tube3}) extends to the elements of the form 
$(\bar J, I)$ by linearity. 
The space of all linear combinations of the pairs (\ref{tube2}) with all values of the
indices $x,a,b,c,d,\alpha,\beta$  equipped with the product $\ast$ thus forms an associative 
algebra called the {\it tube algebra} which we denote ${\cal A}_{M^{U}}$. This construction goes back to the work of Ocneanu  \cite{Ocn1}. In general such an algebra can  be associated with a module category $\mathcal{M}$ over a fusion category $\mathcal{E}$ see   \cite{BBW}, \cite{Lin_Tach}, \cite{BullBar}, \cite{Weak_Hopf} for a recent discussion. In our case $\mathcal{E}={}_{A}\mathcal{C}_{A}$, $\mathcal{M}=\mathcal{C}_{A}$.
The tube algebra is finite dimensional and unital. The identity element is given by the linear combination that involves the identity defect
\be
{\bf 1} = \sum_{a,b}  \left(\bar{\Tilde{\Omega}}_{(0,a)a},\Tilde{\Omega}_{(0,b)b}\right) \, .
\ee  
In the semisimple case relevant for RCFTs, the tube algebra decomposes into a direct sum of its minimal ideals each isomorphic to a complete matrix algebra. Moreover, each such ideal can be associated with an irreducible representation of the tube algebra. The space boundary operators $\Psi_{i}$ labelled by a fixed object $U_{j}$ furnish a representation of the tube algebra which is explicitly described in terms of matrices $\dot{\mathrm{T}}$ 
(\ref{tube_rep}). The mixed pentagon identity (\ref{Tdot_pentagon}) ensures the compatibility with associativity in  ${\cal A}_{M^{U}}$. As  in the bulk case 
discussed in \cite{Lin_Tach}, \cite{BullBar}, the tube algebra ${\cal A}_{M^{U}}$ organises the boundary fields into symmetry multiplets with respect to the topological defects. The fusion of compact open defects was studied in \cite{Koj} for the theories with simple fusion rules (i.e. fusion numbers are 0 or 1)  and the charge conjugation modular invariant  ($A={\bf 1}$).

In general the tube algebra ${\cal A}_{M^{U}}$ is rather large and contains many subalgebras. One way to construct smaller associative algebras 
is by fixing the boundary condition.
Take $N=P=Q=M$ in (\ref{tube1_general}) and assume that 
\begin{equation}\label{tube5}
\quad  M \cong \bigoplus_{a\in \mathcal{J}_A^M}M_a
  \end{equation}
where  $\mathcal{J}^M_A\subseteq \mathcal{J}_A$.  This means that the irreducible modules $M_{a}$ have no multiplicities in  $M$. 
A compact open defect attached to $M$ is specified by choosing a defect $X$ and two junctions 
\be
\bar J\in  {\Hom_{A|\mathbf{1}}}_{(P_{X,M})}(M,X\otimes M)\, , \quad 
I\in   {\Hom_{A|\mathbf{1}}}_{(P_{X,M})}(X\otimes M,M) \, .
\ee
Given this data we consider an element 
\be \label{Mspace}
(\bar J, I) \in \bigoplus_{x\in  \mathcal{K}_{AA}} 
 {\Hom_{A|\mathbf{1}}}_{(P_{X_x,M})}(M,X_{x}\otimes M)\otimes {\Hom_{A|\mathbf{1}}}_{(P_{X_x,M})}(X_{x}\otimes M,M) 
\ee
constructed as in (\ref{compact_vs}). For two compact open defects of this type (i.e. both attached to $M$) the fusion product $\ast$ 
is defined by expanding in elementary pairs (\ref{tube2}) and then using (\ref{tube3}). The fusion product (\ref{tube3}) is defined in such 
a way that only irreducible boundary conditions $M_{a}$ with $a\in \mathcal{J}_A^M$ arise so that the result of the fusion is again an element 
of the space on the right hand side of (\ref{Mspace}). We obtain this way an associative algebra we denote  $\mathcal{A}_{M}$.
For the case of minimal models some concrete calculations of such algebras were presented in \cite{Kon1}. 
In sections \ref{sec4.2} and \ref{sec4.3} we give examples  of the algebras $\mathcal{A}_{M}$ for some rational  free  boson theories.

Note that if we choose in the above construction  
\begin{equation}
     M^U \cong \bigoplus_{a\in \mathcal{J}_A}  M_a
\end{equation}
we obtain the tube algebra $\mathcal{A}_{M^{U}}$.

Before we close this section we would like to discuss generalisations of the above constructions. Firstly, to  allow modules
with non-trivial multiplicities of the irreducible components we  consider embedding maps 
\be
e^{a, \gamma}_{M} \in \operatorname{Hom}_A(M,M_{a}) \, , \quad e_{a, \gamma}^{M} \in \operatorname{Hom}_A(M_{a}, M)
\ee
that establish isomorphisms 
\be
M \equiv \bigoplus_{a \in  \mathcal{J}_A}  n(M,a) M_{a} \, .
\ee
Considering for simplicity the case $N=P=Q=M$ in  (\ref{tube1_general}) we can choose a basis in  the space of compact defects of this type 
labelled by the pairs 
\be
   \left(\Bigl[ \bar{ \Tilde{\Omega}}^\alpha_{(x,b)a}\Bigr]_{\gamma}^{\mu} ,\Bigl[ \Tilde{\Omega}^\beta_{(x,c)d}\Bigr]_{\sigma}^{\nu} \right) 
\ee
where 
\be
\Bigl[ \bar{ \Tilde{\Omega}}^\alpha_{(x,b)a}\Bigr]_{\gamma}^{\mu} = ({\rm id}_{X_{x}}\otimes e_{b,\mu})\circ \bar{ \Tilde{\Omega}}^\alpha_{(x,b)a} \circ e^{a, \gamma}_{M} \, , \quad \Bigl[ \Tilde{\Omega}^\beta_{(x,c)d}\Bigr]_{\sigma}^{\nu} = e^{M}_{d,\nu}\circ  \Tilde{\Omega}^\beta_{(x,c)d} \circ ({\rm id}_{X_{x}}\otimes e_{M}^{c, \sigma} )  \, . 
\ee
The multiplication rule (\ref{tube3}) is modified as 
\bea \label{mult2}
&& \left(\Bigl[\bar{\Tilde{\Omega}}^\alpha_{(x,b^\prime)a}\Bigr]_{\mu_1}^{\nu_1},\Bigl[\Tilde{\Omega}^\delta_{(x,e^\prime)f}\Bigr]_{\mu_2}^{\nu_2}\right)\ast \left(\Bigl[\bar{\Tilde{\Omega}}^\beta_{(y,c)b}\Bigr]_{\mu_3}^{\nu_3}, \Bigl[\Tilde{\Omega}^\gamma_{(y,d)e}\Bigr]_{\mu_4}^{\nu_4} \right) \nonumber \\
&& =\delta_{b,b^\prime}\delta_{e,e^\prime}\delta_{\nu_1, \mu_3}\delta_{\mu_2, \nu_4} \mathlarger{\sum}_{u\in\mathcal{K}_{\! \!AA}}\mathlarger{\sum}_{\rho,\sigma,\varphi}\overline{\mathrm{T}}^{(x\,y\,c)a}_{\rho u\sigma,\alpha b\beta}\,\mathrm{T}^{(x\,y\,d)f}_{\delta e\gamma,\rho u\varphi} \left(\Bigl[\bar{\Tilde{\Omega}}^\sigma_{(u,c)a}\Bigr]_{\mu_1}^{\nu_3},\Bigl[\Tilde{\Omega}^\varphi_{(u,d)f}\Bigr]_{\mu_4}^{\nu_2}\right) \, 
\eea

Secondly, we note that the  fusion multiplication $\ast$ and the constructions of the algebras can be extended to bimodules $X$ between different Frobenius algebras. Each interface $X$ and each pair of junctions should be equipped with two more indices labelling the pair of Frobenius algebras. 
For the  pairs representing matching interfaces the multiplication is given by  formulae (\ref{tube3}), (\ref{mult2})  while in all other cases it should be set to zero.  

\section{Calculating fusing matrices} \label{sec_calculation}

In this section we explain the general approach to evaluating the various fusing matrices presented in subsection \ref{sec2.5}. We assume that the fusing matrices of $\mathcal{C}$, the relevant   Frobenious algebras $A,B,C$ and their representations are given. Our goal is to express the fusing matrices $\mathrm{Y},\mathrm{T}, \dot{\mathrm{T}}$ and $\mathrm{F}[A]$ defined in  \eqref{crossf1}, \eqref{cross4.2}, \eqref{cross2.2} and \eqref{(1)F} in terms of this data. The construction of those fusing matrices involves 3-object morphisms (vertices)  $\tilde{\Lambda}_{(x,y)z}^\alpha$, $\tilde{\Omega}_{(x,a)b}^\beta$ and $\Psi^\gamma_{(a,i)b}$ introduced in \eqref{base3}, \eqref{omegatilde} and \eqref{bcco}. Each of these morphisms can be embedded into the hom-spaces of objects in $\mathcal{C}$. The embeddings are described by sets of linear equations. Solving these equations gives a representation of those vertices as linear combinations of  
$\lambda_{(i,j)k}^\alpha$. The 4-object morphisms involved in the definitions of fusion matrices $\mathrm{Y},\mathrm{T}$ and $\dot{\mathrm{T}}$ 
can  then be expressed in terms of two different bases constructed from $\lambda_{(i,j)k}^\alpha$. As the latter are related by the fusing matrices $\mathrm{F}$ of $\mathcal{C}$ we obtain an expression for the new fusing matrices in terms of $\mathrm{F}$ and the embedding maps.  
We are going to  explain in detail how this is implemented for the $\mathrm{Y}$-matrix first and then, in lesser detail, for the matrices $\mathrm{T}, \dot{\mathrm{T}}$ and $\mathrm{F}[A]$.

\subsection{\texorpdfstring{$\mathrm{Y}$-matrix}{Y-matrix}} \label{sec3.1}

The $\mathrm{Y}$-matrices were defined in  \eqref{crossf1} in terms of  the junctions \eqref{base3}
\begin{equation}
     \Tilde{\Lambda}^\alpha_{(x,y)z}\in{\Hom_{A|C}}_{(P_{x,y})}(X_x\otimes X_y,X_z)\,.
\end{equation}
Here $X_x\in \operatorname{Obj}(\prescript{}{A}{\mathcal{C}}_B),\, X_y\in\operatorname{Obj}(\prescript{}{B}{\mathcal{C}}_C),\, X_z\in\operatorname{Obj}(\prescript{}{A}{\mathcal{C}}_C)$ are simple bimodules and  
\bea
  &&  {\Hom_{A|C}}_{(P_{x,y})}(X_x\otimes X_y,X_z)=\{f\in\Hom(\dot{X}_x\otimes\dot{X}_y,\dot{X}_z)\;|\; f\circ P_{x,y}=f, \nonumber \\
    && \hspace{10mm}f\circ \rho_{x\otimes y}=\rho_{z}\circ\left(\operatorname{id}_{A}\otimes f\right),\, f\circ \Tilde{\rho}_{x\otimes y}=\Tilde{\rho}_{z}\circ\left(f\otimes\operatorname{id}_{C}\right)\}
\eea
where $P$ is the projector introduced in \eqref{proj} while $\rho,\tilde{\rho}$ denote the left and right action of an algebra on a bimodule as in definition \ref{bimoduledef}. Therefore, $\tilde{\Lambda}^{\alpha}_{(x,y)z}$ is an element of $\operatorname{Hom}(\dot{X}_x\otimes\dot{X}_y,\dot{X}_z)$ which must satisfy three constraints:
\begin{subequations}\label{condA}
\begin{align}
\label{cona} \Tilde{\Lambda}^\alpha_{(x,y)z}\circ P_{x,y}&=\Tilde{\Lambda}^\alpha_{(x,y)z}\, , \\
 \label{conb} \Tilde{\Lambda}^\alpha_{(x,y)z}\circ \rho_{x\otimes y}&=\rho_z\circ\left(\operatorname{id}_{A}\otimes \Tilde{\Lambda}^\alpha_{(x,y)z}\right)\, , \\
\label{conc} \Tilde{\Lambda}^\alpha_{(x,y)z}\circ \Tilde{\rho}_{x\otimes y}&=\Tilde{\rho}_z\circ\left(\Tilde{\Lambda}^\alpha_{(x,y)z}\otimes \operatorname{id}_{C}\right) \, .
\end{align}
\end{subequations}
To describe the bimodules $X_{x}$ we introduce as in \eqref{emb1} a complete set of  embeddings and restrictions  
\be
b_{x}^{I}\in \mathrm{Hom}(\dot X_{x}, U_{i})\, , \quad  b_{I}^{x} \in \mathrm{Hom}(U_{i}, \dot X_{x}) 
\ee
where $I=(i, \alpha)$ is a composite index containing a label $i$ of a simple object in ${\mathcal{C}}$ and 
$\alpha$ labels the multiplicities of this object in $\dot X_{x}$. 
From now on we will use the notation 
\begin{equation}
    \mathcal{E}^V_i=\dimh(U_i,V) \enspace \mbox{ for any } V\in \mathrm{Obj}(\mathcal{C}) \, .
\end{equation}
Thus $\alpha = 1, \dots,    \mathcal{E}^{X_{x}}_{i} $. Similarly, we introduce such bases for $X_{y}$ and $X_{z}$. 
Using these maps we  write 
\begin{equation}\label{deco0}
    \tilde{\Lambda}^\alpha_{(x,y)z}=\sum_{I,J,K}\sum_{\beta=1}^{\tensor{N}{_{ij}}{^k}}A^{(x,y)z}_{(I,J)K;\alpha\beta}\,b^z_K\circ \lambda^\beta_{(i,j)k}\circ\left(b_x^I\otimes b_y^J\right)
\end{equation}
or graphically
\begin{equation}\label{deco1}
    \vcenter{\hbox{\hspace{2mm}\begin{tikzpicture}[font=\footnotesize,inner sep=2pt]
  \begin{feynman}
\vertex (i1) at (0,0) [label=below:\(X_x\)];
\vertex [right=1.5cm of i1] (j1) [label=below:\(X_y\)];
\vertex [above=0.8cm of i1] (i2);
\vertex [above =0.8cm of j1] (j2);
\vertex  [right=0.75cm of i1] (k1) ;
\vertex [empty dot][above =2.5cm of k1] (k2)[label=right:\(\alpha\)] {};
\vertex [above=1.7cm of k2] (k3) [label=above:\(X_z\)];
   \draw [middlearrow={latex},thick,rounded corners=1mm] (i1)--(i2)--(k2);
      \draw [middlearrow={latex},thick,rounded corners=1mm] (j1)--(j2)--(k2);
   \diagram*{
(k2)--[thick,middlearrow={latex}] (k3)
  };
  \end{feynman}
\end{tikzpicture}}}=\mathlarger{\sum}_{I,J,K}\mathlarger{\sum}_{\beta=1}^{\tensor{N}{_{ij}}{^k}}A^{(x,y)z}_{(I,J)K;\alpha\beta}
\vcenter{\hbox{\begin{tikzpicture}[font=\footnotesize,inner sep=2pt]
  \begin{feynman}
  \vertex (l) at (0,0)  ;
  \vertex [below=1.5cm of l]  (alpha)  {};
 \vertex  [below right=1.5cm of alpha] [small,orange, dot] (beta) [label=below:\(\beta\)] {};
 \vertex[above=0.9cm of beta] (top1);
 \vertex[right=0.05cm of top1] (top3) [label=above right:\(\tau\)];
 \vertex[above=1cm of top1] (top2)[label=above:\(X_z\)];
 \node[isosceles triangle,
    draw,
    rotate=90,
    fill=aqua,
    minimum size =0.25cm] (T4)at (top1){};
 \vertex [below=5cm of l] (bot1);
 \vertex  [left=1cm of bot1] (i) ;
 \vertex [above=1cm of i] (i2);
 \vertex[above=0.8cm of alpha] (a1);
    \vertex[above=1cm of i2] (i3);
 \vertex [right=0.35cm of bot1] (j)  [label=below:\(X_x\)];
  \vertex [right=1.75cm of bot1] (k) [label=below:\(X_y\)];
  \vertex [above=1cm of j] (j11) ;
  \vertex[left=0.05 of j11] (j111) [label=below left:\(\rho\)];
  \vertex[above=1cm of k] (k11);
  \vertex[right=0.05 of k11] (k111) [label=below right:\(\sigma\)];
  \node[isosceles triangle,
    draw,
    rotate=270,
    fill=aqua,
    minimum size =0.25cm] (T2)at (j11){};
    \node[isosceles triangle,
    draw,
    rotate=270,
    fill=aqua,
    minimum size =0.25cm] (T3)at (k11){};
    \vertex[above=0.8cm of j11] (j22) [label=below right:\(U_i\)];
    \vertex[above=0.8cm of k11]  (k22) [label=below right:\(U_j\)];
  \vertex  (j2) at (0.45,-3.6);
  \vertex [above right=0.15cm of j2] (j3);
  \vertex [above right=0.15cm of j3] (j4);
  \vertex  (k2) at (1.55,-3.6);
  \vertex [above left=0.3cm of k2] (k3);
   \draw [middlearrow={latex},thick,rounded corners=1mm] (j)  -- (T2);
   \draw [middlearrow={latex},thick,rounded corners=1mm] (k)  -- (T3) ;
   \draw [thick,rounded corners=1mm] (T2)  -- (j22) -- (beta);
   \draw [thick,rounded corners=1mm] (T3)  -- (k22) -- (beta);
    \diagram*{
      (beta) -- [thick,edge label=$U_{k}$] (T4),
      (T4)--[thick,middlearrow={latex}] (top2)
    };
  \end{feynman}
\end{tikzpicture}}}
\end{equation}
Here $I=(i,\rho),J=(j,\sigma)$ and $K=(k,\tau)$ where $i,j,k$ run through the simple objects of $\mathcal{C}$ and  $\rho,\sigma$ and $\tau$ are the multiplicity labels  $\rho=1,\ldots,\mathcal{E}_{i}^{X_{x}} \, , \, \sigma=1,\ldots, ,\mathcal{E}_{j}^{X_{y}}\, ,\,\tau=1,\ldots, ,\mathcal{E}_{k}^{X_{z}}$.
The complex numbers  $A^{(x,y)z}_{(I,J)K;\alpha\beta}$ are  to be determined by the conditions \eqref{condA}.

To proceed we need to introduce more notation. Let $V\in \mathrm{Obj}(\mathcal{C})$. A composite label  $I=(i,\rho)$ with $\rho\in\{1,\ldots,\mathcal{E}^V_i\}$ 
labels subobjects of $V$ isomorphic to $U_{i}$. We will use the notation 
\be 
I\prec V
\ee  to indicate that $U_i$ is a subobject of $V$ appearing in the direct sum decomposition of $V$ with the specific choice of embedding and restriction $\rho$. We  shall also write  $i\prec V$ to indicate that $U_i$ is a subobject of $V$ without any consideration regarding the choice of the embedding.

Using the decomposition \eqref{deco1} in the conditions \eqref{condA} we obtain component-wise equations between morphisms which after some manipulations can be used to extract numerical linear equations between the complex numbers $A^{(x,y)z}_{(i,j)k;\alpha\beta}$. 
From  \eqref{cona} using (\ref{proj}) we find a set of linear equations 
\begin{align}\label{conagen}
&\forall I\prec X_x,\forall J\prec X_y,\forall S\prec X_z,\forall \phi\in\{1,\ldots,\tensor{N}{_{ij}}{^s}\} 
\nonumber\\
    &A^{(x,y)z}_{(I,J)S;\alpha\phi}=\nonumber\\ 
    & \hspace{5mm} \sum_{K\prec B}\sum_{\nu=1}^{\mathcal{E}^B_{\bar{k}}}\sum_{\substack{P\prec \dot{X}_x\\Q\prec\dot{X}_y}} \sum_{\gamma=1}^{\tensor{N}{_{pq}}{^s}}\sum_{\delta=1}^{\tensor{N}{_{ik}}{^p}}\sum_{\varepsilon=1}^{\tensor{N}{_{\bar{k}j}}{^q}}\sum_{\eta=1}^{\tensor{N}{_{p\bar{k}}}{^i}} \Delta_0^{(k\mu)(\bar{k}\nu)}\tilde{\rho}^{\hspace{0.15mm}x(P);\delta}_{(I)(K)}\rho^{y(Q);\varepsilon}_{(\bar{k}\nu)(J)}\mathrm{F}^{(p\,\bar{k}\,j)s}_{\gamma q\varepsilon,\eta i\phi}\mathrm{G}^{(i\,k\,\bar{k})i}_{\delta p\eta,0}A^{(x,y)z}_{(P,Q)S;\alpha\gamma}
\end{align}
where $\Delta$ are the comultiplication numbers introduced at \eqref{comultiplication}. In \eqref{conagen} we have used the multi-index notation $I=(i,\rho), J=(j,\sigma),P=(p,\zeta), Q=(q,\xi), S=(s,\eta)$ and $K=(k,\mu)$.

From condition \eqref{conb} we obtain 
\begin{align}\label{conbgen}
    &\forall T\prec A,\forall I\prec X_x,\forall J\prec X_y, \forall q\prec X_x,\forall S\prec X_, \nonumber \\
    &\forall \gamma\in\{1,\ldots,\tensor{N}{_{ti}}{^q}\}, \forall \beta\in\{1,\ldots,\tensor{N}{_{qj}}{^s}\} 
    \nonumber\\
&\hspace{10mm}\sum_{\zeta=1}^{\mathcal{E}^{X_x}_q}\rho^{x(Q);\gamma}_{(T)(I)}A^{(x,y)z}_{(Q,J)S;\alpha\beta}=\sum_{\substack{K\prec X_z\\}}\sum_{\delta=1}^{\tensor{N}{_{ij}}{^k}}\sum_{\varepsilon=1}^{\tensor{N}{_{tk}}{^s}}\rho^{z(S);\varepsilon}_{(T)(K)}\mathrm{F}^{(t\,i\,j)s}_{\varepsilon k\delta,\gamma q\beta}A^{(x,y)z}_{(I,J)K;\alpha\delta}\,.
\end{align}

Finally, from condition \eqref{conc} we find the equations
\begin{align}\label{concgen}
   &\forall T\prec C,\forall I\prec X_x,\forall J\prec X_y, \forall q\prec X_z,\forall \delta\in\{1,\ldots,\tensor{N}{_{ij}}{^q}\}, \nonumber \\
   &\forall S\prec X_z,
   \forall\varepsilon\in\{1,\ldots,\tensor{N}{_{qt}}{^s}\} 
   \nonumber \\
   &\hspace{10mm}\sum_{\substack{K\prec X_y\\}}\sum_{\gamma=1}^{\tensor{N}{_{jt}}{^k}}\sum_{\beta=1}^{\tensor{N}{_{ik}}{^s}}\tilde{\rho}^{y(K);\gamma}_{(J)(T)}\mathrm{F}^{(i\,j\,t)s}_{\beta k\gamma,\delta q\varepsilon}A^{(x,y)z}_{(I,K)S;\alpha\beta}=\sum_{\zeta=1}^{\mathcal{E}^{X_z}_q}\Tilde{\rho}^{z(S);\varepsilon}_{(Q)(T)}A^{(x,y)z}_{(I,J)Q;\alpha\delta}\,.
\end{align}
In equations \eqref{conbgen} and \eqref{concgen} we used the multi-index notation $T=(t,\phi), I=(i,\rho), J=(j,\sigma), Q=(q,\zeta), S=(s,\tau)$ and $K=(k,\eta)$.

Suppose now that we have solved the system of equations given in \eqref{conagen}, \eqref{conbgen} and \eqref{concgen}. In order to solve \eqref{crossf1} for the $\mathrm{Y}$-matrix elements we have to compose that equation with a morphism dual to  one of the diagrams appearing on the right-hand side. To find a suitable dual morphism we introduce a basis
\begin{equation}\label{deco2}
\bar{\Tilde{\Lambda}}^{\gamma}_{(x,y)z}=\sum_{I,J,K}\bar{A}^{(x,y)z}_{(I,J)K;\gamma\delta}\left(b_x^I\otimes b_y^J\right)\circ\bar{\lambda}^\delta_{(i,j)k}\circ b_z^K
\end{equation}
that is dual to the $\Tilde{\Lambda}^\alpha_{(x,y)z}$  introduced in \eqref{deco0}, in the sense that the following equation is satisfied 
\begin{equation}\label{Lambdanorm1}
    \Tilde{\Lambda}^{\alpha}_{(x,y)w}\circ {\bar{\Tilde{\Lambda}}}^\gamma_{(x,y)z}=\delta_{\alpha,\gamma}\delta_{z,w}\operatorname{id}_{X_z}\, .
\end{equation}
Using the two decompositions \eqref{deco0} and \eqref{deco2} in \eqref{Lambdanorm1} we obtain the normalisation condition in terms of the coefficients $A$ and $\bar{A}$
\begin{equation}\label{Lambdanorm2}
    \sum_{I,J}\sum_{\beta=1}^{\tensor{N}{_{ij}}{^k}} \sum_{\varphi=1}^{\mathcal{E}_k^{X_z}}A^{(x,y)w}_{(I,J)(k\tau);\alpha\beta}\bar{A}^{(x,y)z}_{(I,J)(k\varphi);\gamma\beta}=\delta_{\alpha,\gamma}\delta_{z,w} \, \enspace  \forall K\prec X_z \, .
\end{equation}
Considering the numbers $A^{(x,y)z}_{(I,J)K;\alpha\beta}$ known since they can be evaluated by the procedure described earlier in this subsection, we will treat \eqref{Lambdanorm2} as a set of linear equations that one can solve\footnote{The existence of solutions is an assumption that is to be checked in any concrete calculation.} to obtain the numbers $\bar{A}_{(I,J)K;\gamma\beta}^{(x,y)z}$.

To evaluate the  $\mathrm{Y}$-matrices we use  the decomposition \eqref{deco1} in both sides of equation  \eqref{crossf1}, perform the category $\mathcal{C}$ $\mathrm{F}$-move on the left hand side  and then compose both sides with the morphism given by the diagram
\begin{equation}
\mathlarger{\sum}_{I^{\prime\prime},J^{\prime\prime},K^{\prime\prime}}\;\mathlarger{\sum}_{Q^{\prime\prime}\prec X_{v^\prime}}\,\mathlarger{\sum}_{S^{\prime\prime}\prec X_w}\,\mathlarger{\sum}_{\theta^{\prime\prime},\xi^{\prime\prime}} \bar{A}^{(v^\prime,z)w}_{(Q^{\prime\prime},K^{\prime\prime})S^{\prime\prime};\delta^\prime\xi^{\prime\prime}} \bar{A}^{(x,y)z}_{(I^{\prime\prime},J^{\prime\prime})Q^{\prime\prime};\gamma^\prime\theta^{\prime\prime}}  \vcenter{\hbox{\hspace{2mm}\begin{tikzpicture}[font=\footnotesize,inner sep=2pt]
   \begin{feynman}
  \vertex (l) at (0,0)  ;
  \vertex[below=0.8cm of l] (w1) [label=below:\(X_w\)];
  \vertex[above=0.5cm of l] [label=right:\(U_{s^{\prime\prime}}\)];
   \vertex[right=0.1cm of l] (phi) [label=right:\(\phi^{\prime\prime}\)];
  \vertex [small,orange,dot] [above=0.8cm of l] (gam) [label=above:\footnotesize\(\,\,\xi^{\prime\prime}\)] {};
  \vertex [small,orange,dot] [above left=1.25cm of gam] (del) [label=above:\footnotesize\(\,\,\theta^{\prime\prime}\)] {} ;
  \vertex [above right=1cm of gam] (k1);
  \vertex [above=1.4cm of k1] (k) ;
  \vertex [above left=0.7cm of del] (i1);
  \vertex [above=0.7cm of i1] (i) ;
   \vertex [above right=0.7cm of del] (j1);
  \vertex [above=0.7cm of j1] (j) ;
   \draw [thick,rounded corners=1mm] (gam) -- (k1) --(k);
     \draw [thick,rounded corners=1mm] (del) -- (j1) --(j);
       \draw [thick,rounded corners=1mm] (del) -- (i1) --(i);
       \vertex[above=0.8cm of i] (i3) [label=above:\(X_x\)];
       \vertex[above=0.8cm of j] (j3) [label=above:\(X_y\)];
       \vertex[above=0.8cm of k] (k3) [label=above:\(X_z\)];
        \vertex[above right=0.1cm of i] (rho) [label=right:\(\rho^{\prime\prime}\)];
        \vertex[above right=0.1cm of j] (sigma) [label=right:\(\sigma^{\prime\prime}\)];
        \vertex[above right=0.1cm of k] (tau) [label=right:\(\tau^{\prime\prime}\)];
        \vertex[below=0.4cm of i] (i4) [label=left:\(U_{i^{\prime\prime}}\)];
         \vertex[below=0.4cm of j] (j4) [label=left:\(U_{j^{\prime\prime}}\)];
         \vertex[below=0.8cm of k] (k4) [label=right:\(U_{k^{\prime\prime}}\)];
       \node[isosceles triangle,
    draw,
    rotate=270,
    fill=aqua,
    minimum size =0.25cm] (T1)at (l){};
    \node[isosceles triangle,
    draw,
    rotate=90,
    fill=aqua,
    minimum size =0.25cm] (T4)at (k){};
    \node[isosceles triangle,
    draw,
    rotate=90,
    fill=aqua,
    minimum size =0.25cm] (T3)at (j){};
    \node[isosceles triangle,
    draw,
    rotate=90,
    fill=aqua,
    minimum size =0.25cm] (T2)at (i){};
    \diagram*{
     (T1)--[thick] (gam),
     (gam) --[thick,edge label= $U_{q^{\prime\prime}}$] (del),
     (w1)--[thick,middlearrow={latex}] (T1),
     (T2)--[thick,middlearrow={latex}] (i3),
     (T3)--[thick,middlearrow={latex}] (j3),
     (T4)--[thick,middlearrow={latex}] (k3)
    };
  \end{feynman}
\end{tikzpicture}}}
\end{equation}
where we used the multi-index notation $I^{\prime\prime}=(i^{\prime\prime},\rho^{\prime\prime}), J^{\prime\prime}=(j^{\prime\prime},\sigma^{\prime\prime}),K^{\prime\prime}=(k^{\prime\prime},\tau^{\prime\prime}), Q^{\prime\prime}=(q^{\prime\prime},\eta^{\prime\prime})$ and $S^{\prime\prime}=(s^{\prime\prime},\phi^{\prime\prime})$. Performing the  composition and using the normalisation equation \eqref{Lambdanorm2} we obtain the following expression for the $\mathrm{Y}$-matrix elements which holds for all $S\prec X_w$:
\begin{equation}\label{Ysymbol}\small
  \mathrm{Y}^{(x\,y\,z)w}_{\alpha u\beta,\gamma v\delta}=\sum_{\substack{I\prec X_x\\ J\prec X_y\\K\prec X_z}} \, \sum_{\substack{P\prec X_u\\Q\prec X_v}}\sum_{\zeta=1}^{\tensor{N}{_{jk}}{^p}}\sum_{\varepsilon=1}^{\tensor{N}{_{ip}}{^s}} \sum_{\theta=1}^{\tensor{N}{_{ij}}{^q}}\sum_{\xi=1}^{\tensor{N}{_{qk}}{^s}}\sum_{\phi^{\prime\prime}=1}^{\mathcal{E}_s^{X_w}}\mathrm{F}^{(i\,j\,k)s}_{\varepsilon p\zeta,\theta q\xi}A^{(y,z)u}_{(J,K)P;\beta\zeta}A^{(x,u)w}_{(I,P)S;\alpha\varepsilon}\bar{A}^{(x,y)v}_{(I,J)Q;\gamma\theta}\bar{A}^{(v,z)w}_{(Q,K)s\phi^{\prime\prime};\delta\xi}
\end{equation} \normalsize
where $S=(s,\phi)$.

This procedure for finding $\mathrm{Y}$ simplifies dramatically in the following case which is relevant for the free boson theory that we will encounter in the next section. Consider a modular tensor category $\mathcal{C}$ with the fusion rules between simple objects $U_i\otimes U_j \cong U_k$. In other words, the tensor product of $U_i$ and $U_j$ has a direct sum decomposition in terms of simple objects containing a single simple object $U_k$ where $k$ is uniquely determined by the input labels $i,j$. Essentially, this means that every object is a simple current. Similarly, let $A$ be a Frobenius algebra in $\mathcal{C}$ and suppose that the tensor product of simple $A$-$A$-bimodules satisfies $X_x\otimes_A X_y\cong X_z$. Finally, we assume the further property that $\dimh(U_i,\dot{X}_x)\in\{0,1\}$. Let us summarize this special case in a set of equations
\begin{subequations}\label{specialcase}
\begin{align}
&U_i\otimes U_j\cong U_k \\
& X_x\otimes_A X_y\cong X_z\\
&\dimh(U_i,\dot{X}_x)\in\{0,1\}
\end{align}
\end{subequations}
Then, in this case, the definition of the $\mathrm{Y}$-matrix elements reads
\begin{equation}\label{freebos1}
    \vcenter{\hbox{\hspace{-10mm}\begin{tikzpicture}[font=\footnotesize,inner sep=2pt]
  \begin{feynman}
  \vertex (l) at (0,0) [label=above:\(X_w\)] ;
  \vertex [empty dot] (alpha) at (0,-1)  {};
 \vertex  [empty dot] (beta) at (1,-2)  {};
 \vertex  (i) at (-1,-4) [label=below:\(X_x\)];
 \vertex  (left) at (-0.7,-3);
 \vertex [above=1cm of i] (i2);
 \vertex [below left=0.042cm of left] (i3);
 \vertex (j) at (0.4,-4) [label=below:\(X_y\)];
  \vertex (k) at (1.6,-4) [label=below:\(X_z\)];
  \vertex [above=0.8cm of j] (j11);
  \vertex[above=0.8cm of k] (k11);
  \vertex  (j2) at (0.45,-3.6);
  \vertex [above right=0.15cm of j2] (j3);
  \vertex [above right=0.15cm of j3] (j4);
  \vertex  (k2) at (1.55,-3.6);
  \vertex [above left=0.3cm of k2] (k3);
  \draw [middlearrow={latex},thick,rounded corners=1mm] (i) -- (i2) -- (alpha);
   \draw [middlearrow={latex},thick,rounded corners=1mm] (j)  -- (j11) -- (beta);
   \draw [middlearrow={latex},thick,rounded corners=1mm] (k)  -- (k11) -- (beta);
    \diagram*{
      (alpha) --[middlearrow={latex},thick]  (l)   ,
      (beta) -- [middlearrow={latex},thick,edge label=$X_u$] (alpha)
    };
  \end{feynman}
\end{tikzpicture}}}
~\hspace{5mm}=\mathrm{Y}_{ u, v}^{(x\,y\,z)w}
~ 
\vcenter{\hbox{\hspace{1mm}\begin{tikzpicture}[font=\footnotesize,inner sep=2pt]
  \begin{feynman}
  \vertex (l) at (0,0) [label=above:\(X_w\)] ;
  \vertex [empty dot] (del) at (0,-1)  {};
 \vertex  [empty dot] (gamma) at (-1,-2)  {};
 \vertex  (i) at (-1.6,-4) [label=below:\(X_x\)];
 \vertex  (i2) at (-1.55,-2.6);
 \vertex [below left=0.03cm of i2] (i3);
 \vertex (j) at (-0.4,-4) [label=below:\(X_y\)];
  \vertex (k) at (1,-4) [label=below:\(X_z\)];
  \vertex (j2) at (-0.45,-2.6) ;
  \vertex [below right=0.03cm of j2] (j3) ;
  \vertex [above=0.5cm of k] (k2);
  \vertex [above left=0.005cm of k2] (k3);
  \vertex (right) at (0.7,-2);
  \vertex[above=0.7cm of i] (i11);
  \vertex[above=0.7cm of j] (j11);
  \vertex[above=1cm of k] (k11);
   \draw [middlearrow={latex},thick,rounded corners=1mm] (k) -- (k11)  -- (del);
    \draw [middlearrow={latex},thick,rounded corners=1mm] (i) -- (i11)  -- (gamma);
     \draw [middlearrow={latex},thick,rounded corners=1mm] (j) -- (j11) -- (gamma);
    \diagram*{
      (del) --[middlearrow={latex},thick]  (l)   ,
      (gamma) -- [middlearrow={latex},thick, edge label=$X_v$] (del)  
  };
  \end{feynman}
\end{tikzpicture}}}
\end{equation}
where we restrict to the case where all the bimodules involved are $A$-$A$-bimodules, they label  topological defects within the same RCFT. The decomposition \eqref{deco1} also becomes simpler and takes the form
\begin{equation}\label{Aconstants1}
    \vcenter{\hbox{\hspace{2mm}\begin{tikzpicture}[font=\footnotesize,inner sep=2pt]
  \begin{feynman}
\vertex (i1) at (0,0) [label=below:\(X_x\)];
\vertex [right=1.5cm of i1] (j1) [label=below:\(X_y\)];
\vertex [above=0.8cm of i1] (i2);
\vertex [above =0.8cm of j1] (j2);
\vertex  [right=0.75cm of i1] (k1) ;
\vertex [empty dot][above =2.5cm of k1] (k2) {};
\vertex [above=1.7cm of k2] (k3) [label=above:\(X_z\)];
   \draw [middlearrow={latex},thick,rounded corners=1mm] (i1)--(i2)--(k2);
      \draw [middlearrow={latex},thick,rounded corners=1mm] (j1)--(j2)--(k2);
   \diagram*{
(k2)--[thick,middlearrow={latex}] (k3)
  };
  \end{feynman}
\end{tikzpicture}}}=\mathlarger{\sum}_{i,j}A^{(x,y)z}_{(i,j)k}
\vcenter{\hbox{\begin{tikzpicture}[font=\footnotesize,inner sep=2pt]
  \begin{feynman}
  \vertex (l) at (0,0)  ;
  \vertex [below=1.5cm of l]  (alpha)  {};
 \vertex  [below right=1.5cm of alpha] [small,orange, dot] (beta) {};
 \vertex[above=0.9cm of beta] (top1);
 \vertex[right=0.05cm of top1] (top3) ;
 \vertex[above=1cm of top1] (top2)[label=above:\(X_z\)];
 \node[isosceles triangle,
    draw,
    rotate=90,
    fill=aqua,
    minimum size =0.25cm] (T4)at (top1){};
 \vertex [below=5cm of l] (bot1);
 \vertex  [left=1cm of bot1] (i) ;
 \vertex [above=1cm of i] (i2);
 \vertex[above=0.8cm of alpha] (a1);
    \vertex[above=1cm of i2] (i3);
 \vertex [right=0.35cm of bot1] (j)  [label=below:\(X_x\)];
  \vertex [right=1.75cm of bot1] (k) [label=below:\(X_y\)];
  \vertex [above=1cm of j] (j11) ;
  \vertex[left=0.05 of j11] (j111) ;
  \vertex[above=1cm of k] (k11);
  \vertex[right=0.05 of k11] (k111);
  \node[isosceles triangle,
    draw,
    rotate=270,
    fill=aqua,
    minimum size =0.25cm] (T2)at (j11){};
    \node[isosceles triangle,
    draw,
    rotate=270,
    fill=aqua,
    minimum size =0.25cm] (T3)at (k11){};
    \vertex[above=0.8cm of j11] (j22) [label=below right:\(U_i\)];
    \vertex[above=0.8cm of k11]  (k22) [label=below right:\(U_j\)];
  \vertex  (j2) at (0.45,-3.6);
  \vertex [above right=0.15cm of j2] (j3);
  \vertex [above right=0.15cm of j3] (j4);
  \vertex  (k2) at (1.55,-3.6);
  \vertex [above left=0.3cm of k2] (k3);
   \draw [middlearrow={latex},thick,rounded corners=1mm] (j)  -- (T2);
   \draw [middlearrow={latex},thick,rounded corners=1mm] (k)  -- (T3) ;
   \draw [thick,rounded corners=1mm] (T2)  -- (j22) -- (beta);
   \draw [thick,rounded corners=1mm] (T3)  -- (k22) -- (beta);
    \diagram*{
      (beta) -- [thick,edge label=$U_{k}$] (T4),
      (T4)--[thick,middlearrow={latex}] (top2)
    };
  \end{feynman}
\end{tikzpicture}}}
\end{equation}
where recall that $k$ is determined by $i$ and $j$. Using this decomposition in the definition \eqref{freebos1} and performing an $\mathrm{F}$-move on the left-hand side we can compare the resulting morphisms component by component and obtain
\begin{equation}\label{yboson}
    \mathrm{Y}^{(x\,y\,z)w}_{u,v}=\frac{A^{(y,z)u}_{(j,k)p}A^{(x,u)w}_{(i,p)s}}{A^{(x,y)v}_{(i,j)q}A^{(v,z)w}_{(q,k)s}}\mathrm{F}^{(i\,j\,k)s}_{p,q}, \enspace \;  \forall i\prec X_x,\forall j\prec X_y,\forall k\prec X_z
\end{equation}

The matrix elements $\mathrm{Y}^{(x\,y\,z)w}_{u,v}$ are non-zero only if the labels $w,u,v$ are the unique labels that satisfy 
$$
X_w\cong \left( X_x\otimes_AX_y\right)\otimes_A X_z, \quad X_u\cong X_y\otimes_A X_z \quad \text{and}\quad X_v\cong X_x\otimes_A X_y
$$
and $s,p,q$ are the unique labels that satisfy
\begin{equation}\label{allow1}
U_s\cong U_i\otimes U_j\otimes U_k,\quad U_p\cong U_j\otimes U_k \quad \text{and} \quad U_q\cong U_i\otimes U_j
\end{equation}
These selection rules stem from the fusion rules \eqref{specialcase} and they demand that none of the junction spaces appearing in the definition \eqref{freebos1} is zero-dimensional. For these reasons, if we work in the special case of the form \eqref{specialcase} we will usually omit the labels $w,u,v$ when presenting the non-zero elements $\mathrm{Y}^{(x\,y\,z)w}_{u,v}$.

\subsection{\texorpdfstring{$\mathrm{T}$-matrix}{T matrix}}\label{sec3.2}

To calculate the  matrix elements of $\mathrm{T}$ we are going to use the same procedure as the one described in the previous subsection, hence we will omit some of the details that we have already been through. 

The starting point is the definition \eqref{cross4.2} where we notice  the first difference from the previous subsection, namely that it involves one junction of type $\Tilde{\Omega}$ introduced in \eqref{omegatilde}. The second difference is that instead of a $C$-$D$ simple bimodule $X_z$ we now have a simple $C$-module $M_a$. Nevertheless, the fact that $X_z$ has a right module structure was never used  in the previous subsection. The first step is to decompose the $\Tilde{\Omega}$ morphisms in terms of $\lambda$ morphisms. Recall that
\begin{equation}
   \Tilde{\Omega}^\alpha_{(x,a)b}\in \Hom_{A|\mathbf{1}_{(P_{x,a})}}(X_x\otimes M_a,M_b)
\end{equation}
where
\begin{equation}\label{omega1}
     \Hom_{A|\mathbf{1}_{(P_{x,a})}}(X_x\otimes M_a,M_b)=\{f\in\Hom(\dot{X}_x\otimes \dot{M}_a,\dot{M}_b)\;|\; f\circ P_{x,a}=f,\, f\circ \rho_{x\otimes a}= \rho_b\circ (\operatorname{id}_{A}\otimes f)\}\,.
\end{equation}
Hence, the basis $\Tilde{\Omega}^\alpha_{(xa)b}$ of $  \Hom_{A|\mathbf{1}_{(P_{x,a})}}(X_x\otimes M_a,M_b)$ must satisfy two conditions
\begin{subequations}\label{condB}
\begin{align}
 \Tilde{\Omega}^\alpha_{(x,a)b}\circ P_{x,a}&=\Tilde{\Omega}^\alpha_{(x,a)b}\\
  \Tilde{\Omega}^\alpha_{(x,a)b}\circ \rho_{x\otimes a}&=\rho_b\circ\left(\operatorname{id}_{A}\otimes \Tilde{\Omega}^\alpha_{(x,a)b}\right)\,.
\end{align}
\end{subequations}
We can express these bases $\Tilde{\Omega}^\alpha_{(x,a)b}$ in terms of  $\lambda^\beta_{(i,j)k}$ as
\begin{equation}\label{deco4}
    \vcenter{\hbox{\hspace{2mm}\begin{tikzpicture}[font=\footnotesize,inner sep=2pt]
  \begin{feynman}
\vertex (i1) at (0,0) [label=below:\(X_x\)];
\vertex [right=1.5cm of i1] (j1) [label=below:\(M_a\)];
\vertex [above=0.8cm of i1] (i2);
\vertex [above =0.8cm of j1] (j2);
\vertex  [right=0.75cm of i1] (k1) ;
\vertex [above =2.5cm of k1] (k2)[label=right:\(\alpha\)] {};
\path (k2) node [shape=rectangle,draw] {};
\vertex [above=1.7cm of k2] (k3) [label=above:\(M_b\)];
   \draw [middlearrow={latex},thick,rounded corners=1mm] (i1)--(i2)--(k2);
      \draw [thick,rounded corners=1mm] (j1)--(j2)--(k2);
   \diagram*{
(k2)--[thick] (k3)
  };
  \end{feynman}
\end{tikzpicture}}}=\mathlarger{\sum}_{I,J,K}\mathlarger{\sum}_{\beta=1}^{\tensor{N}{_{ij}}{^k}}B^{(x,a)b}_{(I,J)K;\alpha\beta}
\vcenter{\hbox{\begin{tikzpicture}[font=\footnotesize,inner sep=2pt]
  \begin{feynman}
  \vertex (l) at (0,0)  ;
  \vertex [below=1.5cm of l]  (alpha)  {};
 \vertex  [below right=1.5cm of alpha] [small,orange, dot] (beta) [label=below:\(\beta\)] {};
 \vertex[above=0.9cm of beta] (top1);
 \vertex[right=0.05cm of top1] (top3) [label=above right:\(\tau\)];
 \vertex[above=1cm of top1] (top2)[label=above:\(M_b\)];
 \node[isosceles triangle,
    draw,
    rotate=90,
    fill=aqua,
    minimum size =0.25cm] (T4)at (top1){};
 \vertex [below=5cm of l] (bot1);
 \vertex  [left=1cm of bot1] (i) ;
 \vertex [above=1cm of i] (i2);
 \vertex[above=0.8cm of alpha] (a1);
    \vertex[above=1cm of i2] (i3);
 \vertex [right=0.35cm of bot1] (j)  [label=below:\(X_x\)];
  \vertex [right=1.75cm of bot1] (k) [label=below:\(M_a\)];
  \vertex [above=1cm of j] (j11) ;
  \vertex[left=0.05 of j11] (j111) [label=below left:\(\rho\)];
  \vertex[above=1cm of k] (k11);
  \vertex[right=0.05 of k11] (k111) [label=below right:\(\sigma\)];
  \node[isosceles triangle,
    draw,
    rotate=270,
    fill=aqua,
    minimum size =0.25cm] (T2)at (j11){};
    \node[isosceles triangle,
    draw,
    rotate=270,
    fill=aqua,
    minimum size =0.25cm] (T3)at (k11){};
    \vertex[above=0.8cm of j11] (j22) [label=below right:\(U_i\)];
    \vertex[above=0.8cm of k11]  (k22) [label=below right:\(U_j\)];
  \vertex  (j2) at (0.45,-3.6);
  \vertex [above right=0.15cm of j2] (j3);
  \vertex [above right=0.15cm of j3] (j4);
  \vertex  (k2) at (1.55,-3.6);
  \vertex [above left=0.3cm of k2] (k3);
   \draw [middlearrow={latex},thick,rounded corners=1mm] (j)  -- (T2);
   \draw [thick,rounded corners=1mm] (k)  -- (T3) ;
   \draw [thick,rounded corners=1mm] (T2)  -- (j22) -- (beta);
   \draw [thick,rounded corners=1mm] (T3)  -- (k22) -- (beta);
    \diagram*{
      (beta) -- [thick,edge label=$U_{k}$] (T4),
      (T4)--[thick] (top2)
    };
  \end{feynman}
\end{tikzpicture}}}
\end{equation}
for some complex numbers $B^{(x,a)b}_{(I,J)K;\alpha\beta}$ . 
Using this decomposition (\ref{deco4}) in the two conditions \eqref{condB} we obtain two equations involving $B^{(x,a)b}_{(I,J)K;\alpha\beta}$  which are identical to those in the previous subsection, namely \eqref{conagen} and \eqref{conbgen}. This is due to the fact that we never used the right action of $C$ on the bimodule $X_y$ so that all the manipulations we did to arrive at \eqref{conagen} and \eqref{conbgen} are allowed and can  be used in \eqref{condB}. Note that some adjustments to the notation in \eqref{conagen} and \eqref{conbgen} must be made, namely we have to replace  $X_y,X_z$ with $M_a,M_b$ respectively as well as replace the numbers $A$ with $B$. After doing that, the two equations derived from \eqref{condB} have the exact same form as \eqref{conagen} and \eqref{conbgen}. These  provide a system of linear equations to determine the coefficients  $B^{(x,a)b}_{(I,J)K;\alpha\beta}$.

Suppose now that we have obtained $B^{(x,a)b}_{(I,J)K;\alpha\beta}$. To find  the $\mathrm{T}$-matrix elements  we first introduce the numbers $\bar{B}^{(x,a)b}_{(I,J)K;\alpha\beta}$ in the same way as the $\bar{A}$ coefficients are introduced in \eqref{deco2} as the coefficients in the decomposition of the dual morphism $\bar{\Tilde{\Omega}}^\alpha_{(x,a)b}$ in terms of $\bar{\lambda}^\beta_{(i,j)k}$. The numbers $\bar{B}$ can be obtained from the normalisation condition
\begin{equation}
    \Tilde{\Omega}_{(x,a)b}^\alpha\circ \bar{\Tilde{\Omega}}^\beta_{(x,a)b}=\delta_{\alpha,\beta}\operatorname{id}_{M_b} \, .
\end{equation}
Furthermore, using the decomposition \eqref{deco4} in both sides of  \eqref{cross4.2} and composing both sides with a suitable morphism involving the coefficients $\bar{B}^{(x,a)b}_{(I,J)K;\alpha\beta}$ we  find the following expression 
\begin{equation}\small
  \mathrm{T}^{(x\,y\,a)c}_{\alpha b\beta,\gamma v\delta}=\sum_{\substack{I\prec X_x\\ J\prec X_y\\K\prec M_a}} \, \sum_{\substack{P\prec M_b\\Q\prec X_v}}\sum_{\zeta=1}^{\tensor{N}{_{jk}}{^p}}\sum_{\varepsilon=1}^{\tensor{N}{_{ip}}{^s}}\sum_{\theta=1}^{\tensor{N}{_{ij}}{^q}}\sum_{\xi=1}^{\tensor{N}{_{qk}}{^s}}\sum_{\phi^{\prime\prime}=1}^{\mathcal{E}_s^{M_c}}\mathrm{F}^{(i\,j\,k)s}_{\varepsilon p\zeta,\theta q\xi}B^{(y,a)b}_{(J,K)P;\beta\zeta}B^{(x,b)c}_{(I,P)S;\alpha\varepsilon}\bar{A}^{(x,y)v}_{(I,J)Q;\gamma\theta}\bar{B}^{(v,a)c}_{(Q,K)s\phi^{\prime\prime};\delta\xi}
\end{equation} \normalsize
that holds for any  $S=(s,\phi)\prec M_c$.

In the special case described in \eqref{specialcase}, with the additional assumptions
\begin{subequations}\label{specialcase1}
\begin{align}
    &X_x\otimes_A M_a\cong M_b\\
    & \operatorname{dim}\operatorname{Hom}(U_i,M_a)\in\{0,1\}
    \end{align}
\end{subequations}
where $b$ is uniquely determined by the input labels $x$ and $a$ we obtain
\begin{equation}\label{tboson}
 \quad\mathrm{T}^{(x\,y\,a)c}_{b,v}=\frac{B^{(y,a)b}_{(j,k)p}B^{(x,b)c}_{(i,p)s}}{A^{(x,y)v}_{(i,j)q}B^{(v,a)c}_{(q,k)s}}\mathrm{F}^{(i\,j\,k)s}_{p,q},
 \enspace \;    \forall i\prec X_x,\forall j\prec X_y,\forall k\prec M_a\,.
\end{equation}
The matrix elements $\mathrm{T}^{(x\,y\,a)c}_{b,v}$ are non-zero only if the labels $c,b,v$ are the unique labels that satisfy 
$$
M_c\cong \left( X_x\otimes_AX_y\right)\otimes_A M_a, \quad M_b\cong X_y\otimes_A M_a \quad \text{and}\quad X_v\cong X_x\otimes_A X_y
$$
and $s,p,q$ are the unique labels that satisfy
$$
U_s\cong U_i\otimes U_j\otimes U_k,\quad U_p\cong U_j\otimes U_k \quad \text{and} \quad U_q\cong U_i\otimes U_j\,.
$$
For these reasons, if we work in this special case where \eqref{specialcase} and \eqref{specialcase1} hold, we will usually omit the labels $c,b,v$ when presenting the non-zero elements $\mathrm{T}^{(x\,y\,a)c}_{b,v}$.

\subsection{\texorpdfstring{$\dot{\mathrm{T}}$-matrix}{\.T matrix}} \label{sec3.3}

To obtain the fusing matrix $\dot{\mathrm{T}}$ defined in \eqref{cross2.2}  we proceed in the same way as we did in the two previous subsections. We first decompose the junction morphisms of the form $\Psi^\alpha_{(a,j)b}$ introduced in \eqref{bcco} in terms of the morphisms $\lambda^\beta_{(i,j)k}$ as
\begin{equation}\label{deco5}
    \vcenter{\hbox{\hspace{2mm}\begin{tikzpicture}[font=\footnotesize,inner sep=2pt]
  \begin{feynman}
\vertex (i1) at (0,0) [label=below:\(M_a\)];
\vertex [right=1.5cm of i1] (j1) [label=below:\(U_j\)];
\vertex [above=0.8cm of i1] (i2);
\vertex [above =0.8cm of j1] (j2);
\vertex  [right=0.75cm of i1] (k1) ;
\vertex [above =2.5cm of k1][dot] (k2)[label=right:\(\alpha\)] {};
\vertex [above=1.7cm of k2] (k3) [label=above:\(M_b\)];
   \draw [thick,rounded corners=1mm] (i1)--(i2)--(k2);
      \draw [thick,rounded corners=1mm] (j1)--(j2)--(k2);
   \diagram*{
(k2)--[thick] (k3)
  };
  \end{feynman}
\end{tikzpicture}}}=\mathlarger{\sum}_{I,K}\mathlarger{\sum}_{\beta=1}^{\tensor{N}{_{ij}}{^k}}C^{(a,j)b}_{(I)K;\alpha\beta}
\vcenter{\hbox{\begin{tikzpicture}[font=\footnotesize,inner sep=2pt]
  \begin{feynman}
  \vertex (l) at (0,0)  ;
  \vertex [below=1.5cm of l]  (alpha)  {};
 \vertex  [below right=1.5cm of alpha] [small,orange, dot] (beta) [label=below:\(\beta\)] {};
 \vertex[above=0.9cm of beta] (top1);
 \vertex[right=0.05cm of top1] (top3) [label=above right:\(\tau\)];
 \vertex[above=1cm of top1] (top2)[label=above:\(M_b\)];
 \node[isosceles triangle,
    draw,
    rotate=90,
    fill=aqua,
    minimum size =0.25cm] (T4)at (top1){};
 \vertex [below=5cm of l] (bot1);
 \vertex  [left=1cm of bot1] (i) ;
 \vertex [above=1cm of i] (i2);
 \vertex[above=0.8cm of alpha] (a1);
    \vertex[above=1cm of i2] (i3);
 \vertex [right=0.35cm of bot1] (j)  [label=below:\(M_a\)];
  \vertex [right=1.75cm of bot1] (k) [label=below:\(U_j\)];
  \vertex [above=1cm of j] (j11) ;
  \vertex[left=0.05 of j11] (j111) [label=below left:\(\rho\)];
  \vertex[above=1cm of k] (k11);
  \node[isosceles triangle,
    draw,
    rotate=270,
    fill=aqua,
    minimum size =0.25cm] (T2)at (j11){};
    \vertex[above=0.8cm of j11] (j22) [label=below right:\(U_i\)];
    \vertex[above=0.8cm of k11]  (k22) ;
  \vertex  (j2) at (0.45,-3.6);
  \vertex [above right=0.15cm of j2] (j3);
  \vertex [above right=0.15cm of j3] (j4);
  \vertex  (k2) at (1.55,-3.6);
  \vertex [above left=0.3cm of k2] (k3);
   \draw [thick,rounded corners=1mm] (j)  -- (T2);
   \draw [thick,rounded corners=1mm] (k)  -- (T3) ;
   \draw [thick,rounded corners=1mm] (T2)  -- (j22) -- (beta);
   \draw [thick,rounded corners=1mm] (k)  -- (k22) -- (beta);
    \diagram*{
      (beta) -- [thick,edge label=$U_{k}$] (T4),
      (T4)--[thick] (top2)
    };
  \end{feynman}
\end{tikzpicture}}}
\end{equation}
where $C^{(a,j)b}_{(I)K;\alpha\beta}$ are complex numbers to be determined by the condition of $\Psi^\alpha_{(a,j)b}$ being an $A$-module intertwiner:
\begin{equation}
    \Psi^\alpha_{(a,j)b}\circ \rho_{a\otimes j}=\rho_b\circ\left(\operatorname{id}_A\otimes \Psi^\alpha_{(a,j)b}\right)
\end{equation}
Using \eqref{deco5} in the last equation leads to a system of linear equations:
\begin{align}\label{psicongen}
    &\forall T\prec A,\forall I\prec M_a, \forall q\prec M_a,\forall S\prec M_b, \nonumber \\
    &\forall \gamma\in\{1,\ldots,\tensor{N}{_{ti}}{^q}\}, \forall \beta\in\{1,\ldots,\tensor{N}{_{qj}}{^s}\} \quad \text{the following holds:}\nonumber\\
&\hspace{10mm}\sum_{\zeta=1}^{\mathcal{E}^{M_a}_q}\rho^{a(Q);\gamma}_{(T)(I)}C^{(a,j)b}_{(Q)S;\alpha\beta}=\sum_{\substack{K\prec M_b\\}}\sum_{\delta=1}^{\tensor{N}{_{ij}}{^k}}\sum_{\varepsilon=1}^{\tensor{N}{_{tk}}{^s}}\rho^{b(S);\varepsilon}_{(T)(K)}\mathrm{F}^{(t\,i\,j)s}_{\varepsilon k\delta,\gamma q\beta}C^{(a,j)b}_{(I)K;\alpha\delta}
\end{align}
where   $Q=(q,\zeta)$.

Suppose now that we have obtained these numbers $C^{(a,j)b}_{(I)K;\alpha\beta}$. To evaluate the $\dot{\mathrm{T}}$ matrix elements we introduce the numbers $\bar{C}^{(a,j)b}_{(I)K;\alpha\beta}$ in the same way as the $\bar{A}$ coefficients are introduced in \eqref{deco2} as the coefficients in the decomposition of the dual morphism $\bar{\Psi}^\alpha_{(a,j)b}$ in terms of $\bar{\lambda}^\beta_{(i,j)k}$. The numbers  $\bar{C}$ can be obtained from the normalisation condition
\begin{equation}
    \Psi_{(a,j)b}^\alpha\circ \bar{\Psi}^\beta_{(a,j)b}=\delta_{\alpha,\beta}\operatorname{id}_{M_b}
\end{equation}
Subsequently, using the decomposition \eqref{deco5} in both sides of  \eqref{cross2.2} and composing  with a suitable morphism  we find the following expression 
\begin{equation} \small
  \dot{\mathrm{T}}^{(x\,a\,k)c}_{\alpha b\beta,\gamma d\delta}=\sum_{\substack{I\prec X_x\\ J\prec M_a}} \, \sum_{\substack{P\prec M_b\\Q\prec M_d}}\sum_{\zeta=1}^{\tensor{N}{_{jk}}{^p}}\sum_{\varepsilon=1}^{\tensor{N}{_{ip}}{^s}}\sum_{\theta=1}^{\tensor{N}{_{ij}}{^q}}\sum_{\xi=1}^{\tensor{N}{_{qk}}{^s}}\sum_{\phi^{\prime\prime}=1}^{\mathcal{E}_s^{M_c}}\mathrm{F}^{(i\,j\,k)s}_{\varepsilon p\zeta,\theta q\xi}C^{(a,k)b}_{(J)P;\beta\zeta}B^{(x,b)c}_{(I,P)S;\alpha\varepsilon}\bar{B}^{(x,a)d}_{(I,J)Q;\gamma\theta}\bar{C}^{(d,k)c}_{(Q)s\phi^{\prime\prime};\delta\xi}
\end{equation} \normalsize
that holds for all  $S=(s,\phi)\prec M_c$.

In the special case described by
\begin{subequations}\label{specialcase2}
    \begin{align}
        &U_i\otimes U_j\cong U_k \\
& X_x\otimes_A X_y\cong X_z\\
&\dimh(U_i,\dot{X}_x)\in\{0,1\}\\
 &X_w\otimes_A M_a\cong M_b \label{speciald}\\
    & \operatorname{dim}\operatorname{Hom}(U_i,\dot{M}_a)\in\{0,1\}\\
    & M_c\otimes U_i\cong M_c
    \end{align}
\end{subequations}
we find
\begin{equation}\label{tdotboson}
\dot{\mathrm{T}}^{(x\,a\,k)c}_{b,d}=\frac{C^{(a,k)b}_{(j)p}B^{(x,b)c}_{(i,p)s}}{B^{(x,a)d}_{(i,j)q}C^{(d,k)c}_{(q)s}}\mathrm{F}^{(i\,j\,k)s}_{p,q},
  \enspace \;   \forall i\prec X_x,\forall j\prec M_a 
\end{equation}
 Here all possible fusion structures between simple objects are  group-like. For example, while the tensor product $X_w\otimes_A M_a$ is always an $A$-module, the restriction \eqref{speciald} means that it should be irreducible.  Each irreducible defect thus acts on the set of all irreducible modules.

The matrix elements $\mathrm{\dot{T}}^{(x\,a\,k)c}_{b,d}$ are non-zero only if the labels $c,b,d$ are the unique labels that satisfy 
\begin{equation}\label{selec3}
M_c\cong \left( X_x\otimes_BM_a\right)\otimes U_k, \quad M_b\cong M_a\otimes U_k \quad \text{and}\quad M_d\cong X_x\otimes_B M_a
\end{equation}
and $s,p,q$ are the unique labels that satisfy
$$
U_s\cong U_i\otimes U_j\otimes U_k,\quad U_p\cong U_j\otimes U_k \quad \text{and} \quad U_q\cong U_i\otimes U_j.
$$
For these reasons, if we work in the special case of the form \eqref{specialcase2} we will usually omit the labels $c,b,d$ when presenting the non-zero elements $\mathrm{\dot{T}}^{(x\,a\,k)c}_{b,d}$.

\subsection{\texorpdfstring{$\mathrm{F}[A]$-matrix}{\textsuperscript{F[A]-matrix}}}

To calculate the  fusing matrix $\mathrm{F}[A]$ we start from the definition \eqref{(1)F} and use the decomposition \eqref{deco5} of the  junctions $\Psi$. Suppose now that we have obtained the complex coefficients $C^{(a,j)b}_{(I,j)K;\alpha\beta}$ as a solution of the system \eqref{psicongen}. Then the entries of the $\mathrm{F}[A]$ matrices are given by the following expression which holds for any $S\prec M_c$
\begin{equation}
   \mathrm{F}[A]^{(a\,j\,k)c}_{\alpha p\beta,\gamma b\delta}=\sum_{I\prec M_a}\sum_{Q\prec M_b} \sum_{\varepsilon=1}^{\tensor{N}{_{ip}}{^s}}\sum_{\zeta=1}^{\tensor{N}{_{ij}}{q}}\sum_{\eta=1}^{\tensor{N}{_{qk}}{^s}}\sum_{\phi^{\prime\prime}=1}^{\mathcal{E}_s^{M_c}}\mathrm{F}^{(i\,j\,k)s}_{\varepsilon p\beta,\zeta q\eta}C^{(a,p)c}_{(I)S;\alpha\varepsilon}\overbar{C}^{(b,k)c}_{(Q)s\phi^{\prime\prime};\delta\eta}\overbar{C}^{(a,j)b}_{(I)Q;\gamma\zeta}
\end{equation}
where $S=(s,\phi)$.

In the special case described by \eqref{specialcase2} we find
\begin{equation}\label{F(1)boson}
   \mathrm{F}[A]^{(a\,j\,k)c}_{p,b}=\frac{C^{(a,p)c}_{(i)s}}{C^{(a,j)b}_{(i)q}C^{(b,k)c}_{(q)s}}\mathrm{F}^{(i\,j\,k)s}_{p,q}, \enspace \;  \forall i\prec M_a.
\end{equation}
The matrix elements $\mathrm{F}[A]^{(a\,j\,k)c}_{p,b}$ are non-zero only if the labels $c,p,b$ are the unique labels that satisfy
\begin{equation}
    M_c\cong \left(M_a\otimes U_j\right)\otimes U_k,\quad U_p\cong U_j\otimes U_k\quad \text{and} \quad M_b\cong M_a\otimes U_j 
\end{equation}
and $s,q$ are the unique labels that satisfy
$$
U_s\cong U_i\otimes U_j\otimes U_k \quad \text{and} \quad U_q\cong U_i\otimes U_j.
$$
Therefore, whenever we work in this special case we are going to omit the labels $s,p,b$ when presenting the non-zero elements  $\mathrm{F}[A]^{(a\,j\,k)s}_{p,b}$. Moreover, in this case where we have no multiplicities for the relevant junction spaces, the boundary OPE coefficients \eqref{OPEcoefgen} take the form
\begin{equation}\label{OPEcoefbos1}
    C_{j,k}^{(a\,b\,c)p}=\mathrm{G}[A]^{(a\,j\,k)c}_{b,p} \, .
\end{equation}

\section{Free boson compactified on a circle} \label{sec_boson}

In this section, we apply the techniques described earlier in this paper in order to evaluate the various fusing matrices introduced in section \ref{sec2.5} for the conformal field theory of a free boson compactified at a radius of rational square.

\subsection{Generalities} \label{sec4.1}

We will now introduce all the relevant quantities such as the modular tensor category, the symmetric special Frobenius algebras and their representation theory for the free boson theory. The description of these objects will be brief since in this paper we consider them as given data. More details and results are presented in \cite{boson,FRS1,FRS3}.

First, we recall that the chiral symmetry algebra of the free boson $\phi(z)$ is a Heisenberg affine Lie algebra $\hat{\mathfrak{u}}(1)$. For any $N\in \mathbb{Z}_{>0}$ this algebra can be extended by the two vertex operators $W_N^{\pm}(z)=:e^{\pm 2i\sqrt{2N}\phi(z)}:$. The resulting extended chiral algebra, denoted by $\affu(1)_N$, is rational  with $2N$ inequivalent irreducible highest weight representations, labelled $U_0,U_1,\ldots,U_{2N-1}$.

We denote by $\mathcal{U}_N$ the modular tensor category formed by these representations. 
A special feature of this category that will simplify calculations is the fact that the dimensions $\tensor{N}{_{ij}}{^k}$ of all coupling spaces $\Hom(U_i\otimes U_j,U_k)$ are either $0$ or $1$. More specifically the fusion, which furnishes the group $\mathbb{Z}_{2N}$ is found to be
\begin{equation}
   U_k\otimes U_l\cong U_{[k+l]} \quad\;\text{where} \quad[k+l]=k+l\;\text{mod}\; 2N\;\in\mathcal{I}.
\end{equation}
The dual of $k$ is $\bar{k}=2N-k$ and all simple objects have quantum dimension $\operatorname{dim}(U_k)=1$. Under a suitable choice of bases \eqref{bases1}, the braiding and fusing matrices are given by
\begin{equation}\label{Finit}
    \mathrm{R}^{(kl)[k+l]}=e^{-\pi ikl/2N},\qquad \mathrm{F}^{(r\,s\,t)[r+s+t]}_{[s+t][r+s]}=(-1)^{r\sigma(s+t)}
\end{equation}
where $\sigma(s+t)=0$ if $s+t<2N$ and $\sigma(s+t)=1$ otherwise. We denote the bases of $\Hom(U_i\otimes U_j,U_k)$ for which those equalities hold by $\lambda_{(i,j)k}$. Notice that the fusing matrices in \eqref{Finit} only depend on the first three upper indices. Hence, in the following we will usually denote them as $\mathrm{F}^{(r\,s\,t)}$ with the understanding that this stands for the only non-zero matrix element of $\mathrm{F}^{(r\,s\,t)}$. Furthermore, it follows from the explicit form \eqref{Finit} that the $\mathrm{F}$-matrices coincide with their inverse
\begin{equation}
    \mathrm{F}^{(r\,s\,t)}=\mathrm{G}^{(r\,s\,t)}\,.
\end{equation}
\mycomment{
The modular $s$-matrix is found to be
\begin{equation}
    s_{k,l}=e^{-\pi ikl/N} \, .
\end{equation}\\}

Now we turn to describing the Frobenius algebras in $\mathcal{U}_N$. For this free boson theory, we will consider haploid special symmetric Frobenius algebras in $\mathcal{U}_N$. The additional property of haploidity corresponds to the uniqueness of the boundary vacuum state.  Every boundary condition of a full CFT gives rise to a special symmetric Frobenius algebra and all such algebra objects are Morita equivalent. It turns out that there is always a haploid representative of each Morita equivalence class. Furthemore, due to the fact that the quantum dimensions of the simple objects of $\mathcal{U}_N$ are all $1$, the algebra object is additionally of the simple current type. Frobenius algebras which have  all of these properties are called \textit{Schellekens algebras} and they are studied in \cite{FRS3}.

Every haploid special symmetric Frobenius algebra in $\mathcal{U}_N$ is isomorphic to one of the algebras $A_r$ defined as follows (see \cite[Section~3.3]{FRS3}). As an object in $\mathcal{U}_N$ we have
\begin{equation}\label{alg1}
    A_r=\bigoplus_{a=0}^{r-1}U_{2aN/r} \qquad \text{where}\; r\in\mathbb{Z}_{>0} \;\text{divides} \; N\,.
\end{equation}
For $a\in\mathbb{Z}_r$ we denote by $\iota_a\in\Hom(U_{2aN/r},A_r)$ and $\jmath_a\in\Hom(A_r,U_{2aN/r})$ embedding and restriction morphisms for the subobject $U_{2aN/r}$ of $A_r$ with $\jmath_a\circ \iota_a=\operatorname{id}_{U_{2aN/r}}$. We can choose the multiplication and the unit morphism of $A_r$ to be
\begin{equation}\label{mult1}
    m=\sum_{a, b=0}^{r-1} \iota_{a+b} \circ \lambda_{(2 a N / r, 2 b N / r)[2(a+b) N / r]} \circ\left(\jmath_a \otimes \jmath_b\right) \qquad \text { and } \quad \eta=\iota_0\,.
\end{equation}
The compactification radius is related to $r$ and $N$ by
\begin{equation}
    R=\frac{r}{\sqrt{2N}}
\end{equation}
so that different algebra objects correspond to different compactification radii.  A rational  free boson model  is thus specified by the two numbers: $N$ and $r$ which completely characterise the modular tensor category and the Frobenius algebra respectively.

\mycomment{
Conversely, the free boson compactified at a radius of the form $R=\sqrt{P/2Q}$ with $P,Q$ coprime positive integers contains in its subset of holomorphic bulk fields the chiral algebra $\affu(1)_{n^2PQ}$ for any choice of $n\in\mathbb{Z}_{>0}$. The corresponding algebra in $\mathcal{U}_{n^2PQ}$ is then $A_{nP}$. Therefore, the following relations hold between the two different aforementioned parametrizations
\begin{equation}
    r=nP \qquad \text{and}\qquad N=n^2PQ
\end{equation}\\
}

For the left $A_r$-modules one finds that every simple module is isomorphic to an induced module $\operatorname{Ind}_{A_r}(U_k)$, and that the induced modules $\operatorname{Ind}_{A_r}(U_k)$ and $\operatorname{Ind}_{A_r}(U_l)$ are isomorphic if and only if $k\equiv l \;\text{mod}\; 2N/r$. Hence, there are $2N/r$ distinct simple left $A_r$-modules which we denote $M^{(r)}_f,\;f\in \mathcal{J}_{A_r}$ where $\mathcal{J}_{A_r}=\{0,1,\ldots,\frac{2N}{r}-1\}$. Furthermore, there is an additional simplification for this model:  the dimensions of the hom-spaces $\Hom(U_i,A_r\otimes U_k)$ for the subobjects of the induced modules are either $0$ or $1$. The representation numbers \eqref{rep2} are found to be
\begin{equation}\label{rep4}
    \rho^{M_f\,[a+i]}_{(a)(i)}=1
\end{equation}
where $a\prec A_{r}$ and $i\prec M_f$.

Regarding  $A_r$-$A_r$-bimodules, the simple $A_r$-bimodules are labelled by elements of the abelian group 
\begin{equation}
    G^{(r,N)}=H^\ast \times_H \operatorname{Pic}(\mathcal{U}_N)
\end{equation}
where $\operatorname{Pic}(\mathcal{U}_N)=\mathbb{Z}_{2N}$ and $H=\mathbb{Z}_r$ is embedded in $\operatorname{Pic}(\mathcal{U}_{N})$ via $\tilde{\imath} (a)=2aN/r$ (see  \cite[Section~5]{FFRS}). We will label elements of $H^\ast$ by $x\in\mathbb{Z}_r$ via
\begin{equation}
    \psi_x(a)=e^{-2\pi ixa/r}\,.
\end{equation}
The action of $h\in H$ on $k\in\operatorname{Pic}(\mathcal{U}_N)$ is $h.k=k+\tilde{\imath} (h)$. Thus we have
\begin{equation}
    G^{(r,N)}=\mathbb{Z}_r\times_{\mathbb{Z}_r}\mathbb{Z}_{2N}
\end{equation}
where the product over $\mathbb{Z}_r$ implies the identification $(a,k+2N/r)=(a+N/r,k)$.

Now for each $\psi\in H^\ast$ it turns out \cite[Section~5.1]{FFRS} that we can construct an automorphism of $A_r$ as
\begin{equation}\label{twist1}
    t_{\psi}:=\bigoplus_{h\in H}\psi(h)\operatorname{id}_{U_{2hN/r}}\,.
\end{equation}
Then, a simple bimodule corresponding to an element $(\psi,k)\in G^{(r,N)}$ is given by $\alpha_{A_r}^+(U_k)_{t_\psi}$, that is an alpha-induced bimodule for which the right action of $A_r$ is twisted by the automorphism $t_\psi$. More specifically,
the underlying object is $A_r\otimes U_k$ while the left and right actions are 
\begin{equation}\label{act2}
   \rho^{\alpha_{A_r}^+(U_k)_{t_\psi}}=m\otimes \operatorname{id}_{U_k}, \qquad\Tilde{\rho}^{\alpha_{A_r}^+(U_k)_{t_\psi}}=(m\otimes \operatorname{id}_{U_k})\circ \left(\operatorname{id}_{A_r}\otimes c_{U_k,A_r}\right)\circ\left(\operatorname{id}_{A_r\otimes U_k}\otimes t_\psi\right).
\end{equation}
The representation numbers for the left action are of the same form as \eqref{rep4} while for the right action we have $\forall i\prec A_r\otimes U_k$:
\begin{equation}\label{rightaction0}
    \Tilde{\rho}^{M_f\,[i+a]}_{(i)(a)}=\mathrm{R}^{(ka)[k+a]}\tau_\psi(a)
\end{equation}
where $M_f$ denotes a right $A_r$-module with underlying object $A_r\otimes U_k$ and $\tau$ is a number depending on the twist $t_\psi$ and the subobject $a\prec A$. The fusion of two such bimodules is given by the group operation in $G^{(r,N)}$.

It is useful to label the bimodules by another group which is isomorphic to $G^{(r,N)}$, namely
\begin{equation}
    \tilde{G}^{(r,N)}=\left(\mathbb{Z}_{r}\times\mathbb{Z}_{\frac{N}{r}}\times \mathbb{Z}\right)\bigg/\langle(1,1,-2)\rangle
\end{equation}
where $\langle(1,1,-2)\rangle$ denotes the subgroup generated by the element $(1,1,-2)$. Notice that every element of $ \tilde{G}^{(r,N)}$ can be written either as $(a,b,0)$ or $(a,b,1)$ for suitable $a,b$, hence the order of the group is $2N$. The simple $A_r$-bimodules labelled by this group will be denoted $B^{(r)}_{(a,b,\rho)}$ for $(a,b,\rho)\in  \tilde{G}^{(r,N)}$. Let us note that for arbitrary $N,r\in\mathbb{Z}_{>0}$ with $r$ dividing $N$ we find that
\begin{equation}\label{defectgroup}
    \left(\mathbb{Z}_{r}\times\mathbb{Z}_{\frac{N}{r}}\times \mathbb{Z}\right)\bigg/\langle(1,1,-2)\rangle \cong \mathbb{Z}_{\operatorname{gcd}(r,\frac{N}{r})}\times\mathbb{Z}_{2\operatorname{lcm}(r,\frac{N}{r})}\,.
\end{equation}
The fusion of topological defects corresponds to the tensor product over $A_r$ which is given by the group operation in $\tilde{G}^{(r,N)}$:
\begin{equation}\label{bimodfus}
    B_{(a, b, \rho)}^{(r)} \otimes_{A_r} B_{(c, d, \sigma)}^{(r)} \cong B_{(a+c, b+d, \rho+\sigma)}^{(r)}
\end{equation}
while the fusion of a topological defect to a conformal boundary is found to be
\begin{equation}\label{open1}
    B_{(a, b, \rho)}^{(r)} \otimes_{A_r} M_f^{(r)} \cong M_{f+2 b+\rho}^{(r)}\,.
\end{equation}

Notice that the free boson theory described here is a simple current theory \cite{FRS3} and in particular has the properties of the special case described in \eqref{specialcase2}. This implies that we can use the simple formulae \eqref{yboson}, \eqref{tboson}, \eqref{tdotboson} and \eqref{F(1)boson} for the fusing matrices. Furthermore, the equations \eqref{conagen}, \eqref{conbgen} and \eqref{concgen} for calculating the coefficients $A^{(x,y)z}_{(I,J)K;\alpha\beta}$ greatly simplify in this special case. The same can be said for the corresponding equations for calculating the numbers $B^{(x,a)b}_{(I,J)K;\alpha\beta}$ and $C^{(a,j)b}_{(I)K;\alpha\beta}$.

In fact, we can derive a general result for the numbers $C^{(a,j)b}_{(I)K;\alpha\beta}$ as well as the related $\mathrm{F}[A]$ matrices  that holds for any choice of parameters $N$ and $r$. These coefficients are required to satisfy \eqref{psicongen} which in our free boson theory takes the simple form
\begin{equation}\label{Cconstantsbos}
    C^{(a,j)b}_{([t+i])[t+i+j]}=C^{(a,j)b}_{(i)[i+j]},\quad \forall i\prec M_a,\forall t\prec A\,.
\end{equation}
This equation is the result of the explicit form of the Frobenius algebras \eqref{alg1}, the triviality of the representation numbers \eqref{rep4} as well as the $\mathrm{F}$ fusing matrices \eqref{Finit}. Recall that \eqref{Cconstantsbos} provides a linear system of equations for each admissible triple $(a,j,b)$ which satisfy $\operatorname{dim}\operatorname{Hom}(M_a\otimes U_j,M_b)\neq 0$. For each such triple, we find that the solution of this system is
\begin{equation}\label{Cconstantsgen}
     C^{(a,j)b}_{(i)[i+j]}=\begin{cases*}
      \tilde{C}^{(a,j)b} & \text{if}\, $U_i$ \;\text{is a subobject of} \; $M_a$  \\
      $0$        & \text{otherwise}.
    \end{cases*}
\end{equation}
for any $ \tilde{C}^{(a,j)b}\in\mathbb{C}$. In the following, we shall always use the choice of normalisation 
\begin{equation}\label{Cconstantsnorm}
     \tilde{C}^{(a,j)b}=1,\quad \text{for all admissible triples}\; (a,j,b)\,.
\end{equation}
Using this result in \eqref{F(1)boson}, we find that the fusing matrices $\mathrm{F}[A]$ coincide with $\mathrm{F}$ for any choice of $N$ and $r$:
\begin{equation}\label{F[A]boson}
    F[A]^{(a\,j\,k)}=\mathrm{F}^{(i\,j\,k)}=(-1)^{i\sigma(j+k)},\quad\forall i\prec M_a\,.
\end{equation}

Let us briefly discuss the boundary operators in these free boson theories which will be important when we consider boundary RG flows. Firstly, we would like to know which boundary operators can live on a single irreducible conformal boundary. In other words, we want to find the non-trivial morphisms
\begin{equation}\label{bosonbc1}
    \Psi_{(f,i)f}\in\Hom_{A_r}\left(M_f\otimes U_i,M_f\right), \quad i\neq 0
\end{equation}
where $M_f$ is the left $A_r$-module labelling an irreducible conformal boundary condition. Using the fusion rules for the free boson theory we can see that
\begin{equation}
    M_f\otimes U_i\cong M_{f+i\;\text{mod}\; 2N/r},
\end{equation}
hence the hom-space in \eqref{bosonbc1} has non-zero dimension  if
\begin{equation}\label{bosonbc2}
    i=0\;\text{mod}\;\;2N/r.
\end{equation}

Secondly, we are interested in the relevant boundary operators, that is, operators with conformal dimension $\Delta<1$ that can trigger an RG flow. In the free boson theory discussed above, the conformal weight of a boundary operator labelled by $i\in\mathcal{I}$ is
\begin{equation}
    \Delta(i)= \begin{cases}\frac{1}{4 N} i^2 & \text { for } i \leq N \\ \frac{1}{4 N}(2 N-i)^2 & \text { for } i>N\end{cases}
\end{equation}
from which we conclude that a boundary operator $\Psi_{(f,i)f}$ is relevant if $i$ satisfies
\begin{align}
    &i<2\sqrt{N} \qquad \qquad \hspace{-1.5mm} \text { for } i\leq N \nonumber \\
    &i>2N-2\sqrt{N} \quad  \text { for } i>N.
\end{align}
The two lines in the above equation correspond to charge conjugation which does not change the dimension. The only self-conjugate boundary operator is  the one with  $i=N$.

Finally, we remind the reader of our convention for the indices in the fusing matrices. Since the fusion rules for the free boson theory are simple, some indices can be omitted. For example, as explained at the end of section \ref{sec3.1} the matrix elements $\mathrm{Y}^{(x\,y\,z)w}_{u,v}$ will be denoted $\mathrm{Y}^{(x\,y\,z)}$. Similar considerations apply to the other  fusing matrices \eqref{tboson}, \eqref{tdotboson} and \eqref{F(1)boson}. Namely, we will omit the final upper index as well as both lower indices when presenting the non-zero matrix elements of $\mathrm{T},\dot{\mathrm{T}}$ and $^{(1)}\mathrm{F}$.

\mycomment{

We finish with a convention in order to lighten notation which is possible due to the specific structure of the free boson theory having the form summarised in \eqref{specialcase2}. Consider the   matrix elements $\mathrm{Y}^{(x\,y\,z)w}_{u,v}$ as in \eqref{yboson}.  They are non-zero only if the labels $w,u,v$ are the unique labels that satisfy 
$$
X_w\cong \left( X_x\otimes_BX_y\right)\otimes_C X_z, \quad X_u\cong X_y\otimes_C X_z \quad \text{and}\quad X_v\cong X_x\otimes_B X_y
$$
and $s,p,q$ are the unique labels that satisfy
$$
U_s\cong U_i\otimes U_j\otimes U_k,\quad U_p\cong U_j\otimes U_k \quad \text{and} \quad U_q\cong U_i\otimes U_j
$$
For these reasons,  we will omit the labels $w,u,v$ when presenting the non-zero elements $\mathrm{Y}^{(x\,y\,z)w}_{u,v}$. 
}

\subsection{Fusing matrices of the  \texorpdfstring{$N=2, r=2$}{N=2, r=2} model} \label{sec4.2}

In this model, there are four distinct simple objects of the category $\mathcal{U}_2$ denoted $\{U_i,i=0,\ldots,3\}$ and satisfying fusion rules corresponding to the addition in $\mathbb{Z}_4$. The algebra object with $r=2$ is found to be
\begin{equation}
     A_2= U_0\oplus U_2
\end{equation}
with the multiplication given by \eqref{mult1} for $N=2=r$.

There are two distinct simple left $A_2$-modules 
\begin{equation}
    M_0=\left(A_2\otimes U_0,m\otimes \operatorname{id}_{U_0}\right),\qquad\qquad M_1=\left(A_2\otimes U_1,m\otimes \operatorname{id}_{U_1}\right)
\end{equation}
where for the underlying objects we find the following decompositions into simple objects of $\mathcal{U}_2$
\begin{equation}
    \dot{M}_0=A_2\otimes U_0 \cong U_0\oplus U_2=A_2 \qquad \text{and}\qquad \dot{M}_1=A_2\otimes U_1 \cong U_1\oplus U_3.
\end{equation}

For the twist automorphisms we have two possible values of $x\in\mathbb{Z}_2$ which lead to
\begin{equation}
 \psi_0(a)=1 \qquad \text{and} \qquad \psi_1(a)= e^{-\pi i a} .
\end{equation}
Thus, using \eqref{twist1} we find
\begin{equation}
    t_0=\operatorname{id}_{U_0}\oplus \operatorname{id}_{U_2} \qquad \text{and} \qquad  t_1=\operatorname{id}_{U_0}\oplus\left(-\operatorname{id}_{U_2}\right).
\end{equation}
The $t_0$ is trivial so that  there are four distinct simple $A_2$-$A_2$-bimodules with their corresponding labels in $G^{(2,2)}$:
\begin{align}\label{labels}
   & X_0:=\alpha^+_{A_2}(U_0)\; \longleftrightarrow\;(0,0) \qquad \qquad X_1:=\alpha^+_{A_2}(U_1) \; \longleftrightarrow \; (0,1) \nonumber \\
   & X_2:= \alpha^+_{A_2}(U_0)_{t_1} \;\longleftrightarrow \;(1,0) \qquad \qquad X_3:=\alpha^+_{A_2}(U_1)_{t_1}\; \longleftrightarrow\; (1,1).
\end{align}
These bimodules can be labeled by elements of the group
\begin{equation}
    \tilde{G}^{(2,2)} = \left(\mathbb{Z}_{2}\times\mathbb{Z}_{1}\times \mathbb{Z}\right)\bigg/\langle(1,1,-2)\rangle=\left(\mathbb{Z}_{2}\times \mathbb{Z}\right)\bigg/\langle(1,-2)\rangle \cong \mathbb{Z}_4.
\end{equation}
 To make contact with the notation introduced in the previous section that uses $\tilde{G}^{(2,2)}$ as the labelling group we note that
\begin{equation}
    X_0\equiv B^{(2)}_{(0,0,0)},\qquad X_1\equiv B^{(2)}_{(0,0,1)},\qquad X_2\equiv B^{(2)}_{(1,0,0)},\qquad X_3\equiv B^{(2)}_{(1,0,1)}.
\end{equation}
The fusion of topological defects corresponds to the group operation in  $\mathbb{Z}_4$. The element $X_1$ is the generator, so that 
\begin{equation}\label{jun3}
    X_x\otimes_{A_2}X_y\cong X_{[x+y]} \qquad \text{where} \; x,y\in\mathbb{Z}_4.
\end{equation}
We note in passing that this  generalises to the case of $N=r$ for which 
\begin{equation}
    \tilde{G}^{(N,N)}\cong \mathbb{Z}_{2N}.
\end{equation}
This implies that the number of simple topological defects is equal to the number of irreducible representations of the chiral algebra and they both obey the same fusion rules.

 The fusion of a topological defect with a conformal boundary can be written compactly as
\begin{equation}
    X_x\otimes_{A_2}M_a\cong M_{x+a \;\text{mod}\,2}.
\end{equation}
The possible junctions that can occur in our model are restricted  by these fusion rules. Namely, the fusion of two bimodules  describes the allowed triple defect junctions  \eqref{jun1}, while the fusion of a bimodule with a module encodes the allowed open defect junctions  \eqref{jun2}.

Now that we have described all the necessary ingredients of this model, we can proceed with the evaluation of the fusing matrices. To this end, we first need to obtain the complex numbers $A^{(x,y)z}_{(i,j)k},B^{(x,a)b}_{(i,j)k}$ and $C^{(a,j)b}_{(i,j)k}$ introduced in sections \ref{sec3.1}, \ref{sec3.2} and \ref{sec3.3} respectively. Since there are  no multiplicity labels  equation \eqref{conagen} simplifies to
\begin{align}\label{conaboson}
    & A^{(x,y)z}_{(i,j)[i+j]}=\sum_{k\prec A_2}\Delta_0^{(k)(\bar{k})}\tilde{\rho}^{x[i+k]}_{(i)(k)}\mathrm{F}^{([i+k]\,\bar{k}\,j)[i+j]}_{[\bar{k}+j],i}G^{(i\,k\,\bar{k})i}_{[i+k],0}A^{(x,y)z}_{([i+k],[\bar{k}+j])[i+j]}\,.
\end{align}
that holds $\forall i\prec X_x,\forall j\prec X_y$. 
Furthermore, equation \eqref{conbgen} simplifies to 
\be \label{conbboson}
\hspace{10mm}A^{(x,y)z}_{([t+i],j)[t+i+j]}=\mathrm{F}^{(t\,i\,j)[t+i+j]}_{ [i+j], [t+i]}A^{(x,y)z}_{(i,j)[i+j]} \quad \forall t\prec A_2,\forall i\prec X_x,\forall j\prec X_y 
\ee
while \eqref{concgen} becomes
\be\label{concboson}
     \hspace{10mm}\tilde{\rho}^{y[j+t]}_{(j)(t)}\mathrm{F}^{(i\,j\,t)[i+j+t]}_{ [j+t], [i+j]}A^{(x,y)z}_{(i,[j+t])[i+j+t]}=\Tilde{\rho}^{z[i+j+t]}_{([i+j])(t)}A^{(x,y)z}_{(i,j)[i+j]}
\ee
that holds $\forall t\prec A_2,\forall i\prec X_x,\forall j\prec X_y$.

The functions $\tau_\psi$ introduced in \eqref{rightaction0} can be derived by using the explicit form of the twists \eqref{twist1}:
\begin{align}
    &\tau_0(a)=1, \qquad \forall a\prec A_2 \nonumber \\
    & \tau_1(a)=\begin{cases} 
      1,& a=0 \\
      -1, & a=2.
   \end{cases}
\end{align}
Using \eqref{comultnumbers}, (\ref{mult1}), (\ref{Finit})  in  \eqref{conaboson}, \eqref{conbboson} and \eqref{concboson} we obtain a  system of linear equations for the complex numbers $A^{(x,y)z}_{(i,j)[i+j]}$. We present the solution of this system in \eqref{ajun}.

\mycomment{
 This notation encodes all the information of the numbers $A^{(x,y)z}_{(i,j)k}$ introduced earlier since in the free boson theories the label $k$ is uniquely determined by the two labels $i$ and $j$. The label $z$ is also uniquely determined by the $A_2$-$A_2$-bimodule fusion rules, therefore in the following we are going to omit both the labels $k$ and $z$  from the numbers $A_{(i,j)k}^{(x,y)z}$.}

We can now directly use \eqref{yboson} to find the $\mathrm{Y}$-matrices for this model. To this end, consider the $\mathrm{Y}^{(x\,y\,z)}$ matrix as a 3-cochain $Y: (\mathbb{Z}_4)^3\to \mathbb{C}^\times$. Recall that $\mathrm{Y}$ and $\mathrm{F}$ satisfy the pentagon identities which can be rewritten as $d\mathrm{Y}=0$ and $d\mathrm{F}=0$ where $d$ is the coboundary homomorphism. Hence, $\mathrm{Y}$ and $\mathrm{F}$ are 3-cocycles and then \eqref{yboson} is equivalent to the statement that they are in the same cohomology class:
\begin{equation}\label{coho2}
    \mathrm{Y}=\mathrm{F}dA \qquad \text{where}\quad A: \left(\mathbb{Z}_4\right)^2\to \mathbb{C}^\times.
\end{equation}
This implies that $\mathrm{Y}/\mathrm{F}$ must be a coboundary. In  \cite{HLY} it was shown that for any cyclic group $\mathbb{Z}_n$ the 3-cocycles  $\omega\in H^3(\mathbb{Z}_n,\mathbb{C}^\times)$ can be explicitly written as 
\begin{equation}\label{cocycle1}
    \omega(a,b,c)=\operatorname{exp}\left[\frac{2\pi il}{n^2}a(b+c-[b+c])\right]
\end{equation}
for some $l\in\mathbb{Z}_n$, where $[b+c]=b+c\;\text{mod}\;n$. In particular we find that the fusing matrices of $\mathcal{U}_2$ given in \eqref{Finit} 
are obtained from formula  \eqref{cocycle1} with $n=4$, $l=2$.

Note that the system of equations \eqref{conaboson}, \eqref{conbboson} and \eqref{concboson} admits a large freedom in choosing the solution. 
The general solution we give in   \eqref{ajun} is parameterised by  complex numbers $\tilde{A}^{(x,y)}$. Since  $\mathrm{Y}$ and $\mathrm{F}$ are in the same cohomology class we can choose the solution for $A$ to be a closed 2-cochain so that $\mathrm{Y}^{(x\,y\,z)}=\mathrm{F}^{(x\,y\,z)}$. 
We find that this fixes the normalisation constants $\tilde{A}^{(x,y)}$ up to four free parameters. We fix the remaining parameters by imposing an additional 
constraint
\begin{equation} \label{normalisation1}
    \tilde{A}^{(x,x)}=1, \qquad \forall x\in\mathbb{Z}_4.
\end{equation}
that leads to 
\begin{equation}\label{newnorm}
  \Tilde{A}= \left[
\begin{array}{cccc}
 1 & 1 & 1 & 1 \\
 1 & 1 & -\xi^3 & \xi \\
 1 & -\xi & 1 & \xi^3 \\
 1 & -\xi^3 & -\xi & 1 \\
\end{array}
\right]
\end{equation}
where $\xi=e^{i\pi/4}$ and the matrix labels run through the simple bimodule labels $x$ and $y$. With this choice we have
\begin{equation}
   \mathrm{Y}^{(x\,y\,z)}=\mathrm{F}^{(x\,y\,z)}=(-1)^{x\sigma(y+z)}\, .
\end{equation}

To find the other two fusing matrices $\mathrm{T}$ and $\dot{\mathrm{T}}$ we follow a similar procedure. 
The general solution of the equations for the coefficients  $B^{(x,a)b}_{(i,j)k}$ introduced in \ref{sec3.2} is given  in  \eqref{bjun}. It is parameterised by the constants 
$\Tilde{B}^{(x,a)}$ which we fix by imposing 
\begin{equation} \label{normalisation2}
    \Tilde{B}^{(x,a)}=1,\quad \forall x\in\mathbb{Z}_4,\forall a\in\{0,1\}.
\end{equation}

Using  \eqref{tboson} we find the $\mathrm{T}$-matrices for this model. With the solutions for the $A$- and $B$-coefficients chosen as above 
some of  the  entries of the $\mathrm{T}$-matrices are presented graphically below\footnote{ The hollow junctions appearing in these diagrams are the ones introduced in \eqref{cross4.2}.}
\begin{align} 
   &  \vcenter{\hbox{\begin{tikzpicture}[font=\footnotesize,inner sep=2pt]
  \begin{feynman}
  \path[pattern=north east lines,pattern color=ashgrey,very thin] (1.01,4) rectangle (1.6,0);
  \vertex  (i1) at (0,0)  ;
  \vertex (x1) at (0,3.5) ;
  \vertex[above=0.5cm of x1] (top1);
  \vertex[below right=0.1cm of x1] (x5) [label=below left:\(X_1\)];
  \vertex  (x2) at (0.75, 0.65);
  \vertex [above=4cm of i1] (i2);
  \vertex [right=1cm of i1] (j1);
  \vertex[ above=0.4cm of j1] (j3) [label=right:\(M_1\)];
  \vertex [right=1cm of i2] (j2);
  \vertex[ below=0.4cm of j2] (j4) [label=right:\(M_1\)];
  \vertex [below=1.4cm of j2] (j6);
  \vertex   [above=1cm of j1] (j5) ;
  \vertex[ above =1.1cm of j1] (j55);
  \vertex [left=0.15cm of j5]  ;
    \vertex [left=0.1cm of j55]  (j);
  \vertex [left=0.1cm of j6] (j7);
  \vertex (y) at (-0.7,2.6) ;
  \vertex[above=1.4cm of y] (top2);
  \vertex [below right=0.1cm of y] (y1) ;
  \vertex[left=0.8cm of x5] (xx) [label=below left:\(X_1\)];
  \vertex [below=1.4cm of j2] (k1);
  \vertex [below=1.5cm of j2] (k2);
  \vertex [left=0.15cm of k1] (k3);
  \vertex [left=0.15cm of k2] (k4) ;
\draw [thick]   (j7) to[out=90,in=-90] (x1);
   \draw  (1.06,1.1) rectangle (0.8, 1);
   \draw (1.06,2.6) rectangle (0.8, 2.5);
   \draw [thick]  (j) to[out=90,in=-90] (y);
   \vertex[above=0.65 of j55] [label=right:\(M_0\)];
  \diagram*{
  (j2)--[thick] (k1),
  (k2)--[thick] (j55),
  (j5)--[thick] (j1),
  (x1)--[thick,myptr1={latex}] (top1),
   (y)--[thick,myptr1={latex}] (top2)
    };
  \end{feynman}
\end{tikzpicture}}} \hspace{1mm}~=-i
\vcenter{\hbox{\begin{tikzpicture}[font=\footnotesize,inner sep=2pt]
  \begin{feynman}
  \path[pattern=north east lines,pattern color=ashgrey,very thin] (1.01,4) rectangle (1.6,0);
  \vertex  (i1) at (0,0)  ;
  \vertex [empty dot] (x1) at (-0.5,2) {};
  \vertex  (x2) at (0.75, 0.65);
  \vertex [above=4cm of i1] (i2);
  \vertex [right=1cm of i1] (j1);
  \vertex[ above=0.4cm of j1] (j3) [label=right:\(M_1\)];
  \vertex [right=1cm of i2] (j2);
  \vertex[ below=0.4cm of j2] (j4) [label=right:\(M_1\)];
  \vertex   [above=1cm of j1] (j5);
  \vertex   [above=1.1cm of j1] (j6);
  \vertex [left=0.15cm of j5] ;
  \vertex [left=0.1cm of j6] (j7);
  \vertex (x) at (-0.2,1.75) [label=right:\(X_2\)];
\draw [thick] [middlearrow={latex}]   (j7) to[out=90,in=-90] (x1);
   \draw  (1.06,1) rectangle (0.8, 1.1);
   \vertex (k1) at (-1,2.8) ;
   \vertex (k2) at (-1,3.4)  [label=left:\(X_1\)];
   \vertex (k3) at (0,2.8) ;
   \vertex (k4) at (0,3.4) [label=right:\(X_1\)];
   \vertex[above=1.1cm of k2] (k5) ;
   \vertex[above=0.6cm of k4] (k6);
   \vertex[left=0.25cm of k5] (k55);
   \vertex[right=0.25cm of j2] (j11);
    \vertex[below=0.5cm of k5] (k7);
   \draw [myptr={latex},thick,rounded corners=1mm]  (x1) -- (k1) --(k7);
   \draw [myptr={latex},thick,rounded corners=1mm]  (x1) -- (k3) --(k6);
  \diagram*{
  (j1)--[ thick] (j5),
  (j2)--[ thick] (j6)
    };
  \end{feynman}
\end{tikzpicture}}}, \qquad  \vcenter{\hbox{\begin{tikzpicture}[font=\footnotesize,inner sep=2pt]
  \begin{feynman}
  \path[pattern=north east lines,pattern color=ashgrey,very thin] (1.01,4) rectangle (1.6,0);
  \vertex  (i1) at (0,0)  ;
  \vertex (x1) at (0,3.5) ;
  \vertex[above=0.5cm of x1] (top1);
  \vertex[below right=0.1cm of x1] (x5) [label=below left:\(X_3\)];
  \vertex  (x2) at (0.75, 0.65);
  \vertex [above=4cm of i1] (i2);
  \vertex [right=1cm of i1] (j1);
  \vertex[ above=0.4cm of j1] (j3) [label=right:\(M_1\)];
  \vertex [right=1cm of i2] (j2);
  \vertex[ below=0.4cm of j2] (j4) [label=right:\(M_1\)];
  \vertex [below=1.4cm of j2] (j6);
  \vertex   [above=1cm of j1] (j5) ;
  \vertex[ above =1.1cm of j1] (j55);
  \vertex [left=0.15cm of j5]  ;
    \vertex [left=0.1cm of j55]  (j);
  \vertex [left=0.1cm of j6] (j7);
  \vertex (y) at (-0.7,2.6) ;
  \vertex[above=1.4cm of y] (top2);
  \vertex [below right=0.1cm of y] (y1) ;
  \vertex[left=0.8cm of x5] (xx) [label=below left:\(X_3\)];
  \vertex [below=1.4cm of j2] (k1);
  \vertex [below=1.5cm of j2] (k2);
  \vertex [left=0.15cm of k1] (k3);
  \vertex [left=0.15cm of k2] (k4) ;
\draw [thick]   (j7) to[out=90,in=-90] (x1);
   \draw  (1.06,1.1) rectangle (0.8, 1);
   \draw (1.06,2.6) rectangle (0.8, 2.5);
   \draw [thick]  (j) to[out=90,in=-90] (y);
   \vertex[above=0.65 of j55] [label=right:\(M_0\)];
  \diagram*{
  (j2)--[thick] (k1),
  (k2)--[thick] (j55),
  (j5)--[thick] (j1),
  (x1)--[thick,myptr1={latex}] (top1),
   (y)--[thick,myptr1={latex}] (top2)
    };
  \end{feynman}
\end{tikzpicture}}} \hspace{1mm}~=i~
\vcenter{\hbox{\begin{tikzpicture}[font=\footnotesize,inner sep=2pt]
  \begin{feynman}
  \path[pattern=north east lines,pattern color=ashgrey,very thin] (1.01,4) rectangle (1.6,0);
  \vertex  (i1) at (0,0)  ;
  \vertex [empty dot] (x1) at (-0.5,2) {};
  \vertex  (x2) at (0.75, 0.65);
  \vertex [above=4cm of i1] (i2);
  \vertex [right=1cm of i1] (j1);
  \vertex[ above=0.4cm of j1] (j3) [label=right:\(M_1\)];
  \vertex [right=1cm of i2] (j2);
  \vertex[ below=0.4cm of j2] (j4) [label=right:\(M_1\)];
  \vertex   [above=1cm of j1] (j5);
  \vertex   [above=1.1cm of j1] (j6);
  \vertex [left=0.15cm of j5] ;
  \vertex [left=0.1cm of j6] (j7);
  \vertex (x) at (-0.2,1.75) [label=right:\(X_2\)];
\draw [thick] [middlearrow={latex}]   (j7) to[out=90,in=-90] (x1);
   \draw  (1.06,1) rectangle (0.8, 1.1);
   \vertex (k1) at (-1,2.8) ;
   \vertex (k2) at (-1,3.4)  [label=left:\(X_3\)];
   \vertex (k3) at (0,2.8) ;
   \vertex (k4) at (0,3.4) [label=right:\(X_3\)];
   \vertex[above=1.1cm of k2] (k5) ;
   \vertex[above=0.6cm of k4] (k6);
   \vertex[left=0.25cm of k5] (k55);
   \vertex[right=0.25cm of j2] (j11);
    \vertex[below=0.5cm of k5] (k7);
   \draw [myptr={latex},thick,rounded corners=1mm]  (x1) -- (k1) --(k7);
   \draw [myptr={latex},thick,rounded corners=1mm]  (x1) -- (k3) --(k6);
  \diagram*{
  (j1)--[ thick] (j5),
  (j2)--[ thick] (j6)
    };
  \end{feynman}
\end{tikzpicture}}}
\end{align}
Note that the defect labelled by the bimodule $X_2$  together with $X_0$ forms a $\mathbb{Z}_2$ subgroup of the defect fusion ring. It has the following $\mathrm{T}$-matrix values:
\begin{equation}
    \vcenter{\hbox{\begin{tikzpicture}[font=\footnotesize,inner sep=2pt]
  \begin{feynman}
  \path[pattern=north east lines,pattern color=ashgrey,very thin] (1.01,4) rectangle (1.6,0);
  \vertex  (i1) at (0,0)  ;
  \vertex (x1) at (0,3.5) ;
  \vertex[above=0.5cm of x1] (top1);
  \vertex[below right=0.1cm of x1] (x5) [label=below left:\(X_2\)];
  \vertex  (x2) at (0.75, 0.65);
  \vertex [above=4cm of i1] (i2);
  \vertex [right=1cm of i1] (j1);
  \vertex[ above=0.4cm of j1] (j3) [label=right:\(M_a\)];
  \vertex [right=1cm of i2] (j2);
  \vertex[ below=0.4cm of j2] (j4) [label=right:\(M_a\)];
  \vertex [below=1.4cm of j2] (j6);
  \vertex   [above=1cm of j1] (j5) ;
  \vertex[ above =1.1cm of j1] (j55);
  \vertex [left=0.15cm of j5]  ;
    \vertex [left=0.1cm of j55]  (j);
  \vertex [left=0.1cm of j6] (j7);
  \vertex (y) at (-0.7,2.6) ;
  \vertex[above=1.4cm of y] (top2);
  \vertex [below right=0.1cm of y] (y1) ;
  \vertex[left=0.8cm of x5] (xx) [label=below left:\(X_2\)];
  \vertex [below=1.4cm of j2] (k1);
  \vertex [below=1.5cm of j2] (k2);
  \vertex [left=0.15cm of k1] (k3);
  \vertex [left=0.15cm of k2] (k4) ;
\draw [thick]   (j7) to[out=90,in=-90] (x1);
   \draw  (1.06,1.1) rectangle (0.8, 1);
   \draw (1.06,2.6) rectangle (0.8, 2.5);
   \draw [thick]  (j) to[out=90,in=-90] (y);
   \vertex[above=0.65 of j55] [label=right:\(M_a\)];
  \diagram*{
  (j2)--[thick] (k1),
  (k2)--[thick] (j55),
  (j5)--[thick] (j1),
  (x1)--[thick,myptr1={latex}] (top1),
   (y)--[thick,myptr1={latex}] (top2)
    };
  \end{feynman}
\end{tikzpicture}}} \hspace{1mm}~=~
\vcenter{\hbox{\begin{tikzpicture}[font=\footnotesize,inner sep=2pt]
  \begin{feynman}
   \path[pattern=north east lines,pattern color=ashgrey,very thin] (1.01,4) rectangle (1.6,0);
  \vertex  (i1) at (0,0)  ;
  \vertex [empty dot] (x1) at (-0.5,2) {};
  \vertex  (x2) at (0.75, 0.65);
  \vertex [above=4cm of i1] (i2);
  \vertex [right=1cm of i1] (j1);
  \vertex[ above=0.4cm of j1] (j3) [label=right:\(M_{a}\)];
  \vertex [right=1cm of i2] (j2);
  \vertex[ below=0.4cm of j2] (j4) [label=right:\(M_a\)];
  \vertex   [above=1cm of j1] (j5);
  \vertex   [above=1.1cm of j1] (j6);
  \vertex [left=0.15cm of j5] ;
  \vertex [left=0.1cm of j6] (j7);
  \vertex (x) at (-0.05,1.7) [label=right:\(\hspace{-2mm}X_{0}\)];
\draw [thick] [middlearrow={latex}]   (j7) to[out=90,in=-90] (x1);
   \draw  (1.06,1) rectangle (0.8, 1.1);
   \vertex (k1) at (-1,2.8) ;
   \vertex (k2) at (-1,3.4)  [label=left:\(X_2\)];
   \vertex (k3) at (0,2.8) ;
   \vertex (k4) at (0,3.4) [label=right:\(X_2\)];
   \vertex[above=1.1cm of k2] (k5) ;
   \vertex[above=0.6cm of k4] (k6);
   \vertex[left=0.25cm of k5] (k55);
   \vertex[right=0.25cm of j2] (j11);
    \vertex[below=0.5cm of k5] (k7);
   \draw [myptr={latex},thick,rounded corners=1mm]  (x1) -- (k1) --(k7);
   \draw [myptr={latex},thick,rounded corners=1mm]  (x1) -- (k3) --(k6);
  \diagram*{
  (j1)--[ thick] (j5),
  (j2)--[ thick] (j6)
    };
  \end{feynman}
\end{tikzpicture}}} \quad \forall a\in\{0,1\}
\end{equation}
 The remaining values for the fusing matrix $\mathrm{T}$ are presented in \eqref{tcase1.1} and \eqref{tcase1.2}.

We now turn to the $\mathrm{\dot{T}}$-matrices. The general solution to the system of equations \eqref{psicongen} for the constants $C^{(a,j)b}_{(i)k}$ is given in \eqref{junc} . It is parameterised by the coefficients 
$\tilde{C}^{(a,j)}$ which we fix by imposing
\begin{equation}\label{standardnormC}
\tilde{C}^{(a,j)} = 1, \quad \forall a\in\{0,1\},\forall j\in\mathbb{Z_4} \, .
\end{equation}

We are ready to use \eqref{tdotboson} to find the $\mathrm{\dot{T}}$-matrices. The selection rules \eqref{selec3} imply that there are 32 non-zero $\mathrm{\dot{T}}$-matrix elements. When  $x=0$ or $k=0$  the corresponding $\mathrm{\dot{T}}$-matrix element is equal to $1$  
which leaves 18 non-zero $\mathrm{\dot{T}}$-matrix elements. Some of these coefficients are presented on the following diagrams which hold $\forall a\in\{0,1\},\forall j\in\{2,3\}$:
\begin{align}
& \vcenter{\hbox{\hspace{-10mm}\begin{tikzpicture}[font=\footnotesize,inner sep=2pt]
  \begin{feynman}
  \path[pattern=north east lines,pattern color=ashgrey,very thin] (0.01,4) rectangle (0.6,0);
  \vertex (j1) at (0,0);
  \vertex [above=0.8 of j1] (j2);
  \vertex[ above =0.9 of j1] (j3);
  \vertex [small,dot] [ above=2.5cm of j1] (psi)  {};
  \vertex[right=0.2cm of psi] (psi1) ;
  \vertex[below=0.2cm of psi1] (psi2) [label=\(\,\,\,\Psi_2\)];
  \vertex [above=3.4cm of j1] (j4) ;
  \vertex[above=3.5cm of j1] (j5)  ;
  \vertex[ left=0.15cm of  j4] (j7);
  \vertex [left=0.15cm of j3] (j8); 
  \vertex [above =4cm of j1] (j6);
  \draw  (0.06,0.8) rectangle (-0.2, 0.9);
    \vertex[below=0.1cm of j6] (m1) ;
    \vertex[above=1.6cm of j1] (m3) [label=right:\(M_a\)];
    \vertex[below=0.4cm of j6] (m2) [label=right:\(M_a\)];
    \vertex[left=1.2cm of psi] (x) [label=left:\(X_2\)];
    \vertex[above=0.3 of j1] (jj) [label=right:\(M_a\)];
    \vertex [below=0.25cm of x] (x1);
    \vertex [above=1.45cm of x] (x2);
    \draw [thick]    (j8) to[out=90,in=-90] (x1);
    \draw [thick] [middlearrow={latex}]   (x1) to (x2);
   \diagram*{
    (j1) --[thick] (j2),
    (j3) --[thick] (psi),
    (psi) --[thick] (j6)
  };
  \end{feynman}
\end{tikzpicture}}}
= -
\vcenter{\hbox{\hspace{-2mm}\begin{tikzpicture}[font=\footnotesize,inner sep=2pt]
  \begin{feynman}
  \path[pattern=north east lines,pattern color=ashgrey,very thin] (0.01,4) rectangle (0.6,0);
  \vertex (j1) at (0,0);
  \vertex [above=2.4 of j1] (j2);
  \vertex[ above =2.5 of j1] (j3);
  \vertex [small,dot] [ above=0.9cm of j1] (psi)  {};
  \vertex[right=0.2cm of psi] (psi1);
  \vertex[below=0.2cm of psi1] (psi2) [label=\(\,\,\,\Psi_2\)];
  \vertex [above=3.4cm of j1] (j4) ;
  \vertex[above=3.5cm of j1] (j5)  ;
  \vertex[ left=0.15cm of  j4] (j7);
  \vertex [left=0.15cm of j3] (j8); 
  \vertex [above =4cm of j1] (j6);
  \draw  (0.06,2.4) rectangle (-0.2, 2.5);
    \vertex[below=0.1cm of j6] (m1) ;
    \vertex[above=1.6cm of j1] (m3) [label=right:\(M_a\)];
    \vertex[above=1.3cm of j2] (m2) [label=right:\(M_a\)];
    \vertex[above=0.5cm of j4] (xx);
    \vertex[left=1.2cm of xx] (x);
    \vertex[below=0.1cm of x] (x100)  [label=left:\(X_2\)];
    \vertex[above=0.3 of j1] (jj) [label=right:\(M_{a}\)];
    \vertex [below=0.25cm of x] (x1);
    \vertex [above=0.15cm of x] (x2);
    \vertex[left=0.25cm of x2] (x22);
    \vertex[above=0.6cm of j6] (j7);
    \vertex[right=0.25cm of j7] (j77);
    \vertex[left=0.25cm of x2] (x22);
    \vertex[above=0.05cm of x] (x3);
    \vertex[left=0.25cm of x3] (x33);
    \vertex[above=0.1cm of x33] (x333);
    \vertex[above=0.4cm of x2] (x4);
    \draw [thick]    (j8) to[out=90,in=-90] (x1);
    \draw [thick] [middlearrow={latex}]   (x1) to (x2);
   \diagram*{
    (j1) --[thick] (j2),
    (j3) --[thick] (j6)
  };
  \end{feynman}
\end{tikzpicture}}},\quad
 \vcenter{\hbox{\hspace{1mm}\begin{tikzpicture}[font=\footnotesize,inner sep=2pt]
  \begin{feynman}
  \path[pattern=north east lines,pattern color=ashgrey,very thin] (0.01,4) rectangle (0.6,0);
  \vertex (j1) at (0,0);
  \vertex [above=0.8 of j1] (j2);
  \vertex[ above =0.9 of j1] (j3);
  \vertex [small,dot] [ above=2.5cm of j1] (psi)  {};
  \vertex[right=0.2cm of psi] (psi1) ;
  \vertex[below=0.2cm of psi1] (psi2) [label=\(\,\,\,\Psi_j\)];
  \vertex [above=3.4cm of j1] (j4) ;
  \vertex[above=3.5cm of j1] (j5)  ;
  \vertex[ left=0.15cm of  j4] (j7);
  \vertex [left=0.15cm of j3] (j8) ; 
  \vertex [above =4cm of j1] (j6);
  \draw  (0.06,0.8) rectangle (-0.2, 0.9);
    \vertex[below=0.1cm of j6] (m1) ;
    \vertex[above=1.6cm of j1] (m3) [label=right:\(M_j\)];
    \vertex[below=0.4cm of j6] (m2) [label=right:\(M_0\)];
    \vertex[left=1.2cm of psi] (x) [label=left:\(X_1\)];
    \vertex[above=0.3 of j1] (jj) [label=right:\(M_{j+1}\)];
    \vertex [below=0.25cm of x] (x1);
    \vertex [above=1.45cm of x] (x2);
    \draw [thick]    (j8) to[out=90,in=-90] (x1);
    \draw [thick] [middlearrow={latex}]   (x1) to (x2);
   \diagram*{
    (j1) --[thick] (j2),
    (j3) --[thick] (psi),
    (psi) --[thick] (j6)
  };
  \end{feynman}
\end{tikzpicture}}}
= -i 
\vcenter{\hbox{\hspace{-1mm}\begin{tikzpicture}[font=\footnotesize,inner sep=2pt]
  \begin{feynman}
  \path[pattern=north east lines,pattern color=ashgrey,very thin] (0.01,4) rectangle (0.6,0);
  \vertex (j1) at (0,0);
  \vertex [above=2.4 of j1] (j2);
  \vertex[ above =2.5 of j1] (j3);
  \vertex [small,dot] [ above=0.9cm of j1] (psi)  {};
  \vertex[right=0.2cm of psi] (psi1);
  \vertex[below=0.2cm of psi1] (psi2) [label=\(\,\,\,\Psi_j\)];
  \vertex [above=3.4cm of j1] (j4) ;
  \vertex[above=3.5cm of j1] (j5)  ;
  \vertex[ left=0.15cm of  j4] (j7);
  \vertex [left=0.15cm of j3] (j8); 
  \vertex [above =4cm of j1] (j6);
  \draw  (0.06,2.4) rectangle (-0.2, 2.5);
    \vertex[below=0.1cm of j6] (m1) ;
    \vertex[above=1.6cm of j1] (m3) [label=right:\(M_1\)];
    \vertex[above=1.3cm of j2] (m2) [label=right:\(M_0\)];
    \vertex[above=0.5cm of j4] (xx);
    \vertex[left=1.2cm of xx] (x);
    \vertex[below=0.1cm of x] (x100)  [label=left:\(X_1\)];
    \vertex[above=0.3 of j1] (jj) [label=right:\(M_{j+1}\)];
    \vertex [below=0.25cm of x] (x1);
    \vertex [above=0.15cm of x] (x2);
    \vertex[left=0.25cm of x2] (x22);
    \vertex[above=0.6cm of j6] (j7);
    \vertex[right=0.25cm of j7] (j77);
    \vertex[left=0.25cm of x2] (x22);
    \vertex[above=0.05cm of x] (x3);
    \vertex[left=0.25cm of x3] (x33);
    \vertex[above=0.1cm of x33] (x333);
    \vertex[above=0.4cm of x2] (x4);
    \draw [thick]    (j8) to[out=90,in=-90] (x1);
    \draw [thick] [middlearrow={latex}]   (x1) to (x2);
   \diagram*{
    (j1) --[thick] (j2),
    (j3) --[thick] (j6)
  };
  \end{feynman}
\end{tikzpicture}}} 
\end{align}
The rest of the non-trivial $\dot{\mathrm{T}}$ matrix elements can be found in \eqref{tdot1}.

As explained before the fusing matrices  $\mathrm{F}[A_2]$  coincide with $\mathrm{F}$:
\begin{equation}
    \mathrm{F}[A_2]^{(a\,j\,k)}=(-1)^{a\sigma(j+k)}
\end{equation}
where we used the  normalisation \eqref{standardnormC}.

Finally, let us consider the compact open defects introduced in section \ref{sec_tube} and the corresponding associative fusion algebras.  The two smallest algebras  correspond to the two simple $A_2$-modules:
\begin{equation}
    \mathcal{A}_{M_0}=\langle \omega_0^{(0)},\omega_2^{(0)}\rangle_{{\mathbb{C}}} \quad \text{and} \quad   
    \mathcal{A}_{M_1}=\langle \omega_0^{(1)},\omega_2^{(1)}\rangle_{\mathbb{C}}
\end{equation}
 where we introduced the generators 
 \begin{equation} \label{tube_Gen1}
\omega_0^{(0)}=\left(\bar{\tilde{\Omega}}_{(0,0)},\Tilde{\Omega}_{(0,0)}\right),\;\omega_2^{(0)}=\left(\bar{\tilde{\Omega}}_{(2,0)},\Tilde{\Omega}_{(2,0)}\right),\;\omega^{(1)}_0=\left(\bar{\tilde{\Omega}}_{(0,1)},\Tilde{\Omega}_{(0,1)}\right),\;\omega_2^{(1)}=\left(\bar{\tilde{\Omega}}_{(2,1)},\Tilde{\Omega}_{(2,1)}\right)\,.
 \end{equation}
 Here for brevity we omit the third label of the $\bar{\Tilde{\Omega}}$ and $\Tilde{\Omega}$ since it is uniquely determined by the other two labels. 
The product \eqref{tube3} between the compact open defects takes the simple form
\begin{equation}\label{tubeprod1}
    \left(\bar{\Tilde{\Omega}}_{(x,a)},\Tilde{\Omega}_{(x,b)}\right)\ast  \left(\bar{\Tilde{\Omega}}_{(y,c)},\Tilde{\Omega}_{(y,d)}\right)=\delta_{a,y+c\;\text{mod}\;2}\,\delta_{b,y+d\;\text{mod}\;2}\overline{\mathrm{T}}^{(x\,y\,c)}\mathrm{T}^{(x\,y\,d)} \left(\bar{\Tilde{\Omega}}_{([x+y],c)},\Tilde{\Omega}_{([x+y],d)}\right).
\end{equation}
Using this we can see that both of these algebras are isomorphic to the group algebra $\mathbb{C}[\mathbb{Z}_2]\cong {\mathbb C}\oplus {\mathbb C}$.

Let us now turn to the tube algebra $\mathcal{A}_{M^U}$. It is the complex algebra with a basis consisting of   all of the 
sixteen elementary compact open defects. In addition to the generators given in (\ref{tube_Gen1}) we also have 
\begin{align}\label{tube1case1}
& \omega_{1}^{(0)}= \left(\bar{\tilde{\Omega}}_{(1,1)},\Tilde{\Omega}_{(1,1)}\right),\; \omega_{3}^{(0)}=\left(\bar{\tilde{\Omega}}_{(3,1)},\Tilde{\Omega}_{(3,1)}\right) ,\;\omega_{1}^{(1)}=\left(\bar{\tilde{\Omega}}_{(1,0)},\Tilde{\Omega}_{(1,0)}\right),\;\omega_{3}^{(1)}=\left(\bar{\tilde{\Omega}}_{(3,0)},\Tilde{\Omega}_{(3,0)}\right), \nonumber \\
& \gamma^{(1)}_0=\left(\bar{\tilde{\Omega}}_{(0,0)},\Tilde{\Omega}_{(0,1)}\right) ,\;\gamma^{(1)}_2=\left(\bar{\tilde{\Omega}}_{(2,0)},\Tilde{\Omega}_{(2,1)}\right),\;\gamma^{(0)}_0=\left(\bar{\tilde{\Omega}}_{(0,1)},\Tilde{\Omega}_{(0,0)}\right),\;\gamma^{(0)}_2=\left(\bar{\tilde{\Omega}}_{(2,1)},\Tilde{\Omega}_{(2,0)}\right),\nonumber \\
& \gamma_{1}^{(1)}=\left(\bar{\tilde{\Omega}}_{(1,1)},\Tilde{\Omega}_{(1,0)}\right),\;\gamma_{3}^{(1)}=\left(\bar{\tilde{\Omega}}_{(3,1)},\Tilde{\Omega}_{(3,0)}\right),\; \gamma_{1}^{(0)}= \left(\bar{\tilde{\Omega}}_{(1,0)},\Tilde{\Omega}_{(1,1)}\right),\;\gamma_{3}^{(0)}= \left(\bar{\tilde{\Omega}}_{(3,0)},\Tilde{\Omega}_{(3,1)}\right).
\end{align}
The rationale behind the notation is as follows. All compact defects denoted by $\omega_{i}^{(j)}$ have the same boundary condition on the outside (of the attached arc) while those  denoted by $\gamma_{i}^{(j)}$ have different boundary conditions on the outside. The bottom index is the index of the defect while the top index gives the boundary condition which on the diagrams is below the defect arc. It is clear from the notational rule that 
the algebra splits into two ideals 
\be
\mathcal{A}_{M^U} = \omega \oplus \gamma 
\ee
where 
\be
\omega = \langle \omega_{0}^{(0)}, \omega_{0}^{(1)}, \omega_{2}^{(0)}, \omega_{2}^{(1)},  \omega_{1}^{(0)}, \omega_{3}^{(0)},  \omega_{1}^{(1)}, \omega_{3}^{(1)}\rangle_{\mathbb{C}} \, , 
\ee
\be 
 \gamma =\langle \gamma^{(1)}_0, \gamma^{(1)}_2, \gamma^{(0)}_0, \gamma^{(0)}_2,  \gamma_{1}^{(1)}, \gamma_{3}^{(1)}, 
\gamma_{1}^{(0)}, \gamma_{3}^{(0)} \rangle_{\mathbb{C}} \, .
\ee
Each pair of boundary  fields:  $(\Psi_{(0,0)0}, \Psi_{(1,0)1})$, $  (\Psi_{(0,2)0}, \Psi_{(1,2)1})$ spans a two-dimensional representation of the 
subalgebra $\omega$. The kernels of these representations are two-sided ideals which are easy to find using the $\dot {\mathrm T}$ matrices. 
They give a decomposition $\omega = \omega_{+}\oplus \omega_{-} $ with 
\be
 \omega_{+} = \langle \omega_{0}^{(0)} + \omega_{2}^{(0)} , \omega_{0}^{(1)}+ \omega_{2}^{(1)},  \omega_{1}^{(0)}+ \omega_{3}^{(0)},  \omega_{1}^{(1)}+  \omega_{3}^{(1)}\rangle_{\mathbb{C}} 
 \ee
 \be
\omega_{-} = \langle \omega_{0}^{(0)} - \omega_{2}^{(0)} , \omega_{0}^{(1)}- \omega_{2}^{(1)},  \omega_{1}^{(0)}- \omega_{3}^{(0)},  \omega_{1}^{(1)}-  \omega_{3}^{(1)}\rangle_{\mathbb{C}}\,.
\ee
Each of the subalgebras $\omega_{+}$ and $\omega_{-}$ is isomorphic to the matrix algebra $\mathrm{Mat}_{2}({\mathbb C})$. Similarly the second ideal 
$\gamma$ has two two-dimensional representations spanned by the pairs of boundary fields  $(\Psi_{(0,1)1}, \Psi_{(1,1)0})$, 
  $(\Psi_{(0,3)1}, \Psi_{(1,3)0})$. Using these representations we find a decomposition 
  $\gamma = \gamma_{+}\oplus\gamma_{-} $ with 
  \be
  \gamma_{+} = \langle \gamma^{(1)}_0+ \gamma^{(1)}_2, \gamma^{(0)}_0- \gamma^{(0)}_2,  \gamma_{1}^{(1)}- \gamma_{3}^{(1)}, 
\gamma_{1}^{(0)}+ \gamma_{3}^{(0)} \rangle_{\mathbb{C}} \, , 
  \ee
  \be
  \gamma_{-} = \langle \gamma^{(1)}_0- \gamma^{(1)}_2, \gamma^{(0)}_0+ \gamma^{(0)}_2,  \gamma_{1}^{(1)}+ \gamma_{3}^{(1)}, 
\gamma_{1}^{(0)}- \gamma_{3}^{(0)} \rangle_{\mathbb{C}} 
  \ee
where each of $\gamma_{\pm}$ is isomorphic to $\mathrm{Mat}_{2}({\mathbb C})$. The above gives an explicit decomposition of the tube algebra 
into a direct sum of 4 matrix algebras 
\be
\mathcal{A}_{M^U} = \omega_{+}\oplus \omega_{-}\oplus \gamma_{+} \oplus \gamma_{-} \cong \mathrm{Mat}_{2}({\mathbb C})\oplus  \mathrm{Mat}_{2}({\mathbb C})\oplus  \mathrm{Mat}_{2}({\mathbb C})\oplus  \mathrm{Mat}_{2}({\mathbb C}) 
\ee
as expected from semi-simplicity.

\subsection{Fusing matrices of the  \texorpdfstring{$N=4, r=2$}{N=4, r=2} model} \label{sec4.3}

In this model, there are eight distinct simple objects of the category $\mathcal{U}_4$ denoted $\{U_i,\,i=0,\ldots 7\}$ and satisfying fusion rules corresponding to the addition in $\mathbb{Z}_8$. The algebra object with $r=2$ is found to be
\begin{equation}
     \tilde{A}_2= U_0\oplus U_4.
\end{equation}
This algebra has four simple left modules $\{M_a,\,a=0,1,2, 3\}$ with $\dot{M}_a= \tilde{A}_2\otimes U_a$. The labelling group for the $ \tilde{A}_2$-$ \tilde{A}_2$-bimodules is
\begin{equation}
 \tilde{G}^{(2,4)}=\left(\mathbb{Z}_{2}\times\mathbb{Z}_{2}\times \mathbb{Z}\right)\bigg/\langle(1,1,-2)\rangle\cong \mathbb{Z}_2\times \mathbb{Z}_4.
\end{equation}
We notice that again the number of simple topological defects is equal to the number of irreducible representations of the chiral algebra. However, in this model the fusion rules for the two aforementioned objects do not coincide, unlike in the $N=2=r$ case. We denote these bimodules as $X_x$ where the labels $x$ correspond to the labels $(a,b,\rho)$ according to\footnote{Here, the labels $x$ are introduced only for notational convenience and they do not satisfy the $\mathbb{Z}_8$ fusion rules.}
\begin{align}
    &X_0\equiv B_{(0,0,0)}^{(2)},\quad X_1\equiv B_{(0,0,1)}^{(2)} ,\quad X_2\equiv B_{(0,1,0)}^{(2)} ,\quad X_3\equiv B_{(0,1,1)}^{(2)} \nonumber \\
    & X_4\equiv B_{(1,0,0)}^{(2)},\quad X_5\equiv B_{(1,0,1)}^{(2)} ,\quad X_6\equiv B_{(1,1,0)}^{(2)} ,\quad X_7\equiv B_{(1,1,1)}^{(2)}.
\end{align}
The fusion of these bimodules is given by  \eqref{bimodfus} which for this model coincides with the group operation in $\mathbb{Z}_2\times \mathbb{Z}_4$. The fusion of a bimodule  with a module $M_a$ takes the form
\begin{equation}
    X_x\otimes_{\tilde{A}_2}M_a\cong M_{a+x\;\text{mod} \;4}\,.
\end{equation}

We can now proceed in the same way as in \ref{sec4.2} to obtain the complex numbers $A^{(x,y)}_{(i,j)},B^{(x,a)}_{(i,j)}$ and $C^{(a,j)}_{(i)}$ which we present in Appendix \ref{appbos2}. Using  \eqref{yboson} we obtain
\begin{equation}
    \mathrm{Y}^{(x\,y\,z)}=1 \quad \text{if any} \; x,y,z \;\text{is} \; 0.
\end{equation}
The remaining values of the $\mathrm{Y}$ fusing matrices are given in \eqref{Yboson1}.

Furthermore, using \eqref{tboson} we find the  values for the $\mathrm{T}$ fusing matrix. We present some of its entries graphically:
\begin{align}
    &\vcenter{\hbox{\begin{tikzpicture}[font=\footnotesize,inner sep=2pt]
  \begin{feynman}
  \path[pattern=north east lines,pattern color=ashgrey,very thin] (1.01,4) rectangle (1.6,0);
  \vertex  (i1) at (0,0)  ;
  \vertex (x1) at (0,3.5) ;
  \vertex[above=0.5cm of x1] (top1);
  \vertex[below right=0.45cm of x1] (x5) [label=below:\(\hspace{-2mm}B^{(2)}_{(a,b,\rho)}\)];
  \vertex  (x2) at (0.75, 0.65);
  \vertex [above=4cm of i1] (i2);
  \vertex [right=1cm of i1] (j1);
  \vertex[ above=0.4cm of j1] (j3) [label=right:\(M_{2+2b+\rho}\)];
  \vertex [right=1cm of i2] (j2);
  \vertex[ below=0.4cm of j2] (j4) [label=right:\(M_{3}\)];
  \vertex [below=1.4cm of j2] (j6);
  \vertex   [above=1cm of j1] (j5) ;
  \vertex[ above =1.1cm of j1] (j55);
  \vertex [left=0.15cm of j5]  ;
    \vertex [left=0.1cm of j55]  (j);
  \vertex [left=0.1cm of j6] (j7);
  \vertex (y) at (-0.7,2.6) ;
  \vertex[above=1.4cm of y] (top2);
  \vertex [below right=0.1cm of y] (y1) ;
  \vertex[left=1cm of x5] (xx) [label=below left:\(B^{(2)}_{(0,1,1)}\)];
  \vertex [below=1.4cm of j2] (k1);
  \vertex [below=1.5cm of j2] (k2);
  \vertex [left=0.15cm of k1] (k3);
  \vertex [left=0.15cm of k2] (k4) ;
\draw [thick]   (j7) to[out=90,in=-90] (x1);
   \draw  (1.06,1.1) rectangle (0.8, 1);
   \draw (1.06,2.6) rectangle (0.8, 2.5);
   \draw [thick]  (j) to[out=90,in=-90] (y);
   \vertex[above=0.65 of j55] [label=right:\(M_{3+2b+\rho}\)];
  \diagram*{
  (j2)--[thick] (k1),
  (k2)--[thick] (j55),
  (j5)--[thick] (j1),
  (x1)--[thick,myptr1={latex}] (top1),
   (y)--[thick,myptr1={latex}] (top2)
    };
  \end{feynman}
\end{tikzpicture}}} \hspace{1mm}~=-i~
\vcenter{\hbox{\begin{tikzpicture}[font=\footnotesize,inner sep=2pt]
  \begin{feynman}
   \path[pattern=north east lines,pattern color=ashgrey,very thin] (1.01,4) rectangle (1.6,0);
  \vertex  (i1) at (0,0)  ;
  \vertex [empty dot] (x1) at (-0.7,2) {};
  \vertex  (x2) at (0.75, 0.65);
  \vertex [above=4cm of i1] (i2);
  \vertex [right=1cm of i1] (j1);
  \vertex[ above=0.4cm of j1] (j3) [label=right:\(M_{2+2b+\rho}\)];
  \vertex [right=1cm of i2] (j2);
  \vertex[ below=0.4cm of j2] (j4) [label=right:\(M_3\)];
  \vertex   [above=1cm of j1] (j5);
  \vertex   [above=1.1cm of j1] (j6);
  \vertex [left=0.15cm of j5] ;
  \vertex [left=0.1cm of j6] (j7);
  \vertex (x) at (-0.25,1.3) [label=right:\(\hspace{-8mm}B^{(2)}_{(a,b+1,\rho+1)}\)];
\draw [thick] [middlearrow={latex}]   (j7) to[out=90,in=-90] (x1);
   \draw  (1.06,1) rectangle (0.8, 1.1);
   \vertex (k1) at (-1.2,2.8) ;
   \vertex (k2) at (-1.2,3.4)  [label=left:\(B^{(2)}_{(0,1,1)}\)];
   \vertex (k3) at (-0.2,2.8) ;
   \vertex (k4) at (-0.2,3.4) [label=right:\(B^{(2)}_{(a,b,\rho)}\)];
   \vertex[above=1.1cm of k2] (k5) ;
   \vertex[above=0.6cm of k4] (k6);
   \vertex[left=0.25cm of k5] (k55);
   \vertex[right=0.25cm of j2] (j11);
    \vertex[below=0.5cm of k5] (k7);
   \draw [myptr={latex},thick,rounded corners=1mm]  (x1) -- (k1) --(k7);
   \draw [myptr={latex},thick,rounded corners=1mm]  (x1) -- (k3) --(k6);
  \diagram*{
  (j1)--[ thick] (j5),
  (j2)--[ thick] (j6)
    };
  \end{feynman}
\end{tikzpicture}}}
\end{align}
 $\forall(a,b,\rho)\in \tilde{G}^{(2,4)} \mysetminus \{(0,0,0),(1,0,0)\}$. The defect labelled by the bimodule $X_4$ which together with $X_0$ form a $\mathbb{Z}_2$ subgroup of the fusion has the following $\mathrm{T}$-matrix values:
\begin{align}
    &\vcenter{\hbox{\begin{tikzpicture}[font=\footnotesize,inner sep=2pt]
  \begin{feynman}
  \path[pattern=north east lines,pattern color=ashgrey,very thin] (1.01,4) rectangle (1.6,0);
  \vertex  (i1) at (0,0)  ;
  \vertex (x1) at (0,3.5) ;
  \vertex[above=0.5cm of x1] (top1);
  \vertex[below right=0.45cm of x1] (x5) [label=below left:\(X_4\)];
  \vertex  (x2) at (0.75, 0.65);
  \vertex [above=4cm of i1] (i2);
  \vertex [right=1cm of i1] (j1);
  \vertex[ above=0.4cm of j1] (j3) [label=right:\(M_a\)];
  \vertex [right=1cm of i2] (j2);
  \vertex[ below=0.4cm of j2] (j4) [label=right:\(M_{a}\)];
  \vertex [below=1.4cm of j2] (j6);
  \vertex   [above=1cm of j1] (j5) ;
  \vertex[ above =1.1cm of j1] (j55);
  \vertex [left=0.15cm of j5]  ;
    \vertex [left=0.1cm of j55]  (j);
  \vertex [left=0.1cm of j6] (j7);
  \vertex (y) at (-0.7,2.6) ;
  \vertex[above=1.4cm of y] (top2);
  \vertex [below right=0.1cm of y] (y1) ;
  \vertex[left=1cm of x5] (xx) [label=below left:\(X_4\)];
  \vertex [below=1.4cm of j2] (k1);
  \vertex [below=1.5cm of j2] (k2);
  \vertex [left=0.15cm of k1] (k3);
  \vertex [left=0.15cm of k2] (k4) ;
\draw [thick]   (j7) to[out=90,in=-90] (x1);
   \draw  (1.06,1.1) rectangle (0.8, 1);
   \draw (1.06,2.6) rectangle (0.8, 2.5);
   \draw [thick]  (j) to[out=90,in=-90] (y);
   \vertex[above=0.65 of j55] [label=right:\(M_a\)];
  \diagram*{
  (j2)--[thick] (k1),
  (k2)--[thick] (j55),
  (j5)--[thick] (j1),
  (x1)--[thick,myptr1={latex}] (top1),
   (y)--[thick,myptr1={latex}] (top2)
    };
  \end{feynman}
\end{tikzpicture}}} \hspace{1mm}~=~
\vcenter{\hbox{\begin{tikzpicture}[font=\footnotesize,inner sep=2pt]
  \begin{feynman}
   \path[pattern=north east lines,pattern color=ashgrey,very thin] (1.01,4) rectangle (1.6,0);
  \vertex  (i1) at (0,0)  ;
  \vertex [empty dot] (x1) at (-0.5,2) {};
  \vertex  (x2) at (0.75, 0.65);
  \vertex [above=4cm of i1] (i2);
  \vertex [right=1cm of i1] (j1);
  \vertex[ above=0.4cm of j1] (j3) [label=right:\(M_{a}\)];
  \vertex [right=1cm of i2] (j2);
  \vertex[ below=0.4cm of j2] (j4) [label=right:\(M_a\)];
  \vertex   [above=1cm of j1] (j5);
  \vertex   [above=1.1cm of j1] (j6);
  \vertex [left=0.15cm of j5] ;
  \vertex [left=0.1cm of j6] (j7);
  \vertex (x) at (-0.25,1.7) [label=right:\(X_{0}\)];
\draw [thick] [middlearrow={latex}]   (j7) to[out=90,in=-90] (x1);
   \draw  (1.06,1) rectangle (0.8, 1.1);
   \vertex (k1) at (-1,2.8) ;
   \vertex (k2) at (-1,3.4)  [label=left:\(X_4\)];
   \vertex (k3) at (0,2.8) ;
   \vertex (k4) at (0,3.4) [label=right:\(X_4\)];
   \vertex[above=1.1cm of k2] (k5) ;
   \vertex[above=0.6cm of k4] (k6);
   \vertex[left=0.25cm of k5] (k55);
   \vertex[right=0.25cm of j2] (j11);
    \vertex[below=0.5cm of k5] (k7);
   \draw [myptr={latex},thick,rounded corners=1mm]  (x1) -- (k1) --(k7);
   \draw [myptr={latex},thick,rounded corners=1mm]  (x1) -- (k3) --(k6);
  \diagram*{
  (j1)--[ thick] (j5),
  (j2)--[ thick] (j6)
    };
  \end{feynman}
\end{tikzpicture}}} \quad \forall a\in\{0,1,2,3\}
\end{align}
The remaining values are presented in \eqref{tcase2.1} and \eqref{tcase2.2}.\\

Next, we use \eqref{tdotboson} to find the  $\dot{\mathrm{T}}$ fusing matrix. According to the selection rules, there are $256$ non-zero  $\dot{\mathrm{T}}$ matrix elements. Those with $x=0$ or $k=0$ are equal to $1$.  Among the pairs of open defects and boundary operators we find an anticommuting pair: 
$B^{(2)}_{(1,0,0)}$, $\Psi_{(4,a)a}$ inserted on the same conformal boundary condition:
\begin{align}
   \vcenter{\hbox{\hspace{-10mm}\begin{tikzpicture}[font=\footnotesize,inner sep=2pt]
  \begin{feynman}
  \path[pattern=north east lines,pattern color=ashgrey,very thin] (0.01,4) rectangle (0.6,0);
  \vertex (j1) at (0,0);
  \vertex [above=0.8 of j1] (j2);
  \vertex[ above =0.9 of j1] (j3);
  \vertex [small,dot] [ above=2.5cm of j1] (psi)  {};
  \vertex[right=0.2cm of psi] (psi1) ;
  \vertex[below=0.2cm of psi1] (psi2) [label=\(\,\,\,\Psi_4\)];
  \vertex [above=3.4cm of j1] (j4) ;
  \vertex[above=3.5cm of j1] (j5)  ;
  \vertex[ left=0.15cm of  j4] (j7);
  \vertex [left=0.15cm of j3] (j8); 
  \vertex [above =4cm of j1] (j6);
  \draw  (0.06,0.8) rectangle (-0.2, 0.9);
    \vertex[below=0.1cm of j6] (m1) ;
    \vertex[above=1.6cm of j1] (m3) [label=right:\(M_a\)];
    \vertex[below=0.4cm of j6] (m2) [label=right:\(M_a\)];
    \vertex[left=1.2cm of psi] (x) [label=left:\(B^{(2)}_{(1,0,0)}\)];
    \vertex[above=0.3 of j1] (jj) [label=right:\(M_a\)];
    \vertex [below=0.25cm of x] (x1);
    \vertex [above=1.45cm of x] (x2);
    \draw [thick]    (j8) to[out=90,in=-90] (x1);
    \draw [thick] [middlearrow={latex}]   (x1) to (x2);
   \diagram*{
    (j1) --[thick] (j2),
    (j3) --[thick] (psi),
    (psi) --[thick] (j6)
  };
  \end{feynman}
\end{tikzpicture}}}
= -
\vcenter{\hbox{\begin{tikzpicture}[font=\footnotesize,inner sep=2pt]
  \begin{feynman}
  \path[pattern=north east lines,pattern color=ashgrey,very thin] (0.01,4) rectangle (0.6,0);
  \vertex (j1) at (0,0);
  \vertex [above=2.4 of j1] (j2);
  \vertex[ above =2.5 of j1] (j3);
  \vertex [small,dot] [ above=0.9cm of j1] (psi)  {};
  \vertex[right=0.2cm of psi] (psi1);
  \vertex[below=0.2cm of psi1] (psi2) [label=\(\,\,\,\Psi_4\)];
  \vertex [above=3.4cm of j1] (j4) ;
  \vertex[above=3.5cm of j1] (j5)  ;
  \vertex[ left=0.15cm of  j4] (j7);
  \vertex [left=0.15cm of j3] (j8); 
  \vertex [above =4cm of j1] (j6);
  \draw  (0.06,2.4) rectangle (-0.2, 2.5);
    \vertex[below=0.1cm of j6] (m1) ;
    \vertex[above=1.6cm of j1] (m3) [label=right:\(M_a\)];
    \vertex[above=1.3cm of j2] (m2) [label=right:\(M_a\)];
    \vertex[above=0.5cm of j4] (xx);
    \vertex[left=1.2cm of xx] (x);
    \vertex[below=0.1cm of x] (x100)  [label=left:\(B^{(2)}_{(1,0,0)}\)];
    \vertex[above=0.3 of j1] (jj) [label=right:\(M_{a}\)];
    \vertex [below=0.25cm of x] (x1);
    \vertex [above=0.15cm of x] (x2);
    \vertex[left=0.25cm of x2] (x22);
    \vertex[above=0.6cm of j6] (j7);
    \vertex[right=0.25cm of j7] (j77);
    \vertex[left=0.25cm of x2] (x22);
    \vertex[above=0.05cm of x] (x3);
    \vertex[left=0.25cm of x3] (x33);
    \vertex[above=0.1cm of x33] (x333);
    \vertex[above=0.4cm of x2] (x4);
    \draw [thick]    (j8) to[out=90,in=-90] (x1);
    \draw [thick] [middlearrow={latex}]   (x1) to (x2);
   \diagram*{
    (j1) --[thick] (j2),
    (j3) --[thick] (j6)
  };
  \end{feynman}
\end{tikzpicture}}} \quad \forall a\in\{0,1,2,3\}
\end{align}
Some of the other  values of $\dot{\mathrm{T}}$ are given below
\begin{align}
&
 \vcenter{\hbox{\hspace{-10mm}\begin{tikzpicture}[font=\footnotesize,inner sep=2pt]
  \begin{feynman}
  \path[pattern=north east lines,pattern color=ashgrey,very thin] (0.01,4) rectangle (0.6,0);
  \vertex (j1) at (0,0);
  \vertex [above=0.8 of j1] (j2);
  \vertex[ above =0.9 of j1] (j3);
  \vertex [small,dot] [ above=2.5cm of j1] (psi)  {};
  \vertex[right=0.2cm of psi] (psi1) ;
  \vertex[below=0.2cm of psi1] (psi2) [label=\(\,\,\,\Psi_j\)];
  \vertex [above=3.4cm of j1] (j4) ;
  \vertex[above=3.5cm of j1] (j5)  ;
  \vertex[ left=0.15cm of  j4] (j7);
  \vertex [left=0.15cm of j3] (j8); 
  \vertex [above =4cm of j1] (j6);
  \draw  (0.06,0.8) rectangle (-0.2, 0.9);
    \vertex[below=0.1cm of j6] (m1) ;
    \vertex[above=1.6cm of j1] (m3) [label=right:\(M_{j+1}\)];
    \vertex[below=0.4cm of j6] (m2) [label=right:\(M_1\)];
    \vertex[left=1.2cm of psi] (x) [label=left:\(B^{(2)}_{(0,0,1)}\)];
    \vertex[above=0.3 of j1] (jj) [label=right:\(M_{j+2}\)];
    \vertex [below=0.25cm of x] (x1);
    \vertex [above=1.45cm of x] (x2);
    \draw [thick]    (j8) to[out=90,in=-90] (x1);
    \draw [thick] [middlearrow={latex}]   (x1) to (x2);
   \diagram*{
    (j1) --[thick] (j2),
    (j3) --[thick] (psi),
    (psi) --[thick] (j6)
  };
  \end{feynman}
\end{tikzpicture}}}
= -i 
\vcenter{\hbox{\hspace{1mm}\begin{tikzpicture}[font=\footnotesize,inner sep=2pt]
  \begin{feynman}
  \path[pattern=north east lines,pattern color=ashgrey,very thin] (0.01,4) rectangle (0.6,0);
  \vertex (j1) at (0,0);
  \vertex [above=2.4 of j1] (j2);
  \vertex[ above =2.5 of j1] (j3);
  \vertex [small,dot] [ above=0.9cm of j1] (psi)  {};
  \vertex[right=0.2cm of psi] (psi1);
  \vertex[below=0.2cm of psi1] (psi2) [label=\(\,\,\,\Psi_j\)];
  \vertex [above=3.4cm of j1] (j4) ;
  \vertex[above=3.5cm of j1] (j5)  ;
  \vertex[ left=0.15cm of  j4] (j7);
  \vertex [left=0.15cm of j3] (j8); 
  \vertex [above =4cm of j1] (j6);
  \draw  (0.06,2.4) rectangle (-0.2, 2.5);
    \vertex[below=0.1cm of j6] (m1) ;
    \vertex[above=1.6cm of j1] (m3) [label=right:\(M_2\)];
    \vertex[above=1.3cm of j2] (m2) [label=right:\(M_1\)];
    \vertex[above=0.5cm of j4] (xx);
    \vertex[left=1.2cm of xx] (x);
    \vertex[below=0.1cm of x] (x100)  [label=left:\(B^{(2)}_{(0,0,1)}\)];
    \vertex[above=0.3 of j1] (jj) [label=right:\(M_{j+2}\)];
    \vertex [below=0.25cm of x] (x1);
    \vertex [above=0.15cm of x] (x2);
    \vertex[left=0.25cm of x2] (x22);
    \vertex[above=0.6cm of j6] (j7);
    \vertex[right=0.25cm of j7] (j77);
    \vertex[left=0.25cm of x2] (x22);
    \vertex[above=0.05cm of x] (x3);
    \vertex[left=0.25cm of x3] (x33);
    \vertex[above=0.1cm of x33] (x333);
    \vertex[above=0.4cm of x2] (x4);
    \draw [thick]    (j8) to[out=90,in=-90] (x1);
    \draw [thick] [middlearrow={latex}]   (x1) to (x2);
   \diagram*{
    (j1) --[thick] (j2),
    (j3) --[thick] (j6)
  };
  \end{feynman}
\end{tikzpicture}}} \quad \forall j\in\{3,4,5,6\} \nonumber \\[1.5ex]
& \vcenter{\hbox{\hspace{-10mm}\begin{tikzpicture}[font=\footnotesize,inner sep=2pt]
  \begin{feynman}
  \path[pattern=north east lines,pattern color=ashgrey,very thin] (0.01,4) rectangle (0.6,0);
  \vertex (j1) at (0,0);
  \vertex [above=0.8 of j1] (j2);
  \vertex[ above =0.9 of j1] (j3);
  \vertex [small,dot] [ above=2.5cm of j1] (psi)  {};
  \vertex[right=0.2cm of psi] (psi1) ;
  \vertex[below=0.2cm of psi1] (psi2) [label=\(\,\,\,\Psi_j\)];
  \vertex [above=3.4cm of j1] (j4) ;
  \vertex[above=3.5cm of j1] (j5)  ;
  \vertex[ left=0.15cm of  j4] (j7);
  \vertex [left=0.15cm of j3] (j8); 
  \vertex [above =4cm of j1] (j6);
  \draw  (0.06,0.8) rectangle (-0.2, 0.9);
    \vertex[below=0.1cm of j6] (m1) ;
    \vertex[above=1.6cm of j1] (m3) [label=right:\(M_{3+j}\)];
    \vertex[below=0.4cm of j6] (m2) [label=right:\(M_3\)];
    \vertex[left=1.2cm of psi] (x) [label=left:\(B^{(2)}_{(1,0,1)}\)];
    \vertex[above=0.3 of j1] (jj) [label=right:\(M_j\)];
    \vertex [below=0.25cm of x] (x1);
    \vertex [above=1.45cm of x] (x2);
    \draw [thick]    (j8) to[out=90,in=-90] (x1);
    \draw [thick] [middlearrow={latex}]   (x1) to (x2);
   \diagram*{
    (j1) --[thick] (j2),
    (j3) --[thick] (psi),
    (psi) --[thick] (j6)
  };
  \end{feynman}
\end{tikzpicture}}}
= -
\vcenter{\hbox{\hspace{1mm}\begin{tikzpicture}[font=\footnotesize,inner sep=2pt]
  \begin{feynman}
  \path[pattern=north east lines,pattern color=ashgrey,very thin] (0.01,4) rectangle (0.6,0);
  \vertex (j1) at (0,0);
  \vertex [above=2.4 of j1] (j2);
  \vertex[ above =2.5 of j1] (j3);
  \vertex [small,dot] [ above=0.9cm of j1] (psi)  {};
  \vertex[right=0.2cm of psi] (psi1);
  \vertex[below=0.2cm of psi1] (psi2) [label=\(\,\,\,\Psi_j\)];
  \vertex [above=3.4cm of j1] (j4) ;
  \vertex[above=3.5cm of j1] (j5)  ;
  \vertex[ left=0.15cm of  j4] (j7);
  \vertex [left=0.15cm of j3] (j8); 
  \vertex [above =4cm of j1] (j6);
  \draw  (0.06,2.4) rectangle (-0.2, 2.5);
    \vertex[below=0.1cm of j6] (m1) ;
    \vertex[above=1.6cm of j1] (m3) [label=right:\(M_{0}\)];
    \vertex[above=1.3cm of j2] (m2) [label=right:\(M_3\)];
    \vertex[above=0.5cm of j4] (xx);
    \vertex[left=1.2cm of xx] (x);
    \vertex[below=0.1cm of x] (x100)  [label=left:\(B^{(2)}_{(1,0,1)}\)];
    \vertex[above=0.3 of j1] (jj) [label=right:\(M_{j}\)];
    \vertex [below=0.25cm of x] (x1);
    \vertex [above=0.15cm of x] (x2);
    \vertex[left=0.25cm of x2] (x22);
    \vertex[above=0.6cm of j6] (j7);
    \vertex[right=0.25cm of j7] (j77);
    \vertex[left=0.25cm of x2] (x22);
    \vertex[above=0.05cm of x] (x3);
    \vertex[left=0.25cm of x3] (x33);
    \vertex[above=0.1cm of x33] (x333);
    \vertex[above=0.4cm of x2] (x4);
    \draw [thick]    (j8) to[out=90,in=-90] (x1);
    \draw [thick] [middlearrow={latex}]   (x1) to (x2);
   \diagram*{
    (j1) --[thick] (j2),
    (j3) --[thick] (j6)
  };
  \end{feynman}
\end{tikzpicture}}}\; \quad \forall j\in\{5,6,7\}
\end{align}
while the rest are given in \eqref{tdot2}. Here we have used the normalisation given by (\ref{normalisation2}) and (\ref{standardnormC}) for all the possible pairs of labels.

The fusing matrix $\mathrm{F}[ \tilde{A}_2]$ is obtained from \eqref{F[A]boson}:
\begin{equation}
    \mathrm{F}[ \tilde{A}_2]^{(a\,j\,k)}=(-1)^{a\sigma(j+k)}\, .
\end{equation}

Finally, the tube algebra $\mathcal{A}_{M^U}$ with $M^U=M_0\oplus M_1\oplus M_2\oplus M_3$   is found to be a $128$-dimensional associative complex algebra that splits into a direct sum of 8 full matrix algebras $\mathrm{Mat}_{4}({\mathbb C})$  that correspond to the 8 boundary representations spanned by 
the boundary fields with a fixed object $U_{j}$, $j=0,\dots , 7$.

\subsection{Boundary RG flows} \label{RG_section}

In this section we discuss how our calculations can be used to impose constraints on boundary RG flows in the free boson theory. 
All conformal boundary conditions for the compact free boson theory  have been classified \cite{GRW_b}, \cite{GR_b}, \cite{Janik}.
For a generic radius the only conformal boundary conditions with finite boundary entropy are  the Dirichlet and Neumann ones while for 
the radius being a rational number times the self-dual radius there are exceptional boundary conditions which preserve only Virasoro algebra. 
The Dirichlet boundary conditions can be pictured as D0 branes located at points on the circle while the Neumann ones correspond to D1 branes with Wilson lines (or alternatively as D0 branes localised on the T-dual circle). 
In the rational case the exceptional boundary conditions appear for the theories with $N$ being a square.  Since they 
break the extended chiral algebra we do not see them in the TFT formalism. 
We first discuss the simplest case when the conformal boundary condition in the UV is a single irreducible boundary condition $M_{a}$ in a rational theory. 
As we already discussed in section \ref{sec4.1} the non-trivial relevant boundary operators on $M_{a}$ are $\Psi_{(a,j)a}$ with  
\bea
&&j= 0 \enspace {\rm mod} \enspace \frac{2N}{r} \\
&&j<2\sqrt{N} \enspace \mbox{or } j>2(N-\sqrt{N}) \, .
\eea
 We are interested in RG flows triggered by perturbing $M_{a}$ by a boundary action perturbation of the form 
 \be \label{b_pert}
\Delta S =   \int\!\! dt ( \lambda \Psi_{(a,j)a}(t) +   \lambda^\ast  \Psi_{(a,j)a}^{\dagger}(t) ) 
 \ee
 where $\lambda$ is a complex coupling constant with the asterisk standing for complex conjugation and $\Psi_{(a,j)a}^{\dagger}$ is the Hermitian conjugate field that is proportional to $\Psi_{(a,2N-j)a}$.

Since we preserve unitarity the strongest constraint on such flows comes from the $g$-theorem \cite{g1}, \cite{g2}, \cite{g3}.
The $g$-factor $g_{N}$ for a Neumann boundary condition and $g_{D}$ for a Dirichlet one are 
\be
g_{N} = \sqrt{R} \, , \qquad g_{D} = \frac{1}{\sqrt{2R}}
\ee
where $R$ is the radius of the free boson normalised so that the self-dual radius is $R_{\rm s.d.} =\frac{1}{\sqrt{2}}$.  
We observe that when $R\le \sqrt{2}$ we have $2 g_{D} \ge g_{N}$.
Since in the UV our flows start from a single Neumann boundary condition (as all the boundary conditions preserving the extended chiral algebra are Neumann) we see that when $r\le 2\sqrt{N}$ the only possible end point of the flow is a single Dirichlet boundary condition. 
Each Dirichlet boundary condition has the same spectrum so if we just observe the flow of the spectrum putting the same perturbed boundary 
condition on both ends of an interval we cannot distinguish the different Dirichlet boundary conditions at the end point. If, however, we put 
different perturbed boundary conditions on both ends -- those corresponding to different couplings $\lambda_{1}$ and $\lambda_{2}$,  
we can discern from the resulting spectrum the {\it relative} positions of the D0 branes. Information about these relative positions can be 
obtained from topological defects. 

It follows from the fusion rule (\ref{open1}) that the only rational topological defects which can end topologically on $M_{b}$ are 
$B^{(r)}_{(a,0,0)}$.
The fusion  of these defects generates a subring isomorphic to the group ring of  ${\mathbb Z}_{r}$. 
 In the language of \cite{weak_strong} each boundary condition $M_{b}$ is strongly symmetric with respect to the action of this subring.
 By looking at the action of the defects $B^{(r)}_{(a,0,0)}$ on the $U(1)$ bulk primaries we can identify them with rotations of the circle 
 by the angle $\frac{2\pi a}{r}$ that act on the D0 branes by shifting their positions by the same angle. 
For any boundary field $ \Psi_{(b,j)b}$ passing a junction with $B^{(r)}_{(a,0,0)}$ through it results in multiplying the field by a 
number 
\begin{equation}
    \vcenter{\hbox{\hspace{-10mm}\begin{tikzpicture}[font=\footnotesize,inner sep=2pt]
  \begin{feynman}
  \path[pattern=north east lines,pattern color=ashgrey,very thin] (0.01,4) rectangle (0.6,0);
  \vertex (j1) at (0,0);
  \vertex [above=0.8 of j1] (j2);
  \vertex[ above =0.9 of j1] (j3);
  \vertex [small,dot] [ above=2.5cm of j1] (psi)  {};
  \vertex[right=0.2cm of psi] (psi1) ;
  \vertex[below=0.2cm of psi1] (psi2) [label=\(\,\,\,\Psi_j\)];
  \vertex [above=3.4cm of j1] (j4) ;
  \vertex[above=3.5cm of j1] (j5)  ;
  \vertex[ left=0.15cm of  j4] (j7);
  \vertex [left=0.15cm of j3] (j8); 
  \vertex [above =4cm of j1] (j6);
  \draw  (0.06,0.8) rectangle (-0.2, 0.9);
    \vertex[below=0.1cm of j6] (m1) ;
    \vertex[above=1.6cm of j1] (m3) [label=right:\(M_b\)];
    \vertex[below=0.4cm of j6] (m2) [label=right:\(M_b\)];
    \vertex[left=1.2cm of psi] (x) [label=left:\(B^{(r)}_{(a,0,0)}\)];
    \vertex[above=0.3 of j1] (jj) [label=right:\(M_b\)];
    \vertex [below=0.25cm of x] (x1);
    \vertex [above=1.45cm of x] (x2);
    \draw [thick]    (j8) to[out=90,in=-90] (x1);
    \draw [thick] [middlearrow={latex}]   (x1) to (x2);
   \diagram*{
    (j1) --[thick] (j2),
    (j3) --[thick] (psi),
    (psi) --[thick] (j6)
  };
  \end{feynman}
\end{tikzpicture}}}
= \xi_{a,b,j}
\vcenter{\hbox{\hspace{1mm}\begin{tikzpicture}[font=\footnotesize,inner sep=2pt]
  \begin{feynman}
  \path[pattern=north east lines,pattern color=ashgrey,very thin] (0.01,4) rectangle (0.6,0);
  \vertex (j1) at (0,0);
  \vertex [above=2.4 of j1] (j2);
  \vertex[ above =2.5 of j1] (j3);
  \vertex [small,dot] [ above=0.9cm of j1] (psi)  {};
  \vertex[right=0.2cm of psi] (psi1);
  \vertex[below=0.25cm of psi1] (psi2) [label=\(\,\;\Psi_j\)];
  \vertex [above=3.4cm of j1] (j4) ;
  \vertex[above=3.5cm of j1] (j5)  ;
  \vertex[ left=0.15cm of  j4] (j7);
  \vertex [left=0.15cm of j3] (j8); 
  \vertex [above =4cm of j1] (j6);
  \draw  (0.06,2.4) rectangle (-0.2, 2.5);
    \vertex[below=0.1cm of j6] (m1) ;
    \vertex[above=1.6cm of j1] (m3) [label=right:\(M_b\)];
    \vertex[above=1.3cm of j2] (m2) [label=right:\(M_b\)];
    \vertex[above=0.5cm of j4] (xx);
    \vertex[left=1.2cm of xx] (x);
    \vertex[below=0.1cm of x] (x100)  [label=left:\(B^{(r)}_{(a,0,0)}\)];
    \vertex[above=0.3 of j1] (jj) [label=right:\(M_b\)];
    \vertex [below=0.25cm of x] (x1);
    \vertex [above=0.15cm of x] (x2);
    \vertex[left=0.25cm of x2] (x22);
    \vertex[above=0.6cm of j6] (j7);
    \vertex[right=0.25cm of j7] (j77);
    \vertex[left=0.25cm of x2] (x22);
    \vertex[above=0.05cm of x] (x3);
    \vertex[left=0.25cm of x3] (x33);
    \vertex[above=0.1cm of x33] (x333);
    \vertex[above=0.4cm of x2] (x4);
    \draw [thick]    (j8) to[out=90,in=-90] (x1);
    \draw [thick] [middlearrow={latex}]   (x1) to (x2);
   \diagram*{
    (j1) --[thick] (j2),
    (j3) --[thick] (j6)
  };
  \end{feynman}
\end{tikzpicture}}}
\end{equation}
which we can call a charge. The mixed pentagon equation (\ref{Tdot_pentagon}) implies that these charges are all $r$-th roots of unity and that 
$\xi_{a,b,j}=(\xi_{1,b,j})^{a}$ so that it suffices to 
know the charges  $\xi_{1,b,j}$. Our calculations  give the values
\be
\xi_{1,b,j}=e^{-\frac{\pi i j}{N}} \, . 
\ee

We further observe that a general relation 
 \be 
 \xi_{1,b,2N-j} = \xi_{1,b,j}^{*} 
\ee
takes place. It is easy to derive this from the consistency condition between the 
$\dot{\mathrm{T}}$-matrix and the boundary OPE discussed in  section \ref{sec_boundaryope}. This means that passing the  defect 
junction of $M_{b}$ with $B^{(r)}_{(a,0,0)}$ through the perturbation (\ref{b_pert}) results in rotating the complex coupling 
\be \label{coupling_rot}
\lambda \to \lambda (\xi_{1,b,j})^{a} \, .
\ee
General considerations of \cite{Kon1} imply that there exists a topological junction with $B^{(r)}_{(a,0,0)}$ between the endpoints 
of the boundary RG flows for which the couplings differ by the rotation (\ref{coupling_rot}). In the situation when the end points are D0 branes 
they must be spaced accordingly on the circle. When $g$-theorem allows superpositions of two or more D0 branes as the end-points 
they must be again related by the corresponding rotations. 

A more interesting situation arises when the starting point of the flow is described by a superposition of irreducible  modules $M_{a}$. 
 We are going to look at a single example that arises for the $r=N=2$ rational free boson theory we discussed in section \ref{sec4.2}.
 Let us consider the direct sum $M=M_{0}\oplus M_{1}$. The relevant boundary fields are $\Psi_{(0,j)1}$, $\Psi_{(1,j)0}$ for $j=1,3$ and 
 $\Psi_{(0,2)0}$, $\Psi_{(1,2)1}$ so that there is a two-dimensional space of fields for each object $U_{j}$, $j=0,1,2,3$. 
 We introduce coordinates in the corresponding subspaces as 
 \be \label{model_fields1}
 \left[ \begin{array}{c} \alpha \\
 \beta \end{array}\right] = \alpha \Psi_{(0,j)1} + \beta \Psi_{(1,j)0} \, , \enspace j=1, 3 \, , 
 \ee
  \be  \label{model_fields2}
 \left[ \begin{array}{c} \alpha \\
 \beta \end{array}\right] = \alpha \Psi_{(0,2)0} + \beta \Psi_{(1,2)1}  \, .
 \ee
 The elementary topological defects are $X_{j}$, $j=0,1,2,3$.  Their  fusion ring is isomorphic to ${\mathbb Z}_{4}$. Each  defect can be 
 attached to $M$. Moreover the space of junctions for each $X_{j}$ is two-dimensional. We introduce coordinates 
 in each such subspace for fixed $j$ according to 
 \be \label{model_junctions}
 \Omega_{j}^{[\gamma, \delta]} \equiv \gamma\, \Tilde{\Omega}_{(j,1)\, j+1 \,\rm{mod}\, 2} + \delta\,  \Tilde{\Omega}_{(j,0)\, j\, \rm{mod}\, 2} \, , \qquad 
 \gamma, \delta \in {\mathbb C} \, . 
 \ee
Using our results from section \ref{sec4.2} it is straightforward to calculate the matrices that give the action of junctions in the two-dimensional subspaces 
of boundary fields with fixed $i$. We denote the matrices that give the passing of  $\Omega_{j}^{[\gamma, \delta]}$ through the subspace corresponding to 
$U_{i}$  as $\hat X_{j}^{(i)}$. Explicitly we obtain 
\bea \label{mat1}
&& \hat X_{1}^{(1)} = \left[\begin{array}{cc}0&-i\frac{\gamma}{\delta}\\
\frac{\delta}{\gamma} &0\end{array}\right] \, , \quad \hat X_{2}^{(1)} = 
\left[\begin{array}{cc}\frac{\gamma}{\delta}&0\\
0&-\frac{\delta}{\gamma}\end{array}\right] \, , \quad \hat X_{3}^{(1)} =  \left[\begin{array}{cc}0&i\frac{\gamma}{\delta}\\
\frac{\delta}{\gamma} &0\end{array}\right] \,  ,
\eea 
\bea \label{mat2}
&& \hat X_{1}^{(2)} = \left[\begin{array}{cc}0&-i\\
-i &0\end{array}\right] \, , \quad \hat X_{2}^{(2)} = 
\left[\begin{array}{cc}-1&0\\
0&-1\end{array}\right] \, , \quad \hat X_{3}^{(2)} =  \left[\begin{array}{cc}0&i\\
i &0\end{array}\right] \,  ,
\eea 
\bea \label{mat3} 
&& \hat X_{1}^{(3)} = \left[\begin{array}{cc}0&-\frac{\gamma}{\delta}\\
-i\frac{\delta}{\gamma} &0\end{array}\right] \, , \quad \hat X_{2}^{(3)} = 
\left[\begin{array}{cc}-\frac{\gamma}{\delta}&0\\
0&\frac{\delta}{\gamma}\end{array}\right] \, , \quad \hat X_{3}^{(3)} =  \left[\begin{array}{cc}0&-\frac{\gamma}{\delta}\\
i\frac{\delta}{\gamma} &0\end{array}\right] 
\eea 
where we assume that $\gamma \ne 0$, $\delta \ne 0$. 
The fields with $j=1,2,3$ are all relevant. We can draw consequences from the topological junctions for the RG flows provided we construct junctions  
that preserve the subspace of Hermitian boundary fields in the vector space spanned by the relevant fields. 
To describe this subspace we need to know the boundary two-point functions. The complete two-point functions contain the OPE coefficient at the identity field and a factor from the two-point conformal blocks. The two point functions on a circle with insertions at the opposite points are 
\bea
&&   \langle\Psi_{(0,1)1}(1)  \Psi_{(1,3)0}(-1) \rangle = C_{1,3}^{(010)0} B_{13} \langle 1 \rangle_{0} = C_{3,1}^{(101)0}B_{31} \langle 1 \rangle_{1} \, , \label{2p1}\\
&& \langle\Psi_{(0,3)1}(1)  \Psi_{(1,1)0}(-1) \rangle= C_{3,1}^{(010)0} B_{31} \langle 1 \rangle_{0} = C_{1,3}^{(101)0}B_{13} \langle 1 \rangle_{1} \label{2p2}  \, , \\
&& \langle \Psi_{(1,2)1} (1) \Psi_{(1,2)1}(-1) \rangle = C_{2,2}^{(111)0} B_{22} \langle 1\rangle_{1} \, , \\ 
&& \langle \Psi_{(0,2)0}(1)  \Psi_{(0,2)0}(-1) \rangle = C_{2,2}^{(000)0} B_{22}  \langle 1\rangle_{0}
\eea
where $C_{i,j}^{(abc)k}$ are the boundary OPE coefficients \eqref{boundaryOPE}, $\langle 1 \rangle_{a}$ are the disk partition functions for the boundary labelled by $a$, and 
$B_{ij}$ are the normalisation constants for two-point conformal blocks. The latter  are defined as follows. Let 
\be
V_{ij}^{k} : {\mathcal H}_{j} \to {\mathcal H}_{k}
\ee
be the vertex operators corresponding to the primaries of the extended chiral algebra. The two-point conformal block can be written as  
\be
\langle 0| V_{\bar k k}^{0}(z_2) V_{k0}^{k}(z_2) |0\rangle = (z_2-z_1)^{-2\Delta_{k}} B_{\bar k k} \, .
\ee
Note that in (\ref{2p1}), (\ref{2p2}) there are two expressions for each two point function. Those equalities should  hold due to the continuity of the two-point functions \cite{Runkel_A}, \cite{Lewellen} .  
Using the braiding matrix (\ref{Finit}) and the conformal dimension $\Delta_{1}=\Delta_{3}=\frac{1}{8} $ we obtain the relation $B_{13}=-B_{31}$. 
Furthermore as noted in \eqref{F[A]boson}  the values of the boundary OPE coefficients are given by the $\mathrm{F}$-matrix elements. In particular we find 
\be
C_{1,3}^{(010)0}=-C_{3,1}^{(101)0}= 1 \, , \quad C_{3,1}^{(010)0} = -  C_{1,3}^{(101)0} = 1 \, , \quad  C_{2,2}^{(000)0} = - C_{2,2}^{(111)0}= 1 \, .
\ee
Without loss of generality we assume that our fields are normalised so that $B_{13}=1$ and $B_{22}=1$.  The Hermitian conjugation is then given by 
\bea
&& \Psi_{(0,1)1}^{\dagger} = \Psi_{(1,3)0} \, , \qquad \Psi_{(1,1)0}^{\dagger} = -\Psi_{(0,3)1} \, , \nonumber \\
&& \Psi_{(0,2)0}^{\dagger} = \Psi_{(0,2)0} \, , \qquad  \Psi_{(1,2)1}^{\dagger} = -  \Psi_{(1,2)1} \, .
\eea 
Next we introduce a basis of Hermitian operators 
\begin{align}
 v_{1}&= \Psi_{(0,1)1} + \Psi_{(1,3)0} \, ,   &v_{2}&=  \Psi_{(1,1)0} - \Psi_{(0,3)1} \, , \nonumber \\
 v_{3}&= i( \Psi_{(0,1)1} - \Psi_{(1,3)0})  \, ,   &v_{4}& = i( \Psi_{(1,1)0} + \Psi_{(0,3)1}) \, , \nonumber \\
 v_{5}&=  \Psi_{(0,2)0}  \, ,  & v_6 &=  i \Psi_{(1,2)1} \, .
\end{align}
There are two  real subspaces
\be
W^{(1,3)} = \langle  v_{1}, v_2, v_3, v_4 \rangle _{\mathbb R}\, , \qquad W^{(2)}=\langle v_{5}, v_{6}  \rangle _{\mathbb R}\, .
\ee
Choosing $\gamma = i\delta$ in (\ref{model_junctions}) for each junction\footnote{In principle one can choose $\gamma$ and $\delta$ separately for each $j$. Moreover the choice we give is not the only solution. Here however we are not after an exhaustive classification but after a proof of concept example.} with $X_{j}$ we find from (\ref{mat1}),  (\ref{mat2}),  (\ref{mat3})  that on the subspace $W^{(1,3)}$ the corresponding junctions are represented by the matrices 
\be \label{4matrices}
\hat X_{1}^{(1,3)} = \left[ \begin{array}{cccc}
0&1&0&0\\
0&0&1&0\\
0&0&0&1\\
-1&0&0&0
\end{array} \right] \, , \quad 
\hat X_{2}^{(1,3)}= \left[ \begin{array}{cccc}
0&0&-1&0\\
0&0&0&-1\\
1&0&0&0\\
0&1&0&0 
\end{array} \right] \, , \quad 
\hat X_{3}^{(1,3)}= \left[ \begin{array}{cccc}
0&-1&0&0\\
0&0&1&0\\
0&0&0&-1\\
-1&0&0&0
\end{array} \right] \, .
\ee
We see that the matrix elements are all real and hence the chosen junctions preserve the real subspace. 
Consider now a boundary perturbation given by 
\be
\Delta S = \sum_{i=1}^{4} \lambda^{i}\! \int\!\! dt \, v_{i}(t)
\ee
where $\lambda_{i}$ are four real coupling constants which form a vector space with a fixed basis. 
The junctions we consider  act on this vector space by the  transpose of the matrices given in  (\ref{4matrices}). Each orbit of this action is finite. The neighbouring points on the orbit give 
two deformations for which the IR fixed point boundary conditions must admit a topological junction with the corresponding defect $X_{j}$. 

Similarly, in the second real subspace $W^{(2)}$ each junction (\ref{model_junctions}) with any $\gamma \ne 0$, $\delta \ne 0$ acts by 
the matrices 
\be \label{2mat2}
\tilde X_{1}^{(2)} = \left[ \begin{array}{cc}
0&1\\
-1&0 
\end{array}\right] \, , \qquad  \tilde X_{2}^{(2)} 
= \left[ \begin{array}{rr}
-1&0\\
0&-1 
\end{array}\right] \, , \qquad \tilde X_{3}^{(2)} = \left[ \begin{array}{rr}
0&-1\\
1&0 
\end{array}\right] \, .
\ee
These matrices are real so that the subspace  $W_2$ is preserved by these junctions. Considering the deformations 
\be
\Delta S =  \lambda^{5}\! \int\!\! dt \, v_{5}(t) +  \lambda^{6}\! \int\!\! dt \, v_{6}(t)\, ,   \qquad \lambda^5 \in {\mathbb R} \, , \enspace \lambda^{6} \in {\mathbb R} 
\ee
the junctions act on the couplings vector $(\lambda^{5}, \lambda^{6} )$ by the transpose of the matrices in (\ref{2mat2}). 
The corresponding orbits have only two elements so that 
the IR fixed points at the ends of the corresponding flows must have topological junctions with the corresponding defects. 

Notice that there are no real eigenvectors for the above matrices in $W^{(1,3)}$ and $W^{(2)}$ so that we do not obtain any constraints on individual flows but rather a collection of  relative constraints between them. It would be interesting to explore systematically what all of these constraints say about the possible set of IR end points. 

\section{Future directions}
We conclude by mentioning some open problems. 
In this paper  have worked out the details of the linear algebra problems that allow to compute the fusing matrices $\mathrm{Y}$, $\mathrm{T}$, 
$\dot{\mathrm{T}}$ from the RCFT representation theoretic data related to a pair  $({\mathcal C}, A)$. For concrete models like the free boson or the WZW models with non-trivial modular invariants it would be desirable to obtain general analytic expressions for these quantities. We hope this can be done in the future. Although a different approach is needed for this kind of task our method can be always used to check the results. 

In section \ref{RG_section} we outlined possible applications of  the fusing matrices we discussed to boundary RG flows. One can approach these applications more systematically and try to classify all pairs of junctions with boundary conditions and their Hermitian boundary field eigenvectors or subspaces. We hope to address such classification tasks for direct sums of boundary conditions in concrete models in future work. It would be  also interesting to explore 
how restrictive these constraints are in concrete examples of multi-coupling families of flows.  

Another interesting direction would be to understand better  the associative fusion algebras of compact open defects we discussed in section \ref{sec_tube}.
Their applications to boundary and bulk RG flows should be further explored.

\acknowledgments

We are grateful to Ingo Runkel for stimulating discussions, in particular for explaining the 
construction of associators relevant to the fusing matrices we consider in this paper.

\appendix

\section{Associators}

\mycomment{
 The structure of a modular tensor category \cite{Tur1,Kas,KRT,BK} naturally arises from conformal field theory as the representation category of certain rational conformal vertex algebras \cite{FLM,FHL,Kac,Hua2,Lep} which are the mathematical structure encoding physical information of a 2-dimensional chiral RCFT such as its chiral algebra and primary fields. More precisely, the symmetries of a chiral RCFT include conformal symmetries that can be encoded in an conformal vertex algebra $\mathfrak{V}_{\text{Vir}}$ which furnishes in particular a representation of the Virasoro Lie algebra. But, most models have additional symmetries where the  chiral algebra is a larger conformal vertex algebra $\mathfrak{V}\supseteq \mathfrak{V}_{\text{Vir}}$ \cite{Hua}. The spaces of fields of the chiral RCFT are then given by  $\mathfrak{V}$-modules and the category of interest is the representation category $\mathcal{C}:=\operatorname{Rep} (\mathfrak{V})$.

 Firstly,. }
 For reference   a tensor category is defined as follows
 \begin{dfn}\label{def1}
A \textit{tensor} category $\mathcal{C}=(\operatorname{Obj}(\mathcal{C}),\otimes,\mathbf{1},\alpha,\lambda,\rho)$ is a category $\mathcal{C}$, a bifunctor $\otimes:\mathcal{C}\times \mathcal{C}\to \mathcal{C}$, an object $\mathbf{1}\in\operatorname{Obj}(\mathcal{C})$ called the \textit{tensor unit}, and three natural isomorphisms $\alpha,\lambda,\rho$. Explicitly, the natural isomorphism
\begin{equation}\label{aa1}
    \alpha=\alpha_{U,V,W}:U\otimes(V\otimes W)\to(U\otimes V)\otimes W
\end{equation}
called the \textit{associator} is such that the following pentagon diagram commutes
\begin{equation}\label{pentag1}
 \vcenter{\hbox{\hspace{-10mm}\begin{tikzpicture}[inner sep=2pt]
  \begin{feynman}
  \vertex (j1) at (0,0) [label=\(U\otimes(V\otimes(W\otimes Z))\)];
  \vertex [right=4.5cm of j1] (j2) [label=\((U\otimes V)\otimes (W\otimes Z)\)];
  \vertex[right=4.5cm of j2] (j3) [label=\( ((U\otimes V)\otimes W)\otimes Z\)];
  \draw [-stealth] (1.7,0.25)--(2.8,0.25);
   \draw [-stealth] (6.2,0.25)--(7.3,0.25);
   \vertex at (2.2,0.3) [label=\footnotesize\(\alpha\)];
   \vertex at (6.7,0.3) [label=\footnotesize\(\alpha\)];
   \vertex[below=1.5cm of j1] (j4) [label=\(U\otimes((V\otimes W)\otimes Z)\)];
   \vertex [below =1.5cm of j3] (j5) [label=\((U\otimes(V\otimes W))\otimes Z\)];
    \draw [-stealth] (1.7,-1.25)--(7.3,-1.25);
     \draw [-stealth] (0.25,-0.1)--(0.25,-0.8);
      \draw [-stealth] (8.8,-0.1)--(8.8,-0.8);
      \vertex at (0.85,-0.65) [label=\footnotesize\(\operatorname{id}_U\!\otimes\alpha\)];
      \vertex at (9.4,-0.65) [label=\footnotesize\(\alpha\otimes \operatorname{id}_Z\)];
      \vertex at (4,-1.2) [label=\footnotesize\(\alpha\)];
   \diagram*{
    
  };
  \end{feynman}
\end{tikzpicture}}}
\end{equation}
Furthermore, the natural isomorphisms
\begin{equation}
    \lambda_U:\mathbf{1}\otimes U\to U,\qquad \rho_U:U\otimes \mathbf{1}\to U
\end{equation}
called the \textit{left} and \textit{right unitor} respectively, are such that the following triangle diagram commutes
\begin{equation}
 \vcenter{\hbox{\hspace{-10mm}\begin{tikzpicture}[inner sep=2pt]
  \begin{feynman}
  \vertex (j1) at (0,0) [label=\(U\otimes(\mathbf{1}\otimes W)\)];
  \vertex [right=5cm of j1] (j2) [label=\((U\otimes\mathbf{1})\otimes W\)];
  \vertex[below right=3.5cm of j1] (j3) [label=\(U\otimes W\)];
   \draw [-stealth] (1.2,0.25)--node[above]{\footnotesize$\alpha$}(3.9,0.25);
    \draw [-stealth] (0,-0.2)--node[below left]{\footnotesize$\operatorname{id}_U\otimes\lambda_W$}(2.2,-1.9);
    \draw [-stealth] (5,-0.2)--node[below right]{\footnotesize$\rho_U\otimes\operatorname{id}_W$}(2.7,-1.9);
  \diagram*{
    
  };
  \end{feynman}
\end{tikzpicture}}}
\end{equation}
\end{dfn}
Based on coherence theorems \cite[Chapter~VII.2]{Lane}, we will assume without loss of generality that our tensor category is \textit{strict} which means that all the isomorphisms $\alpha,\lambda,\rho$ are identities.

 A module category is defined as 
\begin{dfn}\label{modcat}
    Let $\mathcal{C}$ be a tensor category. A \textit{right module category} over $\mathcal{C}$ is a category $\mathcal{M}$ equipped with a \textit{module product bifunctor} $\otimes: \mathcal{M}\times\mathcal{C} \to M$ and a natural isomorphism with components
\begin{equation}
    m_{M,U,V}: M\otimes (U\otimes V)\to (M\otimes U)\otimes V, \qquad U,V\in\operatorname{Obj}(\mathcal{C}),\, M\in\operatorname{Obj}(\mathcal{M})
\end{equation}
called \textit{module associativity constraint} such that the functor $M\mapsto M \otimes \mathbf{1}$ is an autoequivalence and the following \textit{module pentagon diagram} commutes
\begin{equation}\label{modpent1}
 \vcenter{\hbox{\hspace{-3mm}\begin{tikzpicture}[inner sep=2pt]
  \begin{feynman}
  \vertex (j1) at (0,0) [label=\(M\otimes(U\otimes(V\otimes W))\)];
  \vertex [right=6cm of j1] (j2) [label=\((M\otimes U)\otimes (V\otimes W)\)];
  \vertex[right=6cm of j2] (j3) [label=\( ((M\otimes U)\otimes V)\otimes W\)];
  \draw [-stealth] (1.75,0.25)--(4.3,0.25);
   \draw [-stealth] (7.8,0.25)--(10.3,0.25);
   \vertex at (2.9,0.3) [label=\footnotesize\(m_{M,U,V\otimes W}\)];
   \vertex at (8.9,0.3) [label=\footnotesize\(m_{M\otimes U,V,W}\)];
   \vertex[below=2cm of j1] (j4) [label=\(M\otimes((U\otimes V)\otimes W)\)];
   \vertex [below =2cm of j3] (j5) [label=\((M\otimes(U\otimes V))\otimes W\)];
    \draw [-stealth] (1.75,-1.7)--(10.3,-1.7);
     \draw [-stealth] (0.25,-0.1)--(0.25,-1.4);
      \draw [-stealth] (11.6,-0.1)--(11.6,-1.4);
      \vertex at (1.4,-0.9) [label=\footnotesize\(\operatorname{id}_M\!\otimes \alpha_{U,V,W}\)];
      \vertex at (10.2,-0.9) [label=\footnotesize\(m_{M,U,V}\otimes \operatorname{id}_W\)];
      \vertex at (5.2,-1.6) [label=\footnotesize\(m_{M,U\otimes V,W}\)];
   \diagram*{
    
  };
  \end{feynman}
\end{tikzpicture}}}
\end{equation}
\normalsize  where $\alpha$ is the associator of $\mathcal{C}$ introduced in \eqref{aa1}. There is also a similar definition for a left module category.
\end{dfn}

\mycomment{
Consider a left $A$-module $(\dot{M},\rho^M)$. From the properties of $\mathcal{C}$ we obtain the following completeness relation involving the morphisms introduced in \eqref{emb1}
\begin{equation}
    \sum_{i\in\mathcal{I}}\sum_\alpha b^M_{i,\alpha}\circ b_M^{i,\alpha}=\operatorname{id}_{\dot{M}}
\end{equation}
}

\subsection{Associator of \texorpdfstring{$_A\mathcal{C}_A$}{TEXT}}\label{app3}

Here we prove the pentagon identity for the associator \eqref{associator1} which can be written as
\begin{equation}\label{a1}   \check{\alpha}_{X\otimes_AY,Z,W}\circ\check{\alpha}_{X,Y,Z\otimes_AW}=\left(\check{\alpha}_{X,Y,Z}\otimes_A\operatorname{id}_W\right)\circ \check{\alpha}_{X,Y\otimes_AZ,W}\circ\left(\id_X\otimes_A\check{\alpha}_{Y,Z,W}\right)
\end{equation}
We will proceed by calculating both sides of the last equation and showing that they are identical. Using \eqref{associator1} we find the left-hand side of \eqref{a1} to have the following graphical form
\begin{equation}\label{lhs1}
     \vcenter{\hbox{\hspace{2mm}\scalebox{0.8}{\begin{tikzpicture}[font=\footnotesize,inner sep=2pt]
  \begin{feynman}
  \vertex (x1) at (0,0) [label=below:\(X\)];
  \vertex [right=0.7cm of x1] (y1) [label=below:\(Y\)];
  \vertex[right=0.7cm of y1] (z1) [label=below:\(Z\)];
 \vertex[right=0.7cm of z1] (w1) [label=below:\(W\)];
 \vertex[above=0.5cm of x1] (x2) [label=above right:\(e_{X,Y\otimes_A(Z\otimes_AW)}\)] ;
 \vertex[above=1cm of w1] (w2);
 \vertex[above=0.5cm of y1] (y2);
 \vertex[above=0.5cm of z1] (z2);
 \vertex[above=0.5cm of w1] (w22);
 \vertex[left=0.25cm of x2)] (x3);
 \vertex [right=0.25cm of w2] (w3);
 \draw [fill=yellow] (x3) rectangle (w3);
 \vertex[above=0.5cm of x1] (x2) [label=above right:\(e_{X,Y\otimes_A(Z\otimes_AW)}\)] ;
 \vertex[above=2cm of x2] (x4);
 \vertex[above=1cm of y2] (y3);
 \vertex[left=0.25cm of y3] (y4);
 \vertex[above=1cm of z2] (z3);
 \vertex[above=1cm of w22] (w4);
 \vertex[above=0.5cm of w4] (w5);
 \vertex[right=0.25cm of w5] (w6);
 \draw [fill=yellow] (y4) rectangle (w6);
 \vertex[left=0.25cm of y3] (y4) [label=above right:\(e_{Y,Z\otimes_AW}\)];
 \vertex[above=0.5cm of x2] (x22);
 \vertex[above=0.5cm of y2] (y22);
 \vertex[above=0.5cm of z2] (z22);
 \vertex[above=1.5cm of y22] (y5);
 \vertex[above=2.5cm of z22] (z5);
 \vertex[above=3cm of w22] (w7);
 \vertex[above=0.5cm of y5] (y55);
 \vertex[right=0.25cm of y55] (y6);
 \vertex[left=0.25cm of x4] (x5);
 \draw [fill=yellow] (x5) rectangle (y6);
 \vertex[left=0.25cm of x4] (x5) [label=above right:\(r_{X,Y}\)];
 \vertex[above=0.5cm of x4] (x44);
 \vertex[above=0.5cm of y3] (y33);
 \vertex[above=0.5cm of z3] (z33);
 \vertex[above=0.5cm of x44] (x444);
\vertex[above=0.5cm of y33] (y333);
\vertex[above=0.5cm of y55] (y555);
\vertex[left=0.25cm of x444] (x6);
\vertex[above=0.5cm of w7] (w8);
\vertex[right=0.25cm of w8] (w88);
\draw [fill=yellow] (x6) rectangle (w88);
\vertex[left=0.25cm of x444] (x6) [label=above right:\(P_{X\otimes_AY,Z\otimes_AW}\)];
\vertex[above=0.5cm of x444] (xx1);
\vertex[above=0.5cm of y555] (yy1);
\vertex[above=0.5cm of z5] (zz1);
\vertex[above=0.5cm of w7] (ww1);
\vertex[above=0.5cm of zz1] (zz2);
\vertex[above=0.5cm of ww1] (ww2);
\vertex[above=0.5cm of ww2] (ww3);
\vertex[left=0.25cm of zz2] (zz3);
\vertex[right=0.25cm of ww3] (ww4);
\draw [fill=yellow] (zz3) rectangle (ww4);
\vertex[left=0.25cm of zz2] (zz3) [label=above right:\(e_{Z,W}\)];
\vertex[above=1.5cm of xx1] (xx2);
\vertex[above=1.5cm of yy1] (yy2);
\vertex[above=0.5cm of zz2] (zz4);
\vertex[above=1.5cm of ww3] (ww5);
\vertex[left=0.25cm of xx2] (xx3);
\vertex[above=1cm of zz4] (zz5);
\vertex[above=0.5cm of zz4] (zz55);
\vertex[right=0.25cm of zz5] (zz6);
\draw [fill=yellow] (zz6) rectangle (xx3);
\vertex[left=0.25cm of xx2] (xx3) [label=above right:\(r_{X\otimes_AY,Z}\)];
\vertex[above=0.5cm of xx2] (xx4);
\vertex[above=0.5cm of yy2] (yy3);
\vertex[above=0.5cm of xx4] (xx5);
\vertex[above=0.5cm of yy3] (yy4);
\vertex[above=0.5cm of zz5] (zz7);
\vertex[left=0.25cm of xx5] (xx6);
\vertex[above=0.5cm of ww5] (ww6);
\vertex[right=0.25cm of ww6] (ww7);
\draw [fill=yellow] (ww7) rectangle (xx6);
\vertex[left=0.25cm of xx5] (xx6) [label=above right:\(r_{(X\otimes_AY)\otimes_AZ,W}\)];
\vertex[above=0.5cm of xx5] (xx7);
\vertex[above=0.5cm of yy4] (yy5);
\vertex[above=0.5cm of zz7] (zz8);
\vertex[above=0.5cm of xx7] (xx8) [label=above:\(X\)];
\vertex[above=0.5cm of yy5] (yy6) [label=above:\(Y\)];
\vertex[above=0.5cm of zz8] (zz9) [label=above:\(Z\)];
\vertex[above=0.5cm of ww6] (ww8) [label=above:\(W\)];
   \diagram*{
   (x1)--[thick] (x2),
   (y1)--[thick] (y2),
   (z1)--[thick] (z2),
   (w1)--[thick] (w22),
   (x22)--[thick] (x4),
   (y22)--[thick] (y3),
   (z22)--[thick] (z3),
   (w4)--[thick] (w2),
   (y33)--[thick] (y5),
   (z5)--[thick] (z33),
   (w7)--[thick] (w5),
   (x44)--[thick] (x444),
   (y55)--[thick] (y555),
   (zz1)--[thick] (zz2),
   (ww1)--[thick] (ww2),
   (xx1)--[thick] (xx2),
   (yy1)--[thick] (yy2),
   (ww5)--[thick] (ww3),
   (zz55)--[thick] (zz4),
   (xx5)--[thick] (xx4),
   (yy4)--[thick] (yy3),
   (zz7)--[thick] (zz5),
   (xx7)--[thick] (xx8),
   (yy5)--[thick] (yy6),
   (zz8)--[thick] (zz9),
   (ww6)--[thick] (ww8)
  };
  \end{feynman}
\end{tikzpicture}}}}
\end{equation}
It will prove useful to manipulate the three morphisms in the middle of this expression. To this end, we can use the fact that $r_{X,Y}$ and $e_{Z,W}$ are bimodule intertwiners so they satisfy in particular
\begin{align}\label{a3}
    &e_{Z,W}\circ \rho_{Z\otimes_AW}=\rho_{Z\otimes W}\circ\left(\id_A\otimes e_{Z,W}\right) \qquad \text{and} \nonumber \\
    & r_{X,Y}\circ \Tilde{\rho}_{X\otimes Y}= \Tilde{\rho}_{X\otimes_AY}\circ\left(r_{X,Y}\otimes\id_A\right)
\end{align}
Now using the definition of the projectors \eqref{proj} and \eqref{a3} we can show that for any three $A\text{-}A$ bimodules $X,Y,Z$ we have
\begin{align}\label{a4}
   & P_{X\otimes_AY,Z}\circ(r_{X,Y}\otimes\id_Z)=(r_{X,Y}\otimes\id_Z)\circ P_{X\otimes Y,Z} \qquad \text{and} \nonumber \\
   &(\id_X\otimes e_{Y,Z})\circ P_{X,Y\otimes_AZ}=P_{X,Y\otimes Z}\circ (\id_X\otimes e_{Y,Z})
\end{align}
Using \eqref{a4} into \eqref{lhs1} leads to 
\begin{equation}\label{a5} 
     \vcenter{\hbox{\hspace{2mm}\scalebox{0.8}{\begin{tikzpicture}[font=\footnotesize,inner sep=2pt]
  \begin{feynman}
  \vertex (x1) at (0,0) [label=below:\(X\)];
  \vertex [right=0.7cm of x1] (y1) [label=below:\(Y\)];
  \vertex[right=0.7cm of y1] (z1) [label=below:\(Z\)];
 \vertex[right=0.7cm of z1] (w1) [label=below:\(W\)];
 \vertex[above=0.5cm of x1] (x2) [label=above right:\(e_{X,Y\otimes_A(Z\otimes_AW)}\)] ;
 \vertex[above=1cm of w1] (w2);
 \vertex[above=0.5cm of y1] (y2);
 \vertex[above=0.5cm of z1] (z2);
 \vertex[above=0.5cm of w1] (w22);
 \vertex[left=0.25cm of x2)] (x3);
 \vertex [right=0.25cm of w2] (w3);
 \draw [fill=yellow] (x3) rectangle (w3);
 \vertex[above=0.5cm of x1] (x2) [label=above right:\(e_{X,Y\otimes_A(Z\otimes_AW)}\)] ;
 \vertex[above=2cm of x2] (x4);
 \vertex[above=1cm of y2] (y3);
 \vertex[left=0.25cm of y3] (y4);
 \vertex[above=1cm of z2] (z3);
 \vertex[above=1cm of w22] (w4);
 \vertex[above=0.5cm of w4] (w5);
 \vertex[right=0.25cm of w5] (w6);
 \draw [fill=yellow] (y4) rectangle (w6);
 \vertex[left=0.25cm of y3] (y4) [label=above right:\(e_{Y,Z\otimes_AW}\)];
 \vertex[above=0.5cm of x2] (x22);
 \vertex[above=0.5cm of y2] (y22);
 \vertex[above=0.5cm of z2] (z22);
 \vertex[above=1.5cm of y22] (y5);
 \vertex[above=2.5cm of z22] (z5);
 \vertex[above=3cm of w22] (w7);
 \vertex[above=0.5cm of y5] (y55);
 \vertex[right=0.25cm of y55] (y6);
 \vertex[left=0.25cm of x4] (x5);
 \vertex[above=0.5cm of x4] (x44);
 \vertex[above=0.5cm of y3] (y33);
 \vertex[above=0.5cm of z3] (z33);
 \vertex[above=0.5cm of z33] (zzz1);
 \vertex[above=0.5cm of w5] (www1);
 \vertex[above=0.5cm of www1] (www2);
 \vertex[left=0.25cm of zzz1] (zzz2);
 \vertex[right=0.25cm of www2] (www3);
  \draw [fill=yellow] (zzz2) rectangle (www3);
  \vertex[left=0.25cm of zzz1] (zzz2) [label=above right:\(e_{Z,W}\)];
  \vertex[above=0.5cm of zzz1] (zzz3);
 \vertex[above=0.5cm of x44] (x444);
\vertex[above=0.5cm of y33] (y333);
\vertex[above=0.5cm of y55] (y555);
\vertex[left=0.25cm of x444] (x6);
\vertex[above=0.5cm of w7] (w8);
\vertex[right=0.25cm of w8] (w88);
\draw [fill=yellow] (x6) rectangle (w88);
\vertex[left=0.25cm of x444] (x6) [label=above right:\(P_{X\otimes Y,Z\otimes W}\)];
\vertex[above=0.5cm of x444] (xx1);
\vertex[above=0.5cm of y555] (yy1);
\vertex[above=0.5cm of z5] (zz1);
\vertex[above=0.5cm of w7] (ww1);
\vertex[above=0.5cm of xx1] (xxx1);
\vertex[above=0.5cm of yy1] (yyy1);
\vertex[above=0.5cm of yyy1] (yyy2);
\vertex[right=0.25cm of yyy2] (yyy3);
\vertex[left=0.25cm of xxx1] (xxx2);
\draw [fill=yellow] (yyy3) rectangle (xxx2);
\vertex[left=0.25cm of xxx1] (xxx2) [label=above right:\(r_{X,Y}\)];
\vertex[above=0.5cm of xxx1] (xxx3);
\vertex[above=0.5cm of zz1] (zz2);
\vertex[above=0.5cm of ww1] (ww2);
\vertex[above=0.5cm of ww2] (ww3);
\vertex[left=0.25cm of zz2] (zz3);
\vertex[right=0.25cm of ww3] (ww4);
\vertex[above=1.5cm of xx1] (xx2);
\vertex[above=1.5cm of yy1] (yy2);
\vertex[above=0.5cm of zz2] (zz4);
\vertex[above=1.5cm of ww3] (ww5);
\vertex[left=0.25cm of xx2] (xx3);
\vertex[above=1cm of zz4] (zz5);
\vertex[above=0.5cm of zz4] (zz55);
\vertex[right=0.25cm of zz5] (zz6);
\draw [fill=yellow] (zz6) rectangle (xx3);
\vertex[left=0.25cm of xx2] (xx3) [label=above right:\(r_{X\otimes_AY,Z}\)];
\vertex[above=0.5cm of xx2] (xx4);
\vertex[above=0.5cm of yy2] (yy3);
\vertex[above=0.5cm of xx4] (xx5);
\vertex[above=0.5cm of yy3] (yy4);
\vertex[above=0.5cm of zz5] (zz7);
\vertex[left=0.25cm of xx5] (xx6);
\vertex[above=0.5cm of ww5] (ww6);
\vertex[right=0.25cm of ww6] (ww7);
\draw [fill=yellow] (ww7) rectangle (xx6);
\vertex[left=0.25cm of xx5] (xx6) [label=above right:\(r_{(X\otimes_AY)\otimes_AZ,W}\)];
\vertex[above=0.5cm of xx5] (xx7);
\vertex[above=0.5cm of yy4] (yy5);
\vertex[above=0.5cm of zz7] (zz8);
\vertex[above=0.5cm of xx7] (xx8) [label=above:\(X\)];
\vertex[above=0.5cm of yy5] (yy6) [label=above:\(Y\)];
\vertex[above=0.5cm of zz8] (zz9) [label=above:\(Z\)];
\vertex[above=0.5cm of ww6] (ww8) [label=above:\(W\)];
   \diagram*{
   (x1)--[thick] (x2),
   (y1)--[thick] (y2),
   (z1)--[thick] (z2),
   (w1)--[thick] (w22),
   (x22)--[thick] (x4),
   (y22)--[thick] (y3),
   (z22)--[thick] (z3),
   (w4)--[thick] (w2),
   (y33)--[thick] (y5),
   (zzz1)--[thick] (z33),
   (www1)--[thick] (w5),
   (x44)--[thick] (x444),
   (y55)--[thick] (y555),
   (zz1)--[thick] (zz2),
   (ww1)--[thick] (ww2),
   (xx1)--[thick] (xxx1),
   (yy1)--[thick] (yyy1),
   (ww5)--[thick] (ww3),
   (zz55)--[thick] (zz4),
   (xx5)--[thick] (xx4),
   (yy4)--[thick] (yy3),
   (zz7)--[thick] (zz5),
   (xx7)--[thick] (xx8),
   (yy5)--[thick] (yy6),
   (zz8)--[thick] (zz9),
   (ww6)--[thick] (ww8),
   (x4)--[thick] (x44),
   (y55)--[thick] (y5),
   (zz2)--[thick] (zz4),
   (ww2)--[thick] (ww3),
   (xx2)--[thick] (xxx3),
   (yyy2)--[thick] (yy2),
   (www2)--[thick] (w7),
   (z5)--[thick] (zzz3)
  };
  \end{feynman}
\end{tikzpicture}}}}
\end{equation} 
Using the definition \eqref{associator1}, the right-hand side of \eqref{a1} reads
\begin{equation} \label{a6}
      \vcenter{\hbox{\hspace{2mm}\scalebox{0.8}{\begin{tikzpicture}[font=\footnotesize,inner sep=2pt]
  \begin{feynman}
  \vertex (x1) at (0,0) [label=below:\(X\)];
  \vertex [right=0.7cm of x1] (y1) [label=below:\(Y\)];
  \vertex [right=0.7cm of y1] (z1) [label=below:\(Z\)];
  \vertex [right=0.7cm of z1] (w1) [label=below:\(W\)];
  \vertex[above=1cm of w1] (w2);
  \vertex [above=0.5cm of w1] (w11);
  \vertex[above=0.5cm of x1] (x2);
  \vertex[above=0.5cm of y1] (y2);
  \vertex[above=0.5cm of z1] (z2);
  \vertex[left=0.25cm of x2] (x22);
  \vertex[right=0.25cm of w2] (w22);
  \draw [fill=yellow] (x22) rectangle (w22);
  \vertex[left=0.25cm of x2] (x22) [label=above right:\(e_{X,Y\otimes_A(Z\otimes_AW)}\)];
  \vertex[above=0.5cm of x2] (x3); 
  \vertex[above=0.5cm of y2] (y3);
  \vertex[above=0.5cm of z2] (z3);
  \vertex[above=0.5cm of y3] (y4);
  \vertex[above=0.5cm of z3] (z4);
  \vertex[above=0.5cm of w2] (w3);
  \vertex[above=0.5cm of w3] (w33);
  \vertex[right=0.25cm of w33] (w333);
  \vertex[left=0.25cm of y4] (y44);
   \draw [fill=yellow] (y44) rectangle (w333);
   \vertex[left=0.25cm of y4] (y44) [label=above right:\(e_{Y,Z\otimes_AW}\)];
   \vertex[above=0.5cm of y4] (y5); 
   \vertex[above=0.5cm of z4] (z5);
   \vertex[above=0.5cm of z5] (z6);
   \vertex[above=0.5cm of w33] (w4);
   \vertex[above=0.5cm of w4] (w5); 
   \vertex[right=0.25cm of w5] (w55);
   \vertex[left=0.25cm of z6] (z66);
    \draw [fill=yellow] (z66) rectangle (w55);
    \vertex[left=0.25cm of z6] (z66) [label=above right:\(e_{Z,W}\)];
    \vertex[above=0.5cm of z6] (z7); 
    \vertex[above=0.5cm of z7] (z8);
    \vertex[above=1.5cm of y5] (y6);
    \vertex[above=0.5cm of z8] (z9); 
    \vertex[right=0.25cm of z9] (z99);
    \vertex[left=0.25cm of y6] (y66);
     \draw [fill=yellow] (y66) rectangle (z99);
     \vertex[left=0.25cm of y6] (y66) [label=above right:\(r_{Y,Z}\)];
     \vertex[above=1.5cm of w5] (w6); 
     \vertex[above=0.5cm of w6] (w7);
     \vertex[right=0.25cm of w7] (w77);
     \vertex[above=0.5cm of y6] (y7); 
     \vertex[above=0.5cm of y7] (y8); 
     \vertex[left=0.25cm of y8] (y88);
      \draw [fill=yellow] (y88) rectangle (w77);
      \vertex[left=0.25cm of y8] (y88) [label=above right:\(r_{Y\otimes_AZ,W}\)];
      \vertex[above=0.5cm of z9] (z10);
      \vertex[above=4.5cm of x3] (x4);
      \vertex[left=0.25cm of x4] (x44);
      \vertex[above=0.5cm of y8] (yy1); 
      \vertex[above=0.5cm of z10] (zz1);
      \vertex[above=0.5cm of yy1] (yy2); 
      \vertex[above=0.5cm of zz1] (zz2);
      \vertex[above=0.5cm of w7] (ww1);
      \vertex[above=0.5cm of ww1] (ww2); 
      \vertex[right=0.25cm of ww2] (ww22);
       \draw [fill=yellow] (ww22) rectangle (x44);
       \vertex[left=0.25cm of x4] (x44) [label=above right:\(P_{X,(Y\otimes_AZ)\otimes_AW}\)];
       \vertex[above=0.5cm of x4] (x5); 
       \vertex[above=0.5cm of yy2] (yy3);
       \vertex[above=0.5cm of zz2] (zz3);
       \vertex[above=0.5cm of yy3] (yy4); 
       \vertex[above=0.5cm of zz3] (zz4);
       \vertex[above=0.5cm of ww2] (ww3);
       \vertex[left=0.25cm of yy4] (yy44);
       \vertex[above=0.5cm of ww3] (ww4);
       \vertex[right=0.25cm of ww4] (ww44);
       \draw [fill=yellow] (ww44) rectangle (yy44);
       \vertex[left=0.25cm of yy4] (yy44) [label=above right:\(e_{Y\otimes_AZ,W}\)];
       \vertex[above=0.5cm of yy4] (yy5); 
       \vertex[above=0.5cm of zz4] (zz5); 
       \vertex[above=1.5cm of x5] (x6); 
       \vertex[above=0.5cm of yy5] (yy6); 
       \vertex[above=0.5cm of zz5] (zz6); 
       \vertex[above=0.5cm of zz6] (zz7);
       \vertex[right=0.25cm of zz7] (zz77);
       \vertex[left=0.25cm of x6] (x66);
        \draw [fill=yellow] (x66) rectangle (zz77);
        \vertex[left=0.25cm of x6] (x66)[label=above right:\(r_{X,Y\otimes_AZ}\)];
        \vertex[above=0.5cm of x6] (x7); 
        \vertex[above=0.5cm of yy6] (yy7); 
        \vertex[above=0.5cm of zz6] (zz7); 
        \vertex[above=1.5cm of ww4] (ww5); 
        \vertex[above=0.5cm of ww5] (ww6); 
        \vertex[right=0.25cm of ww6] (ww66);
        \vertex[above=0.5cm of x7] (x8); 
        \vertex[left=0.25cm of x8] (x88);
        \vertex[above=0.5cm of yy7] (yy8); 
        \vertex[above=0.5cm of zz7] (zz8); 
          \draw [fill=yellow] (ww66) rectangle (x88);
          \vertex[left=0.25cm of x8] (x88)[label=above right:\(P_{X\otimes_A(Y\otimes_AZ),W}\)];
          \vertex[above=0.5cm of x8] (xx1); 
          \vertex[above=0.5cm of yy8] (yy9); 
          \vertex[above=0.5cm of zz8] (zz9); 
          \vertex[above=0.5cm of xx1] (xx2); 
          \vertex[above=0.5cm of yy9] (yyy1); 
          \vertex[above=0.5cm of zz9] (zzz1); 
          \vertex[left=0.25cm of xx2] (xx22);
          \vertex[above=0.5cm of zzz1] (zzz2); 
          \vertex[right=0.25cm of zzz2] (zzz22);
           \draw [fill=yellow] (zzz22) rectangle (xx22);
           \vertex[left=0.25cm of xx2] (xx22)[label=above right:\(e_{X,Y\otimes_AZ}\)];
           \vertex[above=0.5cm of xx2] (xx3); 
           \vertex[above=0.5cm of yyy1] (yyy2); 
           \vertex[above=0.5cm of yyy2] (yyy3); 
           \vertex[above=0.5cm of zzz2] (zzz3); 
           \vertex[above=0.5cm of zzz3]  (zzz4); 
           \vertex[left=0.25cm of yyy3] (yyy33);
           \vertex[right=0.25cm of zzz4] (zzz44);
            \draw [fill=yellow] (zzz44) rectangle (yyy33);
             \vertex[left=0.25cm of yyy3] (yyy33)[label=above right:\(e_{Y,Z}\)];
             \vertex[above=0.5cm of yyy3] (yyy4); 
             \vertex[above=1.5cm of xx3] (xx4); 
             \vertex[above=0.5cm of yyy4] (yyy5); 
             \vertex[above=0.5cm of yyy5] (yyy6); 
             \vertex[left=0.25cm of xx4] (xx44);
             \vertex[right=0.25cm of yyy6] (yyy66);
             \draw [fill=yellow] (xx44) rectangle (yyy66);
             \vertex[left=0.25cm of xx4] (xx44)[label=above right:\(r_{X,Y}\)];
             \vertex[above=0.5cm of xx4] (xx5); 
             \vertex[above=1.5cm of zzz4] (zzz5); 
             \vertex[above=0.5cm of xx5] (xx6); 
             \vertex[above=0.5cm of yyy6] (yyy7); 
             \vertex[above=0.5cm of zzz5] (zzz6); 
             \vertex[left=0.25cm of xx6] (xx66);
             \vertex[right=0.25cm of zzz6] (zzz66);
              \draw [fill=yellow] (zzz66) rectangle (xx66);
              \vertex[left=0.25cm of xx6] (xx66)[label=above right:\(r_{X\otimes_AY,Z}\)];
               \vertex[above=0.5cm of xx6] (xx7); 
              \vertex[above=0.5cm of yyy7] (yyy8); 
              \vertex[above=0.5cm of xx7] (xx8); 
              \vertex[above=0.5cm of yyy8] (yyy9); 
              \vertex[above=0.5cm of zzz6] (zzz7); 
              \vertex[above=4.5cm of ww6] (ww7); 
              \vertex[above=0.5cm of ww7] (ww8); 
              \vertex[left=0.25cm of xx8] (xx88);
              \vertex[right=0.25cm of ww8] (ww88);
               \draw [fill=yellow] (ww88) rectangle (xx88);
                \vertex[left=0.25cm of xx8] (xx88)[label=above right:\(r_{(X\otimes_AY)\otimes_AZ,W}\)];
                \vertex[above=0.5cm of xx8] (xx9);
                \vertex[above=0.5cm of yyy9] (yyy0);
                \vertex[above=0.5cm of zzz7] (zzz8);
                \vertex[above=0.5cm of xx9] (xx0) [label=above:\(X\)];
                \vertex[above=0.5cm of yyy0] (y0) [label=above:\(y\)];
                \vertex[above=0.5cm of zzz8] (z0) [label=above:\(Z\)];
                \vertex[above=0.5cm of ww8] (w0)  [label=above:\(W\)];
   \diagram*{
   (x1)--[thick] (x2),
   (y1)--[thick] (y2),
   (z1)--[thick] (z2),
   (w1)--[thick] (w11),
   (y3)--[thick] (y4),
   (z3)--[thick] (z4),
   (w2)--[thick] (w3),
   (z5)--[thick] (z6),
   (w33)--[thick] (w4),
   (y5)--[thick] (y6),
   (z7)--[thick] (z8),
   (z9)--[thick] (z10),
   (y7)--[thick] (y8),
   (w5)--[thick] (w6),
   (x3)--[thick] (x4),
   (yy1)--[thick] (yy2),
   (zz1)--[thick] (zz2),
   (w7)--[thick] (ww1),
   (ww2)--[thick] (ww3),
   (yy3)--[thick] (yy4),
   (zz3)--[thick] (zz4),
   (x5)--[thick] (x6),
   (yy5)--[thick] (yy6),
   (zz5)--[thick] (zz6),
   (x7)--[thick] (x8),
   (yy7)--[thick] (yy8),
   (zz7)--[thick] (zz8),
   (ww4)--[thick] (ww5),
   (xx1)--[thick] (xx2),
   (yy9)--[thick] (yyy1),
   (zz9)--[thick] (zzz1),
   (yyy2)--[thick] (yyy3),
   (zzz2)--[thick] (zzz3),
   (xx3)--[thick] (xx4),
   (yyy4)--[thick] (yyy5),
   (xx5)--[thick] (xx6),
   (yyy6)--[thick] (yyy7),
   (zzz4)--[thick] (zzz5),
   (xx7)--[thick] (xx8),
   (yyy8)--[thick] (yyy9),
   (zzz6)--[thick] (zzz7),
   (ww6)--[thick] (ww7),
   (ww8)--[thick] (w0),
   (xx9)--[thick] (xx0),
   (yyy0)--[thick] (y0),
   (zzz8)--[thick] (z0)
  };
  \end{feynman}
\end{tikzpicture}}}}
\end{equation}
To manipulate this expression we proceed in a similar way as the one described in \eqref{a3} and \eqref{a4}, namely we  use the fact that $e_{X,Y\otimes_AZ}$ and $e_{Y\otimes_AZ,W}$ are bimodule intertwiners. Combining this with the explicit form of the relevant projectors we can show that
\begin{align}\label{a7}
& \left(e_{X,Y\otimes_AZ}\otimes\id_W\right)\circ P_{X\otimes_A(Y\otimes_AZ),W} =P_{X\otimes(Y\otimes_AZ),W}\circ\left(e_{X,Y\otimes_AZ}\otimes\id_W\right) \qquad \text{and}\nonumber \\
& \left(\id_X\otimes e_{Y\otimes_AZ,W}\right)\circ P_{X,(Y\otimes_AZ)\otimes_AW}=P_{X,(Y\otimes_AZ)\otimes W}\circ \left(\id_X\otimes e_{Y\otimes_AZ,W}\right)
\end{align}
Next we  use \eqref{proj11} and also notice that for any three bimodules $X,Y,Z$ we have
\begin{equation}\label{a8}
    P_{X\otimes Y,Z}=\id_X\otimes P_{Y,Z} \qquad \text{and} \qquad P_{X,Y\otimes Z}=P_{X,Y}\otimes \id_Z
\end{equation}
which follows from the explicit form of the projector \eqref{proj}  and from the definition of the right and left action of $A$ on the bimodule $X\otimes Y$ and $Y\otimes Z$ respectively, that is, $\Tilde{\rho}_{X\otimes Y}=\id_X\otimes \tilde{\rho}^Y$ and $\rho_{Y\otimes Z}=\rho^Y\otimes\id_Z$. Finally, we  use the fact that for any $A\text{-}A$-bimodules $X,Y$ the projector $P_{X,Y}$ is an idempotent. Combining all the above, we  put \eqref{a6} into the following form
\begin{equation}\label{a9}
      \vcenter{\hbox{\hspace{2mm}\scalebox{0.8}{\begin{tikzpicture}[font=\footnotesize,inner sep=2pt]
  \begin{feynman}
   \vertex (x1) at (0,0) [label=below:\(X\)];
  \vertex [right=0.7cm of x1] (y1) [label=below:\(Y\)];
  \vertex [right=0.7cm of y1] (z1) [label=below:\(Z\)];
  \vertex [right=0.7cm of z1] (w1) [label=below:\(W\)];
  \vertex[above=1cm of w1] (w2);
  \vertex [above=0.5cm of w1] (w11);
  \vertex[above=0.5cm of x1] (x2);
  \vertex[above=0.5cm of y1] (y2);
  \vertex[above=0.5cm of z1] (z2);
  \vertex[left=0.25cm of x2] (x22);
  \vertex[right=0.25cm of w2] (w22);
  \draw [fill=yellow] (x22) rectangle (w22);
  \vertex[left=0.25cm of x2] (x22) [label=above right:\(e_{X,Y\otimes_A(Z\otimes_AW)}\)];
  \vertex[above=0.5cm of x2] (x3); 
  \vertex[above=0.5cm of y2] (y3);
  \vertex[above=0.5cm of z2] (z3);
  \vertex[above=0.5cm of y3] (y4);
  \vertex[above=0.5cm of z3] (z4);
  \vertex[above=0.5cm of w2] (w3);
  \vertex[above=0.5cm of w3] (w33);
  \vertex[right=0.25cm of w33] (w333);
  \vertex[left=0.25cm of y4] (y44);
   \draw [fill=yellow] (y44) rectangle (w333);
   \vertex[left=0.25cm of y4] (y44) [label=above right:\(e_{Y,Z\otimes_AW}\)];
   \vertex[above=0.5cm of y4] (y5); 
   \vertex[above=0.5cm of z4] (z5);
   \vertex[above=0.5cm of z5] (z6);
   \vertex[above=0.5cm of w33] (w4);
   \vertex[above=0.5cm of w4] (w5); 
   \vertex[right=0.25cm of w5] (w55);
   \vertex[left=0.25cm of z6] (z66);
    \draw [fill=yellow] (z66) rectangle (w55);
    \vertex[left=0.25cm of z6] (z66) [label=above right:\(e_{Z,W}\)];
    \vertex[above=0.5cm of z6] (z7); 
    \vertex[above=0.5cm of z7] (z8);
    \vertex[above=1.5cm of y5] (y6);
    \vertex[above=0.5cm of z8] (z9); 
    \vertex[right=0.25cm of z9] (z99);
    \vertex[left=0.25cm of y6] (y66);
     \draw [fill=yellow] (y66) rectangle (z99);
     \vertex[left=0.25cm of y6] (y66) [label=above right:\(r_{Y,Z}\)];
     \vertex[above=0.5cm of y6] (y7); 
     \vertex[above=0.5cm of y7] (y8); 
     \vertex[above=0.5cm of z9] (zz1); 
     \vertex[above=1.5cm of w5] (w6); 
     \vertex[above=0.5cm of w6] (w7); 
     \vertex[left=0.25cm of y8] (y88);
     \vertex[right=0.25cm of w7] (w77);
      \draw [fill=yellow] (w77) rectangle (y88);
       \vertex[left=0.25cm of y8] (y88)[label=above right:\(P_{Y\otimes_AZ,W}\)];
       \vertex[above=0.5cm of y8] (y9); 
       \vertex[above=0.5cm of zz1] (zz2); 
       \vertex[above=4.5cm of x3] (x4); 
       \vertex[above=0.5cm of y9] (yy1); 
       \vertex[above=0.5cm of zz2] (zz3); 
       \vertex[above=0.5cm of zz3] (zz4); 
       \vertex[left=0.25cm of x4] (x44);
       \vertex[right=0.25cm of zz4] (zz44);
       \draw [fill=yellow] (zz44) rectangle (x44);
       \vertex[left=0.25cm of x4] (x44)[label=above right:\(P_{X,Y\otimes_AZ}\)];
       \vertex[above=0.5cm of x4] (x5); 
       \vertex[above=0.5cm of yy1] (yy2); 
       \vertex[above=0.5cm of yy2] (yy3); 
       \vertex[above=0.5cm of zz4] (zz5); 
       \vertex[above=0.5cm of zz5] (zz6); 
       \vertex[left=0.25cm of yy3] (yy33);
       \vertex[right=0.25cm of zz6] (zz66);
       \draw [fill=yellow] (zz66) rectangle (yy33);
       \vertex[left=0.25cm of yy3] (yy33)[label=above right:\(e_{Y,Z}\)];
       \vertex[above=0.5cm of yy3] (yy4); 
       \vertex[above=1.5cm of x5] (x6); 
       \vertex[above=0.5cm of yy4] (yy5); 
       \vertex[above=0.5cm of yy5] (yy6); 
       \vertex[left=0.25cm of x6] (x66);
       \vertex[right=0.25cm of  yy6] (yy66);
        \draw [fill=yellow] (yy66) rectangle (x66);
         \vertex[left=0.25cm of x6] (x66)[label=above right:\(r_{X,Y}\)];
         \vertex[above=0.5cm of x6] (x7); 
         \vertex[above=0.5cm of x7] (x8); 
         \vertex[above=0.5cm of yy6] (yy7); 
         \vertex[above=1.5cm of zz6] (zz7); 
         \vertex[above=0.5cm of zz7] (zz8); 
         \vertex[left=0.25cm of x8] (x88);
         \vertex[right=0.25cm of zz8] (zz88);
          \draw [fill=yellow] (zz88) rectangle (x88);
          \vertex[left=0.25cm of x8] (x88)[label=above right:\(r_{X\otimes_AY,Z}\)];
          \vertex[above=0.5cm of x8] (x9); 
          \vertex[above=0.5cm of yy7] (yy8); 
          \vertex[above=0.5cm of x9] (xx1); 
          \vertex[above=0.5cm of yy8] (yy9); 
          \vertex[above=0.5cm of zz8] (zz9); 
          \vertex[above=4.5cm of w7] (w8); 
          \vertex[above=0.5cm of w8] (w9); 
          \vertex[left=0.25cm of xx1] (xx11);
          \vertex[right=0.25cm of w9] (w99);
           \draw [fill=yellow] (w99) rectangle (xx11);
            \vertex[left=0.25cm of xx1] (xx11)[label=above right:\(r_{(X\otimes_AY)\otimes_AZ,W}\)];
            \vertex[above=0.5cm of xx1] (xx2); 
            \vertex[above=0.5cm of yy9] (yyy1); 
            \vertex[above=0.5cm of zz9] (zzz1); 
            \vertex[above=0.5cm of xx2] (xx3) [label=above:\(X\)];
            \vertex[above=0.5cm of yyy1] (yyy2) [label=above:\(Y\)];
            \vertex[above=0.5cm of zzz1] (zzz2) [label=above:\(Z\)];
            \vertex[above=0.5cm of w9] (w0) [label=above:\(W\)];
   \diagram*{
 (x1)--[thick] (x2),
   (y1)--[thick] (y2),
   (z1)--[thick] (z2),
   (w1)--[thick] (w11),
   (y3)--[thick] (y4),
   (z3)--[thick] (z4),
   (w2)--[thick] (w3),
   (z5)--[thick] (z6),
   (w33)--[thick] (w4),
   (y5)--[thick] (y6),
   (z7)--[thick] (z8),
   (y7)--[thick] (y8),
   (z9)--[thick] (zz1),
   (w5)--[thick] (w6),
   (x3)--[thick] (x4),
   (y9)--[thick] (yy1),
   (zz2)--[thick] (zz3),
   (yy2)--[thick] (yy3),
   (zz4)--[thick] (zz5),
   (x5)--[thick] (x6),
   (yy4)--[thick] (yy5),
   (x7)--[thick] (x8),
   (yy6)--[thick] (yy7),
   (zz6)--[thick] (zz7),
   (x9)--[thick] (xx1),
   (yy8)--[thick] (yy9),
   (zz8)--[thick] (zz9),
   (w7)--[thick] (w8),
   (w0)--[thick] (w9),
   (xx2)--[thick] (xx3),
   (yyy1)--[thick] (yyy2),
   (zzz1)--[thick] (zzz2)
  };
  \end{feynman}
\end{tikzpicture}}}}
\end{equation}
Now, recalling that $e_{Y,Z}$ is a bimodule intertwiner, we  pull it down through the two projectors in the middle of the last diagram and then use $P_{Y,Z}=e_{Y,Z}\circ r_{Y,Z}$. Doing this we rewrite \eqref{a9} as
\begin{equation}\label{a10}
      \vcenter{\hbox{\hspace{2mm}\scalebox{0.8}{\begin{tikzpicture}[font=\footnotesize,inner sep=2pt]
  \begin{feynman}
   \vertex (x1) at (0,0) [label=below:\(X\)];
  \vertex [right=0.7cm of x1] (y1) [label=below:\(Y\)];
  \vertex [right=0.7cm of y1] (z1) [label=below:\(Z\)];
  \vertex [right=0.7cm of z1] (w1) [label=below:\(W\)];
  \vertex[above=1cm of w1] (w2);
  \vertex [above=0.5cm of w1] (w11);
  \vertex[above=0.5cm of x1] (x2);
  \vertex[above=0.5cm of y1] (y2);
  \vertex[above=0.5cm of z1] (z2);
  \vertex[left=0.25cm of x2] (x22);
  \vertex[right=0.25cm of w2] (w22);
  \draw [fill=yellow] (x22) rectangle (w22);
  \vertex[left=0.25cm of x2] (x22) [label=above right:\(e_{X,Y\otimes_A(Z\otimes_AW)}\)];
  \vertex[above=0.5cm of x2] (x3); 
  \vertex[above=0.5cm of y2] (y3);
  \vertex[above=0.5cm of z2] (z3);
  \vertex[above=0.5cm of y3] (y4);
  \vertex[above=0.5cm of z3] (z4);
  \vertex[above=0.5cm of w2] (w3);
  \vertex[above=0.5cm of w3] (w33);
  \vertex[right=0.25cm of w33] (w333);
  \vertex[left=0.25cm of y4] (y44);
   \draw [fill=yellow] (y44) rectangle (w333);
   \vertex[left=0.25cm of y4] (y44) [label=above right:\(e_{Y,Z\otimes_AW}\)];
   \vertex[above=0.5cm of y4] (y5); 
   \vertex[above=0.5cm of z4] (z5);
   \vertex[above=0.5cm of z5] (z6);
   \vertex[above=0.5cm of w33] (w4);
   \vertex[above=0.5cm of w4] (w5); 
   \vertex[right=0.25cm of w5] (w55);
   \vertex[left=0.25cm of z6] (z66);
    \draw [fill=yellow] (z66) rectangle (w55);
    \vertex[left=0.25cm of z6] (z66) [label=above right:\(e_{Z,W}\)];
    \vertex[above=0.5cm of z6] (z7); 
    \vertex[above=0.5cm of z7] (z8);
    \vertex[above=1.5cm of y5] (y6);
    \vertex[above=0.5cm of z8] (z9); 
    \vertex[right=0.25cm of z9] (z99);
    \vertex[left=0.25cm of y6] (y66);
     \draw [fill=yellow] (y66) rectangle (z99);
     \vertex[left=0.25cm of y6] (y66) [label=above right:\(P_{Y,Z}\)];
     \vertex[above=0.5cm of y6] (y7); 
     \vertex[above=0.5cm of y7] (y8); 
     \vertex[above=0.5cm of z9] (zz1); 
     \vertex[above=1.5cm of w5] (w6); 
     \vertex[above=0.5cm of w6] (w7); 
     \vertex[left=0.25cm of y8] (y88);
     \vertex[right=0.25cm of w7] (w77);
      \draw [fill=yellow] (w77) rectangle (y88);
       \vertex[left=0.25cm of y8] (y88)[label=above right:\(P_{Y\otimes Z,W}\)];
       \vertex[above=0.5cm of y8] (y9); 
       \vertex[above=0.5cm of zz1] (zz2); 
       \vertex[above=4.5cm of x3] (x4); 
       \vertex[above=0.5cm of y9] (yy1); 
       \vertex[above=0.5cm of zz2] (zz3); 
       \vertex[above=0.5cm of zz3] (zz4); 
       \vertex[left=0.25cm of x4] (x44);
       \vertex[right=0.25cm of zz4] (zz44);
       \draw [fill=yellow] (zz44) rectangle (x44);
       \vertex[left=0.25cm of x4] (x44)[label=above right:\(P_{X,Y\otimes Z}\)];
       \vertex[above=0.5cm of x4] (x5); 
       \vertex[above=0.5cm of yy1] (yy2); 
       \vertex[above=0.5cm of yy2] (yy3); 
       \vertex[above=0.5cm of zz4] (zz5); 
       \vertex[above=0.5cm of zz5] (zz6); 
       \vertex[left=0.25cm of yy3] (yy33);
       \vertex[right=0.25cm of zz6] (zz66);
       \vertex[above=0.5cm of yy3] (yy4); 
       \vertex[above=1.5cm of x5] (x6); 
       \vertex[above=0.5cm of yy4] (yy5); 
       \vertex[above=0.5cm of yy5] (yy6); 
       \vertex[left=0.25cm of x6] (x66);
       \vertex[right=0.25cm of  yy6] (yy66);
        \draw [fill=yellow] (yy66) rectangle (x66);
         \vertex[left=0.25cm of x6] (x66)[label=above right:\(r_{X,Y}\)];
         \vertex[above=0.5cm of x6] (x7); 
         \vertex[above=0.5cm of x7] (x8); 
         \vertex[above=0.5cm of yy6] (yy7); 
         \vertex[above=1.5cm of zz6] (zz7); 
         \vertex[above=0.5cm of zz7] (zz8); 
         \vertex[left=0.25cm of x8] (x88);
         \vertex[right=0.25cm of zz8] (zz88);
          \draw [fill=yellow] (zz88) rectangle (x88);
          \vertex[left=0.25cm of x8] (x88)[label=above right:\(r_{X\otimes_AY,Z}\)];
          \vertex[above=0.5cm of x8] (x9); 
          \vertex[above=0.5cm of yy7] (yy8); 
          \vertex[above=0.5cm of x9] (xx1); 
          \vertex[above=0.5cm of yy8] (yy9); 
          \vertex[above=0.5cm of zz8] (zz9); 
          \vertex[above=4.5cm of w7] (w8); 
          \vertex[above=0.5cm of w8] (w9); 
          \vertex[left=0.25cm of xx1] (xx11);
          \vertex[right=0.25cm of w9] (w99);
           \draw [fill=yellow] (w99) rectangle (xx11);
            \vertex[left=0.25cm of xx1] (xx11)[label=above right:\(r_{(X\otimes_AY)\otimes_AZ,W}\)];
            \vertex[above=0.5cm of xx1] (xx2); 
            \vertex[above=0.5cm of yy9] (yyy1); 
            \vertex[above=0.5cm of zz9] (zzz1); 
            \vertex[above=0.5cm of xx2] (xx3) [label=above:\(X\)];
            \vertex[above=0.5cm of yyy1] (yyy2) [label=above:\(Y\)];
            \vertex[above=0.5cm of zzz1] (zzz2) [label=above:\(Z\)];
            \vertex[above=0.5cm of w9] (w0) [label=above:\(W\)];
   \diagram*{
 (x1)--[thick] (x2),
   (y1)--[thick] (y2),
   (z1)--[thick] (z2),
   (w1)--[thick] (w11),
   (y3)--[thick] (y4),
   (z3)--[thick] (z4),
   (w2)--[thick] (w3),
   (z5)--[thick] (z6),
   (w33)--[thick] (w4),
   (y5)--[thick] (y6),
   (z7)--[thick] (z8),
   (y7)--[thick] (y8),
   (z9)--[thick] (zz1),
   (w5)--[thick] (w6),
   (x3)--[thick] (x4),
   (y9)--[thick] (yy1),
   (zz2)--[thick] (zz3),
   (yy2)--[thick] (yy3),
   (zz4)--[thick] (zz5),
   (x5)--[thick] (x6),
   (yy4)--[thick] (yy5),
   (x7)--[thick] (x8),
   (yy6)--[thick] (yy7),
   (zz6)--[thick] (zz7),
   (x9)--[thick] (xx1),
   (yy8)--[thick] (yy9),
   (zz8)--[thick] (zz9),
   (w7)--[thick] (w8),
   (w0)--[thick] (w9),
   (xx2)--[thick] (xx3),
   (yyy1)--[thick] (yyy2),
   (zzz1)--[thick] (zzz2),
   (yy3)--[thick] (yy4),
   (zz5)--[thick] (zz6)
  };
  \end{feynman}
\end{tikzpicture}}}}
\end{equation}
Now recall that $P_{Y\otimes Z,W}=\id_Y\otimes P_{Z,W}$ and that $Z$ is an $A\text{-}A$-bimodule satisfying in particular $\rho_Z\circ\left(\id_A\otimes\Tilde{\rho}_Z\right)=\Tilde{\rho}_Z\circ\left(\rho_Z\otimes \id_A\right)$. Using this and the explicit form of $P_{Z,W}$ and $P_{Y,Z}$ we  see that
\begin{equation}\label{a11}
   \left( \id_Y\otimes P_{Z,W}\right)\circ\left(P_{Y,Z}\otimes\id_W\right)= \left(P_{Y,Z}\otimes\id_W\right)\circ \left( \id_Y\otimes P_{Z,W}\right)
\end{equation}
After using \eqref{a11} into \eqref{a10}, the factor $P_{Z,W}\circ e_{Z,W}$ appears at the bottom part of the diagram $\eqref{a10}$. We find that
\begin{equation}
    P_{Z,W}\circ e_{Z,W}=e_{Z,W}\circ r_{Z,W}\circ e_{Z,W}=e_{Z,W}
\end{equation}
since $r_{Z,W}\circ e_{Z,W}=\id_{Z\otimes_AW}$. Similarly, at the top of the diagram \eqref{a10} after using $P_{X,Y\otimes Z}=P_{X,Y}\otimes \id_Z$, the factor $r_{X,Y}\circ P_{X,Y}$ appears which we find similarly to be equal to $r_{X,Y}$. In the end, we managed to ``eliminate'' the projectors $P_{X,Y\otimes Z}$ and $P_{Y\otimes Z,W}$ from the middle part of diagram \eqref{a10} and we are only left with $P_{Y,Z}$ in the middle part:
\begin{equation}\label{a13}
     \vcenter{\hbox{\hspace{2mm}\scalebox{0.8}{\begin{tikzpicture}[font=\footnotesize,inner sep=2pt]
  \begin{feynman}
  \vertex (x1) at (0,0) [label=below:\(X\)];
  \vertex [right=0.7cm of x1] (y1) [label=below:\(Y\)];
  \vertex[right=0.7cm of y1] (z1) [label=below:\(Z\)];
 \vertex[right=0.7cm of z1] (w1) [label=below:\(W\)];
 \vertex[above=0.5cm of x1] (x2) [label=above right:\(e_{X,Y\otimes_A(Z\otimes_AW)}\)] ;
 \vertex[above=1cm of w1] (w2);
 \vertex[above=0.5cm of y1] (y2);
 \vertex[above=0.5cm of z1] (z2);
 \vertex[above=0.5cm of w1] (w22);
 \vertex[left=0.25cm of x2)] (x3);
 \vertex [right=0.25cm of w2] (w3);
 \draw [fill=yellow] (x3) rectangle (w3);
 \vertex[above=0.5cm of x1] (x2) [label=above right:\(e_{X,Y\otimes_A(Z\otimes_AW)}\)] ;
 \vertex[above=2cm of x2] (x4);
 \vertex[above=1cm of y2] (y3);
 \vertex[left=0.25cm of y3] (y4);
 \vertex[above=1cm of z2] (z3);
 \vertex[above=1cm of w22] (w4);
 \vertex[above=0.5cm of w4] (w5);
 \vertex[right=0.25cm of w5] (w6);
 \draw [fill=yellow] (y4) rectangle (w6);
 \vertex[left=0.25cm of y3] (y4) [label=above right:\(e_{Y,Z\otimes_AW}\)];
 \vertex[above=0.5cm of x2] (x22);
 \vertex[above=0.5cm of y2] (y22);
 \vertex[above=0.5cm of z2] (z22);
 \vertex[above=1.5cm of y22] (y5);
 \vertex[above=2.5cm of z22] (z5);
 \vertex[above=3cm of w22] (w7);
 \vertex[above=0.5cm of y5] (y55);
 \vertex[right=0.25cm of y55] (y6);
 \vertex[left=0.25cm of x4] (x5);
 \vertex[above=0.5cm of x4] (x44);
 \vertex[above=0.5cm of y3] (y33);
 \vertex[above=0.5cm of z3] (z33);
 \vertex[above=0.5cm of z33] (zzz1);
 \vertex[above=0.5cm of w5] (www1);
 \vertex[above=0.5cm of www1] (www2);
 \vertex[left=0.25cm of zzz1] (zzz2);
 \vertex[right=0.25cm of www2] (www3);
  \draw [fill=yellow] (zzz2) rectangle (www3);
  \vertex[left=0.25cm of zzz1] (zzz2) [label=above right:\(e_{Z,W}\)];
  \vertex[above=0.5cm of zzz1] (zzz3);
 \vertex[above=0.5cm of x44] (x444);
\vertex[above=0.5cm of y33] (y333);
\vertex[above=0.5cm of y55] (y555);
\vertex[left=0.25cm of x444] (x6);
\vertex[above=0.5cm of w7] (w8);
\vertex[right=0.25cm of w8] (w88);
\vertex[right=0.7cm of x444] (b1);
\vertex[left=0.25cm of b1] (b11)[label=above right:\(P_{Y,Z}\)];
\vertex[right=0.7cm of b1] (c1);
\vertex[above=0.5cm of c1] (c2);
\vertex[right=0.25cm of c2] (c22);
\draw [fill=yellow] (c22) rectangle (b11);
\vertex[left=0.25cm of b1] (b11)[label=above right:\(P_{Y,Z}\)];
\vertex[above=0.5cm of x444] (xx1);
\vertex[above=0.5cm of y555] (yy1);
\vertex[above=0.5cm of z5] (zz1);
\vertex[above=0.5cm of w7] (ww1);
\vertex[above=0.5cm of xx1] (xxx1);
\vertex[above=0.5cm of yy1] (yyy1);
\vertex[above=0.5cm of yyy1] (yyy2);
\vertex[right=0.25cm of yyy2] (yyy3);
\vertex[left=0.25cm of xxx1] (xxx2);
\draw [fill=yellow] (yyy3) rectangle (xxx2);
\vertex[left=0.25cm of xxx1] (xxx2) [label=above right:\(r_{X,Y}\)];
\vertex[above=0.5cm of xxx1] (xxx3);
\vertex[above=0.5cm of zz1] (zz2);
\vertex[above=0.5cm of ww1] (ww2);
\vertex[above=0.5cm of ww2] (ww3);
\vertex[left=0.25cm of zz2] (zz3);
\vertex[right=0.25cm of ww3] (ww4);
\vertex[above=1.5cm of xx1] (xx2);
\vertex[above=1.5cm of yy1] (yy2);
\vertex[above=0.5cm of zz2] (zz4);
\vertex[above=1.5cm of ww3] (ww5);
\vertex[left=0.25cm of xx2] (xx3);
\vertex[above=1cm of zz4] (zz5);
\vertex[above=0.5cm of zz4] (zz55);
\vertex[right=0.25cm of zz5] (zz6);
\draw [fill=yellow] (zz6) rectangle (xx3);
\vertex[left=0.25cm of xx2] (xx3) [label=above right:\(r_{X\otimes_AY,Z}\)];
\vertex[above=0.5cm of xx2] (xx4);
\vertex[above=0.5cm of yy2] (yy3);
\vertex[above=0.5cm of xx4] (xx5);
\vertex[above=0.5cm of yy3] (yy4);
\vertex[above=0.5cm of zz5] (zz7);
\vertex[left=0.25cm of xx5] (xx6);
\vertex[above=0.5cm of ww5] (ww6);
\vertex[right=0.25cm of ww6] (ww7);
\draw [fill=yellow] (ww7) rectangle (xx6);
\vertex[left=0.25cm of xx5] (xx6) [label=above right:\(r_{(X\otimes_AY)\otimes_AZ,W}\)];
\vertex[above=0.5cm of xx5] (xx7);
\vertex[above=0.5cm of yy4] (yy5);
\vertex[above=0.5cm of zz7] (zz8);
\vertex[above=0.5cm of xx7] (xx8) [label=above:\(X\)];
\vertex[above=0.5cm of yy5] (yy6) [label=above:\(Y\)];
\vertex[above=0.5cm of zz8] (zz9) [label=above:\(Z\)];
\vertex[above=0.5cm of ww6] (ww8) [label=above:\(W\)];
   \diagram*{
   (x1)--[thick] (x2),
   (y1)--[thick] (y2),
   (z1)--[thick] (z2),
   (w1)--[thick] (w22),
   (x22)--[thick] (x4),
   (y22)--[thick] (y3),
   (z22)--[thick] (z3),
   (w4)--[thick] (w2),
   (y33)--[thick] (y5),
   (zzz1)--[thick] (z33),
   (www1)--[thick] (w5),
   (x44)--[thick] (x444),
   (y55)--[thick] (y555),
   (zz1)--[thick] (zz2),
   (ww1)--[thick] (ww2),
   (xx1)--[thick] (xxx1),
   (yy1)--[thick] (yyy1),
   (ww5)--[thick] (ww3),
   (zz55)--[thick] (zz4),
   (xx5)--[thick] (xx4),
   (yy4)--[thick] (yy3),
   (zz7)--[thick] (zz5),
   (xx7)--[thick] (xx8),
   (yy5)--[thick] (yy6),
   (zz8)--[thick] (zz9),
   (ww6)--[thick] (ww8),
   (x4)--[thick] (x44),
   (y55)--[thick] (y5),
   (zz2)--[thick] (zz4),
   (ww2)--[thick] (ww3),
   (xx2)--[thick] (xxx3),
   (yyy2)--[thick] (yy2),
   (www2)--[thick] (w7),
   (z5)--[thick] (zzz3),
   (x444)--[thick] (xx1),
   (w8)--[thick] (w7)
  };
  \end{feynman}
\end{tikzpicture}}}}
\end{equation}
Now we can compare the left and right side of our initial pentagon equation \eqref{a1}, namely compare \eqref{a5} with \eqref{a13}. Using \eqref{a8} twice for $P_{X\otimes Y,Z\otimes W}$ we see that the two expressions are identically equal, proving the pentagon identity for the tensor category $_A\mathcal{C}_A$ with associator given by \eqref{associator1}.

\subsection{Associator of \texorpdfstring{$_A\mathcal{C}_B\times \prescript{}{B}{\mathcal{C}}_C\times\mathcal{C}_C$}{TEXT}} \label{app4}

Considering that in our applications we are mainly interested in the situation where all the Frobenius algebras are the same $A=B=C$, we will restrict to this case. Nevertheless, the considerations presented here can be straightforwardly generalised.

The mixed pentagon identity that the associator \eqref{associator3} must satisfy is
\begin{equation}\label{b1}
\Tilde{\alpha}_{X\otimes_AY,Z,M}\circ\Tilde{\alpha}_{X,Y,Z\otimes_AM}=\left(\check{\alpha}_{X,Y,Z}\otimes_A\id_M\right)\circ\Tilde{\alpha}_{X,Y\otimes_AZ,M}\circ\left(\id_X\otimes_A\Tilde{\alpha}_{Y,Z,M}\right)
\end{equation}
where $\check{\alpha}$ is defined in \eqref{associator1}, $X,Y,Z$ are $A\text{-}A$-bimodules while $M$ is a left $A$-module. The only difference between the two pentagons \eqref{a1} and \eqref{b1} is that in the latter we have $M$ which is an $A$-module instead of $W$ which is an $A\text{-}A$-bimodule. However, the same proof presented in appendix \ref{app3} still holds, since all the steps used are valid. This stems from the fact that there was no point in the proof where the right action of $A$ on $W$ was used, treating it essentially as a left $A$-module.

More expicitly, \eqref{a7} can be used as is, since  $e_{X,Y\otimes_AZ}$ and $e_{Y\otimes_AZ,W}$ are again bimodule intertwiners. Furthermore, formulas like \eqref{a8} still hold even if $Z$ is only a left $A$-module. Next, $e_{Y,Z}$ is still a bimodule intertwiner, thus the steps leading to \eqref{a10} can be repeated in our current case of proving \eqref{b1}. Finally, it is clear that the remaining steps used in appendix \ref{app3} are also valid here.

\subsection{Associator of \texorpdfstring{$_A\mathcal{C} _B\times\mathcal{C} _B\times\mathcal{C}$}{TEXT}} \label{app5}

We restrict ourselves to the case $A=B$ noting that all considerations can be  easily extended to the general case.
The mixed pentagon identity that the associator \eqref{associator4} must satisfy is
\begin{equation}\label{c1}
    \dot{\alpha}_{X\otimes_AY,M,U}\circ\Tilde{\alpha}_{X,Y,M\otimes U}=\left(\Tilde{\alpha}_{X,Y,M}\otimes\id_U\right)\circ \alpha_{X,Y\otimes_AM,U}\circ\left(\id_X\otimes_A\alpha_{Y,M,U}\right)
\end{equation}
where $\tilde{\alpha}$ is defined in \eqref{associator3}, $X,Y$ are $A\text{-}A$-bimodules, $M$ is a left $A$-module and $U\in\operatorname{Obj}(\mathcal{C})$. We will proceed by calculating both sides of \eqref{c1} and showing that they are identical. Using the definitions of the relevant associators we find that the left-hand side of \eqref{c1} is represented graphically as

\begin{equation}\label{c2}
      \vcenter{\hbox{\hspace{2mm}\scalebox{0.8}{\begin{tikzpicture}[font=\footnotesize,inner sep=2pt]
  \begin{feynman}
  \vertex (x1) at (0,0) [label=below:\(X\)];
  \vertex [right=0.7cm of x1] (y1) [label=below:\(Y\)];
  \vertex[right=0.7cm of y1] (m1) [label=below:\(M\)];
 \vertex[right=0.7cm of m1] (u1) [label=below:\(U\)];
 \vertex[above=0.5cm of x1] (x2); 
 \vertex[above=0.5cm of y1] (y2); 
 \vertex[above=0.5cm of m1] (m2); 
 \vertex[above=0.5cm of u1] (u2); 
 \vertex[above=0.5cm of u2] (u3); 
 \vertex[left=0.25cm of x2] (x22);
 \vertex[right=0.25cm of u3] (u33);
 \draw [fill=yellow] (u33) rectangle (x22);
 \vertex[above=0.5cm of x1] (x2) [label=above right:\(e_{X,Y\otimes_A(M\otimes U)}\)];
 \vertex[above=0.5cm of x2] (x3); 
 \vertex[above=0.5cm of y2] (y3); 
 \vertex[above=0.5cm of m2] (m3); 
 \vertex[above=0.5cm of y3] (y4); 
 \vertex[above=0.5cm of m3] (m4); 
 \vertex[above=0.5cm of u3] (u4); 
\vertex[above=0.5cm of u4] (u5); 
\vertex[left=0.25cm of y4] (y44);
\vertex[right=0.25cm of u5] (u55);
 \draw [fill=yellow] (u55) rectangle (y44);
 \vertex[above=0.5cm of y3] (y4)[label=above right:\(e_{Y,M\otimes U}\)];
 \vertex[above=0.5cm of y4] (y5); 
 \vertex[above=0.5cm of m4] (m5);
 \vertex[above=1.5cm of x3] (x4); 
 \vertex[above=0.5cm of y5] (y6); 
 \vertex[above=0.5cm of y6] (y7); 
 \vertex[left=0.25cm of x4] (x44);
 \vertex[right=0.25cm of y7] (y77);
 \draw [fill=yellow] (y77) rectangle (x44);
 \vertex[above=1.5cm of x3] (x4)[label=above right:\(r_{X,Y}\)];
 \vertex[above=0.5cm of x4] (x5); 
 \vertex[above=0.5cm of x5] (x6); 
 \vertex[above=0.5cm of y7] (y8); 
 \vertex[above=1.5cm of m5] (m6) ; 
 \vertex[above=1.5cm of u5] (u6); 
 \vertex[above=0.5cm of u6] (u7); 
 \vertex[left=0.25cm of x6] (x66);
 \vertex[right=0.25cm of u7] (u77);
 \draw [fill=yellow] (x66) rectangle (u77);
 \vertex[above=0.5cm of x5] (x6)[label=above right:\(P_{X\otimes_AY,M\otimes U}\)];
 \vertex[above=0.5cm of x6] (x7); 
 \vertex[above=0.5cm of y8] (y9); 
 \vertex[above=0.5cm of m6] (m7); 
 \vertex[above=0.5cm of x7] (x8); 
 \vertex[above=0.5cm of y9] (yy1); 
 \vertex[above=0.5cm of m7] (m8); 
 \vertex[above=0.5cm of m8] (m9); 
 \vertex[left=0.25cm of x8] (x88);
 \vertex[right=0.25cm of m9] (m99);
 \draw [fill=yellow] (m99) rectangle (x88);
 \vertex[above=0.5cm of x7] (x8)[label=above right:\(r_{X\otimes_AY,M}\)];
 \vertex[above=0.5cm of x8] (x9); 
 \vertex[above=0.5cm of yy1] (yy2); 
 \vertex[above=1.5cm of u7] (u0) [label=above:\(U\)];
 \vertex[above=0.5cm of x9] (x0)[label=above:\(X\)];
 \vertex[above=0.5cm of yy2] (y0)[label=above:\(Y\)];
 \vertex[above=0.5cm of m9] (m0)[label=above:\(M\)];
   \diagram*{
  (x1)--[thick] (x2),
  (y1)--[thick] (y2),
  (m1)--[thick] (m2),
  (u1)--[thick] (u2),
  (y3)--[thick] (y4),
  (m3)--[thick] (m4),
  (u3)--[thick] (u4),
  (x3)--[thick] (x4),
  (y5)--[thick] (y6),
  (x6)--[thick] (x5),
  (y7)--[thick] (y8),
  (m5)--[thick] (m6),
  (u5)--[thick] (u6),
  (x7)--[thick] (x8),
  (y9)--[thick] (yy1),
  (m7)--[thick] (m8),
  (u7)--[thick] (u0),
  (x9)--[thick] (x0),
  (yy2)--[thick] (y0),
  (m9)--[thick] (m0)
  };
  \end{feynman}
\end{tikzpicture}}}}
\end{equation}
Next we  use $P_{X\otimes_AY,M\otimes U}=P_{X\otimes_AY,M}\otimes \id_U$  yielding the term $r_{X\otimes_A,Y,M}\circ P_{X\otimes_AY,M}$ on the top part of \eqref{c2}. Using the definition \eqref{proj11} we find that 
\begin{equation}
r_{X\otimes_A,Y,M}\circ P_{X\otimes_AY,M}=r_{X\otimes_AY,M}\circ e_{X\otimes_AY,M}\circ r_{X\otimes_AY,M}= r_{X\otimes_AY,M}
\end{equation}
since
\begin{equation}
    r_{X\otimes_AY,M}\circ e_{X\otimes_AY,M}=\id_{(X\otimes_AY)\otimes_AM}
    \end{equation}
    Therefore, the left-hand side of \eqref{c1} becomes
\begin{equation}\label{c3}
      \vcenter{\hbox{\hspace{2mm}\scalebox{0.8}{\begin{tikzpicture}[font=\footnotesize,inner sep=2pt]
  \begin{feynman}
  \vertex (x1) at (0,0) [label=below:\(X\)];
  \vertex [right=0.7cm of x1] (y1) [label=below:\(Y\)];
  \vertex[right=0.7cm of y1] (m1) [label=below:\(M\)];
 \vertex[right=0.7cm of m1] (u1) [label=below:\(U\)];
 \vertex[above=0.5cm of x1] (x2); 
 \vertex[above=0.5cm of y1] (y2); 
 \vertex[above=0.5cm of m1] (m2); 
 \vertex[above=0.5cm of u1] (u2); 
 \vertex[above=0.5cm of u2] (u3); 
 \vertex[left=0.25cm of x2] (x22);
 \vertex[right=0.25cm of u3] (u33);
 \draw [fill=yellow] (u33) rectangle (x22);
 \vertex[above=0.5cm of x1] (x2) [label=above right:\(e_{X,Y\otimes_A(M\otimes U)}\)];
 \vertex[above=0.5cm of x2] (x3); 
 \vertex[above=0.5cm of y2] (y3); 
 \vertex[above=0.5cm of m2] (m3); 
 \vertex[above=0.5cm of y3] (y4); 
 \vertex[above=0.5cm of m3] (m4); 
 \vertex[above=0.5cm of u3] (u4); 
\vertex[above=0.5cm of u4] (u5); 
\vertex[left=0.25cm of y4] (y44);
\vertex[right=0.25cm of u5] (u55);
 \draw [fill=yellow] (u55) rectangle (y44);
 \vertex[above=0.5cm of y3] (y4)[label=above right:\(e_{Y,M\otimes U}\)];
 \vertex[above=0.5cm of y4] (y5); 
 \vertex[above=0.5cm of m4] (m5);
 \vertex[above=1.5cm of x3] (x4); 
 \vertex[above=0.5cm of y5] (y6); 
 \vertex[above=0.5cm of y6] (y7); 
 \vertex[left=0.25cm of x4] (x44);
 \vertex[right=0.25cm of y7] (y77);
 \draw [fill=yellow] (y77) rectangle (x44);
 \vertex[above=1.5cm of x3] (x4)[label=above right:\(r_{X,Y}\)];
 \vertex[above=0.5cm of x4] (x5); 
 \vertex[above=0.5cm of x5] (x6); 
 \vertex[above=0.5cm of y7] (y8); 
 \vertex[above=1.5cm of m5] (m6) ; 
 \vertex[above=1.5cm of u5] (u6); 
 \vertex[above=0.5cm of u6] (u7); 
 \vertex[left=0.25cm of x6] (x66);
 \vertex[above=0.5cm of x5] (a1);
 \vertex[above=0.5cm of y7] (b1);
 \vertex[above=1.5cm of m5] (c1);
 \vertex[above=0.5cm of c1] (c2);
 \vertex[left=0.25cm of a1] (a11);
 \vertex[right=0.25cm of c2] (c22);
  \draw [fill=yellow] (a11) rectangle (c22);
  \vertex[left=0.25cm of a1] (a11)[label=above right:\(r_{X\otimes_AY,M}\)];
  \vertex[above=0.5cm of a1] (a2);
  \vertex[above=0.5cm of b1] (b2);
  \vertex[above=0.5cm of a2] (a0) [label=above:\(X\)];
  \vertex[above=0.5cm of b2] (b0) [label=above:\(Y\)];
  \vertex[above=0.5cm of c2] (c0) [label=above:\(M\)];
  \vertex[above=2.5cm of u5] (d0) [label=above:\(U\)];
 \vertex[right=0.25cm of u7] (u77);
 \vertex[above=0.5cm of x6] (x7); 
 \vertex[above=0.5cm of y8] (y9); 
 \vertex[above=0.5cm of m6] (m7); 
 \vertex[above=0.5cm of x7] (x8); 
 \vertex[above=0.5cm of y9] (yy1); 
 \vertex[above=0.5cm of m7] (m8); 
 \vertex[above=0.5cm of m8] (m9); 
 \vertex[left=0.25cm of x8] (x88);
 \vertex[right=0.25cm of m9] (m99);
 \vertex[above=0.5cm of x8] (x9); 
 \vertex[above=0.5cm of yy1] (yy2); 
   \diagram*{
   (u6)--[thick] (d0),
  (x1)--[thick] (x2),
  (y1)--[thick] (y2),
  (m1)--[thick] (m2),
  (u1)--[thick] (u2),
  (y3)--[thick] (y4),
  (m3)--[thick] (m4),
  (u3)--[thick] (u4),
  (x3)--[thick] (x4),
  (y5)--[thick] (y6),
  (x6)--[thick] (x5),
  (y7)--[thick] (y8),
  (m5)--[thick] (m6),
  (u5)--[thick] (u6),
  (x7)--[thick] (x8),
  (y9)--[thick] (yy1),
  (m7)--[thick] (m8)
  };
  \end{feynman}
\end{tikzpicture}}}}
\end{equation}
We proceed with the right-hand side of \eqref{c1} which takes the form
\begin{equation}\label{c4}
     \vcenter{\hbox{\hspace{2mm}\scalebox{0.8}{\begin{tikzpicture}[font=\footnotesize,inner sep=2pt]
  \begin{feynman}
   \vertex (x1) at (0,0) [label=below:\(X\)];
  \vertex [right=0.7cm of x1] (y1) [label=below:\(Y\)];
  \vertex [right=0.7cm of y1] (z1) [label=below:\(M\)];
  \vertex [right=0.7cm of z1] (w1) [label=below:\(U\)];
  \vertex[above=1cm of w1] (w2);
  \vertex [above=0.5cm of w1] (w11);
  \vertex[above=0.5cm of x1] (x2);
  \vertex[above=0.5cm of y1] (y2);
  \vertex[above=0.5cm of z1] (z2);
  \vertex[left=0.25cm of x2] (x22);
  \vertex[right=0.25cm of w2] (w22);
  \draw [fill=yellow] (x22) rectangle (w22);
  \vertex[left=0.25cm of x2] (x22) [label=above right:\(e_{X,Y\otimes_A(M\otimes U)}\)];
  \vertex[above=0.5cm of x2] (x3); 
  \vertex[above=0.5cm of y2] (y3);
  \vertex[above=0.5cm of z2] (z3);
  \vertex[above=0.5cm of y3] (y4);
  \vertex[above=0.5cm of z3] (z4);
  \vertex[above=0.5cm of w2] (w3);
  \vertex[above=0.5cm of w3] (w33);
  \vertex[right=0.25cm of w33] (w333);
  \vertex[left=0.25cm of y4] (y44);
   \draw [fill=yellow] (y44) rectangle (w333);
   \vertex[left=0.25cm of y4] (y44) [label=above right:\(e_{Y,M\otimes U}\)];
   \vertex[above=0.5cm of y4] (y5); 
   \vertex[above=0.5cm of z4] (z5);
   \vertex[above=0.5cm of z5] (z6);
   \vertex[above=0.5cm of w33] (w4);
   \vertex[above=0.5cm of w4] (w5); 
   \vertex[right=0.25cm of w5] (w55);
   \vertex[left=0.25cm of z6] (z66);
    \vertex[above=0.5cm of z6] (z7); 
    \vertex[above=0.5cm of z7] (z8);
    \vertex[above=1.5cm of y5] (y6);
    \vertex[above=0.5cm of z8] (z9); 
    \vertex[right=0.25cm of z9] (z99);
    \vertex[left=0.25cm of y6] (y66);
     \draw [fill=yellow] (y66) rectangle (z99);
     \vertex[left=0.25cm of y6] (y66) [label=above right:\(r_{Y,M}\)];
     \vertex[above=0.5cm of y6] (y7); 
     \vertex[above=0.5cm of y7] (y8); 
     \vertex[above=0.5cm of z9] (zz1); 
     \vertex[above=1.5cm of w5] (w6); 
     \vertex[above=0.5cm of w6] (w7); 
     \vertex[above=4.5cm of x3] (x4); 
     \vertex[below=1cm of x4] (x444);
     \vertex[left=0.25cm of x444] (xx444);
     \vertex[left=0.25cm of y8] (y88);
     \vertex[right=0.25cm of w7] (w77);
     \vertex[above=0.5cm of x444] (x4444);
      \draw [fill=yellow] (w77) rectangle (xx444);
      \vertex[left=0.25cm of x444][label=above right:\(P_{X,(Y\otimes_AM)\otimes U}\)];
       \vertex[above=0.5cm of y8] (y9); 
       \vertex[above=0.5cm of zz1] (zz2); 
       \vertex[above=4.5cm of x3] (x4); 
       \vertex[above=0.5cm of y9] (yy1); 
       \vertex[above=0.5cm of zz2] (zz3); 
       \vertex[above=0.5cm of zz3] (zz4); 
       \vertex[left=0.25cm of x4] (x44);
       \vertex[right=0.25cm of zz4] (zz44);
       \draw [fill=yellow] (zz44) rectangle (x44);
       \vertex[left=0.25cm of x4] (x44)[label=above right:\(P_{X,Y\otimes_AM}\)];
       \vertex[above=0.5cm of x4] (x5); 
       \vertex[above=0.5cm of yy1] (yy2); 
       \vertex[above=0.5cm of yy2] (yy3); 
       \vertex[above=0.5cm of zz4] (zz5); 
       \vertex[above=0.5cm of zz5] (zz6); 
       \vertex[left=0.25cm of yy3] (yy33);
       \vertex[right=0.25cm of zz6] (zz66);
       \draw [fill=yellow] (zz66) rectangle (yy33);
       \vertex[left=0.25cm of yy3] (yy33)[label=above right:\(e_{Y,M}\)];
       \vertex[above=0.5cm of yy3] (yy4); 
       \vertex[above=1.5cm of x5] (x6); 
       \vertex[above=0.5cm of yy4] (yy5); 
       \vertex[above=0.5cm of yy5] (yy6); 
       \vertex[left=0.25cm of x6] (x66);
       \vertex[right=0.25cm of  yy6] (yy66);
        \draw [fill=yellow] (yy66) rectangle (x66);
         \vertex[left=0.25cm of x6] (x66)[label=above right:\(r_{X,Y}\)];
         \vertex[above=0.5cm of x6] (x7); 
         \vertex[above=0.5cm of x7] (x8); 
         \vertex[above=0.5cm of yy6] (yy7); 
         \vertex[above=1.5cm of zz6] (zz7); 
         \vertex[above=0.5cm of zz7] (zz8); 
         \vertex[left=0.25cm of x8] (x88);
         \vertex[right=0.25cm of zz8] (zz88);
          \draw [fill=yellow] (zz88) rectangle (x88);
          \vertex[left=0.25cm of x8] (x88)[label=above right:\(r_{X\otimes_AY,M}\)];
          \vertex[above=0.5cm of x8] (x9); 
          \vertex[above=0.5cm of yy7] (yy8); 
          \vertex[above=0.5cm of x9] (xx1); 
          \vertex[above=0.5cm of yy8] (yy9); 
          \vertex[above=0.5cm of zz8] (zz9); 
          \vertex[above=4.0cm of w7] (w8); 
          \vertex[above=0.5cm of w8] (w9); 
          \vertex[left=0.25cm of xx1] (xx11);
          \vertex[right=0.25cm of w9] (w99);
            \vertex[above=0.5cm of x9] (xx2); 
            \vertex[above=0.5cm of yy8] (yyy1); 
            \vertex[above=0.5cm of zz8] (zzz1); 
            \vertex[above=0.0001cm of xx2] (xx3) [label=above:\(X\)];
            \vertex[above=0.0001cm of yyy1] (yyy2) [label=above:\(Y\)];
            \vertex[above=0.0001cm of zzz1] (zzz2) [label=above:\(M\)];
            \vertex[above=0.0001cm of w9] (w0) [label=above:\(U\)];
   \diagram*{
   (xx1)--[thick] (xx2),
   (yy9)--[thick] (yyy1),
   (zz9)--[thick] (zzz1),
   (w8)--[thick] (w9),
   (x4)--[thick] (x4444),
   (z6)--[thick] (z7),
   (w4)--[thick] (w5),
 (x1)--[thick] (x2),
   (y1)--[thick] (y2),
   (z1)--[thick] (z2),
   (w1)--[thick] (w11),
   (y3)--[thick] (y4),
   (z3)--[thick] (z4),
   (w2)--[thick] (w3),
   (z5)--[thick] (z6),
   (w33)--[thick] (w4),
   (y5)--[thick] (y6),
   (z7)--[thick] (z8),
   (y7)--[thick] (y8),
   (z9)--[thick] (zz1),
   (w5)--[thick] (w6),
   (x3)--[thick] (x444),
   (y9)--[thick] (yy1),
   (zz2)--[thick] (zz3),
   (yy2)--[thick] (yy3),
   (zz4)--[thick] (zz5),
   (x5)--[thick] (x6),
   (yy4)--[thick] (yy5),
   (x7)--[thick] (x8),
   (yy6)--[thick] (yy7),
   (zz6)--[thick] (zz7),
   (x9)--[thick] (xx1),
   (yy8)--[thick] (yy9),
   (zz8)--[thick] (zz9),
   (w7)--[thick] (w8),
   (w0)--[thick] (w9),
   (xx2)--[thick] (xx3),
   (yyy1)--[thick] (yyy2),
   (zzz1)--[thick] (zzz2)
  };
  \end{feynman}
\end{tikzpicture}}}}
\end{equation}
 We next use the identity $P_{X,(Y\otimes_AM)\otimes U}=P_{X,Y\otimes_AM}\otimes\id_U$ and the fact that the projector is an idempotent $P_{X,Y\otimes_AM}\circ P_{X,Y\otimes_AM}=P_{X,Y\otimes_AM}$ in the middle part of the diagram \eqref{c4}. Then we recall that $e_{Y,M}$ is a left $A$-module intertwiner  between the $A$-modules $Y\otimes_AM$ and $Y\otimes M$ satisfying
\begin{equation}
    e_{Y,M}\circ \rho_{Y\otimes_A M}=\rho_{Y\otimes M}\circ \left(\id_A\otimes e_{Y,M}\right)
\end{equation}
Using this and the explicit form of the projector $P_{X.Y\otimes_AM}$ we  show that
\begin{equation}
    \left( \id_X\otimes e_{Y,M}\otimes \id_U\right)\circ \left( P_{X,Y\otimes_AM}\otimes \id_U\right)=\left( P_{X,Y\otimes M}\otimes \id_U\right)\circ \left( \id_X\otimes e_{Y,M}\otimes \id_U\right)
\end{equation}
to pull $e_{Y,M}$ below $P_{X,Y\otimes_AM}$ in the middle part of \eqref{c4}. In the end, \eqref{c4} becomes
\begin{equation}\label{c7}
     \vcenter{\hbox{\hspace{2mm}\scalebox{0.8}{\begin{tikzpicture}[font=\footnotesize,inner sep=2pt]
  \begin{feynman}
   \vertex (x1) at (0,0) [label=below:\(X\)];
  \vertex [right=0.7cm of x1] (y1) [label=below:\(Y\)];
  \vertex [right=0.7cm of y1] (z1) [label=below:\(M\)];
  \vertex [right=0.7cm of z1] (w1) [label=below:\(U\)];
  \vertex[above=1cm of w1] (w2);
  \vertex [above=0.5cm of w1] (w11);
  \vertex[above=0.5cm of x1] (x2);
  \vertex[above=0.5cm of y1] (y2);
  \vertex[above=0.5cm of z1] (z2);
  \vertex[left=0.25cm of x2] (x22);
  \vertex[right=0.25cm of w2] (w22);
  \draw [fill=yellow] (x22) rectangle (w22);
  \vertex[left=0.25cm of x2] (x22) [label=above right:\(e_{X,Y\otimes_A(M\otimes U)}\)];
  \vertex[above=0.5cm of x2] (x3); 
  \vertex[above=0.5cm of y2] (y3);
  \vertex[above=0.5cm of z2] (z3);
  \vertex[above=0.5cm of y3] (y4);
  \vertex[above=0.5cm of z3] (z4);
  \vertex[above=0.5cm of w2] (w3);
  \vertex[above=0.5cm of w3] (w33);
  \vertex[right=0.25cm of w33] (w333);
  \vertex[left=0.25cm of y4] (y44);
   \draw [fill=yellow] (y44) rectangle (w333);
   \vertex[left=0.25cm of y4] (y44) [label=above right:\(e_{Y,M\otimes U}\)];
   \vertex[above=0.5cm of y4] (y5); 
   \vertex[above=0.5cm of z4] (z5);
   \vertex[above=0.5cm of z5] (z6);
   \vertex[above=0.5cm of w33] (w4);
   \vertex[above=0.5cm of w4] (w5); 
   \vertex[right=0.25cm of w5] (w55);
   \vertex[left=0.25cm of z6] (z66);
    \vertex[above=0.5cm of z6] (z7); 
    \vertex[above=0.5cm of z7] (z8);
    \vertex[above=1.5cm of y5] (y6);
    \vertex[above=0.5cm of z8] (z9); 
    \vertex[right=0.25cm of z9] (z99);
    \vertex[left=0.25cm of y6] (y66);
     \draw [fill=yellow] (y66) rectangle (z99);
     \vertex[left=0.25cm of y6] (y66) [label=above right:\(P_{Y,M}\)];
     \vertex[above=0.5cm of y6] (y7); 
     \vertex[above=0.5cm of y7] (y8); 
     \vertex[above=0.5cm of z9] (zz1); 
     \vertex[above=1.5cm of w5] (w6); 
     \vertex[above=0.5cm of w6] (w7); 
     \vertex[above=4.5cm of x3] (x4); 
     \vertex[below=1cm of x4] (x444);
     \vertex[left=0.25cm of x444] (xx444);
     \vertex[left=0.25cm of y8] (y88);
     \vertex[right=0.25cm of w7] (w77);
     \vertex[above=0.5cm of x444] (x4444);
       \vertex[above=0.5cm of y8] (y9); 
       \vertex[above=0.5cm of zz1] (zz2); 
       \vertex[above=4.5cm of x3] (x4); 
       \vertex[above=0.5cm of y9] (yy1); 
       \vertex[above=0.5cm of zz2] (zz3); 
       \vertex[above=0.5cm of zz3] (zz4); 
       \vertex[left=0.25cm of x4] (x44);
       \vertex[right=0.25cm of zz4] (zz44);
       \draw [fill=yellow] (zz44) rectangle (x44);
       \vertex[left=0.25cm of x4] (x44)[label=above right:\(P_{X,Y\otimes M}\)];
       \vertex[above=0.5cm of x4] (x5); 
       \vertex[above=0.5cm of yy1] (yy2); 
       \vertex[above=0.5cm of yy2] (yy3); 
       \vertex[above=0.5cm of zz4] (zz5); 
       \vertex[above=0.5cm of zz5] (zz6); 
       \vertex[left=0.25cm of yy3] (yy33);
       \vertex[right=0.25cm of zz6] (zz66);
       \vertex[above=0.5cm of yy3] (yy4); 
       \vertex[above=1.5cm of x5] (x6); 
       \vertex[above=0.5cm of yy4] (yy5); 
       \vertex[above=0.5cm of yy5] (yy6); 
       \vertex[left=0.25cm of x6] (x66);
       \vertex[right=0.25cm of  yy6] (yy66);
        \draw [fill=yellow] (yy66) rectangle (x66);
         \vertex[left=0.25cm of x6] (x66)[label=above right:\(r_{X,Y}\)];
         \vertex[above=0.5cm of x6] (x7); 
         \vertex[above=0.5cm of x7] (x8); 
         \vertex[above=0.5cm of yy6] (yy7); 
         \vertex[above=1.5cm of zz6] (zz7); 
         \vertex[above=0.5cm of zz7] (zz8); 
         \vertex[left=0.25cm of x8] (x88);
         \vertex[right=0.25cm of zz8] (zz88);
          \draw [fill=yellow] (zz88) rectangle (x88);
          \vertex[left=0.25cm of x8] (x88)[label=above right:\(r_{X\otimes_AY,M}\)];
          \vertex[above=0.5cm of x8] (x9); 
          \vertex[above=0.5cm of yy7] (yy8); 
          \vertex[above=0.5cm of x9] (xx1); 
          \vertex[above=0.5cm of yy8] (yy9); 
          \vertex[above=0.5cm of zz8] (zz9); 
          \vertex[above=4.0cm of w7] (w8); 
          \vertex[above=0.5cm of w8] (w9); 
          \vertex[left=0.25cm of xx1] (xx11);
          \vertex[right=0.25cm of w9] (w99);
            \vertex[above=0.5cm of x9] (xx2); 
            \vertex[above=0.5cm of yy8] (yyy1); 
            \vertex[above=0.5cm of zz8] (zzz1); 
            \vertex[above=0.0001cm of xx2] (xx3) [label=above:\(X\)];
            \vertex[above=0.0001cm of yyy1] (yyy2) [label=above:\(Y\)];
            \vertex[above=0.0001cm of zzz1] (zzz2) [label=above:\(M\)];
            \vertex[above=0.0001cm of w9] (w0) [label=above:\(U\)];
   \diagram*{
   (zz1)--[thick] (zz2),
   (w6)--[thick] (w7),
   (y8)--[thick] (y9),
   (x444)--[thick] (x4444),
   (yy3)--[thick] (yy4),
   (zz5)--[thick] (zz6),
   (xx1)--[thick] (xx2),
   (yy9)--[thick] (yyy1),
   (zz9)--[thick] (zzz1),
   (w8)--[thick] (w9),
   (x4)--[thick] (x4444),
   (z6)--[thick] (z7),
   (w4)--[thick] (w5),
 (x1)--[thick] (x2),
   (y1)--[thick] (y2),
   (z1)--[thick] (z2),
   (w1)--[thick] (w11),
   (y3)--[thick] (y4),
   (z3)--[thick] (z4),
   (w2)--[thick] (w3),
   (z5)--[thick] (z6),
   (w33)--[thick] (w4),
   (y5)--[thick] (y6),
   (z7)--[thick] (z8),
   (y7)--[thick] (y8),
   (z9)--[thick] (zz1),
   (w5)--[thick] (w6),
   (x3)--[thick] (x444),
   (y9)--[thick] (yy1),
   (zz2)--[thick] (zz3),
   (yy2)--[thick] (yy3),
   (zz4)--[thick] (zz5),
   (x5)--[thick] (x6),
   (yy4)--[thick] (yy5),
   (x7)--[thick] (x8),
   (yy6)--[thick] (yy7),
   (zz6)--[thick] (zz7),
   (x9)--[thick] (xx1),
   (yy8)--[thick] (yy9),
   (zz8)--[thick] (zz9),
   (w7)--[thick] (w8),
   (w0)--[thick] (w9),
   (xx2)--[thick] (xx3),
   (yyy1)--[thick] (yyy2),
   (zzz1)--[thick] (zzz2)
  };
  \end{feynman}
\end{tikzpicture}}}}
\end{equation}
Now we can use $P_{X,Y\otimes M}=P_{X,Y}\otimes\id_M$ and then the term $r_{X,Y}\circ P_{X,Y}$ appears at the middle left part of \eqref{c7}. We find that 
\begin{equation}
r_{X,Y}\circ P_{X,Y}=r_{X,Y}\circ e_{X,Y}\circ r_{X,Y}=r_{X,Y}
\end{equation}
 Similarly, in the middle right part of \eqref{c7} we use $P_{Y,M}\otimes\id_U=P_{Y,M\otimes U}$ so that the term $P_{Y,M\otimes U}\circ e_{Y,M\otimes U}$ appears. We calculate that term as 
 \begin{equation}
 P_{Y,M\otimes U}\circ e_{Y,M\otimes U}=e_{Y,M\otimes U}\circ r_{Y,M\otimes U}\circ e_{Y,M\otimes U}=e_{Y,M\otimes U}
 \end{equation}
 In the end, \eqref{c7} becomes

\begin{equation}\label{c8}
      \vcenter{\hbox{\hspace{2mm}\scalebox{0.8}{\begin{tikzpicture}[font=\footnotesize,inner sep=2pt]
  \begin{feynman}
  \vertex (x1) at (0,0) [label=below:\(X\)];
  \vertex [right=0.7cm of x1] (y1) [label=below:\(Y\)];
  \vertex[right=0.7cm of y1] (m1) [label=below:\(M\)];
 \vertex[right=0.7cm of m1] (u1) [label=below:\(U\)];
 \vertex[above=0.5cm of x1] (x2); 
 \vertex[above=0.5cm of y1] (y2); 
 \vertex[above=0.5cm of m1] (m2); 
 \vertex[above=0.5cm of u1] (u2); 
 \vertex[above=0.5cm of u2] (u3); 
 \vertex[left=0.25cm of x2] (x22);
 \vertex[right=0.25cm of u3] (u33);
 \draw [fill=yellow] (u33) rectangle (x22);
 \vertex[above=0.5cm of x1] (x2) [label=above right:\(e_{X,Y\otimes_A(M\otimes U)}\)];
 \vertex[above=0.5cm of x2] (x3); 
 \vertex[above=0.5cm of y2] (y3); 
 \vertex[above=0.5cm of m2] (m3); 
 \vertex[above=0.5cm of y3] (y4); 
 \vertex[above=0.5cm of m3] (m4); 
 \vertex[above=0.5cm of u3] (u4); 
\vertex[above=0.5cm of u4] (u5); 
\vertex[left=0.25cm of y4] (y44);
\vertex[right=0.25cm of u5] (u55);
 \draw [fill=yellow] (u55) rectangle (y44);
 \vertex[above=0.5cm of y3] (y4)[label=above right:\(e_{Y,M\otimes U}\)];
 \vertex[above=0.5cm of y4] (y5); 
 \vertex[above=0.5cm of m4] (m5);
 \vertex[above=1.5cm of x3] (x4); 
 \vertex[above=0.5cm of y5] (y6); 
 \vertex[above=0.5cm of y6] (y7); 
 \vertex[left=0.25cm of x4] (x44);
 \vertex[right=0.25cm of y7] (y77);
 \draw [fill=yellow] (y77) rectangle (x44);
 \vertex[above=1.5cm of x3] (x4)[label=above right:\(r_{X,Y}\)];
 \vertex[above=0.5cm of x4] (x5); 
 \vertex[above=0.5cm of x5] (x6); 
 \vertex[above=0.5cm of y7] (y8); 
 \vertex[above=1.5cm of m5] (m6) ; 
 \vertex[above=1.5cm of u5] (u6); 
 \vertex[above=0.5cm of u6] (u7); 
 \vertex[left=0.25cm of x6] (x66);
 \vertex[above=0.5cm of x5] (a1);
 \vertex[above=0.5cm of y7] (b1);
 \vertex[above=1.5cm of m5] (c1);
 \vertex[above=0.5cm of c1] (c2);
 \vertex[left=0.25cm of a1] (a11);
 \vertex[right=0.25cm of c2] (c22);
  \draw [fill=yellow] (a11) rectangle (c22);
  \vertex[left=0.25cm of a1] (a11)[label=above right:\(r_{X\otimes_AY,M}\)];
  \vertex[above=0.5cm of a1] (a2);
  \vertex[above=0.5cm of b1] (b2);
  \vertex[above=0.5cm of a2] (a0) [label=above:\(X\)];
  \vertex[above=0.5cm of b2] (b0) [label=above:\(Y\)];
  \vertex[above=0.5cm of c2] (c0) [label=above:\(M\)];
  \vertex[above=2.5cm of u5] (d0) [label=above:\(U\)];
 \vertex[right=0.25cm of u7] (u77);
 \vertex[above=0.5cm of x6] (x7); 
 \vertex[above=0.5cm of y8] (y9); 
 \vertex[above=0.5cm of m6] (m7); 
 \vertex[above=0.5cm of x7] (x8); 
 \vertex[above=0.5cm of y9] (yy1); 
 \vertex[above=0.5cm of m7] (m8); 
 \vertex[above=0.5cm of m8] (m9); 
 \vertex[left=0.25cm of x8] (x88);
 \vertex[right=0.25cm of m9] (m99);
 \vertex[above=0.5cm of x8] (x9); 
 \vertex[above=0.5cm of yy1] (yy2); 
   \diagram*{
   (u6)--[thick] (d0),
  (x1)--[thick] (x2),
  (y1)--[thick] (y2),
  (m1)--[thick] (m2),
  (u1)--[thick] (u2),
  (y3)--[thick] (y4),
  (m3)--[thick] (m4),
  (u3)--[thick] (u4),
  (x3)--[thick] (x4),
  (y5)--[thick] (y6),
  (x6)--[thick] (x5),
  (y7)--[thick] (y8),
  (m5)--[thick] (m6),
  (u5)--[thick] (u6),
  (x7)--[thick] (x8),
  (y9)--[thick] (yy1),
  (m7)--[thick] (m8)
  };
  \end{feynman}
\end{tikzpicture}}}}
\end{equation}
which is identical to the left-hand side shown in \eqref{c3}, proving the pentagon identity \eqref{c1}.

\section{Matrices $A, B, C$ and fusing matrices for rational free  boson}

\subsection{\texorpdfstring{$N=2, r=2$}{N=2, r=2}} \label{appbos1}

Let us first introduce a convenient notational convention regarding the three sets of coefficients $A,B,C$  in \eqref{Aconstants1}, \eqref{deco4} and \eqref{deco5} respectively. Consider for instance $A^{(x,y)z}_{(i,j)k}$ as in \eqref{Aconstants1}. Due to the structure of the free boson theory which is summarized in \eqref{specialcase2}, the labels $z$ and $k$ are uniquely determined by the remaining four labels. More specifically, the coefficients $A^{(x,y)z}_{(i,j)k}$ are non zero only if $z$ and $k$ are the unique labels that satisfy $X_z\cong X_x\otimes_B X_y$ and $U_k\cong U_i\otimes U_j$. Therefore, we will omit these labels when working in the free boson setting. A similar argument holds for the last upper and last lower indices of $B^{(x,a)b}_{(i,j)k}$ and $C^{(a,j)b}_{(i)k}$.

At this point we note the following index sets for the model with $N=2, r=2$. The index set for the simple objects of the category $\mathcal{U}_2$ is $\mathcal{I}=\{0,1,2,3\}$, the index set for the $A_2$-modules is $\mathcal{J}_{A_2}=\{0,1\}$ while for the $A_2$-$A_2$-bimodules we have $\mathcal{K}_{A_2A_2}=\{0,1,2,3\}$.   Using the results from section \ref{sec4.1} we can construct the system of linear equations from \eqref{conaboson}, \eqref{conbboson} and \eqref{concboson} in order to find the complex numbers $A^{(x,y)}_{(i,j)}$.  This system is solved by the following set of $\lvert \mathcal{K}_{A_2A_2}\rvert^2$ matrices of dimension $|\mathcal{I}|\times |\mathcal{I}|$:
\begin{align}\label{ajun}
&A^{(0,0)}= \Tilde{A}^{(0,0)}
    \begin{bmatrix}
1 & 0 & 1 &0\\
0 & 0 & 0 &0\\
1 &0 & 1 &0 \\
0 &0&0&0
\end{bmatrix},\;
  &A^{(0,1)}=\Tilde{A}^{(0,1)}  \left[
\begin{array}{cccc}
 0 & 1 & 0 & 1 \\
 0 & 0 & 0 & 0 \\
 0 & 1 & 0 & 1 \\
 0 & 0 & 0 & 0 \\
\end{array}
\right],\quad &A^{(0,2)}=\Tilde{A}^{(0,2)} \left[
\begin{array}{cccc}
 1 & 0 & 1 & 0 \\
 0 & 0 & 0 & 0 \\
 1 & 0 & 1 & 0 \\
 0 & 0 & 0 & 0 \\
\end{array}
\right], \nonumber \\ 
 &A^{(0,3)}=\Tilde{A}^{(0,3)} \left[
\begin{array}{cccc}
 0 & 1 & 0 & 1 \\
 0 & 0 & 0 & 0 \\
 0 & 1 & 0 & 1 \\
 0 & 0 & 0 & 0 \\
\end{array}
\right],
\; &A^{(1,0)}=\Tilde{A}^{(1,0)}\left[
\begin{array}{cccc}
 0 & 0 & 0 & 0 \\
 1 & 0 & -i  & 0 \\
 0 & 0 & 0 & 0 \\
 1 & 0 & -i  & 0 \\
\end{array}
\right],\quad &A^{(1,1)}=\Tilde{A}^{(1,1)} \left[
\begin{array}{cccc}
 0 & 0 & 0 & 0 \\
 0 & 1 & 0 & -i  \\
 0 & 0 & 0 & 0 \\
 0 & 1& 0 & -i \\
\end{array}
\right], \nonumber\\
&A^{(1,2)}=\Tilde{A}^{(1,2)} \left[
\begin{array}{cccc}
 0 & 0 & 0 & 0 \\
 1 & 0 & -i  & 0 \\
 0 & 0 & 0 & 0 \\
 1 & 0 & -i  & 0 \\
\end{array}
\right],\; &A^{(1,3)}=\Tilde{A}^{(1,3)} \left[
\begin{array}{cccc}
 0 & 0 & 0 & 0 \\
 0 & 1 & 0 & -i  \\
 0 & 0 & 0 & 0 \\
 0 & 1 & 0 & -i  \\
\end{array}
\right],\quad
 &A^{(2,0)}=\Tilde{A}^{(2,0)} \left[
\begin{array}{cccc}
 1 & 0 & -1 & 0 \\
 0 & 0 & 0 & 0 \\
 1 & 0 & -1& 0 \\
 0 & 0 & 0 & 0 \\
\end{array}
\right], \nonumber\\
& A^{(2,1)}=\Tilde{A}^{(2,1)} \left[
\begin{array}{cccc}
 0 &1 & 0 & -1\\
 0 & 0 & 0 & 0 \\
 0 &1 & 0 & -1 \\
 0 & 0 & 0 & 0 \\
\end{array}
\right],\; &A^{(2,2)}=\Tilde{A}^{(2,2)} \left[
\begin{array}{cccc}
 1 & 0 & -1& 0 \\
 0 & 0 & 0 & 0 \\
 1 & 0 & -1 & 0 \\
 0 & 0 & 0 & 0 \\
\end{array}
\right],\quad &A^{(2,3)}=\Tilde{A}^{(2,3)} \left[
\begin{array}{cccc}
 0 & 1 & 0 & -1 \\
 0 & 0 & 0 & 0 \\
 0 & 1 & 0 & -1 \\
 0 & 0 & 0 & 0 \\
\end{array}
\right], \nonumber \\
&A^{(3,0)}=\Tilde{A}^{(3,0)} \left[
\begin{array}{cccc}
 0 & 0 & 0 & 0 \\
 1 & 0 & i  & 0 \\
 0 & 0 & 0 & 0 \\
 1 & 0 & i  & 0 \\
\end{array}
\right],\; &A^{(3,1)}=\Tilde{A}^{(3,1)} \left[
\begin{array}{cccc}
 0 & 0 & 0 & 0 \\
 0 &1 & 0 & i  \\
 0 & 0 & 0 & 0 \\
 0 &1 & 0 & i  \\
\end{array}
\right], \quad &A^{(3,2)}=\Tilde{A}^{(3,2)} \left[
\begin{array}{cccc}
 0 & 0 & 0 & 0 \\
 1 & 0 & i  & 0 \\
 0 & 0 & 0 & 0 \\
 1 & 0 & i  & 0 \\
\end{array}
\right], \nonumber \\
& A^{(3,3)}=\Tilde{A}^{(3,3)} \left[
\begin{array}{cccc}
 0 & 0 & 0 & 0 \\
 0 &1 & 0 & i  \\
 0 & 0 & 0 & 0 \\
 0 & 1 & 0 & i  \\
\end{array}
\right]
\end{align}
for any complex numbers $\tilde{A}^{(x,y)}$. Here we have used the matrix notation with matrix entries $A^{(x,y)}_{(i,j)}=\Tilde{A}^{(x,y)}\Tilde{A}_{(i,j)}^{(x,y)}$ where  the row label $i$ and the column label $j$ run through the index set $\mathcal{I}$.

Next, we present the solution for the coefficients $B^{(x,a)}_{(i,j)}$. The solution consists of $\lvert \mathcal{K}_{A_2A_2}\rvert \lvert \mathcal{J}_{A_2}\rvert$ matrices of dimension $\lvert \mathcal{I}\rvert\times \lvert \mathcal{I}\rvert$ with matrix elements
\begin{align*}
    &B^{(0,0)}=\Tilde{B}^{(0,0)}\left[
\begin{array}{cccc}
 1 & 0 & 1 & 0 \\
 0 & 0 & 0 & 0 \\
 1 & 0 & 1 & 0 \\
 0 & 0 & 0 & 0 \\
\end{array}
\right],& B^{(0,1)}=\Tilde{B}^{(0,1)}\left[
\begin{array}{cccc}
 0 & 1 & 0 & 1 \\
 0 & 0 & 0 & 0 \\
 0 &1 & 0 & 1 \\
 0 & 0 & 0 & 0 \\
\end{array}
\right], \quad& B^{(1,0)}=\Tilde{B}^{(1,0)} \left[
\begin{array}{cccc}
 0 & 0 & 0 & 0 \\
 i  & 0 & 1 & 0 \\
 0 & 0 & 0 & 0 \\
 i & 0 & 1 & 0 \\
\end{array}
\right], \nonumber \\
& B^{(1,1)}=\Tilde{B}^{(1,1)} \left[
\begin{array}{cccc}
 0 & 0 & 0 & 0 \\
 0 & i & 0 & 1 \\
 0 & 0 & 0 & 0 \\
 0 & i  & 0 & 1 \\
\end{array}
\right], &
B^{(2,0)}=\Tilde{B}^{(2,0)} \left[
\begin{array}{cccc}
 -1 & 0 & 1 & 0 \\
 0 & 0 & 0 & 0 \\
 -1 & 0 & 1 & 0 \\
 0 & 0 & 0 & 0 \\
\end{array}
\right],\quad& B^{(2,1)}=\Tilde{B}^{(2,1)} \left[
\begin{array}{cccc}
 0 & -1 & 0 & 1 \\
 0 & 0 & 0 & 0 \\
 0 & -1 & 0 & 1 \\
 0 & 0 & 0 & 0 \\
\end{array}
\right],
\end{align*}
\begin{align}\label{bjun}
&B^{(3,0)}=\Tilde{B}^{(3,0)} \left[
\begin{array}{cccc}
 0 & 0 & 0 & 0 \\
 -i  & 0 & 1 & 0 \\
 0 & 0 & 0 & 0 \\
 -i  & 0 & 1 & 0 \\
\end{array}
\right], 
&B^{(3,1)}=\Tilde{B}^{(3,1)} \left[
\begin{array}{cccc}
 0 & 0 & 0 & 0 \\
 0 & -i  & 0 & 1 \\
 0 & 0 & 0 & 0 \\
 0 & -i & 0 & 1\\
\end{array}
\right]
\end{align}
where $\Tilde{B}^{(x,a)}\in\mathbb{C}$ are arbitrary and the row and column labels run through the simple objects of the category $\mathcal{U}_2$. Using these results in \eqref{tboson} we obtain the following for the $\mathrm{T}$-matrices:
\begin{equation}\label{tcase1.1}
      \mathrm{T}^{(x\,y\,a)}=1 \qquad \text{if any} \;x,y \;\text{is} \;0 .
\end{equation}
Some of the remaining values are
\begin{align}\label{tcase1.2}
& \mathrm{T}^{(1\,1\,1)}=-i,\qquad  \mathrm{T}^{(3\,3\,1)}=i \nonumber \\ 
&  \mathrm{T}^{(1\,1\,0)}= \mathrm{T}^{(2\,2\,0)}= \mathrm{T}^{(3\,3\,0)}= \mathrm{T}^{(2\,2\,1)}=1 \nonumber \\
&  \mathrm{T}^{(1\,2\,0)}= \mathrm{T}^{(1\,2\,1)}= \mathrm{T}^{(3\,1\,0)}= \mathrm{T}^{(2\,3\,1)}=\xi \nonumber \\
& \mathrm{T}^{(1\,3\,1)}= \mathrm{T}^{(2\,3\,0)}=-\xi \nonumber \\
&  \mathrm{T}^{(2\,1\,0)}= \mathrm{T}^{(3\,2\,0)}= \mathrm{T}^{(3\,2\,1)}= \mathrm{T}^{(3\,1\,1)}=\xi^3 \nonumber \\
&  \mathrm{T}^{(2\,1\,1)}= \mathrm{T}^{(1\,3\,0)}=-\xi^3
\end{align}
while the remaining of the non-zero values are  equal to $1$. Here, we used the  normalisation \eqref{normalisation2} for the constants $\tilde{B}^{(x,a)}$.

We now present the results for the coefficients $C^{(a,j)}_{(i)}$. The solution of  system \eqref{psicongen} consists of the following $\lvert \mathcal{J}_{A_2}\rvert \lvert \mathcal{I}\rvert$ vectors of dimension $\lvert \mathcal{I}\rvert$:
\begin{align}\label{junc}
    &C^{(0,0)}=\Tilde{C}^{(0,0)} \begin{bmatrix}
        1 \\
        0\\
        1\\
        0
    \end{bmatrix},\quad C^{(0,1)}=\Tilde{C}^{(0,1)} \begin{bmatrix}
        1 \\
        0\\
        1\\
        0
    \end{bmatrix},\quad 
    C^{(0,2)}=\Tilde{C}^{(0,2)}  \begin{bmatrix}
        1 \\
        0\\
        1\\
        0
    \end{bmatrix},\quad C^{(0,3)}=\Tilde{C}^{(0,3)}  \begin{bmatrix}
        1 \\
        0\\
        1\\
        0
    \end{bmatrix} \nonumber \\
    & C^{(1,0)}=\Tilde{C}^{(1,0)}  \begin{bmatrix}
        0 \\
        1\\
        0\\
        1
    \end{bmatrix},\quad  C^{(1,1)}=\Tilde{C}^{(1,1)}  \begin{bmatrix}
        0 \\
        1\\
        0\\
        1
    \end{bmatrix},\quad 
     C^{(1,2)}=\Tilde{C}^{(1,2)}  \begin{bmatrix}
        0 \\
        1\\
        0\\
        1
    \end{bmatrix},\quad  C^{(1,3)}=\Tilde{C}^{(1,3)}  \begin{bmatrix}
        0 \\
        1\\
        0\\
        1
    \end{bmatrix}
\end{align}
where $\Tilde{C}^{(a,j)}\in\mathbb{C}$ are arbitrary. They were obtained by using \eqref{Cconstantsgen} as well as the normalisation \eqref{Cconstantsnorm}.  Here,  the row label $i$ is running through the simple objects of $\mathcal{U}_2$. Finally, we obtain the following results for the $\mathrm{\dot{T}}$-matrices
\begin{align}\label{tdot1}
   & \dot{\mathrm{T}}^{(1\,0\,2)}= \dot{\mathrm{T}}^{(1\,0\,3)}=\dot{\mathrm{T}}^{(1\,1\,1)}=\dot{\mathrm{T}}^{(1\,1\,2)}=-i \nonumber \\
   & \dot{\mathrm{T}}^{(1\,1\,3)}=\dot{\mathrm{T}}^{(2\,0\,3)}=\dot{\mathrm{T}}^{(2\,1\,1)}=\dot{\mathrm{T}}^{(3\,1\,3)}=-1 \nonumber \\
   & \dot{\mathrm{T}}^{(3\,0\,2)}=\dot{\mathrm{T}}^{(3\,0\,3)}=\dot{\mathrm{T}}^{(3\,1\,1)}=\dot{\mathrm{T}}^{(3\,1\,2)}=i
\end{align}
while the remaining of the non-zero values are equal to $1$. Here we used the normalisation \eqref{standardnormC} for the boundary field junctions.

\subsection{\texorpdfstring{$N=4, r=2$}{N=4, r=2}} \label{appbos2}

In this appendix, since the dimensions of the matrices $A^{(x,y)}_{(i,j)},B^{(x,a)}_{(i,j)}$ and $C^{(a,j)}_{(i)}$ are quite large, we  present only the non-zero matrix elements. Recall that for $N=4, r=2$, the index set of simple objects of $\mathcal{U}_4$ is $\tilde{\mathcal{I}}=\{0,1,\ldots,7\}$ while the index set for $\tilde{A}_2$-modules is $\mathcal{J}_{\tilde{A}_2}=\{0,1,2,3\}$ and the one for $\tilde{A}_2$-$\tilde{A}_2$-bimodules is $\mathcal{K}_{\tilde{A}_2\tilde{A}_2}=\{0,1,\ldots,7\}$.

We first present the solution to the linear system constructed from \eqref{conaboson}, \eqref{conbboson} and \eqref{concboson}. The solution consists of $\lvert \mathcal{K}_{\tilde{A}_2\tilde{A}_2}\rvert ^2$ matrices of dimension $\lvert \tilde{\mathcal{I}}\rvert\times \lvert \tilde{\mathcal{I}}\rvert$ with matrix elements
\begin{equation}\label{amatrices1}
    A^{(x,y)}_{(i,j)}=\tilde{A}^{(x,y)}\tilde{A}^{(x,y)}_{(i,j)}
\end{equation}
for any complex numbers $\tilde{A}^{(x,y)}$.  The non-zero matrix elements  $\tilde{A}^{(x,y)}_{(i,j)}$ are found to be
\begin{align}
    &\tilde{A}^{(0,0)}_{(0,0)}= \tilde{A}^{(0,0)}_{(4,0)}= \tilde{A}^{(0,0)}_{(0,4)}= \tilde{A}^{(0,0)}_{(4,4)}=\tilde{A}^{(0,1)}_{(0,1)}=\tilde{A}^{(0,1)}_{(4,1)}=\tilde{A}^{(0,1)}_{(0,5)}=\tilde{A}^{(0,1)}_{(4,5)}=1 ,\nonumber \\
    & \tilde{A}^{(0,2)}_{(0,2)}= \tilde{A}^{(0,2)}_{(4,2)}= \tilde{A}^{(0,2)}_{(0,6)}= \tilde{A}^{(0,2)}_{(4,6)}=\tilde{A}^{(0,3)}_{(0,3)}=\tilde{A}^{(0,3)}_{(4,3)}=\tilde{A}^{(0,3)}_{(0,7)}=\tilde{A}^{(0,3)}_{(4,7)}=1 , \nonumber \\
    &\tilde{A}^{(1,0)}_{(1,0)}= \tilde{A}^{(1,0)}_{(5,0)}= \tilde{A}^{(1,1)}_{(1,1)}= \tilde{A}^{(1,1)}_{(5,1)}=1,\quad \tilde{A}^{(1,0)}_{(1,4)}= \tilde{A}^{(1,0)}_{(5,4)}=\tilde{A}^{(1,1)}_{(1,5)}=\tilde{A}^{(1,1)}_{(5,5)}=-i ,\nonumber \\
    & \tilde{A}^{(1,2)}_{(1,2)}= \tilde{A}^{(1,2)}_{(5,2)}=\tilde{A}^{(1,3)}_{(1,3)}=\tilde{A}^{(1,3)}_{(5,3)}=1,\quad  \tilde{A}^{(1,2)}_{(1,6)}= \tilde{A}^{(1,2)}_{(5,6)}=\tilde{A}^{(1,3)}_{(1,7)}=\tilde{A}^{(1,3)}_{(5,7)}=-i,\nonumber \\
    & \tilde{A}^{(2,0)}_{(0,0)}=\tilde{A}^{(2,0)}_{(4,0)}=\tilde{A}^{(2,1)}_{(0,1)}=\tilde{A}^{(2,1)}_{(4,1)}=1,\quad \tilde{A}^{(2,0)}_{(0,4)}=\tilde{A}^{(2,0)}_{(4,4)}=\tilde{A}^{(2,1)}_{(0,5)}=\tilde{A}^{(2,1)}_{(4,5)}=-1 ,\nonumber \\
    & \tilde{A}^{(2,2)}_{(0,2)}=\tilde{A}^{(2,2)}_{(4,2)}=\tilde{A}^{(2,3)}_{(0,3)}=\tilde{A}^{(2,3)}_{(4,3)}=1 ,\quad \tilde{A}^{(2,2)}_{(0,6)}=\tilde{A}^{(2,2)}_{(4,6)}=\tilde{A}^{(2,3)}_{(0,7)}=\tilde{A}^{(2,3)}_{(4,7)}=-1 ,\nonumber \\
    & \tilde{A}^{(3,0)}_{(1,0)}=\tilde{A}^{(3,0)}_{(5,0)}=\tilde{A}^{(3,1)}_{(1,1)}=\tilde{A}^{(3,1)}_{(5,1)}=1,\quad \tilde{A}^{(3,0)}_{(1,4)}=\tilde{A}^{(3,0)}_{(5,4)}=\tilde{A}^{(3,1)}_{(1,5)}=\tilde{A}^{(3,1)}_{(5,5)}=i,\nonumber \\
&
\tilde{A}^{(3,2)}_{(1,2)}=\tilde{A}^{(3,2)}_{(5,2)}=\tilde{A}^{(3,3)}_{(1,3)}=\tilde{A}^{(3,3)}_{(5,3)}=1,\quad \tilde{A}^{(3,2)}_{(1,6)}=\tilde{A}^{(3,2)}_{(5,6)}=\tilde{A}^{(3,3)}_{(1,7)}=\tilde{A}^{(3,3)}_{(5,7)}=i,\nonumber \\
    &\tilde{A}^{(4,0)}_{(2,0)}=\tilde{A}^{(4,0)}_{(6,0)}=\tilde{A}^{(4,0)}_{(2,4)}=\tilde{A}^{(4,0)}_{(6,4)}=\tilde{A}^{(4,1)}_{(2,1)}=\tilde{A}^{(4,1)}_{(6,1)}=\tilde{A}^{(4,1)}_{(2,5)}=\tilde{A}^{(4,1)}_{(6,5)}=1,\nonumber \\
    &\tilde{A}^{(4,2)}_{(2,2)}=\tilde{A}^{(4,2)}_{(6,2)}=\tilde{A}^{(4,2)}_{(2,6)}=\tilde{A}^{(4,2)}_{(6,6)}=\tilde{A}^{(4,3)}_{(2,3)}=\tilde{A}^{(4,3)}_{(6,3)}=\tilde{A}^{(4,3)}_{(2,7)}=\tilde{A}^{(4,3)}_{(6,7)}=1,\nonumber \\
    & \tilde{A}^{(5,0)}_{(3,0)}=\tilde{A}^{(5,0)}_{(7,0)}=\tilde{A}^{(5,1)}_{(3,1)}=\tilde{A}^{(5,1)}_{(7,1)}=1,\quad \tilde{A}^{(5,0)}_{(3,4)}=\tilde{A}^{(5,0)}_{(7,4)}=\tilde{A}^{(5,1)}_{(3,5)}=\tilde{A}^{(5,1)}_{(7,5)}=-i,\nonumber \\
    & \tilde{A}^{(5,2)}_{(3,2)}=\tilde{A}^{(5,2)}_{(7,2)}=\tilde{A}^{(5,3)}_{(3,3)}=\tilde{A}^{(5,3)}_{(7,3)}=1,\quad \tilde{A}^{(5,2)}_{(3,6)}=\tilde{A}^{(5,2)}_{(7,6)}=\tilde{A}^{(5,3)}_{(3,7)}=\tilde{A}^{(5,3)}_{(7,7)}=-i,\nonumber \\
    & \tilde{A}^{(6,0)}_{(2,0)}=\tilde{A}^{(6,0)}_{(6,0)}=\tilde{A}^{(6,1)}_{(2,1)}=\tilde{A}^{(6,1)}_{(6,1)}=1,\quad \tilde{A}^{(6,0)}_{(2,4)}=\tilde{A}^{(6,0)}_{(6,4)}=\tilde{A}^{(6,1)}_{(2,5)}=\tilde{A}^{(6,1)}_{(6,5)}=-1,\nonumber\\
    & \tilde{A}^{(6,2)}_{(2,2)}=\tilde{A}^{(6,2)}_{(6,2)}=\tilde{A}^{(6,3)}_{(2,3)}=\tilde{A}^{(6,3)}_{(6,3)}=1,\quad \tilde{A}^{(6,2)}_{(2,6)}=\tilde{A}^{(6,2)}_{(6,6)}=\tilde{A}^{(6,3)}_{(2,7)}=\tilde{A}^{(6,3)}_{(6,7)}=-1,\nonumber \\
    & \tilde{A}^{(7,0)}_{(3,0)}=\tilde{A}^{(7,0)}_{(7,0)}=\tilde{A}^{(7,1)}_{(3,1)}=\tilde{A}^{(7,1)}_{(7,1)}=1,\quad \tilde{A}^{(7,0)}_{(3,4)}=\tilde{A}^{(7,0)}_{(7,4)}=\tilde{A}^{(7,1)}_{(3,5)}=\tilde{A}^{(7,1)}_{(7,5)}=i,\nonumber\\
    &\tilde{A}^{(7,2)}_{(3,2)}=\tilde{A}^{(7,2)}_{(7,2)}=\tilde{A}^{(7,3)}_{(3,3)}=\tilde{A}^{(7,3)}_{(7,3)}=1,\quad \tilde{A}^{(7,2)}_{(3,6)}=\tilde{A}^{(7,2)}_{(7,6)}=\tilde{A}^{(7,3)}_{(3,7)}=\tilde{A}^{(7,3)}_{(7,7)}=i.
    \end{align}
Notice that these are only 32 out of the 64 matrices $\tilde{A}^{(x,y)}_{(i,j)}$ that appear on \eqref{amatrices1}. The remaining matrices are obtained by
\begin{equation}
    \tilde{A}^{(x,y)}_{(i,j)}=\tilde{A}^{(x,y\;\text{mod}\;4)}_{(i,j)},\quad \forall x,y,i,j
\end{equation}
\mycomment{
\begin{equation}
    \tilde{A}^{(x,4)}_{(i,j)}=\tilde{A}^{(x,0)}_{(i,j)},\quad \tilde{A}^{(x,5)}_{(i,j)}=\tilde{A}^{(x,1)}_{(i,j)},\quad \tilde{A}^{(x,6)}_{(i,j)}=\tilde{A}^{(x,2)}_{(i,j)},\quad \tilde{A}^{(x,7)}_{(i,j)}=\tilde{A}^{(x,3)}_{(i,j)},\quad \forall x,i,j.
\end{equation}
}

Using these we can compute the $\mathrm{Y}$ fusing matrices from \eqref{yboson} and obtain
\begin{align}\label{Yboson1}
&\mathrm{Y}^{(1\,1\,3)}=\mathrm{Y}^{(1\,1\,7)}=\mathrm{Y}^{(1\,5\,3)}=\mathrm{Y}^{(1\,5\,7)}= \mathrm{Y}^{(1\,2\,2)}= \mathrm{Y}^{(1\,2\,3)}= \mathrm{Y}^{(1\,2\,6)}= \mathrm{Y}^{(1\,2\,7)}= -i \nonumber \\
&  \mathrm{Y}^{(1\,3\,1)}= \mathrm{Y}^{(1\,3\,5)}= \mathrm{Y}^{(1\,3\,2)}= \mathrm{Y}^{(1\,3\,3)}= \mathrm{Y}^{(1\,3\,6)}= \mathrm{Y}^{(1\,3\,7)}= \mathrm{Y}^{(1\,6\,2)}= \mathrm{Y}^{(1\,6\,3)}=-i \nonumber \\
& \mathrm{Y}^{(1\,6\,6)}=\mathrm{Y}^{(1\,6\,7)}= \mathrm{Y}^{(1\,7\,1)}= \mathrm{Y}^{(1\,7\,5)}= \mathrm{Y}^{(1\,7\,2)}= \mathrm{Y}^{(1\,7\,3)}=\mathrm{Y}^{(1\,7\,6)}= \mathrm{Y}^{(1\,7\,7)} =-i \nonumber \\
&   \mathrm{Y}^{(3\,1\,3)} = \mathrm{Y}^{(3\,1\,7)}= \mathrm{Y}^{(3\,5\,3)}= \mathrm{Y}^{(3\,5\,7)}= \mathrm{Y}^{(3\,2\,2)}=\mathrm{Y}^{(3\,2\,3)} = \mathrm{Y}^{(3\,2\,6)} = \mathrm{Y}^{(3\,2\,7)}  =-i \nonumber \\
&   \mathrm{Y}^{(3\,3\,1)} = \mathrm{Y}^{(3\,3\,5)}= \mathrm{Y}^{(3\,3\,2)}= \mathrm{Y}^{(3\,3\,3)}= \mathrm{Y}^{(3\,3\,6)}= \mathrm{Y}^{(3\,3\,7)} = \mathrm{Y}^{(3\,6\,2)}= \mathrm{Y}^{(3\,6\,3)} =-i \nonumber\\
&  \mathrm{Y}^{(3\,6\,6)} = \mathrm{Y}^{(3\,6\,7)} = \mathrm{Y}^{(3\,7\,1)}= \mathrm{Y}^{(3\,7\,5)}= \mathrm{Y}^{(3\,7\,2)}= \mathrm{Y}^{(3\,7\,3)}= \mathrm{Y}^{(3\,7\,6)}= \mathrm{Y}^{(3\,7\,7)}=-i \nonumber \\
& \mathrm{Y}^{(4\,1\,3)} = \mathrm{Y}^{(4\,1\,7)} = \mathrm{Y}^{(4\,5\,3)} = \mathrm{Y}^{(4\,5\,7)} = \mathrm{Y}^{(4\,2\,2)} = \mathrm{Y}^{(4\,2\,3)}   = \mathrm{Y}^{(4\,2\,6)} = \mathrm{Y}^{(4\,2\,7)} =-1 \nonumber \\
&   \mathrm{Y}^{(4\,3\,1)}= \mathrm{Y}^{(4\,3\,5)}= \mathrm{Y}^{(4\,3\,2)}= \mathrm{Y}^{(4\,3\,3)} = \mathrm{Y}^{(4\,3\,6)} = \mathrm{Y}^{(4\,3\,7)} = \mathrm{Y}^{(4\,6\,2)} = \mathrm{Y}^{(4\,6\,3)} =-1 \nonumber \\
& \mathrm{Y}^{(4\,6\,6)}= \mathrm{Y}^{(4\,6\,7)} = \mathrm{Y}^{(4\,7\,1)}  = \mathrm{Y}^{(4\,7\,5)}   = \mathrm{Y}^{(4\,7\,2)}   = \mathrm{Y}^{(4\,7\,3)}   = \mathrm{Y}^{(4\,7\,6)}     = \mathrm{Y}^{(4\,7\,7)}  =-1\nonumber \\
&  \mathrm{Y}^{(6\,1\,3)}   =\mathrm{Y}^{(6\,1\,7)} =\mathrm{Y}^{(6\,5\,3)}    =\mathrm{Y}^{(6\,5\,7)}    =\mathrm{Y}^{(6\,2\,2)}   =\mathrm{Y}^{(6\,2\,3)}   =\mathrm{Y}^{(6\,2\,6)}    =\mathrm{Y}^{(6\,2\,7)}  =-1 \nonumber \\
& \mathrm{Y}^{(6\,3\,1)}   =\mathrm{Y}^{(6\,3\,5)} =\mathrm{Y}^{(6\,3\,2)}  =\mathrm{Y}^{(6\,3\,3)}  =\mathrm{Y}^{(6\,3\,6)}  =\mathrm{Y}^{(6\,3\,7)} =\mathrm{Y}^{(6\,6\,2)} =\mathrm{Y}^{(6\,6\,3)} =-1\nonumber \\
& \mathrm{Y}^{(6\,6\,6)} =\mathrm{Y}^{(6\,6\,7)}   =\mathrm{Y}^{(6\,7\,1)} =\mathrm{Y}^{(6\,7\,5)} =\mathrm{Y}^{(6\,7\,2)}  =\mathrm{Y}^{(6\,7\,3)}  =\mathrm{Y}^{(6\,7\,6)}  =\mathrm{Y}^{(6\,7\,7)}  =-1 \nonumber \\
& \mathrm{Y}^{(5\,1\,3)} =\mathrm{Y}^{(5\,1\,7)} =\mathrm{Y}^{(5\,5\,3)}   =\mathrm{Y}^{(5\,5\,7)}  =\mathrm{Y}^{(5\,2\,2)} =\mathrm{Y}^{(5\,2\,3)} =\mathrm{Y}^{(5\,2\,6)} =\mathrm{Y}^{(5\,2\,7)} =i\nonumber \\
& \mathrm{Y}^{(5\,3\,1)} =\mathrm{Y}^{(5\,3\,5)} =\mathrm{Y}^{(5\,3\,2)} =\mathrm{Y}^{(5\,3\,3)} =\mathrm{Y}^{(5\,3\,6)} =\mathrm{Y}^{(5\,3\,7)} =\mathrm{Y}^{(5\,6\,2)} =\mathrm{Y}^{(5\,6\,3)}=i\nonumber \\
&\mathrm{Y}^{(5\,6\,6)} =\mathrm{Y}^{(5\,6\,7)} =\mathrm{Y}^{(5\,7\,1)}  =\mathrm{Y}^{(5\,7\,5)} =\mathrm{Y}^{(5\,7\,2)} =\mathrm{Y}^{(5\,7\,3)} =\mathrm{Y}^{(5\,7\,6)} =\mathrm{Y}^{(5\,7\,7)} =i \nonumber \\
&\mathrm{Y}^{(7\,1\,3)} =\mathrm{Y}^{(7\,1\,7)} =\mathrm{Y}^{(7\,5\,3)}  =\mathrm{Y}^{(7\,5\,7)}  =\mathrm{Y}^{(7\,2\,2)} =\mathrm{Y}^{(7\,2\,3)} =\mathrm{Y}^{(7\,2\,6)} =\mathrm{Y}^{(7\,2\,7)} =i\nonumber \\
& \mathrm{Y}^{(7\,3\,1)} =\mathrm{Y}^{(7\,3\,5)} =\mathrm{Y}^{(7\,3\,2)} =\mathrm{Y}^{(7\,3\,3)} =\mathrm{Y}^{(7\,3\,6)} =\mathrm{Y}^{(7\,3\,7)} =\mathrm{Y}^{(7\,6\,2)} =\mathrm{Y}^{(7\,6\,3)} =i\nonumber \\
& \mathrm{Y}^{(7\,6\,6)} =\mathrm{Y}^{(7\,6\,7)} =\mathrm{Y}^{(7\,7\,1)} =\mathrm{Y}^{(7\,7\,5)} =\mathrm{Y}^{(7\,7\,2)} =\mathrm{Y}^{(7\,7\,3)} =\mathrm{Y}^{(7\,7\,6)} =\mathrm{Y}^{(7\,7\,7)} =i
\end{align}
while the remaining of the non-zero values are equal to $1$. Here we used the  normalisation  $\tilde{A}^{(x,y)}=1$ in \eqref{amatrices1} for all pairs of bimodule labels $x,y$.

To find the complex numbers $B^{(x,a)}_{(i,j)}$ we proceed in the way described in \ref{sec3.2} and construct a system of linear equations for these numbers.  The solution consists of $\lvert \mathcal{K}_{\tilde{A}_2\tilde{A}_2}\rvert \lvert \mathcal{J}_{\tilde{A}_2}\rvert$ matrices of dimension $\lvert \tilde{\mathcal{I}}\rvert\times \lvert \tilde{\mathcal{I}}\rvert$ with matrix elements
\begin{equation}\label{bmatrices1}
    B^{(x,a)}_{(i,j)}=\tilde{B}^{(x,a)}\tilde{B}^{(x,a)}_{(i,j)}
\end{equation}
for any complex numbers $\tilde{B}^{(x,a)}$.  The non-zero matrix elements  $\tilde{B}^{(x,y)}_{(i,j)}$ are found to be
\begin{align}\label{bmatrices2}
   &\tilde{B}^{(0,0)}_{(0,0)}= \tilde{B}^{(0,0)}_{(4,0)}= \tilde{B}^{(0,0)}_{(0,4)}= \tilde{B}^{(0,0)}_{(4,4)}=\tilde{B}^{(0,1)}_{(0,1)}=\tilde{B}^{(0,1)}_{(4,1)}=\tilde{B}^{(0,1)}_{(0,5)}=\tilde{B}^{(0,1)}_{(4,5)}=1 ,\nonumber \\
   & \tilde{B}^{(0,2)}_{(0,2)}=\tilde{B}^{(0,2)}_{(4,2)}=\tilde{B}^{(0,2)}_{(0,6)}=\tilde{B}^{(0,2)}_{(4,6)}=\tilde{B}^{(0,3)}_{(0,3)}=\tilde{B}^{(0,3)}_{(4,3)}=\tilde{B}^{(0,3)}_{(0,7)}=\tilde{B}^{(0,3)}_{(4,7)}=1,\nonumber \\
   & \tilde{B}^{(1,0)}_{(1,0)}=\tilde{B}^{(1,0)}_{(5,0)}=\tilde{B}^{(1,1)}_{(1,1)}=\tilde{B}^{(1,1)}_{(5,1)}=1,\quad \tilde{B}^{(1,0)}_{(1,4)}=\tilde{B}^{(1,0)}_{(5,4)}=\tilde{B}^{(1,1)}_{(1,5)}=\tilde{B}^{(1,1)}_{(5,5)}=-i,\nonumber \\
&
\tilde{B}^{(1,2)}_{(1,2)}=\tilde{B}^{(1,2)}_{(5,2)}=\tilde{B}^{(1,3)}_{(1,3)}=\tilde{B}^{(1,3)}_{(5,3)}=1,\quad \tilde{B}^{(1,2)}_{(1,6)}=\tilde{B}^{(1,2)}_{(5,6)}=\tilde{B}^{(1,3)}_{(1,7)}=\tilde{B}^{(1,3)}_{(5,7)}=-i,\nonumber\\
   & \tilde{B}^{(4,0)}_{(0,0)}=\tilde{B}^{(4,0)}_{(4,0)}=\tilde{B}^{(4,1)}_{(0,1)}=\tilde{B}^{(4,1)}_{(4,1)}=1,\quad \tilde{B}^{(4,0)}_{(0,4)}=\tilde{B}^{(4,0)}_{(4,4)}=\tilde{B}^{(4,1)}_{(0,5)}=\tilde{B}^{(4,1)}_{(4,5)}=-1,\nonumber \\
   & \tilde{B}^{(4,2)}_{(0,2)}=\tilde{B}^{(4,2)}_{(4,2)}=\tilde{B}^{(4,3)}_{(0,3)}=\tilde{B}^{(4,3)}_{(4,3)}=1,\quad \tilde{B}^{(4,2)}_{(0,6)}=\tilde{B}^{(4,2)}_{(4,6)}=\tilde{B}^{(4,3)}_{(0,7)}=\tilde{B}^{(4,3)}_{(4,7)}=-1,\nonumber \\
   & \tilde{B}^{(5,0)}_{(1,0)}=\tilde{B}^{(5,0)}_{(5,0)}=\tilde{B}^{(5,1)}_{(1,1)}=\tilde{B}^{(5,1)}_{(5,1)}=1,\quad \tilde{B}^{(5,0)}_{(1,4)}=\tilde{B}^{(5,0)}_{(5,4)}=\tilde{B}^{(5,1)}_{(1,5)}=\tilde{B}^{(5,1)}_{(5,5)}=i,\nonumber\\
   & \tilde{B}^{(5,2)}_{(1,2)}=\tilde{B}^{(5,2)}_{(5,2)}=\tilde{B}^{(5,3)}_{(1,3)}=\tilde{B}^{(5,3)}_{(5,3)}=1,\quad \tilde{B}^{(5,2)}_{(1,6)}=\tilde{B}^{(5,2)}_{(5,6)}=\tilde{B}^{(5,3)}_{(1,7)}=\tilde{B}^{(5,3)}_{(5,7)}=i,\nonumber\\
   & \tilde{B}^{(2,0)}_{(2,0)}= \tilde{B}^{(2,0)}_{(6,0)}= \tilde{B}^{(2,0)}_{(2,4)}= \tilde{B}^{(2,0)}_{(6,4)}= \tilde{B}^{(2,1)}_{(2,1)}= \tilde{B}^{(2,1)}_{(6,1)}= \tilde{B}^{(2,1)}_{(2,5)}= \tilde{B}^{(2,1)}_{(6,5)}=1,\nonumber\\
   &  \tilde{B}^{(2,2)}_{(2,2)}=\tilde{B}^{(2,2)}_{(6,2)}=\tilde{B}^{(2,2)}_{(2,6)}=\tilde{B}^{(2,2)}_{(6,6)}=\tilde{B}^{(2,3)}_{(2,3)}=\tilde{B}^{(2,3)}_{(6,3)}=\tilde{B}^{(2,3)}_{(2,7)}=\tilde{B}^{(2,3)}_{(6,7)}=1,\nonumber\\
   & \tilde{B}^{(3,0)}_{(3,0)}=\tilde{B}^{(3,0)}_{(7,0)}=\tilde{B}^{(3,1)}_{(3,1)}=\tilde{B}^{(3,1)}_{(7,1)}=1,\quad \tilde{B}^{(3,0)}_{(3,4)}=\tilde{B}^{(3,0)}_{(7,4)}=\tilde{B}^{(3,1)}_{(3,5)}=\tilde{B}^{(3,1)}_{(7,5)}=-i,\nonumber\\
   &\tilde{B}^{(3,2)}_{(3,2)}=\tilde{B}^{(3,2)}_{(7,2)}=\tilde{B}^{(3,3)}_{(3,3)}=\tilde{B}^{(3,3)}_{(7,3)}=1,\quad \tilde{B}^{(3,2)}_{(3,6)}=\tilde{B}^{(3,2)}_{(7,6)}=\tilde{B}^{(3,3)}_{(3,7)}=\tilde{B}^{(3,3)}_{(7,7)}=-i,\nonumber\\
   & \tilde{B}^{(6,0)}_{(2,0)}=\tilde{B}^{(6,0)}_{(6,0)}=\tilde{B}^{(6,1)}_{(2,1)}=\tilde{B}^{(6,1)}_{(6,1)}=1,\quad \tilde{B}^{(6,0)}_{(2,4)}=\tilde{B}^{(6,0)}_{(6,4)}=\tilde{B}^{(6,1)}_{(2,5)}=\tilde{B}^{(6,1)}_{(6,5)}=-1,\nonumber \\
   & \tilde{B}^{(6,2)}_{(2,2)}=\tilde{B}^{(6,2)}_{(6,2)}=\tilde{B}^{(6,3)}_{(2,3)}=\tilde{B}^{(6,3)}_{(6,3)}=1,\quad \tilde{B}^{(6,2)}_{(2,6)}=\tilde{B}^{(6,2)}_{(6,6)}=\tilde{B}^{(6,3)}_{(2,7)}=\tilde{B}^{(6,3)}_{(6,7)}=-1,\nonumber \\
   &\tilde{B}^{(7,0)}_{(3,0)}=\tilde{B}^{(7,0)}_{(7,0)}=\tilde{B}^{(7,1)}_{(3,1)}=\tilde{B}^{(7,1)}_{(7,1)}=1,\quad \tilde{B}^{(7,0)}_{(3,4)}=\tilde{B}^{(7,0)}_{(7,4)}=\tilde{B}^{(7,1)}_{(3,5)}=\tilde{B}^{(7,1)}_{(7,5)}=i,\nonumber\\
   & \tilde{B}^{(7,2)}_{(3,2)}=\tilde{B}^{(7,2)}_{(7,2)}=\tilde{B}^{(7,3)}_{(3,3)}=\tilde{B}^{(7,3)}_{(7,3)}=1 ,\quad \tilde{B}^{(7,2)}_{(3,6)}=\tilde{B}^{(7,2)}_{(7,6)}=\tilde{B}^{(7,3)}_{(3,7)}=\tilde{B}^{(7,3)}_{(7,7)}=i.
\end{align}

Now we can use \eqref{tboson} to obtain the $\mathrm{T}$-matrices:
\begin{equation}\label{tcase2.1}
     \mathrm{T}^{(x\,y\,a)}=1 \quad \text{if any} \;x,y,a \;\text{is} \;0.
\end{equation}
Some of the remaining values are
\begin{align}\label{tcase2.2}
   & \mathrm{T}^{(1\,1\,3)} = \mathrm{T}^{(1\,5\,3)} =\mathrm{T}^{(1\,2\,2)} =\mathrm{T}^{(1\,2\,3)} = \mathrm{T}^{(1\,3\,1)} =\mathrm{T}^{(1\,3\,2)} =\mathrm{T}^{(1\,3\,3)} =\mathrm{T}^{(1\,6\,2)}= -i \nonumber \\
   &  \mathrm{T}^{(1\,6\,3)} =\mathrm{T}^{(1\,7\,1)}=\mathrm{T}^{(1\,7\,2)}=\mathrm{T}^{(1\,7\,3)} =\mathrm{T}^{(3\,1\,3)} =\mathrm{T}^{(3\,5\,3)} =\mathrm{T}^{(3\,2\,2)} =\mathrm{T}^{(3\,2\,3)}=-i \nonumber \\
   &  \mathrm{T}^{(3\,3\,1)}= \mathrm{T}^{(3\,3\,2)}= \mathrm{T}^{(3\,3\,3)}=  \mathrm{T}^{(3\,6\,2)}= \mathrm{T}^{(3\,6\,3)}= \mathrm{T}^{(3\,7\,1)}= \mathrm{T}^{(3\,7\,2)}=\mathrm{T}^{(3\,7\,3)}= -i\nonumber \\
   & \mathrm{T}^{(4\,1\,3)}=  \mathrm{T}^{(4\,5\,3)}=  \mathrm{T}^{(4\,2\,2)}= \mathrm{T}^{(4\,2\,3)}=  \mathrm{T}^{(4\,3\,1)}= \mathrm{T}^{(4\,3\,2)}=\mathrm{T}^{(4\,3\,3)}= \mathrm{T}^{(4\,6\,2)}=-1 \nonumber \\
   & \mathrm{T}^{(4\,6\,3)}= \mathrm{T}^{(4\,7\,1)}=\mathrm{T}^{(4\,7\,2)}=\mathrm{T}^{(4\,7\,3)}= \mathrm{T}^{(6\,1\,3)}= \mathrm{T}^{(6\,5\,3)}= \mathrm{T}^{(6\,2\,2)}= \mathrm{T}^{(6\,2\,3)}= -1\nonumber \\
   & \mathrm{T}^{(6\,3\,1)}= \mathrm{T}^{(6\,3\,2)}= \mathrm{T}^{(6\,3\,3)}=  \mathrm{T}^{(6\,6\,2)}= \mathrm{T}^{(6\,6\,3)}=  \mathrm{T}^{(6\,7\,1)}= \mathrm{T}^{(6\,7\,2)}= \mathrm{T}^{(6\,7\,3)}= -1\nonumber \\
   & \mathrm{T}^{(5\,1\,3)}=  \mathrm{T}^{(5\,5\,3)}=  \mathrm{T}^{(5\,2\,2)}= \mathrm{T}^{(5\,2\,3)}=  \mathrm{T}^{(5\,3\,1)}= \mathrm{T}^{(5\,3\,2)}= \mathrm{T}^{(5\,3\,3)}= \mathrm{T}^{(5\,6\,2)}= i\nonumber \\
   & \mathrm{T}^{(5\,6\,3)}=  \mathrm{T}^{(5\,7\,1)}= \mathrm{T}^{(5\,7\,2)}= \mathrm{T}^{(5\,7\,3)}= \mathrm{T}^{(7\,1\,3)}= \mathrm{T}^{(7\,5\,3)}=  \mathrm{T}^{(7\,2\,2)}= \mathrm{T}^{(7\,2\,3)}= i\nonumber \\
   & \mathrm{T}^{(7\,3\,1)}= \mathrm{T}^{(7\,3\,2)}= \mathrm{T}^{(7\,3\,3)}=  \mathrm{T}^{(7\,6\,2)}= \mathrm{T}^{(7\,6\,3)}=  \mathrm{T}^{(7\,7\,1)}= \mathrm{T}^{(7\,7\,2)}= \mathrm{T}^{(7\,7\,3)}= i
\end{align}
while the remaining of the non-zero values are equal to $1$. Here, we used the normalisation $\tilde{B}^{(x,a)}=1,\,\forall x,a$ for the open defect junctions.\\

To find the complex numbers $C^{(a,j)}_{(i)}$ we proceed in the way described in \ref{sec3.3} and construct a system of linear equations for these numbers. The solution consists of $\lvert \mathcal{J}_{\tilde{A}_2}\rvert \lvert \mathcal{I}\rvert$ vectors of dimension $\lvert \tilde{\mathcal{I}}\rvert$ of the form
\begin{equation}\label{cmatrices1}
    C^{(a,j)}_{(i)}=\tilde{C}^{(a,j)}\tilde{C}^{(a,j)}_{(i)}
\end{equation}
for any complex numbers $\tilde{C}^{(a,j)}$.  The  entries $\tilde{C}^{(a,j)}_{(i)}$ are obtained by \eqref{Cconstantsgen}:
\begin{equation}
    \tilde{C}^{(a,j)}_{(i)}=\begin{cases*}
      1 & \text{if}\, $U_i$ \;\text{is a subobject of} \; $M_a$  \\
      $0$        & \text{otherwise}.
    \end{cases*}
\end{equation}
We use these results in \eqref{tdotboson} to obtain the following values for the $\mathrm{\dot{T}}$-matrices:
\begin{align}\label{tdot2}
   & \dot{\mathrm{T}}^{(0\,0\,4)}=\dot{\mathrm{T}}^{(0\,0\,5)}= \dot{\mathrm{T}}^{(0\,0\,6)}=\dot{\mathrm{T}}^{(0\,0\,7)}= \dot{\mathrm{T}}^{(1\,1\,3)}=  \dot{\mathrm{T}}^{(1\,1\,4)}=\dot{\mathrm{T}}^{(1\,1\,5)}=\dot{\mathrm{T}}^{(1\,1\,6)}=-i \nonumber \\
    & \dot{\mathrm{T}}^{(1\,2\,2)}=\dot{\mathrm{T}}^{(1\,2\,3)}=\dot{\mathrm{T}}^{(1\,2\,4)}=\dot{\mathrm{T}}^{(1\,2\,5)}= \dot{\mathrm{T}}^{(1\,3\,1)}=\dot{\mathrm{T}}^{(1\,3\,2)}=\dot{\mathrm{T}}^{(1\,3\,3)}=\dot{\mathrm{T}}^{(1\,3\,4)}= -i \nonumber \\
    & \dot{\mathrm{T}}^{(3\,0\,4)}=\dot{\mathrm{T}}^{(3\,0\,5)}=\dot{\mathrm{T}}^{(3\,0\,6)}=\dot{\mathrm{T}}^{(3\,0\,7)}= \dot{\mathrm{T}}^{(3\,1\,3)}= \dot{\mathrm{T}}^{(3\,1\,4)}= \dot{\mathrm{T}}^{(3\,1\,5)}= \dot{\mathrm{T}}^{(3\,1\,6)}= -i \nonumber \\
      &\dot{\mathrm{T}}^{(3\,2\,2)}= \dot{\mathrm{T}}^{(3\,2\,3)}= \dot{\mathrm{T}}^{(3\,2\,4)}= \dot{\mathrm{T}}^{(3\,2\,5)}=   \dot{\mathrm{T}}^{(3\,3\,1)}= \dot{\mathrm{T}}^{(3\,3\,2)}= \dot{\mathrm{T}}^{(3\,3\,3)}=  \dot{\mathrm{T}}^{(3\,3\,4)}= -i \nonumber \\
      & \dot{\mathrm{T}}^{(1\,1\,7)}=  \dot{\mathrm{T}}^{(1\,2\,6)}= \dot{\mathrm{T}}^{(1\,2\,7)}=  \dot{\mathrm{T}}^{(1\,3\,5)}=  \dot{\mathrm{T}}^{(1\,3\,6)}=  \dot{\mathrm{T}}^{(1\,3\,7)}=  \dot{\mathrm{T}}^{(4\,0\,4)}= \dot{\mathrm{T}}^{(4\,0\,5)}= -1\nonumber \\
      & \dot{\mathrm{T}}^{(4\,0\,6)}= \dot{\mathrm{T}}^{(4\,0\,7)} =\dot{\mathrm{T}}^{(4\,1\,3)}=  \dot{\mathrm{T}}^{(4\,1\,4)}=  \dot{\mathrm{T}}^{(4\,1\,5)}=  \dot{\mathrm{T}}^{(4\,1\,6)}=  \dot{\mathrm{T}}^{(4\,2\,2)}= \dot{\mathrm{T}}^{(4\,2\,3)}=   -1\nonumber \\
      &\dot{\mathrm{T}}^{(4\,2\,4)}=  \dot{\mathrm{T}}^{(4\,2\,5)}=  \dot{\mathrm{T}}^{(4\,3\,1)}=   \dot{\mathrm{T}}^{(4\,3\,2)}=   \dot{\mathrm{T}}^{(4\,3\,3)}=   \dot{\mathrm{T}}^{(4\,3\,4)}=\dot{\mathrm{T}}^{(5\,1\,7)}= \dot{\mathrm{T}}^{(5\,2\,6)}=   -1\nonumber \\
      & \dot{\mathrm{T}}^{(5\,2\,7)}= \dot{\mathrm{T}}^{(5\,3\,5)}= \dot{\mathrm{T}}^{(5\,3\,6)}= \dot{\mathrm{T}}^{(5\,3\,7)}=  \dot{\mathrm{T}}^{(3\,1\,7)}= \dot{\mathrm{T}}^{(3\,2\,6)}= \dot{\mathrm{T}}^{(3\,2\,7)}=\dot{\mathrm{T}}^{(3\,3\,5)}= -1\nonumber \\
      &  \dot{\mathrm{T}}^{(3\,3\,6)}= \dot{\mathrm{T}}^{(3\,3\,7)}=  \dot{\mathrm{T}}^{(6\,0\,4)}= \dot{\mathrm{T}}^{(6\,0\,5)}= \dot{\mathrm{T}}^{(6\,0\,6)}= \dot{\mathrm{T}}^{(6\,0\,7)}= \dot{\mathrm{T}}^{(6\,1\,3)}= \dot{\mathrm{T}}^{(6\,1\,4)}=-1\nonumber \\
      & \dot{\mathrm{T}}^{(6\,1\,5)}= \dot{\mathrm{T}}^{(6\,1\,6)}= \dot{\mathrm{T}}^{(6\,2\,2)}= \dot{\mathrm{T}}^{(6\,2\,3)}=\dot{\mathrm{T}}^{(6\,2\,4)}=\dot{\mathrm{T}}^{(6\,2\,5)}= \dot{\mathrm{T}}^{(6\,3\,1)}=\dot{\mathrm{T}}^{(6\,3\,2)}= -1\nonumber \\
      & \dot{\mathrm{T}}^{(6\,3\,3)}=\dot{\mathrm{T}}^{(6\,3\,4)}= \dot{\mathrm{T}}^{(7\,1\,7)}= \dot{\mathrm{T}}^{(7\,2\,6)}=\dot{\mathrm{T}}^{(7\,2\,7)}= \dot{\mathrm{T}}^{(7\,3\,5)}=\dot{\mathrm{T}}^{(7\,3\,6)}=\dot{\mathrm{T}}^{(7\,3\,7)}= -1 \nonumber \\
      & \dot{\mathrm{T}}^{(5\,0\,4)}=\dot{\mathrm{T}}^{(5\,0\,5)}=\dot{\mathrm{T}}^{(5\,0\,6)}=\dot{\mathrm{T}}^{(5\,0\,7)}= \dot{\mathrm{T}}^{(5\,1\,3)}=\dot{\mathrm{T}}^{(5\,1\,4)}=\dot{\mathrm{T}}^{(5\,1\,5)}=\dot{\mathrm{T}}^{(5\,1\,6)}= i\nonumber \\
      & \dot{\mathrm{T}}^{(5\,2\,2)}=\dot{\mathrm{T}}^{(5\,2\,3)}=\dot{\mathrm{T}}^{(5\,2\,4)}=\dot{\mathrm{T}}^{(5\,2\,5)}= \dot{\mathrm{T}}^{(5\,3\,1)}=\dot{\mathrm{T}}^{(5\,3\,2)}=\dot{\mathrm{T}}^{(5\,3\,3)}=\dot{\mathrm{T}}^{(5\,3\,4)}=i\nonumber \\
     & \dot{\mathrm{T}}^{(7\,0\,4)}=\dot{\mathrm{T}}^{(7\,0\,5)}=\dot{\mathrm{T}}^{(7\,0\,6)}=\dot{\mathrm{T}}^{(7\,0\,7)}= \dot{\mathrm{T}}^{(7\,1\,3)}=\dot{\mathrm{T}}^{(7\,1\,4)}=\dot{\mathrm{T}}^{(7\,1\,5)}=\dot{\mathrm{T}}^{(7\,1\,6)}= i\nonumber \\
     & \dot{\mathrm{T}}^{(7\,2\,2)}= \dot{\mathrm{T}}^{(7\,2\,3)}=\dot{\mathrm{T}}^{(7\,2\,4)}=\dot{\mathrm{T}}^{(7\,2\,5)}= \dot{\mathrm{T}}^{(7\,3\,1)}=\dot{\mathrm{T}}^{(7\,3\,2)}=\dot{\mathrm{T}}^{(7\,3\,3)}=\dot{\mathrm{T}}^{(7\,3\,4)}= i
\end{align}
while the rest of the non-zero values are equal to $1$. Here, we used the normalisation \eqref{Cconstantsnorm}.


 \bibliographystyle{JHEP}


\end{document}